\documentclass[12pt]{article}
\pdfoutput=1
\usepackage{jheppub}

\usepackage{amssymb}
\usepackage{graphicx}
\usepackage{subcaption}
\usepackage{hyperref}
\usepackage{dsfont}
\usepackage{braket}
\usepackage{slashed}
\usepackage{bm}
\usepackage{pifont}
\usepackage{appendix}
\usepackage[table]{xcolor}
\usepackage{multicol}
\usepackage{mathrsfs}
\usepackage{amsmath}
\usepackage{ulem}
\usepackage{float}

\newcommand{\roughly}[1]{\mathrel{\raise.3ex\hbox{$#1$\kern-0.85em
\lower1ex\hbox{$\sim$}}}}

\def\ol#1{\overline{#1}}
\def\exd{{\hbox{d}}}

\def\QED{{\scriptscriptstyle QED}}
\def\NQED{{\Phi\,{\scriptscriptstyle QED}}}
\def\NS{{\scriptscriptstyle NS}}

\def\bea{\begin{eqnarray}}
\def\eea{\end{eqnarray}}
\def\be{\begin{equation}}
\def\ee{\end{equation}}

\def\bfe{{\bf e}}

\def\bfp{{\bf p}}
\def\bfr{{\bf r}}
\def\bfv{{\bf v}}
\def\bfx{{\bf x}}
\def\bfy{{\bf y}}

\def\bfA{{\bf A}}
\def\bfB{{\bf B}}
\def\bfE{{\bf E}}
\def\bfF{{\bf F}}
\def\bfI{{\bf I}}
\def\bfJ{{\bf J}}
\def\bfL{{\bf L}}
\def\bfS{{\bf S}}

\def\ssA{{\scriptscriptstyle A}}
\def\ssB{{\scriptscriptstyle B}}
\def\ssD{{\scriptscriptstyle D}}
\def\ssF{{\scriptscriptstyle F}}

\def\ssI{{\scriptscriptstyle I}}

\def\ssL{{\scriptscriptstyle L}}

\def\ssN{{\scriptscriptstyle N}}
\def\ssP{{\scriptscriptstyle P}}

\def\ssS{{\scriptscriptstyle S}}

\def\ssZ{{\scriptscriptstyle Z}}
\def\\mfK \mfK {{\scriptscriptstyle \mfK \mfK }}

\def\cA{\mathcal{A}}
\def\cB{\mathcal{B}}
\def\cC{\mathcal{C}}
\def\cD{\mathcal{D}}

\def\cF{\mathcal{F}}

\def\cI{\mathcal{I}}
\def\c\mfK {\mathcal{\mfK }}
\def\cL{\mathcal{L}}
\def\cM{\mathcal{M}}
\def\cN{\mathcal{N}}
\def\cO{\mathcal{O}}

\def\cR{\mathcal{R}}
\def\cS{\mathcal{S}}

\def\cV{\mathcal{V}}
\def\cW{\mathcal{W}}

\def\cY{\mathcal{Y}}
\def\cZ{\mathcal{Z}}

\def\mfc{\mathfrak{c}}
\def\mff{\mathfrak{f}}
\def\mfg{\mathfrak{g}}
\def\mfi{\mathfrak{i}}
\def\mfj{\mathfrak{j}}
\def\mfm{\mathfrak{m}}

\def\mfz{\mathfrak{z}}
\def\mfB{\mathfrak{B}}
\def\mfD{\mathfrak{D}}
\def\mfH{\mathfrak{H }}
\def\mfK{\mathfrak{K }}
\def\mfN{\mathfrak{N}}

\def\msC{\mathscr{C}}
\def\msD{\mathscr{D}}
\def\msP{\mathscr{P}}

\def\za{Z\alpha}
\def\zas{Z\alpha\,}

\def\nn{\nonumber}

\def\({\left(}
\def\){\right)}

\def\pref#1{(\ref{#1})}

\def\exx{{\mathfrak{s}}}

\usepackage{graphicx}% Include figure files
\usepackage{dcolumn}% Align table columns on decimal point

\title{Precision Nuclear-Spin Effects in Atoms: \\
EFT Methods for Reducing %Nuclear 
Theory Errors}

\author{L.~Zalavari,}
\author{C.P.~Burgess,}
\author{P.~Hayman}
\author{and M.~Rummel}
\affiliation[a]{Department of Physics \& Astronomy, McMaster University, Hamilton, Ontario, L8S 4M1, Canada}
\affiliation[b]{Perimeter Institute for Theoretical Physics, Waterloo, Ontario, N2L 2Y5, Canada }
 \emailAdd{zalavarl@mcmaster.ca}
\emailAdd{cburgess@perimeterinstitute.ca}
\emailAdd{haymanpf@mcmaster.ca}
\emailAdd{rummelm@mcmaster.ca}

\date{}

\abstract{We use effective field theory to compute the influence of nuclear structure on precision calculations of atomic energy levels. As usual, the EFT's effective couplings correspond to the various nuclear properties (such as the charge radius, nuclear polarizabilities, Friar and Zemach moments {\it etc.}) that dominate its low-energy electromagnetic influence on its surroundings. By extending to spinning nuclei the arguments developed for spinless ones in {\tt arXiv:1708.09768}, we use the EFT to show -- to any fixed order in $Z\alpha$ (where $Z$ is the atomic number and $\alpha$ the fine-structure constant) and the ratio of nuclear to atomic size  --  that nuclear properties actually contribute to electronic energies through fewer parameters than the number of these effective nuclear couplings naively suggests. Our result is derived using a position-space method for matching effective parameters to nuclear properties in the EFT, that more efficiently exploits the simplicity of the small-nucleus limit in atomic systems. By showing that precision calculations of atomic spectra depend on fewer nuclear uncertainties than naively expected, this observation allows the construction of many nucleus-independent combinations of atomic energy differences whose measurement can be used to test fundamental physics (such as the predictions of QED) because their theoretical uncertainties are not limited by the accuracy of nuclear calculations. We provide several simple examples of such nucleus-free predictions for Hydrogen-like atoms. 
}

\begin{document}
\maketitle

\newpage

\section{Introduction}

Modern experimental techniques allow exquisitely accurate measurements of atomic transition frequencies. For simple atoms it is hoped that these measurements can be turned into a test of fundamental theory by comparing with equally precise predictions; modern alternatives to the classic comparisons between theory and experiment for the Lamb shift in Hydrogen \cite{wijngaarden2000} -- \cite{hessels2019}. 

The sad fact that atomic nuclei are not point charges is a major obstruction to this program, because atomic energy shifts due to nuclear structure can be larger than the fundamental corrections to be measured. Furthermore, the intricacies of the strong interactions  make {\it ab initio} predictions of nuclear properties necessarily inaccurate, often making nuclear uncertainties the dominant theoretical error when predicting atomic energy levels \cite{kks} -- \cite{ji2020}. 

In this paper we provide the details for (and provide broader applications of the results of) \cite{PPEFT-HydLet}, which aims to push past this floor in theoretical error by systematically identifying combinations of energy differences from which all of the effects of nuclear physics cancel.  Because the accuracy with which such observables can be predicted is not limited by nuclear uncertainties (by construction), their measurement can provide a potentially telling test of fundamental theory. Given $N_{\rm exp}$ well-measured transitions involving a specific type of atom and $N_{\rm nuc}$ parameters governing nuclear contributions to atomic energies, there are $N_{\rm exp} - N_{\rm nuc}$ independent observables for which a nucleus-free prediction can be made. 

While this counting is not so remarkable an observation in itself, what is perhaps more surprising is how small $N_{\rm nuc}$ turns out to be. In the applications made here, for instance, precisely two nuclear parameters suffice even though our predictions are accurate enough to include the effects of the nuclear charge radius, nuclear Friar and Zemach moments, nuclear polarizabilities, effects of multiple photon exchange, and so on. Only two parameters turn out to capture all of these different nuclear effects in our examples because atomic energy shifts only sample nuclear physics at extremely low energies relative to typical nuclear scales. 

The precise value for $N_{\rm nuc}$ depends on the accuracy that is required. This accuracy is most precisely specified when atomic energy levels are expressed as a perturbative expansion about the leading Bohr formula (for the binding energy of a nonrelativistic point-like lepton\footnote{For simplicity of language we speak in the main text about ordinary atoms -- {\it i.e.}~electrons orbiting nuclei -- but our analysis applies equally well to muonic atoms. At various points in the text we point out how the larger muon mass changes the relative size of different contributions.} to a static, charge $Ze$ point nucleus), 
\be
  \varepsilon_n = - \frac{(Z\alpha)^2m_r }{2n^2} \,,
\ee
where $m_r = mM/(m+M) = m + \cO(m^2/M)$ is the reduced mass, with $M$ the nuclear mass and $m$ the relevant lepton mass, while $n = 1,2,\cdots$ is the usual principal quantum number. This expansion comes in powers of four small quantities: the lepton speed $\mathrm{v}_e \sim Z \alpha$; the fine-structure constant\footnote{Unless otherwise stated we use fundamental units, for which $\hbar = c = k_\ssB = 1$.} $\alpha = e^2/4\pi$ (which enters without $Z$ in QED radiative corrections, for instance); the ratio of lepton to nuclear mass, $m/M$; and the ratio $R/a_\ssB \sim mRZ \alpha$ of nuclear and atomic length scales, where $R \sim 1$ fm is a measure of nuclear radius (more about which below) and $a_\ssB \sim (Z \alpha\,m)^{-1}$ is the Bohr radius. 

In terms of this expansion Ref.~\cite{ppeftA} found when working out to order $m^3 R^2 (Z\alpha)^6$ and $m^4 R^3 (Z\alpha)^5$ that only $N_{\rm nuc} = 1$ parameter was required (for each type of nucleus) to capture {\it all}~dependence of leptonic energy levels on nuclear substructure, at least for spinless nuclei.\footnote{More precisely, although one parameter captures all nuclear effects for either an electron or a muon orbiting a spinless nucleus, the parameter differs for the two so measurements of electrons cannot be used to infer nuclear contributions to muonic atoms without additional information from nuclear models.}  In this paper we extend this analysis to include nuclear-spin effects to the same order, and find all effects of nuclear substructure require only a single new parameter (so $N_{\rm nuc} =2$). Furthermore, since this accuracy provides a precision of about 0.01 kHz it is sufficient (with two exceptions) for current measurement precision with elemental Hydrogen. The corresponding precision for muonic Hydrogen is about $10^{-3}$ meV, and so is close to (but, as discussed at length in \S\ref{muonPunchline}, not quite) accurate enough for these measurements as well. A summary of this parameter counting as a function of expansion order can be found in Table \ref{table1}. 

\begin{center}
\begin{table}[h!]
\hfil
\begin{tabular}{|c|c|c|c|c|c|}
\hline
Abs. order & Rel. order & $j = 1/2$ & $j = 3/2$ &  H (kHz) &  muonic H (meV) \\ 
\hline
$m^3R^2(Z\alpha)^4$ & chg. rad. & \cellcolor{cyan} \ding{51} & \ding{53} &  $\cellcolor{green} 1.7 \times 10^{3}$  &  \cellcolor{green} $6.1\times10^1$ \\ 
\hline
$m^4R^3(Z\alpha)^5$ & $(mRZ\alpha)$ & \ding{51} & \ding{53} &  $\cellcolor{green} 2.6 \times 10^{-2}$  &  $\cellcolor{green} 2.0\times10^{-1}$ \\
\hline
$m^5R^4(Z\alpha)^6$ & $(mRZ\alpha)^2$ & \ding{51} & \cellcolor{cyan} \ding{51} &  $4.2\times10^{-7}$ &  \cellcolor{yellow} $6.6\times10^{-4}$ \\
\hline
$m^3R^2(Z\alpha)^6$ & $(Z\alpha)^2$ & \ding{51} & \ding{53} &  \cellcolor{green} $8.9\times10^{-2}$ &  \cellcolor{green} $3.2\times10^{-3}$ \\
\hline
$m^4R^3(Z\alpha)^7$ & $(mRZ\alpha)(Z\alpha)^2$ & \ding{51} & \ding{53} &  $1.4\times10^{-6}$ &  $1.1\times10^{-5}$ \\
\hline
$m^3R^2(Z\alpha)^8$ & $(Z\alpha)^4$ & \ding{51} & \ding{53} &  $4.7\times10^{-6}$ &  $1.7\times10^{-7}$ \\
\hline
\hline
$m \exx  (Z\alpha)^3$ & hfs & \ding{51} & \ding{51} &  $5.3\times10^5$ &  $9.4\times10^1$ \\
\hline
$m^2R \exx  (Z\alpha)^4$ & LO & \cellcolor{cyan} \ding{51} & \ding{53} &  \cellcolor{green}$1.1\times10^{1}$ &  \cellcolor{green} $4.0\times10^{-1}$ \\
\hline
$m^3R^2 \exx  (Z\alpha)^5$ & $(mRZ\alpha)$ & \ding{51} &  \ding{53} &  $2.2\times10^{-4}$ &  \cellcolor{green} $1.7\times10^{-3}$ \\
\hline
$m^4R^3 \exx  (Z\alpha)^6$ & $(mRZ\alpha)^2$ & \ding{51} & \cellcolor{cyan} \ding{51} &  $4.6\times10^{-9}$ &  $7.1\times10^{-6}$ \\
\hline
$m^2R \exx  (Z\alpha)^6$ & $(Z\alpha)^2$ & \ding{51} & \ding{53} &  $5.8\times10^{-4}$ &  $2.1\times10^{-5}$ \\
\hline
$m^3R^2 \exx  (Z\alpha)^7$ & $(mRZ\alpha)(Z\alpha)^2$ & \ding{51} & \ding{53} &  $1.2\times10^{-8}$ &  $9.0\times10^{-8}$ \\
\hline
$m^2R \exx  (Z\alpha)^8$ & $(Z\alpha)^4$ & \ding{51} & \ding{53} &  $3.1\times10^{-8}$ &  $1.1\times10^{-9}$ \\
\hline
\hline
$m \exx ^2 (Z\alpha)^4$ & $\mathrm{hfs}^2$ & \ding{51} & \ding{51} & \cellcolor{green} $4.3\times10^{-2}$  & \cellcolor{green}$1.6\times10^{-3}$ \\
\hline
$m^2 R \exx ^2 (Z\alpha)^5$ & LO & \cellcolor{cyan}\ding{51} & \ding{53} & $8.8\times10^{-7}$ &   $6.7\times10^{-6}$ \\
\hline
$m^3R^2 \exx ^2 (Z\alpha)^6$ & $(mRZ\alpha)$ & \ding{51} &  \ding{53} &  $1.8\times10^{-11}$ &  $2.8\times10^{-8}$ \\
\hline
$m^4R^3 \exx ^2 (Z\alpha)^7$ & $(mRZ\alpha)^2$ & \ding{51} & \cellcolor{cyan} \ding{51} &  $3.7\times10^{-16}$ &  $1.2\times10^{-10}$ \\
\hline
$m^2 R \exx ^2 (Z\alpha)^7$ & $(Z\alpha)^2$ & \ding{51} & \ding{53} &  $4.7\times10^{-11}$ &  $3.6\times10^{-10}$ \\
\hline
$m^3 R^2 \exx ^2 (Z\alpha)^8$ & $(mRZ\alpha)(Z\alpha)^2$ & \ding{51} & \ding{53} &  $9.6\times10^{-16}$ &  $1.5\times10^{-12}$ \\
\hline
$m^2 R \exx ^2 (Z\alpha)^9$ & $(Z\alpha)^4$ & \ding{51} & \ding{53} &  $2.5\times10^{-15}$ &  $1.9\times10^{-14}$ \\
\hline
\end{tabular}
\caption{\label{table1}The order of magnitude of various nuclear contributions to atomic energy shifts. Double lines separate blocks involving different powers of nuclear moments, $\mu_\ssN$, where $\exx \sim \cO(m e \mu_\ssN/4\pi) \sim \cO(mR \,Z \alpha)$ -- with $m$ the lepton mass, $R$ a measure of nuclear size and $\alpha$ the fine-structure constant -- is more precisely defined in eq.~\pref{exxdef}. The checks and crosses indicate if a term of the given order actually arises (excluding recoil effects and QED radiative corrections, which do not introduce new parameters) for lepton states with $j=\frac12$ and $j=\frac32$. New parameters enter when integration constants for new modes are required, and cyan shading indicates the order where this first arises, for different choices for $j$ and powers of $\exx$. The final two columns evaluate the numerical size implied by the powers of $\exx$, $Z\alpha$ and $mRZ\alpha$, for ordinary Hydrogen (electrons) and for muonic Hydrogen, using assumptions spelled out in the text below. Green boxes flag terms required to achieve an accuracy of order 0.001 kHz for Hydrogen (or $10^{-3}$ meV for muonic Hydrogen). The yellow square flags a term not computed here, which is likely to be relevant to muonic Hydrogen experiments.}
\end{table}
\end{center}

\vspace{-1.5cm}
Our construction of nucleus-free combinations relies on identifying combinations of observables for which the $R/a_\ssB$ corrections cancel order-by-order in $\alpha$ and $Z\alpha$, and our tool for finding these exploits the fact that the expansion in $R/a_\ssB$ is most efficiently captured using an appropriate effective field theory (EFT), described in detail below.

\subsection{The EFT framework}

It has long been known that these expansions are efficiently organized using EFT methods, since both the expansion in powers of $\mathrm{v}_e$ and $R/a_\ssB$ arise as low-energy approximations. In quantum field theory the expansion in powers of $\mathrm{v}_e \sim Z \alpha$ can be systematized using non-relativistic quantum electrodynamics (NRQED) \cite{caswell}, and efficiently allows the inclusion of second-quantized radiative corrections with the standard Schr\"odinger treatment of Coulomb bound states. One way to think about our formalism is as a version of NRQED where the projection onto the single-nucleus sector is achieved using first-quantized methods. This allows the matching of the nuclear-size effective couplings to be performed very efficiently, using a near-nucleus  boundary condition.  

We do {\it not} explicitly start from NRQED here (though our formalism is also easily adapted to the non-relativistic fields of NRQED), since it is equally easy to treat the light lepton relativistically. So we instead choose to perturb around the relativistic Dirac-Coulomb system, rather than the nonrelativistic Schr\"odinger-Coulomb system. We do so because our main EFT focus is on those interactions that capture the $R/a_\ssB$ expansion. 

If the nucleus\footnote{For concreteness this discussion proceeds assuming a spin-half nucleus but our effective theory works for arbitrary nuclear spin.} were a point particle its leading relativistic electromagnetic interactions with leptons and photons would be described by the renormalizable QED lagrangian
\begin{equation} \label{QEDlepnuc}
  S_\NQED = -\int   \exd^4 x \left\{ \frac{1}{4} F_{\mu\nu} F^{\mu\nu} + \ol{\Psi} \left[ \slashed{D} + m \right] \Psi + \ol{\Phi} \left[ \slashed{D} + M \right] \Phi  \right\},
\end{equation}%
for electromagnetic field strength $F_{\mu\nu} = \partial_\mu A_\nu - \partial_\nu A_\mu$, Dirac lepton field $\Psi$ and spin-half nucleus field $\Phi$ (with $D_\mu \Psi = \partial_\mu \Psi + i e A_\mu \Psi$ and $D_\mu \Phi = \partial_\mu \Phi - i Z e A_\mu \Phi$). Here $M \gg m$ is the mass of the nucleus where (as above) $m$ is the lepton mass. In reality \pref{QEDlepnuc} gets supplemented by additional renormalizable terms involving other light fields (such as any other light leptons), as well as by nonrenormalizable terms describing shorter-wavelength physics that is already integrated out (such as those describing the weak interactions and so on). 

From an EFT perspective the $R/a_\ssB$ expansion is captured by a subset of the higher-dimensional terms not listed explicitly in \pref{QEDlepnuc}. The ones that are relevant are those --- see {\it e.g.} \cite{hill, pineda1997, paz2015, pineda2004} --- consistent with the symmetries of the strong interactions (like electromagnetic gauge invariance, parity, and so on), that couple $\Phi$ to $\Psi$ and $A_\mu$ nonminimally, such as
\be \label{QEDnuc}
  S_{\rm nuc} = - \int \exd^4x \left\{ \frac{\widetilde c_d}{2}( \ol{\Phi} \gamma^{\mu\nu} \Phi )\, F_{\mu\nu} + \widetilde c_s (\ol{\Psi}  \Psi )\, (\ol{\Phi} \Phi) + \widetilde c_v (\ol{\Psi} \gamma^\mu \Psi )\, (\ol{\Phi} \gamma_\mu\Phi) + \cdots \right\} \,.
\ee
Here $\gamma^{\mu\nu} := -\frac{i}{4} \left[ \gamma^\mu, \gamma^\nu \right]$ and effective couplings like $\widetilde c_d$ and $\widetilde c_v$ have dimensions (length)${}^p$ for positive $p$. EFT methods exploit the fact that such interactions capture the low-energy effects obtained by integrating out {\it any} kinds of nuclear degrees of freedom besides the nucleus' overall spin and position. 

The effective couplings obtained in this way dominate the low-energy interactions of finite-sized nuclei at wavelengths much longer than nuclear size. Of the couplings given above $\widetilde c_d \propto R$ captures the anomalous nuclear magnetic moment, $\widetilde c_v \propto R^2$ is related to its electromagnetic charge radius and so on. The precise interpretation of parameters like $\widetilde c_d$, $\widetilde c_s$ and $\widetilde c_v$ is found using matching calculations that compare the predictions of $S_\NQED + S_{\rm nuc}$ for lepton-nucleus and photon-nucleus scattering, either with measurements or (in principle) with {\it ab initio} predictions of the Standard Model (in practice this is where calculations using nuclear models come in).

The arguments of this paper are easiest to make using a variant of the above EFT that is better adapted to atomic calculations. Rather than treating the nucleus using the second-quantized field $\Phi$ the variant we prefer instead uses a first-quantized description (described in more detail in refs.~\cite{ppeft1, ppeft2, ppeft3}). A first-quantized EFT description of the nucleus is more efficient because an atom includes only a single nucleus, making the rest of the multi-particle Fock space accessed by $\Phi$ unnecessary. In principle one integrates out all multi-particle degrees of freedom to arrive at an EFT that contains only nuclear collective coordinates: its centre-of-mass position and nuclear spin, interacting with the second-quantized fields $\Psi$ and $A_\mu$.

The resulting EFT is worked out for spinless nuclei in refs. \cite{ppeft3, ppeftA}, where the only relevant nuclear degree of freedom is the position operator that describes the nucleus' world-line, $\msP$: $x^\mu = y^\mu(s)$. In this effective theory the remaining second-quantized degrees of freedom are described by
\be \label{QEDlep}
  S_\QED = -\int   \exd^4 x \left\{ \frac{1}{4} F_{\mu\nu} F^{\mu\nu} + \ol{\Psi} \left[ \slashed{D} + m \right] \Psi  \right\},
\ee
to which one adds the action describing couplings to the nuclear degrees of freedom: 
\bea \label{Spdefnospin}
S_p &=& - \int_\msP   \exd s \left\{ \sqrt{-\dot{y}^2} \; M  -Ze\, \dot{y}^\mu A_\mu  + c_s \sqrt{-\dot{y}^2}  \; ( \ol{\Psi}  \Psi )  + i  c_v\, \dot{y}^\mu\, (\ol{\Psi} \gamma_\mu \Psi) + \cdots \right\}   \nn\\
&=& -\int \exd^4 x \int_\msP   \exd s \left\{ \sqrt{-\dot{y}^2} \; M  -Ze\, \dot{y}^\mu A_\mu  + c_s \sqrt{-\dot{y}^2}  \; ( \ol{\Psi}  \Psi ) \phantom{\frac12} \right.\\
&&\qquad\qquad\qquad\qquad\qquad\qquad \left.\phantom{\frac12} + i  c_v\, \dot{y}^\mu\, (\ol{\Psi} \gamma_\mu \Psi) + \cdots \right\}  \, \delta^4[x - y(s)],,\nn
\eea
where the second line emphasizes that in the first line the `bulk' fields $\Psi(x)$ and $A_\mu(x)$ are evaluated along the nuclear world-line $x^\mu = y^\mu(s)$. $\dot y^\mu$ here represents $\exd y^\mu/\exd s$ where $s$ is an arbitrary parameter along the world-line. Quantization proceeds by evaluating the functional integral of $\exp[i(S_\QED +S_p)]$ with respect to $y^\mu(s)$ (which in particular captures the response of nuclear recoil -- more about which below) as well as $\Psi(x)$ and $A_\mu(x)$. 

In this language the $\Phi$-dependent part of eqs.~\pref{QEDlepnuc} and \pref{QEDnuc} are replaced by $S_p$, and the spin-independent part of effective couplings --- {\it e.g.}~$\widetilde{c_s}, \widetilde{c_v}$ of \pref{QEDnuc} --- are captured by the effective couplings --- {\it e.g.} $c_s, c_v$ of \pref{Spdefnospin} --- within $S_p$. For this to be possible $S_p$ is again required to be the most general local action consistent with the field content and symmetries, which in this case now also include arbitrary reparametrizations of the world-line parameter $s$.  On dimensional grounds the effective couplings $c_s$ and $c_v$ are of order $R^2$, and the ellipses in \pref{Spdefnospin} contain interactions with couplings having dimension (length)${}^p$ for $p > 2$, with those -- like the ones describing nuclear polarizability -- arising at order $R^3$ (but not involving nuclear spin) considered in detail in \cite{ppeftA}.

Physical interpretation is simpler if the world-line parameter $s$ is chosen to be proper time, $\tau$ (in which case $\dot y^2 = \eta_{\mu\nu} \, \dot y^\mu \dot y^\nu = -1$). With this choice then $\dot y^\mu = \gamma ( 1, \bfv )$ with $\gamma := \exd t/\exd \tau = (1- \mathbf{v}^2)^{-1/2}$ and $\bfv := \exd \bfy/\exd t$, so once evaluated in the atomic center-of-mass frame \pref{Spdefnospin} becomes 
\bea \label{Spdefnospin2}
S_p &=& - \int_\msP   \exd \tau \left\{ M  -Ze\, \gamma (A_0 + \bfv \cdot \bfA)  + c_s  \; ( \ol{\Psi}  \Psi ) \phantom{\frac12} \right.\nn\\
&&\qquad\qquad\qquad\qquad \left.\phantom{\frac12} + i  c_v\, \gamma \Bigl[   (\ol{\Psi} \gamma_0 \Psi) + \bfv \cdot (\ol{\Psi} \boldsymbol{\gamma} \Psi) \Bigr] + \cdots \right\}    \nn\\
&=& - \int   \exd^4x \,\left\{ \sqrt{1 - \bfv^2} \Bigl[ M + c_s  \; ( \ol{\Psi}  \Psi ) \Bigr] -Ze\,  (A_0 + \bfv \cdot \bfA)  \phantom{\frac12} \right.\\
&&\qquad\qquad\qquad\qquad \left.\phantom{\frac12} + i  c_v\, \Bigl[   (\ol{\Psi} \gamma_0 \Psi) + \bfv \cdot (\ol{\Psi} \boldsymbol{\gamma} \Psi) \Bigr] + \cdots \right\}  \, \delta^3[\bfx - \bfy(\tau)]\,.\nn
\eea
The terms of \pref{Spdefnospin} and \pref{Spdefnospin2} involving $M$ and $Ze$ are recognizable as describing the rest mass and Coulomb coupling of a point nucleus, while the subsequent terms carry the leading information about nuclear substructure. 

The $\bfv$-dependent terms of \pref{Spdefnospin2} contain the nuclear recoil effects, with $M \sqrt{1-\bfv^2}$ capturing the usual relativistic kinematics $E = \sqrt{\bfp^2 + M^2}$. $\bfv$-dependent terms that also involve $c_s$ or $c_v$ contain mixed nuclear-size/recoil contributions. Their size can be estimated given that, in atoms, the nuclear momenta are of order $|\bfp| \sim Z \alpha m$, and so $|\bfv| \sim Z\alpha m/M$. Similarly the leptonic matrix element   $\langle \ol\Psi \boldsymbol{\gamma} \Psi \rangle \sim |\psi(0)|^2 \bfv_e$ is of order $(Z \alpha m)^3 (Z \alpha)$ and so $c_v \bfv \cdot \ol{\Psi} \boldsymbol{\gamma} \Psi$ contributes to energies a shift of order $c_v (Z\alpha)^5 m^4/M$. Given that the non-recoil term implies $c_v \sim Z \alpha R^2$ is a measure of the nuclear charge radius, the energy shift is $m^4 R^2 (Z\alpha)^6/M$. Since we work here to an accuracy of order $m^4 R^3 (Z\alpha)^5$ and $m^3 R^2 (Z\alpha)^6$, we must keep the leading mixed recoil/nuclear-size effects in what follows.\footnote{Recoil effects for point nuclei must of course also be included, as we do in what follows.} Because recoil effects do not change the parameter counting given below, we put them aside temporarily and return to them in \S\ref{energies}.

Dropping recoil effects, in the nuclear rest frame (with the nucleus situated at the origin) one finds
\be \label{Spdefnospinrest}
S_p = - \int   \exd^4 x \, \Bigl\{ M  -Ze\, A_0  +c_s \, ( \ol{\Psi}  \Psi )+ i  c_v\, (\ol{\Psi} \gamma_0 \Psi) + \cdots \Bigr\} \delta^3(\bfx) \,.
\ee
This first-quantized, point-particle EFT (or PPEFT) treatment of the central object has been tested in a variety of other systems with compact central sources \cite{ppeftA, falltocenter, echoes, twospecies}, and found to reproduce in a simpler way many standard results. In particular, ref.~\cite{ppeftA} shows that its application to spinless nuclei -- including some of the order-$R^3$ terms shown as ellipses in \pref{Spdefnospin} and \pref{Spdefnospinrest} -- correctly captures the nuclear charge radius, Friar moment, polarizability {\it etc.} that provide the leading contributions to shifts in atomic energy levels. 

A major purpose of the present paper is to extend the action \pref{Spdefnospin2} to include nuclear spin --- leading {\it e.g.}~to eq.~\pref{SpForm} --- and to explore the consequences of new terms for atomic energies.

\subsection{Parameter counting for atomic energy shifts}

In principle, the above discussion allows existing calculations of atomic energy shifts due to nuclear size to be described in two steps. First use an explicit nuclear model to compute effective couplings like $c_s$ or $c_v$ (which can be mapped to the various nuclear moments encountered in the literature). Second, compute the dependence of atomic energies on these effective couplings. Ref.~\cite{ppeftA} elaborates on this process in the special case of spinless nuclei. 

A virtue of proceeding in this way is its economy of effort: for a specific nuclear model the first step need be taken only once, with the results usable in the second step for any number of different type of observables (provided these are at low enough energies to allow the assumed expansion in powers of $R$). Alternatively the second step can be done once to obtain how a specific observable depends on the effective couplings, with only the first step needing to be repeated to calculate these couplings using a variety of different nuclear models. 

For the present purposes, however, we focus specifically on step two (which we  generalize to include nuclei with spin). We do so because the PPEFT formulation of the nucleus provides insight into the number of independent ways that nuclear structure (or, equivalently, the effective couplings, $c_s$, $c_v$ and so on) can contribute to atomic energy shifts. In particular it provides a systematic way to identify observables that do not depend at all on the nuclear effective couplings.  

Tracing how effective couplings appear in low-energy atomic observables is particularly transparent in the first-quantized language. This is because in this formalism observables like energy shifts turn out to acquire their dependence on nuclear properties purely through the near-nucleus boundary conditions satisfied by the external fields $\Psi$ and $A_\mu$, which in turn depends on the effective couplings like $c_s,c_v$. This boundary condition is found by integrating the equations of motion for these fields over a small ball centred at the nuclear position (possibly weighted by spherical harmonics). It is {\it only} because these boundary conditions depend on couplings like $c_s$ and $c_v$ that observables like atomic energy levels are sensitive to nuclear structure. 

For a familar example of this general argument, consider the field equations for $A_0$ obtained from the action $S_\QED+S_p$ as given in \pref{QEDlep} and \pref{Spdefnospinrest}. The field $A_0$ satisfies (in Coulomb gauge: $\nabla \cdot \bfA = 0$)
\be \label{A0eq}
   \nabla^2 A_0   - ie \, \ol{\Psi} \gamma_0 \Psi =  Ze \, \delta^3(\bfx) \,.
\ee
Treating the $A_\mu \ol{\Psi} \gamma^\mu \Psi$ term of $S_\QED$ --- as well as all but the first two terms on the right-hand side of \pref{Spdefnospinrest} --- as perturbations then allows (to leading order) the dropping of the $\ol{\Psi} \gamma_0 \Psi$ term in \pref{A0eq}. Integrating what's left over a small Gaussian pillbox centred at the origin gives the standard Gaussian boundary condition\footnote{In our metric conventions the usual electrostatic potential is $A^0 = - A_0$.}
\be \label{A0BCeq}
  \int \exd^2 \Omega \Big( \bfe_r \cdot \nabla A_0 \Big)\Big|_{r = 0^+} = 4\pi \lim_{r\to 0^+} r^2\partial_r A_0  =  Ze \,,
\ee
where $\bfe_r$ denotes the outward-pointing radial unit normal. This boundary condition fixes an integration constant of the Coulomb solution in the usual way to ensure $A^0 = Ze/(4\pi r)$. (Higher electromagnetic multipole moments are similarly found from higher-derivative terms in $S_p$ that are linear in $A_\mu$ \cite{EFTBook}.)

Now comes the main point. The argument just given applies equally well for the field $\Psi$, and also once interactions are included in the equations of motion.\footnote{As explored in more detail in \cite{ppeft1,ppeft2,ppeft3,EFTBook}, once $S_p$ is not linear in the bulk fields this procedure necessarily involves regularizing coincident divergences at the source position, and renormalizing effective couplings in $S_p$. (See also \cite{Goldberger:2001tn, Goldberger:2004jt, Agashe:2002bx, deRham:2007mcp, Burgess:2008yx, Bayntun:2009im} for similar discussions in another context.)} By changing the boundary conditions at the origin the presence of the interactions in $S_p$ necessarily alters the mode functions and energy eigenvalues of the fields $\Psi$ and $A_\mu$, and this is how effective couplings like $c_s$ or $c_v$ end up affecting electronic properties. This observation is useful because it means that many effective couplings can only appear in observables in a limited way, as is now argued.

To see why parameters like $c_s$ or $c_v$ (and by extension the various nuclear moments) only appear in specific combinations in observables, imagine finding the mode functions for $\Psi$ by separating variables in its field equation in spherical coordinates. One then seeks a basis of solutions of the form $u_{\ssL}(t,r,\theta,\varphi) = \cR_{\ssL}(\kappa r) Y_{\ssL}(\theta, \varphi) \, e^{-i \omega t}$, where $\kappa$ is a function of the mode's energy, $\omega$, while $Y_\ssL(\theta,\varphi)$ represent appropriate spherical harmonics for which $L$ denotes the collection of angular-momentum labels relevant to the problem ({\it e.g.}~for spinless fields $L = \{ l , l _z \}$ is a pair of integers in three spatial dimensions). The radial mode then satisfies an ordinary second-order differential equation, whose general solution has the form
\be
   \cR_{\ssL} (\kappa r) = \msC_\ssL \cR^{\scriptscriptstyle \msC}_{\ssL}(\kappa r) + \msD_\ssL\cR^{\scriptscriptstyle \msD}_{\ssL}(\kappa r) \,,
\ee
with two integration constants, $\msC_\ssL$ and $\msD_\ssL$, corresponding to the two basis solutions, $\cR^{\scriptscriptstyle \msC}_{\ssL}(\kappa r)$ and $\cR^{\scriptscriptstyle \msD}_\ssL(\kappa r)$, for each value of $\kappa$ and each choice of angular-momentum quantum numbers $L$. The boundary condition at the origin fixes the value of $\msD_{\ssL}/\msC_{\ssL}$ and once the interactions in \pref{Spdefnospinrest} are included in the action this value also depends on effective couplings like $c_s$ and $c_v$. This is in practice how effective interactions localized at the nucleus ultimately modify physics far from the nucleus, and in particular influence the shapes and energies of the electronic energy eigenmodes. Furthermore, it is {\it  only} through such boundary conditions that these  eigenmodes `learn' about non-pointlike nuclear-physics effects. 

The above discussion is important because it shows that effective couplings typically only enter into electronic energy shifts through the values they imply for the ratio $\msD_{\ssL}/\msC_{\ssL}$   for each choice for $L$. Furthermore, at low orders in the $R/a_\ssB$ expansion only a few angular momenta contribute at all, because of the suppression near the nucleus of higher-$L$ wave-functions. This leads to one of our main points: 

\begin{quote}{\it If there are more effective couplings in $S_p$ than there are relevant integration constants $\msC_{\ssL}, \msD_{\ssL}$, then the effective couplings cannot all appear independently in observables: all that matters for experiments are the values of} $\msD_{\ssL}/\msC_{\ssL}$. 
\end{quote}

For spinless nuclei ref.~\cite{ppeftA} shows that out to orders $m^4 R^3 (Z\alpha)^5$ and $m^3 R^2 (Z\alpha)^6$ (but not including order $m^5 R^4 (Z\alpha)^6$) all nuclear-size contributions --- {\it i.e.}~the charge radius, the Friar moment \cite{friar}, the electromagnetic nuclear polarizabilities, leading recoil corrections and a few others --- contribute to atomic energy shifts only through their contributions to one parameter: the value of $\msD_{\ssL}/\msC_{\ssL}$   for $S$-wave modes. This is only one parameter because this ratio has a predictable dependence on $\kappa$ (and so also on the principal quantum number $n = 1,2,\cdots$). Because the couplings $c_i$ capture {\it all} possible nuclear-size effects, it follows that (to this order in $Z\alpha$ and $R/a_\ssB = mRZ\alpha$) {\it all} nuclear finite-size effects can enter into atomic energy levels only through a single independent parameter: the $S$-wave value of $\msD_{\ssL}/\msC_{\ssL}$   (for  details see \cite{ppeftA}). 

A similar statement applies at higher orders. Starting at order $m^5 R^4 (Z\alpha)^6$ the ratio $\msD_{\ssL}/\msC_{\ssL}$   for the $P$-wave modes is also required when computing atomic energies, so to this order (for spinless nuclei) all nuclear effects only enter atomic energies through {\it two} independent parameters. One can continue on in this way to any order and identify the number of independent nuclear contributions that can robustly arise. In all cases there are fewer contributions than would be naively expected by counting nuclear effective couplings (like $c_s$, $c_v$, {\it etc}) allowed at this order in $R$ (or by counting nuclear `moments' within specific nuclear models, as is the usual calculational practice).

\subsubsection*{This paper}

With the above logic in mind, we now can state what the present paper achieves. We first extend the analysis of \cite{ppeftA} to include nonzero nuclear spin. This allows us to broaden the applicability of these conclusions to general single-lepton atoms and ions, and in particular to those of most practical interest: Hydrogen (and muonic Hydrogen). We find that the inclusion of spin introduces additional constants, but to any fixed order it remains true that nuclear-size effects enter into atomic energies through fewer parameters than might be expected based on nuclear modelling.
Our more detailed conclusions regarding the number of relevant parameters at any given order are summarized in Table \ref{table1}, which contains a list of contributions to atomic energy shifts due to nuclear finite-size effects. The Table is organized into three blocks, each containing terms that involve specific powers of the nuclear magnetic moment, $\mu_\ssN$ (which enters through the small dimensionless parameter $\exx \propto m e \mu_\ssN$). Successive rows in each block list higher-order terms additionally suppressed by powers of $Z\alpha$ or $R/a_\ssB = mR Z \alpha$. (The first block -- independent of magnetic moments -- reproduces in particular the results of \cite{ppeftA} for spinless nuclei.) 

The Table's 2nd column gives the suppression of each term relative to the leading-order contribution that shares the same power of $\exx$. The leading contribution independent of $\exx$ is the usual charge-radius term. For terms involving at least one power of $\exx$ the entry marked `LO' contains the leading dependence on finite nuclear size $R$ at this order in $\exx$. (The term before this one depends on nuclear size only through $\mu_\ssN$, which need not vanish even for a point nucleon). 

The third and fourth columns of Table \ref{table1} indicate with a check or a cross whether or not this size a contribution actually arises when computed using Dirac-Coulomb wavefunctions (and neglecting QED radiative corrections\footnote{These radiative corrections are included within our formalism, as are recoil corrections, because all of $\Psi$, $A_\mu$ and $y^\mu$ are fully dynamic quantum operators. We discuss radiative and recoil corrections in more detail in \S\ref{energies}, but what is important is this: they do not introduce any new parameters (though they do of course depend on the bulk-field mode functions, and so on the existing parameters, $\msD_{\ssL}/\msC_{\ssL}$).} that are suppressed by additional powers of $\alpha$). These two columns differ in the value they assume for the lepton's total angular momentum quantum number, $j$, with column 3 giving the dominant result for $j = \frac12$ and column 4 giving the same for $j = \frac32$. Comparing these columns shows the angular-momentum suppression expected for nuclear effects due to the suppression of the wave-function outside the nucleus near the origin. For instance $j= \frac32$ states are irrelevant for contributions lower-order than $m^5 R^4 (Z\alpha)^6$, but once this order is reached a new constant enters because the value of $\msD_{\ssL}/\msC_{\ssL}$   for $j=\frac32$ contributes observably.

The final two columns give an indication of the numerical size of each term, with column 5 providing the numbers for ordinary (electronic) Hydrogen while column 6 does so for muonic Hydrogen. Electronic and muonic Hydrogen differ only in the value of the lepton mass that is used in the corresponding estimate, which implies that the combination $mR$ is not particularly small for muonic Hydrogen (and so $R/a_\ssB \sim Z\alpha$, rather than being much smaller). For illustration purposes squares are shaded green if the resulting estimate is of order $0.001$ kHz or larger (for Hydrogen) or of order $10^{-3}$ meV or more for muonic Hydrogen. The given numerical values evaluate $R$ using the charge radius, $R = r_p = 0.84087$ fm, for $\exx$-independent contributions, but use the Zemach radius, $R = r_z = 1.082$ fm for $\exx$-dependent contributions \cite{aldo2013exp}. The yellow square flags a term not computed here that might also contribute observably for muonic Hydrogen.

Only terms on rows shaded green are required if one works only to this accuracy, and Table \ref{table1} shows that in this case nuclear finite-size effects enter through just two independent constants. One controls the contributions independent of nuclear spin (and contains in particular the contributions of the charge radius, Friar moment, nuclear polarizabilities, leading recoil corrections and more \cite{ppeftA}) and the second, spin-dependent, parameter captures the nuclear Zemach moment. The nuclear magnetic moment also enters into atomic energies at this order, though we do not count this as an unknown nuclear parameter because its value is accurately determined by other means. 

We see from the Table that the conclusions of \cite{ppeftA} also apply for spinning nuclei, though with one additional constant required at the accuracy discussed above. This is so despite there being many more parameters apparently relevant in the first-quantized action, $S_p$, and in explicit calculations using nuclear models \cite{friar}. Of course, this does not mean that the parameters of nuclear models are not intrinsically independent; what it says is that only a small number of combinations of them ever appear in observables {\it at the extremely low energies relevant to atomic energy levels}. As discussed below, the two constants through which they appear to this order for spin-half nuclei can be captured by $\msD_{\ssL}/\msC_{\ssL}$  for $S$-wave modes for the two different values of total atomic angular momentum quantum number: $F$ (where $\bfF = \bfI + \bfJ$ as usual combines nuclear spin and electronic total angular momentum). 

Why care that the many moments of nuclear models (or parameters in the action $S_p$) can only enter atomic energies through their contributions to the (comparatively fewer) mode constants $\msD_{\ssL}/\msC_{\ssL}$? This observation is useful because theoretical calculations of nuclear moments are notoriously difficult and in some cases introduce the dominant theoretical uncertainties in calculations of atomic energy levels. These uncertainties can be larger than the size of other small effects whose measurement might ultimately provide new tests of fundamental physics. Robustly knowing that these uncertainties only enter atomic levels in a small number of independent ways opens up ways to remove nuclear uncertainties from some precision atomic measurements. This can be done in several ways: simply use experiments to determine the relevant nuclear parameters (which, as argued above, are fewer than the naive number of nuclear moments); or combine observables in such a way that nuclear contributions completely cancel, whose results are not subject at all to uncertainties associated with nuclear (or other short-distance) physics.

We illustrate how this can be done in practice by identifying the values of the two independent nuclear parameters using two particularly well-measured atomic energy-level differences, with results for these parameters summarized in Table \ref{sat} for atomic Hydrogen and Table \ref{mat} for muonic Hydrogen. These values are then used to predict the nuclear component of other transition frequencies working at the order indicated by green entries in Table \ref{table1}. These predictions are presented for the best-measured transitions of atomic Hydrogen in Table \ref{ahprecise}, while results for the much broader list of transitions compiled in \cite{kramida} are given in Tables \ref{table2} through \ref{table4}. Each of these tables  gives the finite-nuclear-size contribution as computed for these transitions, together with an estimate of the errors involved, at the orders listed in green in Table \ref{table1}, as summarized in \S\ref{energies}.   

As these tables show, at present the largest uncertainty comes from the error in the theoretical prediction for the two reference transitions as computed using a point-like nucleus (or the `point-like' theory, for short). What is important is that the size of all of these errors can improve in a way that does not depend on nuclear physics. Although the required calculations are challenging in practice, improved computations for a point-like nucleus are in principle straightforward to perform. The same is true for the `truncation' error given in Column 5 of the Tables, and of course experimental errors do not require improvements in nuclear-physics calculations. As these errors improve, so do the overall theoretical errors in the predictions like those in Column 2 of the Tables. Errors in nuclear calculations play no role in our predictions because all relevant nuclear parameters are taken directly from atomic observations. 

We find in this way a broad and robust class of predictions whose intrinsic error is not set by our ability to compute with nuclear models. In the language of earlier paragraphs, given precise measurements for $N_{\rm exp}$ leptonic energy differences (for a specific nucleus), we make $N_{\exp}-2$ model-independent predictions for how nuclei shift atomic energies, essentially by eliminating the two independent nuclear parameters. What is important is that the error in these predictions is controlled only by the accuracy of the experiments used to determine the two parameters, plus the error implied by working only to a fixed order in the small quantities $Z\alpha$ and $\exx \sim R/a_\ssB = mRZ\alpha$ in both the `point-like' theory contributions and the effective couplings.  

The rest of this paper is organized as follows. \S\ref{ppeft} derives the point-particle effective action for central sources with spin, keeping effective interactions with couplings out to dimension (length)${}^2$.  To the order we work the main difference relative to \cite{ppeftA} is the presence of the dipole nuclear magnetic field, and we use standard  perturbation theory to establish its consequences for atomic energy levels. 

In particular, \S\ref{running} then explores the near-nucleus boundary conditions implied for the electron field by the relevant contact interactions at the nucleus, \S \ref{ssec:Renorm} gives a discussion of the associated divergences in these calculations in the near-nucleus limit, and derives the renormalization-group (RG) evolution of the new effective couplings. These contact interactions also shift electronic energy levels in a way that competes with the effects of the nuclear magnetic dipole field. These sections show in detail why (for each nuclear isotope) only two parameters (plus the nuclear magnetic moment) are required to describe all nucleus-dependent shifts to the order we work. 

Next, \S\ref{energies} collects expressions for nucleus-dependent atomic energy shifts. \S\ref{electronPunchline} uses the existing experimental data for Hydrogen to fit the two relevant nuclear parameters, and applies these to make predictions for nuclear-size effects for other transitions, with prediction errors that are independent of the limitations of nuclear models. \S\ref{muonPunchline} briefly discusses the same steps for muonic Hydrogen. Finally, \S\ref{conc} summarizes our conclusions and comments on possible future directions. Several  appendices outline useful calculational details, and in particular Appendix \ref{los} provides a list of notation used.

\section{PPEFT for sources with spin}
\label{ppeft}

To set up the position-space point-particle effective theory (PPEFT) for spinning nuclei we couple second-quantized electron and photon fields to the first-quantized nuclear centre-of-mass and spin degrees of freedom. The resulting action is the sum of a `bulk' part and a `point-particle' part, $S= S_\ssB + S_p$, where the bulk part consists of standard quantum electrodynamics (QED), as in \pref{QEDlep} (repeated here, for convenience),
\begin{equation} \label{QEDlep2}
S_\ssB = -\int   \exd^4 x \left\{ \frac{1}{4} F_{\mu\nu} F^{\mu\nu} + \ol{\Psi} \Bigl( \slashed{D} + m \Bigr) \Psi \right\} \,.
\end{equation}
As before, $F_{\mu\nu}$ is the field strength for the electromagnetic gauge potential $A_\mu(x)$, and $\Psi(x)$ is the spin-half Dirac field of the orbiting lepton with mass $m$ and charge $q = -e$ (with covariant derivative $D_\mu = \partial_\mu + ieA_\mu$). 

Our main focus here is in the formulation of $S_p$ for the first-quantized nucleus, with the new feature relative to refs.~\cite{ppeft1, ppeft2, ppeft3, ppeftA} being the inclusion of the nuclear spin degrees of freedom in the first-quantized nuclear action, $S_p$. 

\subsection{Spin on a world-line}

The classical and quantum dynamics of first-quantized spinless relativistic particles propagating in spacetime is discussed in many textbooks  \cite{gitman1990, polchinski, zwiebach2009}. The extension to first-quantized spinning particles started in the early days of supersymmetry where it was found that first-quantized supersymmetric systems built using Grassmann (classically anti-commuting) fields described spinning particles \cite{casalbuoni1975, barducci1976, berezin1977, brink, vecchia, ravndal1980}. 

Classical Grassman variables naturally arise when describing spin because on quantization they furnish finite-dimensional representations of rotations in the quantum Hilbert space (as is seen explicitly below). The particle's total spin quantum number, $s$, is then fixed in terms of the dimension, $2s+1$, of this representation. We here follow this lead and use such a Grassmann field to describe nuclear spin, introducing a 4-vector of new Grassmann fields, $\xi^\mu(s)$, on the nuclear world-line, which at the classical level satisfies $\left\{ \xi^\mu, \xi^\nu \right\} = 0$.

\subsubsection*{Kinematics}

Supplementing the unperturbed action for the centre-of-mass motion of the nucleus with the free action for $\xi^\mu$ gives
\be \label{Sp0def}
  S_{p0} = - \int   \exd s \,  \left\{ \sqrt{-\dot{y}^2}\; M + i\xi^\mu \dot{\xi}_\mu -  (Ze) \dot{y}^\mu A_\mu  \right\}  \,,
\ee
where $s$ is an arbitrary world-line parameter. 

Once quantized, the classical anticommutation relation becomes modified\footnote{See Appendix \ref{AppendixA} for more details and for our Dirac matrix conventions.} to become \cite{brink}:
\begin{equation}
\label{clifford}
\left\{ \hat{\xi}^\mu, \hat{\xi}^\nu \right\} = -\frac{1}{2}\, \eta^{\mu\nu} \,.
\end{equation}
A technical complication arises when quantizing because this is a constrained classical system, whose canonical positions and momenta are not all independent of one another. Quantization requires the toolkit put together by Dirac in \cite{dirac} and others \cite{dirac1964, teitelboim, gitman1990, henneaux} for constrained systems, combined with standard techniques for anti-commuting objects summarised (for instance) in \cite{casalbuoni, DeWittSuperMan}.

The system's Hilbert space (as usual) furnishes a representation of this algebra, and we choose the spin of the nucleus when we choose the dimension of the representation that is of interest. For spin-$\frac{1}{2}$ fermions, we use a 4-dimensional\footnote{This becomes two-dimensional once antiparticle states are projected out.}  representation in terms of Dirac matrices, 
\begin{equation} \label{xirep}
\hat{\xi}^\mu = \frac{i}{2} \,\Gamma^\mu,
\end{equation}
since the Clifford algebra identity $\{\Gamma^\mu, \Gamma^\nu\} = 2 \eta^{\mu\nu}$ then ensures \eqref{clifford} is satisfied. We use the notation $\Gamma^\mu$ (rather than $\gamma^\mu$) here to emphasize that these matrices act in the spin-space of the nucleus and reserve $\gamma^\mu$ for the matrices that act on the bulk electron field $\Psi$.

\subsubsection*{Interactions}

The EFT program for a first-quantized and spinning nucleus then asks for all possible local interactions on the world-line that can be written using the fields $y^\mu(s)$, $\xi^\mu(s)$ as well as the `bulk' fields $A_\mu(x=y(s))$ and $\Psi(x = y(s))$. One must write down all allowed operators to a given order to capture all spin-dependent effects consistent with the assumed symmetries (which for the applications below we take, as before, to be the symmetries of the strong and electromagnetic interactions). 

We require the point-particle localized interactions to be hermitian (this is not always required for localized sources, see {\it e.g.}~\cite{falltocenter}, but is appropriate for the present application); to be Grassmann-even; to be invariant under Poincare transformations; to be electromagnetic gauge-invariant; to preserve separately C, P and T transformations; and arbitrary reparameterizations of the nuclear world-line: $s \to s' = f(s)$. Then, keeping only interactions out to order (length)${}^2$, the most general interactions work out to be
\begin{eqnarray}\label{SpForm}
S_p &=& S_{p0}  +\int \exd s \; \left\{  i\mu_\ssN \sqrt{-\dot{y}^2} \;\xi^\mu \xi^\nu F_{\mu\nu}  +  i c_{\rm em} \dot{y}^\mu \xi^\rho \xi^\sigma \partial_\mu F_{\rho\sigma} \right. \notag \\
&& \qquad\qquad -\ol{\Psi} \left[ \sqrt{-\dot{y}^2} \left( c_s + i c_2 \epsilon_{\alpha\beta\gamma\delta}\xi^\alpha \xi^\beta \xi^\gamma \xi^\delta \gamma_5 + i c_\ssF \xi^\mu \xi^\nu \gamma_{\mu\nu} \right) \right.   \\
&& \qquad\qquad\qquad + i \dot{y}^\mu \left( c_v \gamma_\mu + c_3  \epsilon_{\alpha\beta\gamma\delta}\xi^\alpha \xi^\beta \xi^\gamma \xi^\delta \gamma_5\gamma_\mu \right) \Big] \Psi + \cdots  \Big\} \,,\nn
\end{eqnarray}
where $S_{p0}$ is as given in \pref{Sp0def} and (as above) $y^\mu(s)$ is the bosonic centre-of-mass position of the source, $\xi^\mu(s)$ is the Grassmann coordinate representing nuclear spin, while overdots denote derivatives with respect to the world-line parameter.  The quantities $\mu_\ssN$, $c_{\rm em}$, $c_s$, $c_2$, $c_v$, $c_\ssF$, $c_3$ and so on are the effective couplings that arise to this order, where $\mu_\ssN$ has dimension (length) and the rest have dimension (length)${}^2$.

We next turn to the boundary conditions for $A_\mu$ and $\Psi$ that are implied by this action, starting first with the electromagnetic field. We specialize when doing so to spin-half nuclei (both for concreteness's sake and with a view to applications to Hydrogen).

\subsection{Implications for the electromagnetic field}

Varying $A_\mu$ in the action $S_\ssB + S_p$ yields the field equation  
\begin{eqnarray} \label{EMeqs}
\partial_\nu F^{\mu\nu} &=& -ie \ol{\Psi} \gamma^\mu \Psi + Ze \int   \exd s \, \dot{y}^\mu \delta^4[x-y(s)] + i\mu_\ssN  \, \partial_\nu \int   \exd s \, \sqrt{-\dot y^2} \; \left[ \hat\xi^\mu, \hat\xi^\nu \right]  \delta^4[x-y(s)] \notag \\
&& \qquad \qquad - i c_{\rm em} \partial_\sigma \partial_\nu \int   \exd s \, \left[ \hat\xi^\mu, \hat\xi^\nu \right] \dot{y}^\sigma \delta^4[x-y(s)] \\
 &=& -ie \ol{\Psi} \gamma^\mu \Psi + Ze\, \delta^\mu_0 \, \delta^3(\bfx)  + \mu_\ssN \, \Gamma^{\mu\nu} \, \partial_\nu  \delta^3(\bfx)  - c_{\rm em} \Gamma^{\mu\nu}\partial_0 \partial_\nu   \delta^3(\bfx) \,, \nn
\end{eqnarray}
where the second line uses the spin-half version of \pref{xirep} as well as the definition $\Gamma^{\mu\nu} := -\frac{i}{4} \left[ \Gamma^\mu, \Gamma^\nu \right]$; we specialize to the nuclear rest-frame with nucleus situated at the origin -- {\it i.e.}~$\dot{\bfy}(s) = \bfy(s) = 0$ -- and we parameterize the world-line using proper time ({\it i.e.}~$s = \tau$, with $- \dot y^2 = 1$ and $\dot y^\mu = \delta^\mu_0$). Specialization to the rest frame simplifies the discussion by excluding nuclear recoil effects that are suppressed by inverse powers of the nuclear mass, since these are not required for the applications we have in mind. But there is no reason why such effects cannot also be included as corrections to the second equality of \pref{EMeqs}, which would instead be obtained by evaluating the first equation in the atomic centre-of-mass frame.

Neglect of inverse powers of nuclear mass also simplifies the above by allowing the removal of its antiparticle states, leaving the two spin states of the non-relativisitic nucleus at the origin, familiar from atomic physics. This is achieved by projecting out the anti-particle solutions from the nuclear states (as described in Appendix \ref{AppendixA}). Together with dropping nuclear recoil this also means the only matrices to survive unsuppressed by nuclear velocity are
\be \label{Gammavis0}
   \Gamma^0 \to -i\, \mathds{1} \quad \hbox{and} \quad 
   \Gamma_5 \Gamma_k \to -i \tau_k \quad \hbox{and} \quad
   \Gamma^{ij} \to \frac{1}{2} \epsilon^{ijk} \tau_k \,,
\ee
which are $2\times 2$ matrices acting in nuclear spin-space, for which our conventions use $\tau_k$ to denote the Pauli matrices (with the same matrices acting in electron-spin space being denoted $\sigma_k$).

With these choices the electromagnetic field equations \pref{EMeqs} become
\begin{eqnarray} \label{EMptEqs}
\nabla \cdot \bfE = \partial_\nu F^{0\nu} &=& -ie \ol{\Psi} \gamma^0 \Psi + Ze\,  \delta^3(\bfx)\,, \notag \\
(- \partial_t \bfE + \nabla \times \bfB)^i = \partial_\nu F^{i \nu} &=& -ie \ol{\Psi} \gamma^i \Psi + \mu_\ssN \, \epsilon^{ilk} I_k \partial_l  \delta^3(\bfx)\,,
\end{eqnarray}
where $\bfI := \frac12 \, \boldsymbol{\tau}$ denotes the nuclear spin vector, which is an operator in the space of nuclear spins. These show that the nuclear part of the electromagnetic current 4-vector is:
\begin{equation} \label{nuclearA0vecA}
j^0 = Ze\,\delta^3(\bfx)\,, \hspace{16pt} j^i =  \mu_\ssN \, \epsilon^{ilk} I_k \,\partial_l \delta^3(\bfx)\,.
\end{equation}

Following standard practice, we work perturbatively in quantum-field interactions like $eA_\mu \ol{\Psi} \gamma^\mu \Psi$, whose contributions can be tracked by evaluating the appropriate Feynman graphs for QED. In principle we would like also {\it not} to perturb in the nucleus-generated electromagnetic fields, and so include these in the evolution of interaction-picture fields. The boundary conditions for the interaction-picture fields therefore are derived by following the arguments leading to \pref{A0BCeq}, but using the contributions of both the currents $j^0$ and $j^i$ of \pref{nuclearA0vecA} on the right-hand side of eqs.~\pref{EMptEqs}.

In interaction picture (and in Coulomb gauge) the solution to the Maxwell equations that satisfy the nucleus-dependent boundary conditions generated by the right-hand side of \pref{EMptEqs} gives \pref{A0BCeq} as before (for the electrostatic potential), while use of the $j^i$ boundary condition\footnote{In detail, the $A_0$ boundary condition is obtained as before by straight-up integration of the field equation over a small spherical Gaussian pillbox of radius $\epsilon$, while the boundary condition for $\bfA$ comes from a similar integration, but weighted by an $l = 1$ spherical harmonic \cite{EFTBook}.} generates the standard magnetic dipole field \cite{jackson, griffithsem}, 
\be \label{NucEMFs}
   A^0 = A^0_{\rm nuc} = \frac{Ze}{4\pi r} , \hspace{18pt} \bfA = \bfA_{\rm nuc} + \bfA_{\rm rad} =  \frac{\boldsymbol{\mu} \times \bfr}{4\pi r^3} + \bfA_{\rm rad}\,,
\ee
where\footnote{Notice $\mu_\ssN$ here denotes the nuclear magnetic moment ({\it not} the nuclear magneton) including the nuclear $g$-factor.} $\boldsymbol{\mu} :=  \mu_\ssN \bfI$ is the nuclear magnetic moment\footnote{That is, \pref{NucEMFs} is the classical solution obtained using the boundary condition that formally follows from \pref{EMptEqs} with $\bfI$ regarded as a specified function. For first-quantized nuclei we may treat $\bfI$ as an operator in the knowledge that $\bfA_{\rm nuc}$ ultimately appears within an expectation value between two nuclear spins (as we see explicitly below).} and $\bfA_{\rm rad}(\bfr,t)$ denotes the operator-valued radiation component of the interaction-picture electromagnetic field (whose boundary conditions are the standard, nucleus-independent, ones).

\subsection{Lepton mode functions}
\label{section:lepModes}

The previous sections show that the $A_\mu$-dependent terms in the action \pref{SpForm} alter the interaction-picture electromagnetic field only by capturing the nuclear magnetic moment (and by doing so give a physical interpretation for the effective coupling $\mu_\ssN$). Repeating the above exercise for the electron field reveals more information, however, leading to the field equation: 
\begin{eqnarray}
0 &=& \left( \slashed{D} + m \right) \Psi + \int \exd s \, \delta^4[x-y(s)]  \left\{ \sqrt{-\dot{y}^2} \left(  c_s + i c_2 \epsilon_{\mu\nu\rho\sigma}\hat\xi^\mu \hat\xi^\nu \hat\xi^\rho \hat\xi^\sigma \gamma_5 + i c_\ssF\, \hat\xi^\mu \hat\xi^\nu \gamma_{\mu\nu} \right) \right.  \notag \\
&&\qquad\qquad\qquad\qquad\qquad\qquad + \Big. i \dot{y}^\mu \left( c_v \gamma_\mu + c_3 \epsilon_{\mu\nu\rho\sigma}\hat\xi^\mu \hat\xi^\nu \hat\xi^\rho \hat\xi^\sigma \gamma_5\gamma_\mu \right) \Big\} \Psi,
\end{eqnarray}
where, as before, the lepton covariant derivative is $D_\mu\Psi =(\partial_\mu +ie A_\mu)\Psi$. Using again the representation \pref{xirep} and specializing to the nuclear rest frame (and parameterizing using proper time) then gives
\begin{eqnarray} \label{Psieq}
0 &=& \left[ \gamma^0 \left( \partial_0 + i e A_0 \right) + \gamma^i \left( \partial_i + ie A_i \right) + m \right] \Psi + \delta^3(\bfx) \left[ c_s - i c_v \gamma^0 + \frac{c_\ssF}{2} \epsilon^{ijk} I_k \gamma_{ij}   \right] \Psi \,,\nn\\
\end{eqnarray}
where the terms involving $c_2$ and $c_3$ are proportional to $\Gamma_5 := -i \Gamma^0 \Gamma^1 \Gamma^2 \Gamma^3$ and so vanish in the nuclear rest frame -- {\it c.f.}~eqs.~\pref{Gammavis0} -- and (as above) $\bfI := \frac12 \, \boldsymbol{\tau}$ is the nuclear spin (acting in nuclear-spin space).

As above, our perturbative treatment of quantum-field interactions allows the term $\bfA_{\rm rad} \cdot \ol{\Psi} \boldsymbol{\gamma} \Psi$ to be dropped in the interaction-picture evolution of the fields, though the nucleus-generated Coulomb and magnetic-dipole fields do appear in the interaction-picture evolution of the field operator $\Psi$. Away from the nuclear position eq.~\pref{Psieq} then boils down to the Dirac equation in the presence of a Coulomb potential and a dipole magnetic field:
\begin{equation} \label{modeEQ}
0 = \left[ -i\gamma^0\left( \omega - eA_0^{\rm nuc} \right) + \boldsymbol{\gamma} \cdot \nabla  + m \right] \psi + ie \boldsymbol{\gamma} \cdot \bfA_{\rm nuc} \psi \,,
\end{equation}
where $A_0^{\rm nuc}$ and $\bfA_{\rm nuc}$ are the Coulomb and magnetic-dipole contributions given in \pref{NucEMFs}. 

The delta-function terms in \pref{Psieq} contribute once the equation is integrated over a small sphere of radius $\epsilon$ that includes the nucleus (possibly weighted by spherical harmonics\footnote{For the $S$-wave modes of later interest no weighting by spherical harmonics is necessary.}). For example (as described in more detail in Appendix \ref{AppendixFBC}), for $S$-wave modes integrating over a small sphere (of radius $\epsilon$) about the position of the delta function implies a boundary condition
\be 
\label{fermionbc1}
  0 = \int \exd^2 \Omega_2 \, \epsilon^2 \left[ \gamma^r + \hat{c}_s - i \gamma^0 \hat{c}_v + \hat{c}_\ssF \,\bfI \cdot  \boldsymbol{\Sigma} \right] \psi(\epsilon) \,,
\ee
for fields at $r = \epsilon$, where the couplings $\hat c_i$ are related to the couplings appearing in $S_p$ by  $\hat{c}_i = {c_i}/({4\pi \epsilon^2})$ and $\Sigma_k$ is defined by $\gamma^{ij} = \epsilon^{ijk} \Sigma_k$, where $\gamma^{\mu\nu} := -\frac{i}4 [\gamma^\mu, \gamma^\nu]$. This near-nucleus boundary condition fixes some of the integration constants that arise when integrating \pref{modeEQ}, and thereby allows them to depend on $c_s$, $c_v$ and $c_\ssF$. It is through this dependence that nuclear properties alter atomic energy levels. 

Two new steps are required in order to compute the effects of the nucleus on electronic levels. First the Dirac equation \pref{modeEQ} must be solved away from the nucleus in the presence of the dipole magnetic field, which is done in this section by perturbing in the magnetic moment $\mu_\ssN$. The second step (performed in \S\ref{running} below) takes these solutions and imposes the near-nucleus boundary conditions implied by \pref{fermionbc1} to determine some of the integration constants found when solving \pref{modeEQ}. In particular these solutions are {\it not} assumed to be bounded at the origin, and indeed the boundary conditions following from \pref{fermionbc1} are only consistent with boundedness when the effective couplings $c_s$, $c_v$, {\it etc.}~all vanish.  The full effect of spin-dependent finite-size nuclei on atomic energy levels receives contributions from both of these two steps, as we now see.

\subsubsection{Dirac-Coulomb modes}

We start by reviewing the properties of modes, $u(\bfx,t)$, of the Dirac equation in the presence of a Coulomb potential but without a dipole magnetic field. Denoting mode energy by $\omega$ one seeks solutions of the form $u(\bfx, t) = e^{-i\omega t} \psi(\bfx)$. Away from the origin the function $\psi$ satisfies:
\begin{equation} \label{modeEQDC}
0 = \Bigl[ -i\gamma^0\left( \omega - eA_0^{\rm nuc} \right) + \boldsymbol{\gamma} \cdot \nabla  + m \Bigr] \psi  \,.
\end{equation}
where (as above) $A_0^{\rm nuc}$ is the Coulomb potential of eq.~\pref{NucEMFs}.  

The standard Dirac-Coulomb Hamiltonian mode functions separate in polar coordinates, and are labelled by the quantum numbers $|njj_z\varpi\rangle$ where $n$ is the principal quantum number, $j = \frac12, \frac32, \cdots$ and $j_z = -j, -j+1, \cdots, j-1, j$, stand for total electronic angular momentum, and $\varpi = \pm$ is the parity quantum number.\footnote{Strictly speaking the parity of a state is $(-)^l $ where $l := j - \frac12\,\varpi$, so $\varpi$ {\it determines} the parity, but need not be equal to it for all $j$. This distinction does not matter in practice for the states of most interest, for which $j = \frac12$.} Working in a basis for which $\gamma^0$ is diagonal, the corresponding mode functions are Dirac spinors \cite{ppeft3, ll}:
\begin{equation} \label{OmegaDef1}
\psi = \left( \begin{array}{c} \Omega_{jlj_z\varpi}(\theta,\phi) \; \mff_{nj}(r) \\ i\Omega_{jl'j_z\varpi}(\theta,\phi) \; \mfg_{nj}(r) \end{array} \right),
\end{equation}
where $\Omega_{jlj_z\varpi}$ denotes a 2-component spinor spherical harmonic,
\begin{equation} \label{OmegaDef2}
\Omega_{jlj_z\varpi} := \left( \begin{array}{c} \varpi \sqrt{\frac{l +\varpi\, j_z + \frac{1}{2}}{2l + 1}} \;Y_{l, j_z-\frac{1}{2}} (\theta, \phi) \\ \\
\sqrt{\frac{l -\varpi\, j_z + \frac{1}{2}}{2l+1}} \;Y_{l, j_z + \frac{1}{2}}(\theta, \phi) \end{array} \right) \,,
\end{equation}
with total, orbital and projected total angular momentum quantum numbers $(j, l, j_z)$. Here the orbital quantum numbers $l$ and $l'$ are related to $j$ and parity by $l =  j - \frac12 \varpi $ and $l' = j + \frac12 \varpi$. $Y_{l l_z}(\theta, \phi)$ are the usual scalar spherical harmonics. 

The functions $\mff_{nj}(r)$ and $\mfg_{nj}(r)$ are the solutions to the radial part of the Dirac equation (more about which below). Skipping details -- see \cite{ppeftA} for an enumeration of more steps using much the same formalism, but for spinless nuclei -- the radial functions are given by  
\begin{eqnarray}
\label{dc}
\mff_{nj}(r)   &=& \sqrt{m+\omega} \; e^{-{\rho}/{2}}  \left\{ \mathscr{C}  \rho^{\zeta-1}\left[ \mathcal{M}_1 - \left( \frac{a}{c} \right) \mathcal{M}_2 \right] + \mathscr{D}  \rho^{-\zeta-1} \left[ \mathcal{M}_3 -  \left( \frac{a'}{c} \right) \mathcal{M}_4 \right] \right\}, \notag \\
\mfg_{nj}(r)   &=&     - \sqrt{m-\omega} \; e^{-{\rho}/{2}} \left\{\msC \rho^{\zeta-1}  \left[ \mathcal{M}_1 + \left( \frac{a}{c} \right) \mathcal{M}_2 \right] + \msD \rho^{-\zeta-1}  \left[ \mathcal{M}_3 +  \left( \frac{a'}{c} \right) \mathcal{M}_4 \right] \right\}, \nn\\
\end{eqnarray}
where $\msC$ and $\msD$ are integration constants  and the functions $\mathcal{M}_\mathfrak{i}$ are given in terms of confluent hypergeometric functions -- defined in \eqref{chfseries} -- $\mathcal{M} (\beta, \gamma; z) := {}_1\mathcal{F}_1[\beta; \gamma; z]$ with different arguments:
\begin{eqnarray}\label{hypergeos}
&&\mathcal{M}_1 :=  \mathcal{M}\left(a, b; \rho \right)\,, \;\; \mathcal{M}_2 :=  \mathcal{M} \left( a+1, b;\rho \right)\,,\nn\\ \;\;
&&\mathcal{M}_3 :=  \mathcal{M}\left( a', b'; \rho\right)\,, \;\;  \mathcal{M}_4 := \mathcal{M}\left( a'+1, b'; \rho \right) \,.
\end{eqnarray}
The various parameters appearing in \pref{dc} and \pref{hypergeos} are defined by
\begin{eqnarray}
\label{hyperparam}
a &:=& \zeta-\frac{Z\alpha \omega}{\kappa}, \hspace{12pt} a' :=  -\left( \zeta + \frac{Z\alpha\omega}{\kappa} \right), \hspace{12pt} b := 1+ 2\zeta, \hspace{12pt} b' := 1-2\zeta, \hspace{12pt}  \notag \\
c &:=& \mfK  - \frac{Z\alpha m}{\kappa}, \quad \rho := 2\kappa r, \hspace{24pt} \kappa := \sqrt{m^2 - \omega^2}, \hspace{24pt} \zeta := \sqrt{\mfK ^2 - (Z\alpha)^2} \,,
\end{eqnarray}
where $\mfK$ is the Dirac quantum number, defined by
\be \label{mfKdef}
   \mfK   := -\varpi \left( j + \frac{1}{2} \right) = \mp \left( j + \frac{1}{2} \right) 
   \quad \hbox{for parity $\pm$ states}\,.
\ee
For later purposes we note that only $\mathcal{M}_1$ and $\mathcal{M}_2$ are bounded at the origin, and so the the radial functions are bounded at the origin only when $\msD = 0$, as is usually chosen when working with a point-like spinless nucleus. 

Bound states have $\omega < m$ and for these normalizability at large $r$ requires the integration constants to be related by 
\begin{equation}
\label{normal}
-\left( \frac{\msD}{\msC} \right) = \frac{\Gamma[1+2\zeta]\; \Gamma\left[ -\zeta - ({Z\alpha \omega}/{\kappa}) \right]}{\Gamma[1-2\zeta]\; \Gamma\left[\zeta - ({Z\alpha \omega}/{\kappa}) \right] } \,.
\end{equation}
The Dirac-Coulomb bound-state energies are then determined by choosing $\omega$ to ensure that \pref{normal} is consistent with the condition on $\msD/\msC$ that comes from the near-nucleus boundary condition (described in more detail in \S\ref{running}). 

As mentioned earlier, for a point-like spinless nucleus -- {\it i.e.}~in the absence of the nucleus-dependent $\delta^3(\bfx)$ terms in \pref{Psieq} -- these boundary conditions simply state that the solution is bounded at the origin, which implies $\msD = 0$. Using $\msD = 0$ in \pref{normal} then implies the standard point-nucleus Dirac-Coulomb energy spectrum $\omega = \omega^\ssD_{nj}$, where
\be  \label{Diracomega}
\omega^{\ssD}_{nj} = m \sqrt{1 -  \left( \frac{\kappa^\ssD_{nj}}{m}\right)^2}
\ee
with
\be  \label{Diracomega2}
 \kappa^{\ssD}_{nj} = \frac{m Z\alpha}{\cN} \quad \hbox{and} \quad
  \cN = n \sqrt{1 - \frac{2(n-|\mfK  |) (Z\alpha)^2}{n^2(\zeta + |\mfK  |)}} \,.
\ee
Here $n = 1,2,\cdots$ is the usual principal quantum number. 

More generally, when $\msD/\msC$ is nonzero but small the solution obtained by solving \pref{normal} for $\omega$ becomes\footnote{The subscript $F$ anticipates that $\delta \omega$ depends on the total atomic angular momentum quantum number, $F$, through its dependence on the quantity $\msD/\msC$.} $\omega_{n\ssF j\varpi} = \omega^\ssD_{nj} + \delta \omega_{n\ssF j \varpi}$ with $\delta \omega_{n\ssF j\varpi}$ given \cite{ppeftA} by eq.~\pref{normshift} of Appendix \ref{AppendixC}. 

In what follows it is important to keep in mind that the mode energy, $\omega$, is {\it not} the same as the physical single-particle energy measured in atomic systems. The entire energy relevant to experiments includes many corrections, and (to the order required here) takes the form
\be \label{energylevelanswer}
   \omega_{n\ssF j \varpi} = \omega^{\ssD}_{nj} + \delta \omega_{n\ssF j \varpi}  + \varepsilon_{n\ssF j \varpi}^{{\rm mag}}  + \varepsilon^{\QED}_{n\ssF j\varpi}  + \varepsilon_{n\ssF j \varpi}^{{\rm rec}} \,.
\ee
The first of these is the single-particle Dirac-Coulomb spectrum of \pref{Diracomega}, while $\delta \omega_{n\ssF j \varpi}$ denotes the shift in the mode spectrum coming from having $\msD/\msC \neq 0$ when solving eq.~\pref{normal}. As we see explicitly below, nonzero values for $\msD/\msC$ arise when nuclei are not point-like, and so provide part of the influence of nuclear structure on atomic spectra. 

The contribution $\varepsilon_{n\ssF j \varpi}^{{\rm mag}} =  \varepsilon_{n\ssF j \varpi}^{(1)} + \varepsilon_{n\ssF j \varpi}^{(ho)}$ contains the influence of the nuclear magnetic field, $\bfA_{\rm nuc}$, which to the accuracy desired here can be computed perturbatively. The first-order effects we denote by $\varepsilon_{n\ssF j \varpi}^{(1)}$, whose calculation is described at length below. This term contains both spin-dependent point-nucleus contributions (such as the hyperfine splitting) and spin-dependent finite-size nuclear effects. Higher-order contributions, denoted $\varepsilon_{n\ssF j \varpi}^{(ho)}$, are also relevant \cite{aldo2013th}, though for current precision their form for point nuclei is sufficient. 

The next contribution arises when perturbing in the radiation component of the electromagnetic field, $\varepsilon^{\QED} = \varepsilon^{{\rm pt}-\QED}_{n\ssF j\varpi} + \varepsilon^{\ssN-\QED}_{n\ssF j\varpi}$, with $\varepsilon^{{\rm pt}-\QED}_{n\ssF j\varpi}$ describing standard QED corrections (such as the Lamb shift) as computed for point nuclei, and $\varepsilon^{\ssN-\QED}$ describing nuclear-size effects in these QED corrections. 

The final contribution in \pref{energylevelanswer} contains recoil corrections (those terms suppressed by powers of $m/M$ that are not simply the result of using the reduced mass in the non-relativistic problem). As described above -- {\it c.f.}~the discussion surrounding eq.~\pref{Spdefnospin2} -- this also divides into point-nucleus and a nuclear-structure piece, $ \varepsilon_{n\ssF j \varpi}^{{\rm rec}}  = \varepsilon_{n\ssF j \varpi}^{{\rm pt-rec}} + \varepsilon_{n\ssF j \varpi}^{{\rm N-rec}}$, both of which contribute at the order we work.

The next two sections compute the energy shifts $\delta \omega$ and $\varepsilon^{(1)}$ in some detail, with a view to counting systematically the number of relevant nuclear parameters. While both types of QED corrections are relevant to modern experiments, as are recoil corrections, we argue in \S\ref{energies} why these contribute only to predictions for the {\it value} of the two nuclear parameters, rather than introducing new independent parameters themselves. As such they are not conceptual obstacles to identifying nucleus-free observables.

\subsubsection{Effects of the nuclear magnetic dipole}

This section reviews the form of Dirac mode functions in the presence of the nuclear dipole magnetic field, with the magnetic field treated perturbatively. Discussions of this perturbation expansion are given in the literature \cite{schwartz, borie1982, schwinger}, though  for point-like nuclei ({\it i.e.}~where the unperturbed radial-mode integration constants satisfy $\msD/\msC=0$). We redo these calculations here explicitly however because nonzero $\msD/\msC$ is required by finite-size nuclear effects. Because this is conceptually straightforward (though tedious) only the main features of the calculations are described here, with more details given in Appendix \ref{AppendixB}. 

As described below, for the present purposes we need work only to linear order in the nuclear spin-dependent effects. For the point nucleus it would therefore suffice to compute the linear-order energy shift without also needing the first-order change to the Dirac mode functions. An important change relative to the point-nucleus problem is that the determination of $\msD/\msC$ to first-order also requires knowing the leading perturbative corrections to the mode functions as well.

We choose a basis of zeroth-order energy eigenstates, $|nFF_z j \varpi \rangle$, that also diagonalize total atomic angular momentum, $\bfF = \bfJ + \bfI$, that sums nuclear spin $\bfI$ with the total leptonic angular momentum $\bfJ$. We do so by combining the Dirac-Coulomb states described above with nuclear spin states to make states that take definite values for $\bfF^2$ and $F_z$,
\begin{equation}\label{FModeFuncs}
\psi_{n\ssF j \varpi}(r,\theta,\phi) := \langle r, \theta, \phi | n F \, f_z; I, j ; \varpi \rangle_{0} = \left( \begin{array}{r} \mathcal{Y}_{F f_z}^{j, \varpi} \, \mff_{nj \varpi}(r) \\ i \mathcal{Y}_{F f_z}^{j, -\varpi} \, \mfg_{nj \varpi}(r) \end{array} \right) \,,
\end{equation}
where the functions $\cY_{Ff_z}^{j, \pm \varpi}$ are defined in eq.\pref{cYdef2} to be 
\begin{equation}\label{cYdef2x}
\mathcal{Y}_{F = j +  \frac{\alpha}{2}, f_z}^{j, \varpi} = \left[ \begin{array}{c}
\alpha \sqrt{\frac{j + \frac{1}{2} + \alpha f_z}{2j + 1}} \; \Omega_{j, l, f_z - \frac{1}{2},\varpi} \\
\sqrt{\frac{j + \frac{1}{2} -\alpha f_z}{2j + 1}}\;  \Omega_{j, l, f_z + \frac{1}{2},\varpi} \end{array} \right] \,.
\end{equation}
with $\Omega_{jlj_z\varpi}$ defined in \pref{OmegaDef2} and the nuclear 2-component spinors defined by
\be \label{app:etadefsx}
  \eta_{\frac{1}{2}, +\frac{1}{2}} = \left[ \begin{array}{c} 1 \\ 0 \end{array} \right] \quad \hbox{and} \quad
  \eta_{\frac{1}{2}, -\frac{1}{2}} = \left[ \begin{array}{c} 0 \\ 1 \end{array} \right] \,.
\ee
To avoid confusion we use square brackets to denote spinors in nuclear-spin space and round brackets to denote the same in electron-spin space.

\subsubsection*{First-order energy shift}

States with different $F_z$ are degenerate at zeroth-order in the magnetic-moment field, necessitating the use of degenerate perturbation theory. Consequently one seeks a basis that diagonalizes the perturbing interaction $\cL_{\rm int} = -e\gamma^0 \boldsymbol{\gamma} \cdot \bfA_{\rm nuc}$ within the degenerate subspace of interest. For a degenerate eigenspace with fixed $j$ the states with definite values of $F$ and $F_z$ provide precisely the required basis.\footnote{This is not to say that eigenstates with fixed $F, F_z, j, j_z$ diagonalize the entire perturbing Hamiltonian, since mixing between opposite parity states that share the same values of $F, F_z$ but with different $j$ quantum numbers, can still occur, as has been known for some time \cite{brodsky, hh}. This mixing first appears in the energy at second order in the magnetic moment, and at first-order in the corrections to the wave-functions (as we describe in more detail later). Because its contributions to nuclear finite-size energy shifts are smaller than the precision to which we work in this paper, we do not calculate them in detail.}

The first-order energy shift for these states becomes
\bea \label{E1explexpr}
\varepsilon^{(1)}_{n \ssF j \varpi} &=& -\left(\frac{e}{4\pi}\right) \frac{\int \exd^3x \; r^{-2} \, \psi^\dagger \gamma^0 \boldsymbol{\gamma} \cdot \left( \boldsymbol{\mu} \times \hat{\bfr} \right) \psi }{\int \exd^3x \, \psi^\dagger \psi},  \\
&=& \left(\frac{e\mu_\ssN}{4\pi}\right) \frac{1}{\cD} \int  \exd^2 \Omega_2 \,  \left[ \(\mathcal{Y}_{\ssF, f_z}^{j, \varpi}\)^\dagger \Sigma \,\mathcal{Y}_{\ssF, f_z}^{j, -\varpi}  - \left( \mathcal{Y}_{\ssF, f_z}^{j, -\varpi}\right)^\dagger \Sigma\, \mathcal{Y}_{\ssF, f_z}^{j, \varpi} \right] \int \exd r\; \mff\cdot \mfg  \,,\nn
\eea
where
\be\label{SigmamfDdefs}
 \Sigma :=   i \left(\bfI \times \hat{\bfr} \right) \cdot  \boldsymbol{\sigma} 
 \quad \hbox{and} \quad
  \cD :=  \int \exd r \, r^2 \left( \mff^2 + \mfg^2 \right) \,,
\ee
and we suppress the quantum numbers $\{n,F,j,\varpi\}$ on $\psi$ and the radial functions to avoid notational clutter.

This can be further simplified using the property $\mathcal{Y}_{\ssF, f_z}^{j, -\varpi} = - \sigma^r \mathcal{Y}_{\ssF, f_z}^{j, \varpi}$, where $\sigma^r := \hat \bfr \cdot \boldsymbol{\sigma}$ is the radially-pointing Pauli matrix acting on the leptonic spin space, and a $(2I +1) \times (2I+1) = 2 \times 2$ unit matrix acting in nuclear-spin space is not written explicitly. Additionally, using
\begin{align} 
&\boldsymbol{\sigma} \sigma^r = \left\{  \mathds{1} \,,  - i \sigma^\phi \,,  i\sigma^\theta \right\} \,, \quad 
\sigma^r \boldsymbol{\sigma} = \left\{  \mathds{1} \,,  i \sigma^\phi \,, -i\sigma^\theta \right\} \,, \notag \\
& \quad \left(\bfI \times \hat{\bfr} \right) \cdot \left(  \boldsymbol{\sigma} \sigma^r - \sigma^r  \boldsymbol{\sigma} \right) = -2i \left( I^\theta \sigma^\theta + I^\phi \sigma^\phi \right),
\end{align}
the first-order mode-energy shift simplifies to  
\be
\label{E1partial}
 \varepsilon^{(1)}_{n \ssF j \varpi} =  \left(\frac{e \mu_\ssN}{4\pi}\right) \mfK   X_\ssF\left\{ \frac{ \int \exd r \, \mff\cdot \mfg}{\int \exd r \, r^2 \left( \mff^2 + \mfg^2 \right)} \right\} .
\ee
This expression evaluates the angular integration as in the literature \cite{borie1982} 
\be \label{I.Sintegral}
   2 \int \exd^2 \Omega_2 \( \mathcal{Y}_{\ssF, f_z}^{j, \varpi}\)^\dagger \left( I^\theta \sigma^\theta + I^\phi \sigma^\phi \right) \mathcal{Y}_{\ssF, f_z}^{j, \varpi}  = - \mfK   X_\ssF \,,
\ee
where the variable $X_\ssF$ is defined by
\be
\label{amatrix}
   X_\ssF :=     \frac{F(F+1) - j(j+1) - I(I+1)}{j(j+1)} =\left\{ {(j+1)^{-1}\quad \hbox{if $F = j+\frac12$}\atop  - j^{-1} \quad \hbox{if $F = j-\frac12$}} \right.\,,
\ee
and the final equality specializes to $I = \frac12$. 

The numerator of \pref{amatrix} arises ubiquitously in what follows because it is the eigenvalue of $2\,  \bfI \cdot \bfJ = (\bfI + \bfJ)^2 - \bfJ^2 - \bfI^2$ evaluated in a state with definite nuclear, electronic and total atomic angular momentum quantum numbers $I$, $j$ and $F$. In all of the energy shifts discussed below the dependence on $F$ appears through this combination, as is ultimately required by rotational invariance.

It is convenient to extract the dimensionless combination 
\be \label{exxdef}
  \exx := \frac{m e \mu_\ssN}{4\pi} \ll 1 \,,
\ee
where $m$ is (as usual) the lepton mass, because this is the small quantity that controls the size of nuclear-spin effects. Because our focus is on nuclear finite-size effects, and because current experimental precision for both atomic and muonic Hydrogen is insensitive to finite-size effects at order $\exx^2$, for our purposes it suffices in what follows to work to linear order in $\exx$. At this order \pref{E1partial} implies the energy shift is
\be
\label{E1partial2}
 \varepsilon^{(1)}_{n \ssF j \varpi} =  \frac{\exx \mfK X_\ssF}{m}   \left\{ \frac{ \int \exd r \, \mff\cdot \mfg}{\int \exd r \, r^2 \left( \mff^2 + \mfg^2 \right)} \right\}   =:  -4\exx \mfK X_\ssF \left( \frac{\kappa^3}{m^2}\right)  \left( \frac{\mfN}{\mfD} \right).
\ee

The last equality of \pref{E1partial2} evaluates the radial matrix elements inside the braces, for which both numerator and denominator naturally divide up into three parts. That is, defining
\be
  \int_0^\infty  \exd r \, \mff\cdot \mfg =: -\frac{\msC^2}{2}  \, \mfN 
\ee
one finds
\be \label{rmatrix1}
  \mfN = \mfN_{\rm pt} + \left( \frac{\msD}{\msC} \right) \, \mfN_1 + \left( \frac{\msD}{\msC} \right)^2 \mfN_2  \,,
\ee
and defining $\mfD$ as in \pref{SigmamfDdefs},
\be  \label{cDvsmfD}
   \cD  :=   \int \exd r \, r^2 \left( \mff^2 + \mfg^2 \right) =  \frac{\msC^2 m}{(2\kappa)^3} \; \mfD  \,,
\ee
implies it can be written
\be \label{rmatrix2}  
  \mfD = \mfD_{\rm pt} + \left( \frac{\msD}{\msC} \right) \, \mfD_1 + \left( \frac{\msD}{\msC} \right)^2 \mfD_2  \,.
\ee
In these expressions the subscript `pt' labels the contribution of a point-like nucleus ({\it i.e.}~one for which the radial mode functions have $\msD = 0$) and the remaining $\msD$-dependent terms represent the nuclear-size dependent contributions (described in more detail below). An explicit factor of the integration constant $\msC^2$ is factorized out of these definitions to emphasize how the energy shift depends only on the ratio $\msD/\msC$, and not on each of these constants separately.  

The contributions to \pref{rmatrix1} and \pref{rmatrix2} are given in terms of a basic class of integrals of the form,
\begin{equation}\label{IpijDef}
\cI_{\mathfrak{i} \mathfrak{j}}^{(p)} := \int_0^\infty \exd \rho \, e^{-\rho} \rho^{p} \cM_\mathfrak{i} \cM_\mathfrak{j},
\end{equation}
where $\mathfrak{i}, \mathfrak{j} = 1,2,3,4$, corresponding to the functions $\cM_\mfi$ defined in \pref{hypergeos}, and $p$ is a real number that depends on which of the $\cM_\mfi$ appearing in \pref{dc} are relevant.  For some of the choices of $p$ encountered below the integrals $\cI_{\mfi\mfj}^{(p)}$ diverges at the $\rho \to 0$ limit, a divergence that below gets renormalized into the effective coupling $c_\ssF$. 

Explicit formulae for $\mfN$ and $\mfD$ obtained by performing these integrals are given in eqs.~\pref{Nforms} and \pref{Dforms} of Appendix \S\ref{Appssec:EShift}, which state
\be
 \mfN_{\rm pt}  = \left[\cI_{11}^{(2\zeta-2)} - \left( \frac{a}{c}\right)^2 \cI_{22}^{(2\zeta-2)} \right] \,,
\ee
and
\bea \label{NformsTxT}
  \mfN_1 &=& 2  \left[ \cI_{13}^{(-2)} - \left(\frac{aa'}{c^2}\right) \cI_{24}^{(-2)} \right] \nn\\
 \mfN_2 &=& \cI_{33}^{(-2\zeta-2)} - \left( \frac{a'}{c} \right)^2  \cI_{44}^{(-2\zeta-2)}  \,,
\eea
while
\be 
 \mfD_{\rm pt} = \left[2\cI_{11}^{(2\zeta)} - \frac{4\omega}{m}  \left( \frac{a}{c} \right) \cI_{12}^{(2\zeta)} + 2\left( \frac{a}{c} \right)^2 \cI_{22}^{(2\zeta)} \right]\,,
\ee
and
\bea \label{DformsTxT}
 \mfD_1 &=& 2  \left[ 2\cI_{13}^{(0)} -  \frac{2\omega}{m}  \left( \frac{a'}{c} \right) \cI_{14}^{(0)} -  \frac{2\omega}{m}  \left( \frac{a}{c} \right) \cI_{23}^{(0)} +  2\left(\frac{aa'}{c^2} \right) \cI_{24}^{(0)} \right]   \\
  \mfD_2 &=&  2  \left[ \cI_{33}^{(-2\zeta)}  - \frac{2\omega}{m}  \left( \frac{a'}{c} \right)  \cI_{34}^{(-2\zeta)} + \left( \frac{a'}{c}\right)^2 \cI_{44}^{(-2\zeta)} \right]  \,.\nn
\eea
The integrals appearing in $\mfN_1$, $\mfN_2$, $\mfD_1$ and $\mfD_2$ are the ones that can diverge as $\rho \to 0$, and when present this divergence is regularized by restricting the integration to $\rho > \eta$ (or, more simply, using dimensional regularization) as described in Appendix \ref{ssec:MatrixElementApp}.

As a check, consider first the point-nucleus contributions, $\mfN_{\rm pt}$ and $\mfD_{\rm pt}$. These involve only the confluent hypergeometric profiles, $\mathcal{M}_1$ and $\mathcal{M}_2$ and converge in the limit $\rho \to 0$, making them easy to evaluate (for details see Appendix \ref{ssec:MatrixElementApp}), leading to
\bea
\label{rmatrixpoint}
\mfN_{\rm pt} &=&  \frac{(-2)\left[ \Gamma[1 + 2\zeta] \right]^2 \Gamma \left[ 1 - \zeta + \frac{Z\alpha\omega}{\kappa} \right] \left( \frac{Z\alpha m}{\kappa}\right)}{(4\zeta^2 -1) (2\zeta) \Gamma\left[ 1 +\zeta + \frac{Z\alpha\omega}{\kappa}\right]\left( \mfK   - \frac{Z\alpha m}{\kappa} \right)} \left( 1 -    \frac{2\mfK \omega}{m}\right) \,, \nn \\
 \mfD_{\rm pt} &=& -\frac{4\left[ \Gamma(1+2\zeta) \right]^2 \Gamma\left( 1 - \zeta + \frac{Z\alpha \omega}{\kappa} \right) \left( \frac{Z\alpha m}{\kappa} \right)}{\Gamma\left(1 + \zeta + \frac{Z\alpha \omega}{\kappa} \right) \left( \mfK    - \frac{Z\alpha m}{\kappa} \right)} \,.
\eea
Using these, the energy shift for a point-like spin-half nucleus obtained from \pref{E1partial2} by using \pref{rmatrixpoint} and $\msD = 0$ in \eqref{rmatrix1} and \eqref{rmatrix2} is
\bea\label{Efermi}
\varepsilon^{\rm{hfs}}_{n\ssF j \varpi} &=&  -\exx  \mfK X_\ssF \left( \frac{ \kappa^3}{m^2} \right)  \frac{\left( 1 - 2\mfK \omega/m\right)}{\zeta\left( 4\zeta^2 -1 \right)} \\
&=&  -\mfK    X_\ssF\; \frac{  g_p m^2(Z\alpha)^4}{2M}   \left( \frac{m_r}{m} \right)^3   \left[ \frac{ 1- 2\mfK    \sqrt{1- {(Z\alpha)^2}/{\mathcal{N}^2}}}{\mathcal{N}^3 \zeta\left( 4\zeta^2 -1 \right)} \right]  \qquad \mathrm{(Hydrogen)},\nn
\eea
where the second line specializes to Hydrogen and evaluates $\kappa$ using $\kappa = m_r Z\alpha/\cN$ with reduced mass $m_r = mM/(m+M)$ (where $M$ is the nuclear mass) and $\cN$ as defined in \pref{Diracomega2} with $\mfK$ given in \pref{mfKdef}. Also used are the definition \pref{exxdef} of $\exx$ and $\mu_p =  g_p(Ze/2M)$ with $g_p$ the proton's $g$-factor. When evaluated to compute the energy difference between the $nS_{j=1/2}^{\ssF = 1} $ and $nS_{j=1/2}^{\ssF = 0}$ states in Hydrogen, this expression agrees with standard results for relativistic hyperfine splitting \cite{borie1982, schwinger, eides}.

So far so good, but what about non-point-like nuclei? To capture the finite-size effects we must use the modified near-nucleus boundary condition for $\Psi$ implied by the nuclear effective interactions in $S_p$ (which imply $\msD/\msC \neq 0$). This also requires dealing with the divergent integrals that appear in expressions \pref{NformsTxT} and \pref{DformsTxT} for the magnetic-moment dependent energy shift, whose explicit form becomes
\bea
\label{E1}
\varepsilon^{(1)}_{n \ssF j \varpi} &=& \varepsilon^{\rm{hfs}}_{n\ssF j \varpi} \left[ \frac{1 + (\msD/\msC)(\mfN_1/\mfN_{\rm pt}) + (\msD/\msC)^2 (\mfN_2/\mfN_{\rm pt} )}{1 + (\msD/\msC)(\mfD_1/\mfD_{\rm pt}) + (\msD/\msC)^2 (\mfD_2/\mfD_{\rm pt} )}  \right] \\
&\simeq&- \exx \mfK X_\ssF \left( \frac{ \kappa^3}{m^2} \right) \frac{\left( 1 - 2\mfK {\omega}/{m} \right)}{\zeta\left( 4\zeta^2 -1 \right)}  \left[ 1 + C_\eta - \frac{\mfc}{n} +\cdots \right] \,,\nn
\eea
and in the second line $\mfc$ and $C_\eta$ are $n$-independent constants. Of these $\mfc$ is defined below in \pref{ca+0def} and so contains the various nuclear effective couplings. Unlike $\mfc$, the constant $C_\eta$ depends on the regularization parameter, $\eta$, associated with the near-nucleus divergences described above. In practice the detailed form of $C_\eta$ does not matter in what follows because it gets absorbed into the nuclear effective coupling $c_\ssF$. 

What makes possible the absorption of $C_\eta$ into a counterterm is the fact that neither $\mfc$ nor $C_\eta$ depend on the principal quantum number and so the $n$-dependence in \pref{E1} is either explicit or contained within the standard expressions \pref{Diracomega} and \pref{Diracomega2} for $\kappa$ and $\omega$. As shown in detail in \S\ref{ssec:Renorm}, it is only because $C_\eta$ does not come together with additional $n$ dependence that its contribution to the energy is proportional to $1/n^3$ and so can be absorbed into a counterterm like $c_\ssF$ for an interaction localized at the nucleus' position (whose contribution to the energy is proportional to $|\psi(0)|^2$ and so is also $\propto 1/n^3$). 

The same is not true of the term $\mfc/n$ in \pref{E1}, whose $n$-dependence is a genuine prediction. As argued below (see also Appendix \ref{ssec:MatrixElementApp}) matching to nuclear properties implies $\mfc \sim \cO[(mR\za)^2]$ and so given that $\varepsilon^{\rm{hfs}} \sim \cO[(\za)^4(m^2/M) \sim \cO[(\za)^4m^2R$ -- with $M \sim 1/R$ being the nuclear mass, see {\it e.g.} \pref{E1partial2x}  -- the constant $\mfc$ turns out to contribute to the energy at order $m(\za)^3(mR\za)^3$. For electrons this is smaller than the $\cO[m(\za)^4(mR\za)^2]$ and $\cO[m(\za)^2(mR\za)^3]$ contributions computed here.\footnote{This also makes this contribution competitive with the $\cO[m(\za)^2(mR\za)^4]$ contributions that are also not computed here, but which can be important for muonic Hydrogen.}

\subsubsection*{First-order mode-function correction}
 
As described above, eq.~\pref{E1} is not the whole story. Previously we have mentioned -- {\it c.f.}~\pref{energylevelanswer} -- that at the accuracy of interest here finite nuclear size contributes to electron energies in two different ways: through the contributions of $\msD/\msC$ to\footnote{Because we compute by perturbing using zeroth-order Coulomb bound-state wave-functions, each of which satisfies \pref{normal}, the new eigenstates found by perturbing with $\bfA_{\rm nuc}$ are automatically normalizable assuming only that \eqref{normal} is satisfied.}  $\delta \omega_{n\ssF j \varpi}$ and to $\varepsilon_{n\ssF j \varpi}^{(1)}$. 

In both of these contributions nuclear properties enter through the values implied for $\msD/\msC$ by the near-nucleus boundary conditions -- such as \pref{fermionbc1}. Since it turns out that calculating the implications of \pref{fermionbc1} for $\msD/\msC$ requires knowing the first-order correction to the radial wave-functions, $\mff$ and $\mfg$, due to the magnetic moment interaction, we now pause to compute this.

Using standard first-order Rayleigh-Schr\"odinger perturbation theory, we find the following leading correction to the relativistic Dirac state due to the nuclear magnetic field:
\be \label{1stOrdStShift}
   \ket{nFF_z j\varpi}_1 =  \sum_{\widetilde{n} \not = n}  \frac{\cC_{\widetilde{n} n F F_z j \varpi}}{ E_{n \ssF j \varpi}^{(0)} - E_{\widetilde{n} \ssF j \varpi}^{(0)}}   \ket{\widetilde{n}FF_z j \varpi}_0 + \hbox{($\widetilde j$ terms)}  \,,
\ee
where `($\widetilde j$ terms)' denote contributions coming from summing states with $\widetilde j \neq j$; terms that can be neglected in what follows as explained in Appendix \ref{sec:StateShiftApp}. The displayed sum is only over principal quantum numbers that differ from that of the state being perturbed, and the coefficients are
\bea
\label{state1partial}
  \cC_{\widetilde{n} n F F_z  j  \varpi} &=& -\left( \frac{e}{4\pi \widetilde \cD} \right)   \int \exd^3x \, r^{-2} \,  \widetilde{\psi}_{\widetilde{n}\ssF \widetilde{j} \widetilde{\varpi}}^\dagger \gamma^0 \boldsymbol{\gamma} \cdot \left(\boldsymbol{\mu} \times \hat{\mathbf{r}} \right) \psi_{n\ssF j \varpi},  \\
&=& - \left(\frac{\exx}{m \widetilde \cD} \right) \int \exd \Omega_2 \left( \cY_{\ssF, f_z}^{\widetilde{j} \widetilde{\varpi}} \right)^\dagger \left( I^\theta \sigma^\theta + I^\phi \sigma^\phi \right) \cY_{\ssF, f_z}^{j \varpi}  \int  \exd r\, \left( \widetilde{\mff} \, \mfg + \widetilde{\mfg} \, \mff \right)  \,, \nn
\eea
with
\be 
 \widetilde \cD :=   \int  \exd r\, r^2 \left(\widetilde{\mff}^2 + \widetilde{\mfg}^2\right) = \frac{\widetilde{\msC}^2 m}{(2\widetilde{\kappa})^3} \widetilde{\mfD} \,.
\ee
defined in the same way as is $\cD$ -- in eq.~\pref{SigmamfDdefs} -- but evaluated for the state $\widetilde\psi$ (for more detail see Appendix \ref{AppendixB}).

Notice that the integrals $\widetilde{\mfD}_{\rm pt}$, $\widetilde{\mfD}_1$ and $\widetilde{\mfD}_2$ appearing in $\widetilde{\mfD}$, are defined in terms of $\widetilde\cD$ using \pref{cDvsmfD} and \pref{rmatrix2} -- {\it i.e.}~with $\msC \to \widetilde \msC$, $\kappa \to \widetilde \kappa$, $n \to \widetilde{n}$ and so on -- the only new quantity here is the radial integral in the numerator. Defining 
\bea
 \label{mfNsDef}
 \mfN^s &:=&- \frac{1}{\msC \widetilde{\msC} } \int_0^\infty \exd r \, \left( \widetilde{\mff} \,\mfg + \widetilde{\mfg} \,\mff \right) \nn\\
&=&  m (2\widetilde{\kappa})^{\widetilde{\zeta} -1} (2\kappa)^{\zeta-1} \left( \widetilde{\kappa} + \kappa \right)^{1-\widetilde{\zeta} - \zeta} \\
&& \qquad \qquad \qquad \times \left\{ \mfN^s_{\rm pt}  +\left( \frac{\msD}{\msC} \right) \mfN^s_1 + \left( \frac{\widetilde{\msD}}{\widetilde{\msC}} \right) \widetilde{\mfN}^s_1 + \left( \frac{\widetilde{\msD} \msD}{\widetilde{\msC} \msC} \right) \mfN^s_2 \right\} \,,\nn
\eea
and evaluating the angular integral for $(\widetilde{j}, \widetilde{\varpi}) = (j, \varpi)$ the first-order state correction given in \pref{1stOrdStShift} becomes  
\bea 
\label{rmatrixstate}
  \ket{nFF_z j \varpi}_1 &=& - \sum_{\widetilde{n} \neq n}   \frac{ \exx (2\widetilde{\kappa})^3 \mfK X_\ssF}{2m^2  \left( E_{n \ssF j \varpi}^{(0)} - E_{\widetilde{n} \ssF j \varpi}^{(0)} \right)} \left( \frac{\msC \mfN^s}{\widetilde{\msC} \widetilde{\mfD} } \right)\ket{\widetilde{n}FF_z j \varpi}_0 + \hbox{($\widetilde j$ terms)} \nn\\
  &=:&  \exx X_\ssF \widehat{\sum_{\widetilde{n}}} \ket{\widetilde{n}FF_z j \varpi}_0 + \hbox{($\widetilde j$ terms)} \,,
\eea
where the last equality defines the $\widehat{\sum}$ operator, which is therefore given by  
\bea
\label{SigmaHatDef}
    \widehat{\sum_{\widetilde{n}}} \ket{\widetilde{n}FF_z j \varpi}_0 &:=& -\sum_{\widetilde{n} \neq n}  \frac{4 \widetilde{\kappa}^3 \mfK    (4\widetilde{\kappa} \kappa)^{\zeta-1} \left( \widetilde{\kappa} + \kappa \right)^{1 - 2\zeta} }{m \left( E_{n \ssF j \varpi}^{(0)} - E_{\widetilde{n} \ssF j \varpi}^{(0)} \right) }   \left( \frac{ \msC \mfN_{\rm pt}^{s}}{ \widetilde{\msC} \widetilde{\mfD}_{\rm pt}} \right) \\
  && \times \left\{ \frac{1 + \left(\frac{\msD}{\msC} \right) \mfN^s_1 / \mfN^s_{\rm pt} + \left(\frac{\widetilde{\msD}}{ \widetilde{\msC}} \right) \widetilde{\mfN}^s_1/ \mfN^s_{\rm pt} +  \left( \frac{\widetilde{\msD} \msD}{ \widetilde{\msC} \msC} \right) \mfN^s_2 / \mfN^s_{\rm pt}}{1 + \left(\frac{\widetilde{\msD}}{\widetilde{\msC}} \right) \widetilde{\mfD}_1/\widetilde{\mfD}_{\rm pt} + \left(\frac{\widetilde{\msD}}{\widetilde{\msC}} \right)^2 \widetilde{\mfD}_2/\widetilde{\mfD}_{\rm pt}} \right\} \ket{\widetilde{n}FF_z j \varpi}_0\,.\nn
\eea
The integrals $\mfN^s_{\rm pt}, \mfN^s_1, \widetilde{\mfN}^s_1$ and $\mfN^s_2$ appearing here are given explicitly in terms of integrals similar to $\cI^{(p)}_{\mfi\mfj}$ in Appendix \ref{sec:StateShiftApp}, in eqs.~\pref{type2} and \pref{Nsform} and subsequent paragraphs.

The solution to the leptonic equations of motion correct to first order in $\exx$ then is,
\begin{align}
\label{psi1}
&\psi_{n \ssF j \varpi} = \left[ \begin{array}{c} \mathcal{Y}_{\ssF, f_z}^{j, \varpi} \left( \mff_{n j \varpi}^{(0)}(r) + \exx X_\ssF \, \mff_{n j \varpi}^{(1)}(r) + \cdots \right) \\ i \mathcal{Y}_{\ssF, f_z}^{j, -\varpi} \left( \mfg_{n j \varpi}^{(0)}(r) + \exx X_\ssF\, \mfg_{n j \varpi}^{(1)}(r) + \cdots \right) \end{array} \right]
\end{align}
with the ellipses representing terms of order $\mathcal{O}(\exx ^2)$ or higher and the first-order function corrections are given by
\begin{align}
\label{fg1}
&\mff_{n j \varpi}^{(1)}(r) := \widehat{\sum_{\widetilde{n}}}  \; \mff_{\widetilde{n} j \varpi}^{(0)}(r)  \quad \hbox{and} \quad  \mfg_{n j \varpi}^{(1)}(r) := \widehat{\sum_{\widetilde{n}}}  \; \mfg_{\widetilde{n} j \varpi}^{(0)}(r) \,.
\end{align}
This concludes our perturbative calculations of the lepton modes to linear order in $\exx$.

\subsection{Near-nucleus fermion boundary conditions}
\label{running}
 
We next determine the values of $\msD/\msC$ required by the fermionic boundary conditions, obtained by a more careful treatment of the delta-function terms in the fermionic field equation \eqref{Psieq}. This section quotes the main results, with more details given in Appendix \ref{AppendixFBC}. Because it happens that the dominant effects arise from boundary conditions for $j=\frac12$ modes, we focus here on these. The relevant near-nucleus boundary condition -- applied at distance $r = \epsilon$ from the nucleus -- is given in \pref{fermionbc1}, repeated here for convenience 
\be 
\label{fermionbc1x}
  0 = \int \exd^2 \Omega_2 \, \epsilon^2 \left[ \gamma^r + \hat{c}_s - i \gamma^0 \hat{c}_v + \hat{c}_\ssF \,\bfI \cdot  \boldsymbol{\Sigma} \right] \psi_{n \ssF j \varpi}(\epsilon) \,,
\ee
where (as before) $\hat{c}_i = {c_i}/({4\pi \epsilon^2})$.

As applied to the positive parity $j=\frac12$ state, performing the angular integration implies the boundary condition for the radial function becomes
\begin{align}
\label{bc++}
\hat{c}_s - \hat{c}_v + \cZ_{\ssF} \,\hat{c}_\ssF = \frac{ \mfg_{n \frac{1}{2} +}(\epsilon)}{ \mff_{n \frac{1}{2} +}(\epsilon)},
\end{align}
where $\cZ_\ssF := \frac12 \left[F(F+1) - \frac32\right] = \cZ_{\ssF+}$ where $\cZ_{\ssF\varpi}$ is defined for $ j = \frac12$ states as (see Appendix \ref{AppendixFBC}) 
\be \label{ZFvsXF}
   \cZ_{\ssF\varpi} := \frac{ 2\varpi+1}{6} \left[ F(F+1)  - I(I+1) -\frac34 \right]  =\frac{ 2\varpi+1}{8}\; X_\ssF  \,,
\ee
where the last equality uses the definition \pref{amatrix} of $X_\ssF$. Repeating the same exercise for the negative parity, $j=\frac12$ state similarly gives (see Appendix \ref{AppendixFBC}) 
\be 
\label{bc--}
\hat{c}_s + \hat{c}_v +\mathcal{Z}_{\ssF} \, \hat{c}_\ssF = \frac{\mff_{n \frac{1}{2} -}(\epsilon)}{\mfg_{n \frac{1}{2} -}(\epsilon)}.
\ee
As elaborated in Appendix \ref{AppendixFBC}, both eqs.~\pref{bc++} and \pref{bc--} use a compact notation that suppresses an implicit dependence of the couplings on both $F$ and $\varpi$ (see {\it e.g.}~eq.~\pref{coeffexp}).

As usual, there are two equivalent ways to read these last two equations. The simplest way is to evaluate the right-hand side of these equations using the solutions \pref{dc} to the radial equation, and regard them as being solved for $\msD/\msC$ as a function of the $\hat c_i$. This shows explicitly how the integration constants are determined by the effective nuclear couplings. Because physical quantities (like leptonic energy levels) can be expressed as functions of $\msD/\msC$ they also acquire a dependence on the $\hat c_i$. 

The other way to interpret these equations is as renormalization-group equations that define the running of the renormalized couplings, $\hat c_i$. That is, if the value of $\epsilon$ is to be changed without modifying physical quantities (like electron energy levels), then the explicit $\epsilon$-dependence visible in \pref{bc++} and \pref{bc--} must cancel against an $\epsilon$-dependence that is implicit in the couplings $\hat c_i$ \cite{ppeft1, ppeft2, ppeft3, ppeftA}.

The remainder of this section focuses on the first of these two points of view, and we return to the second approach in \S\ref{ssec:Renorm} below.

\subsubsection{Solution for $\msD/\msC$}

The goal is to solve eqs.~\pref{bc++} and \pref{bc--} for the integration constant $\msD/\msC$, to linear order in $\exx$. Because the explicitly calculable terms of eqs.~\pref{bc++} and \pref{bc--} depend on nuclear spin only through the spin-dependance of $X_\ssF$ this suggests that the same should also be true for the couplings $\hat{c}_{s, v}$ and integration constants $\msD/\msC$ at $\cO(\exx)$. This leads to the ansatz
\be
\label{coeffexp}
\hat{c}_{s, v} = \hat{c}_{s, v}^{(0)} + \exx X_\ssF \, \hat{c}_{s, v}^{(1)} + \cO(\exx^2) \,,
\ee
where the first terms are the couplings found in \cite{ppeftA} for spinless nuclei that are independent of the total atomic angular momentum, $F$. The second terms are the first-order corrections whose $F$-dependence is guaranteed by rotation invariance to be proportional to $\left[ F(F+1) - j(j+1) - I(I+1) \right]$. The integration constants are then solved perturbatively in $\exx$, using
\begin{equation}
\label{caexp}
\left( \frac{\msD}{\msC} \right) = \left( \frac{\msD}{\msC} \right)^{(0)} + \exx  X_\ssF \left( \frac{\msD}{\msC} \right)^{(1)} + \cdots \, ,
\end{equation}
where the $F$-independent part of eqs.~\pref{bc++} and \pref{bc--} determine $\(\msD/\msC\)^{(0)}$ and $\(\msD/\msC\)^{(1)}$ is fixed by their $X_\ssF$-dependent terms.

With this in mind we also expand the right-hand side of these boundary conditions to linear order in $\exx$, using the state-correction result from first-order perturbation theory in \eqref{psi1}, to write it as,
\be
\label{gfratio1}
\frac{\mfg_\varpi }{\mff_\varpi } = \frac{\mfg_\varpi ^{(0)} + \exx X_\ssF \, \mfg_\varpi^{(1)} + \cdots}{\mff_\varpi^{(0)} + \exx X_\ssF\, \mff_\varpi^{(1)} + \cdots} \simeq \frac{\mfg_\varpi^{(0)}}{\mff_\varpi^{(0)}} \left[ 1 + \exx X_\ssF \left( \frac{\mfg_\varpi^{(1)}}{\mfg_\varpi^{(0)}} -  \frac{\mff_\varpi^{(1)}}{\mff_\varpi^{(0)}} \right) + \cO \( \exx^2 \) \right] \,,
\ee
where $\varpi = \pm$ is the electron state's parity, $\mfg_\varpi^{(0)}, \mff_\varpi^{(0)}$ are given by \eqref{dc} and $\mfg_\varpi^{(1)}, \mff_\varpi^{(1)}$ are given by \eqref{fg1}. Because the functions $\mfg_\varpi^{(0)}, \mff_\varpi^{(0)}$ and $\mfg_\varpi^{(1)}, \mff_\varpi^{(1)}$ are themselves functions of the ratio $\msD_\varpi /\msC_\varpi$, which itself can depend on nuclear spin -- {\it c.f.}  \eqref{caexp} -- to find all terms that appear at $\cO(\exx)$ requires using \pref{caexp} in \pref{gfratio1}, expanding in powers of $\exx$ and grouping terms. 

What is important when doing so is this: because all of the $\cO(\exx)$ terms in \eqref{gfratio1} are proportional to $X_\ssF$ both sides of eqs.~\pref{bc++} and \pref{bc--} share the same dependence on nuclear spin out to linear order in $\exx$. This is no accident because, to linear order, rotation invariance implies the nuclear spin appears only through the combination $\bfI \cdot \bfJ$, whose matrix elements give the spin-dependence in both $X_\ssF$ and $\cZ_{\ssF\varpi} \propto X_\ssF$. This shows how $(\msD/\msC)^{(0)}$ is determined in terms of the coefficients $\hat{c}_s^{(0)}$ and $\hat{c}_v^{(0)}$ and by $\mff_\varpi^{(0)}/\mfg_\varpi^{(0)}$ -- as in \cite{ppeftA}. Similarly $(\msD/\msC)^{(1)}$ is given in terms of the $\cO(\exx)$ parts of $\hat{c}_\ssF, \hat{c}_s^{(1)}$ and $\hat{c}_v^{(1)}$ together with $(\mff_\varpi^{(1)}/\mff_\varpi^{(0)}) - (\mfg_\varpi^{(1)}/\mfg_\varpi^{(0)})$ and $\mff_\varpi^{(0)}/\mfg_\varpi^{(0)}$.

The details of this calculation can be found at the end of Appendix \ref{AppendixF} and here we only quote the results, separately for each parity choice $\varpi = \pm$.

\subsubsection*{Positive parity states}

Using the small-$r$ asymptotic form for the radial solutions of eqs.~\pref{dc} in the parity-even boundary condition \pref{bc++} then gives, at zeroth order in $\exx$,
\be \label{running+0}
 \hat{c}_s^{(0)} - \hat{c}_v^{(0)}  =   -\chi \left[  \frac{\left(c + a\right) + \left(c + a'\right)\left( \msD_+ / \msC_+ \right)^{(0)} (2\kappa \epsilon)^{-2\zeta}}{\left(c - a\right) + \left(c - a'\right)\left( \msD_+ / \msC_+ \right)^{(0)}  (2\kappa\epsilon)^{-2\zeta}}\right]  \,, % \bar{\lambda}_+^{(0)} :=
\ee
where the parameters on the right-hand side are given in eqs.~\pref{hyperparam} and 
\be\label{chidef}
  \chi := \sqrt{ \frac{m - \omega}{m + \omega}} \,.
\ee
Inverting \pref{running+0} then gives the integration constants in terms of $\hat{c}_s^{(0)} - \hat{c}^{(0)}_v$:
\be \label{DC0vscscv+}
  \left( \frac{\msD_+}{\msC_+} \right)^{(0)} = - \left[ \frac{\(\hat{c}_s^{(0)} - \hat{c}_v^{(0)}\)(c-a) + \chi (c + a) }{ \(\hat{c}_s^{(0)} - \hat{c}_v^{(0)} \) (c - a') + \chi(c+a')} \right] (2\kappa \epsilon)^{2\zeta} \,,
\ee
as is also found in \cite{ppeftA} for spinless nuclei.

Next, consider the $\cO(\exx)$ terms on both sides of the boundary condition \pref{bc++} which reads -- using $\cZ_\ssF = \cZ_{\ssF+} =  \frac38 \, X_\ssF$,
\bea \label{cFfromBC}
 &&\exx\(\hat{c}_s^{(1)} - \hat{c}_v^{(1)}\) + \frac38\, \hat{c}_\ssF \\
 && \qquad\qquad\qquad =  \exx \left\{  \frac{(-2) \chi  c\( a' - a \)   (2\kappa \epsilon)^{-2 \zeta}}{\left[ \left(c - a \right) +  \left(c  - a' \right)\( {\msD_+}/{\msC_+} \)^{(0)}  (2\kappa \epsilon)^{-2\zeta}\right]^2}  \(\frac{\msD_+}{\msC_+} \)^{(1)} + \Lambda_+ \right\} \,,\nn
\eea
where $\Lambda_+$ is given by 
\bea \label{Lambda+form}
 \Lambda_+ &=&  - \sqrt{\frac{m+\widetilde{\omega}}{m+\omega}}\;\widehat{\sum_{\widetilde n}}  \left[ \frac{\widetilde{\msC}_+ {(2\widetilde\kappa\epsilon)}^{\zeta-1} c}{\msC_+  (2\kappa\epsilon)^{\zeta-1} \widetilde{c}} \right] \left[ \frac{\left(\widetilde{c}  - \widetilde{a} \right) + \left( \widetilde{c} - \widetilde{a}' \right)\( {\widetilde{\msD}_+}/{\widetilde{\msC}_+} \)^{(0)} {(2\widetilde \kappa\epsilon)}^{-2\zeta} }{\left( c - a \right) + \left( c - a' \right) \( {\msD_+}/{\msC_+} \)^{(0)} (2\kappa\epsilon)^{-2\zeta} } \right]  \notag \\
&&  \times \left\{ \widetilde{\chi} \left[ \frac{\left(\widetilde{c} + \widetilde{a} \right) +  \left( \widetilde{c} + \widetilde{a}' \right)\( {\widetilde{\msD}_+}/{\widetilde{\msC}_+} \)^{(0)} {(2\widetilde\kappa\epsilon)}^{-2\zeta}}{\left(\widetilde{c} - \widetilde{a} \right) +\left( \widetilde{c} - \widetilde{a}' \right) \( {\widetilde{\msD}_+}/{\widetilde{\msC}_+} \)^{(0)} {(2\widetilde\kappa\epsilon)}^{-2\zeta} } \right]\right. \\
&& \qquad\qquad\qquad\qquad\qquad \left.\phantom{\frac{1^2}{2^2}} - \chi \left[\frac{\left(c + a \right) + \left( c + a' \right) \( {\msD_+}/{\msC_+} \)^{(0)} (2\kappa\epsilon)^{-2\zeta}}{\left(c - a  \right) + \left( c - a' \right) \( {\msD_+}/{\msC_+} \)^{(0)} (2\kappa\epsilon)^{-2\zeta}} \right]\right\},\nn
\eea
and comes from evaluating the terms involving ${\mfg_+^{(1)}}$ and ${\mff_+^{(1)}}$ in the square bracket of \pref{gfratio1}. In particular, $\Lambda_+$ does not depend on $(\msD_+/\msC_+)^{(1)}$.

As shown in detail in the paragraph surrounding \pref{ca+0def} below, the two quantities in the final braces in \pref{Lambda+form} cancel one another -- at least to leading order in $(Z\alpha)^2$ -- which in turn ensures that $\Lambda_+$ vanishes to the order we require. This simplifies enormously the above boundary condition, whose solution for $\msD_+/\msC_+$ becomes
\bea \label{DC1vscF+}
&& \exx \left( \frac{\msD_+}{\msC_+} \right)^{(1)}  = - \;  \frac{ \left[ \exx\(\hat{c}_s^{(1)} - \hat{c}_v^{(1)}\)+ \frac38 \,\hat{c}_\ssF \right]}{4\chi c \left(a'-a\right)} \\
&&\qquad\qquad\qquad \qquad\qquad \times \left[ \left(c - a \right) + \left(c  - a' \right)  \( \frac{\msD_+}{\msC_+} \)^{(0)} (2\kappa \epsilon)^{-2\zeta}\right]^2 (2\kappa\epsilon)^{2\zeta}\, .\nn
\eea

Eqs.~\pref{DC0vscscv+} and \pref{DC1vscF+} solve the problem of obtaining the integration constant ratio $\msD/\msC$ as functions of the effective couplings, $c_s$, $c_v$ and $c_\ssF$, for positive-parity $j=\frac12$ states. The next step, in prinicple, is to use these expressions in formulae like \pref{E1partial2} and \pref{normshift} for atomic energy shifts to predict how these are influenced by finite nuclear size. Before taking this step we first repeat the above exercise for parity-odd $j=\frac12$ states.

\subsubsection*{Negative parity states}
 
Returning now to the boundary condition \pref{bc--}, repeating the same steps as before ({\it i.e.}~expanding the radial functions and integration constants to linear order in $\exx$) leads to two separate relations that determine $(\msD_-/\msC_-)^{(0)}$ and $(\msD_-/\msC_-)^{(1)}$ in terms of $\hat c_s$, $\hat c_v$ and $\hat c_\ssF$. The parity-odd counterpart to \pref{running+0} is given by
\be \label{running-0}
\hat{c}_s^{(0)} + \hat{c}_v^{(0)}   =  -\frac{1}{\chi}\left[ \frac{\left(c  - a \right) +\left( c - a' \right) \( {\msD_-}/{\msC_-} \)^{(0)} (2\kappa\epsilon)^{-2\zeta} }{\left(c + a  \right) +\left( c  + a'  \right) \( {\msD_-}/{\msC_-} \)^{(0)} (2\kappa\epsilon)^{-2\zeta} } \right] \,,
\ee
which, when solved for the integration constants, gives 
\be \label{DC0vscscv-}
  \left( \frac{\msD_-}{\msC_-} \right)^{(0)} = - \left[ \frac{ \chi \(\hat{c}_s^{(0)} + \hat{c}_v^{(0)}\)  (c+a) + (c - a) }{ \chi \(\hat{c}_s^{(0)} + \hat{c}_v^{(0)}\) (c + a') + (c-a')} \right] (2\kappa \epsilon)^{2\zeta} \,.
\ee
as the counterpart to eq.~\pref{DC0vscscv+}. 

Similarly, the $\cO(\exx)$ terms of the parity-odd boundary condition \pref{bc--} are, again using $\cZ_\ssF = \frac38\, X_\ssF$,
\be  \label{cFfromBC-}
  \exx \( \hat{c}_s^{(1)} + \hat{c}_v^{(1)} \) + \frac{ 3\hat{c}_\ssF}8 =   \exx \left\{  \frac{2\chi^{-1}  c\( a' - a \)   (2\kappa \epsilon)^{-2 \zeta}}{\left[ \left(c + a \right) +  \left(c  + a' \right) \( {\msD_-}/{\msC_-} \)^{(0)} (2\kappa \epsilon)^{-2\zeta}\right]^2}  \(\frac{\msD_-}{\msC_-} \)^{(1)} \!\!\! \!\! + \Lambda_- \right\} \,,
\ee
with $\Lambda_-$ given by 
\bea \label{Lambda-form}
 \Lambda_- &=&  - \sqrt{\frac{m-\widetilde{\omega}}{m-\omega}}\;\widehat{\sum_{\widetilde n}}  \left[ \frac{\widetilde{\msC}_- e^{-\widetilde\kappa \epsilon} {(2\widetilde\kappa\epsilon)}^{\zeta-1} c}{\msC_- e^{-\kappa\epsilon} (2\kappa\epsilon)^{\zeta-1} \widetilde{c}} \right] \left[ \frac{\left(\widetilde{c}  + \widetilde{a} \right) +\left( \widetilde{c} + \widetilde{a}' \right) \( {\widetilde{\msD}_-}/{\widetilde{\msC}_-} \)^{(0)} {(2\widetilde \kappa\epsilon)}^{-2\zeta} }{\left( c + a \right) + \left( c + a' \right) \( {\msD_-}/{\msC_-} \)^{(0)} (2\kappa\epsilon)^{-2\zeta}} \right]  \notag \\
&&  \times \left\{ \frac{1}{\widetilde{\chi}} \left[ \frac{\left(\widetilde{c} - \widetilde{a} \right) + \left( \widetilde{c} - \widetilde{a}' \right)\( {\widetilde{\msD}_-}/{\widetilde{\msC}_-} \)^{(0)} {(2\widetilde\kappa\epsilon)}^{-2\zeta} }{\left(\widetilde{c} + \widetilde{a} \right) +\left( \widetilde{c} + \widetilde{a}' \right) \( {\widetilde{\msD}_-}/{\widetilde{\msC}_-} \)^{(0)} {(2\widetilde\kappa\epsilon)}^{-2\zeta} } \right]\right. \\
&& \qquad\qquad\qquad\qquad\qquad \left.\phantom{\frac{1^2}{2^2}} -\frac{1}{ \chi} \left[ \frac{\left(c - a \right) + \left( c - a' \right)\( {\msD_-}/{\msC_-} \)^{(0)} (2\kappa\epsilon)^{-2\zeta} }{\left(c + a  \right) +  \left( c + a' \right)\( {\msD_-}/{\msC_-} \)^{(0)} (2\kappa\epsilon)^{-2\zeta}} \right]\right\},\nn
\eea
coming from evaluating the terms involving ${\mfg_-^{(1)}}$ and ${\mff_-^{(1)}}$ in the square bracket of \pref{gfratio1}. As before, $\Lambda_-$ does not depend on $(\msD_-/\msC_-)^{(1)}$ and, as argued in the discussion surrounding \pref{lambda-0x}, the two terms in the final braces of \pref{Lambda-form} cancel to leading order, ensuring that $\Lambda_-$ vanishes to the order we require. This allows the solution 
\bea \label{DC1vscF-}
&&\exx \left( \frac{\msD_-}{\msC_-} \right)^{(1)}  =  \frac{ \chi \left[ \exx \( \hat{c}_s^{(1)} + \hat{c}_v^{(1)} \) + \frac38\, \hat{c}_\ssF \right] }{2c \left(a'-a\right)}   \\ 
&& \qquad \qquad \qquad\qquad \qquad \qquad \times \left[ \left(c + a \right) + \left(c  + a' \right)  \( \frac{\msD_-}{\msC_-} \)^{(0)} (2\kappa \epsilon)^{-2\zeta}\right]^2 (2\kappa\epsilon)^{2\zeta}\, .\nn
\eea

\section{Renormalization}
\label{ssec:Renorm}

So far so good. But as things stand it looks like all predictions of nucleus-induced shifts on atomic energy levels depend explicitly on the arbitrary parameters $\epsilon$ (the position where the boundary conditions \pref{bc++} and \pref{bc--} are imposed) and $\eta$ (the regularization scale associated with the divergent integrals $\mfN_1$, $\mfN_2$, $\mfD_1$ and $\mfD_2$ -- introduced {\it e.g.} below eq.~\pref{type1ayy22}). We now address how sensible predictions are possible despite the presence of these arbitrary scales. 

Physical predictions are possible because all of the dependence on these arbitrary scales can be renormalized into the definitions of effective couplings like $c_s$, $c_v$ and $c_\ssF$. That is to say: what counts are physical predictions that relate observables to other observables, and effective couplings just play a role in intermediate steps when making these relations. For instance, in practice measurements of some observables are usually used to determine the values of the effective couplings, and any real physical content only emerges once these values are used to infer the numerical size of other observables (that can themselves be measured). What is important is that all of the arbitrary dependence on $\epsilon$ and $\eta$ cancels out once observables are related to observables. In detail, this cancellation happens because any explicit dependence on $\epsilon$ and $\eta$ cancels with an implicit dependence that is hidden in the values that are used for $\hat c_s(\epsilon, \eta)$, $\hat c_v(\epsilon, \eta)$ and $\hat c_\ssF(\epsilon,\eta)$. If $\epsilon$ and $\eta$ were varied then the inferred values these couplings acquire on comparison to measurements also change, and they do so (by construction) in precisely the way that is required to keep physical observables fixed.

\subsection{Cancellation of $\epsilon$-dependence}
\label{epsCancel}
 
To see how this works, start first with the cancellation of $\epsilon$-dependence. We do so first for those nuclear finite-size contributions that do not depend on nuclear spin and then repeat the exercise at linear order in nuclear spin.

\subsubsection{Contributions independent of nuclear spin}

At one level the dependence on $\epsilon$ that is required of the couplings $\hat{c}_s^{(0)}$ and $\hat{c}_v^{(0)}$ is simply given by the boundary condition itself: eq.~\pref{bc++}, provided this is interpreted as giving the left-hand side as a function of $\epsilon$, with the parameter $(\msD_+/\msC_+)^{(0)}$ held fixed. In this point of view most of the information contained in \pref{bc++} specifies the class of trajectories along which any couplings like $\hat{c}_s^{(0)}$ and $\hat{c}_v^{(0)}$ must evolve in order to keep observables independent of $\epsilon$. The constant $(\msD_+/\msC_+)^{(0)}$ is then regarded as specifying precisely which trajectory within this class the couplings of a particular nucleus lie.   

\begin{figure}[h!]
\centering
\includegraphics{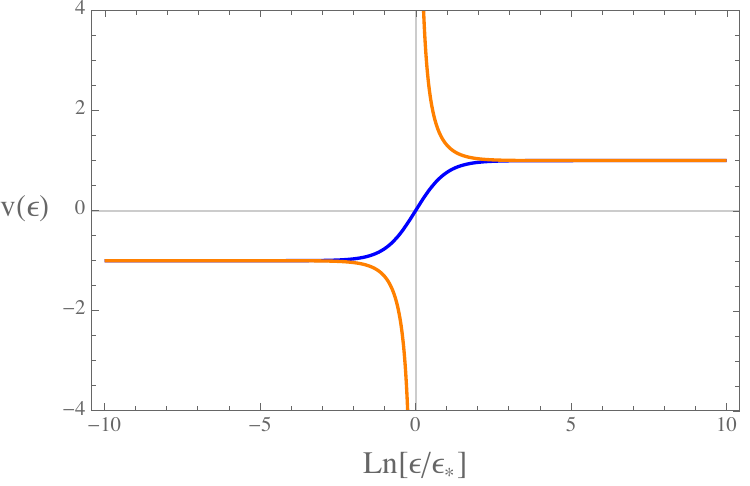}
\caption{Illustration of the two categories of RG flow described by the solutions \pref{appRGsolution} to the evolution equation \pref{appuniversalDE}. This figure plots the universal variable $v(\epsilon)$ against the logarithmic variable $\ln(\epsilon/\epsilon_\star)$. An example of each of the two categories of flow is shown. }
\label{figureOldFlow}
\end{figure}

This picture is laid out in more detail in \cite{ppeft1, ppeft2, ppeft3, falltocenter, ppeftA} and briefly summarized for convenience in Appendix \ref{ssec:AppEvoEq}. The upshot is that relations like \pref{running+0} and \pref{running-0} can all be regarded as special cases of equations of the form:
\be \label{gvseps}
  g(\epsilon) = \frac{A (2\kappa \epsilon)^{2\zeta} + B}{C (2\kappa\epsilon)^{2\zeta} + D} \,,
\ee
where  $A,B,C$ and $D$ are known parameters and $g$ is a representative coupling ({\it i.e.}~a specific combination of the couplings $c_s$, $c_v$ and $c_\ssF$). For {\it any} such evolution it is possible to define a universal coupling $v(\epsilon)$, such that
\be\label{gvsvdef}
  g (\epsilon)  =\frac12 \left( \frac{A}{C} - \frac{B}{D} \right) v(\epsilon) + \frac12 \left( \frac{A}{C} + \frac{B}{D} \right) \,.
\ee
and for which the evolution of $v$ with $\epsilon$ is simple. As is easily shown ({\it e.g.}~in Appendix \ref{ssec:AppEvoEq}), eqs.~\pref{gvseps} and \pref{gvsvdef} ensure that $v$ satisfies
\be \label{universalDE}
  \epsilon \, \frac{\exd v}{\exd \epsilon} =  \zeta (1-v^2) \,.
\ee
whose general solution is 
\be \label{RGsolution}
  v(\epsilon) = \frac{ (v_0 + 1)(\epsilon/\epsilon_0)^{2\zeta} + (v_0 - 1)}{(v_0 + 1)(\epsilon/\epsilon_0)^{2\zeta} - (v_0 - 1)} =  \frac{(\epsilon/\epsilon_\star)^{2\zeta} + y_\star}{(\epsilon/\epsilon_\star)^{2\zeta} - y_\star}  \,.
\ee
Here the first equality chooses the integration constant to ensure $v(\epsilon_0) = v_0$, and the second equality instead chooses $v(\epsilon_\star) = 0$ (if $y_\star = -1$) or $v(\epsilon_\star) = \infty$ (if $y_\star = +1$), where $y_\star = \hbox{sign}(|v|-1) = \pm 1$ is a universal constant along the trajectory (in the sense that it does not depend on $\epsilon$). 

For $\zeta > 0$ eq.~\pref{RGsolution} describes a universal flow that runs from $v_0 = -1$ to $v_\infty = +1$ as $\epsilon$ flows from 0 to $\infty$, corresponding to the initial variable flowing from $g_0 = B/D$ when $\epsilon = 0$ to $g_\infty = A/C$ as $\epsilon \to \infty$. Plots of these flows for each choice of $y_\star = \pm 1$ are given explicitly in Fig.~\ref{figureOldFlow}. Physically, this flow describes the crossover between the two independent solutions to the radial mode equation with increasing distance from the source at the origin \cite{falltocenter} ({\it i.e.}~in the present case, from the nucleus). This crossover happens because the two radial solutions have different small-$r$ asymptotic forms (typically power laws, with powers related to the two fixed points of the above flow equation), with one solution or the other dominating at large or small radius.  

To keep things concrete we next show how things work for positive-parity $j=\frac12$ states, and then return to describe the extension to negative-parity states.

\subsubsection*{Positive parity}
 
For the specific case of $j= \frac12$ parity-even states the evolution equation predicted by \pref{bc++} has the form of \pref{gvseps} if we identify $g=-\(\hat{c}_s^{(0)} - \hat{c}_v^{(0)}\)/\chi$ and use $\zeta = \sqrt{1-(Z\alpha)^2}$ and
\bea
   \frac{A}{C} &=& \frac{c+a}{c-a} = \frac{-1+\zeta -(m+\omega)Z\alpha/\kappa}{-1-\zeta - (m-\omega)Z\alpha/\kappa} \simeq n + \cdots  \nn\\
   \frac{B}{D} &=& \frac{c+a'}{c-a'} =  \frac{-1-\zeta -(m+\omega)Z\alpha/\kappa}{-1+\zeta - (m-\omega)Z\alpha/\kappa}  \simeq \frac{4n}{(Z\alpha)^2} + \cdots \,.
\eea
The approximate equalities here specialize to the leading Coulomb expression $m - \omega \simeq (Z\alpha)^2m/(2n^2) + \cdots$ and so $\kappa \simeq Z\alpha m/n + \cdots$ as well as $\zeta \simeq 1 - \frac12(Z\alpha)^2 + \cdots$, with ellipses describing contributions suppressed by additional powers of $(Z\alpha)^2$.  

Applying these expressions -- as well as $\chi = \sqrt{(m-\omega)/(m+ \omega)} \simeq Z\alpha/(2n) + \cdots$ -- to \pref{gvsvdef} then shows that the evolution of the quantity
\be 
\label{running+0x}
\bar{\lambda}_+^{(0)} := \hat{c}_s^{(0)} - \hat{c}_v^{(0)}  = -\chi \left[ \frac{\left(c + a\right) + \left(c + a'\right)\left( \msD_+ / \msC_+ \right)^{(0)} (2\kappa\epsilon)^{-2\zeta}}{\left(c - a\right) + \left(c - a'\right)\left( \msD_+ / \msC_+ \right)^{(0)}  (2\kappa\epsilon)^{-2\zeta}} \right],
\ee
has the equivalent form
\bea
\label{lambda+0}
  \bar{\lambda}_+^{(0)} &=&- \frac\chi2 \left( \frac{c+a}{c-a} - \frac{c+a'}{c-a'} \right) v_+^{(0)}(\epsilon) - \frac\chi2 \left( \frac{c+a}{c-a} + \frac{c+a'}{c-a'} \right)   \nn\\
 &=&- \frac\chi2 \left( \frac{c+a}{c-a} - \frac{c+a'}{c-a'} \right)\left[ \frac{\left( \epsilon /\epsilon_{\star+} \right)^{2\zeta} + y_{\star+}}{\left(\epsilon / \epsilon_{\star+}\right)^{2\zeta} - y_{\star+}} \right]- \frac\chi2 \left( \frac{c+a}{c-a} + \frac{c+a'}{c-a'} \right)  .
\eea
Using the values given in \pref{hyperparam} for the parameters $a$, $a'$ and $c$ then gives
\be 
\label{lambda+0x}
 \bar{\lambda}_+^{(0)} \simeq \frac{1}{Z\alpha} \Bigl(v_+^{(0)} - 1 \Bigr) 
 = \frac{1}{Z\alpha} \left\{ \left[ \frac{\left( \epsilon /\epsilon_{\star+} \right)^{2\zeta} + y_{\star+}}{\left(\epsilon / \epsilon_{\star+}\right)^{2\zeta} - y_{\star+}} \right]- 1 \right\} \,,
\ee
which drops terms suppressed by $(Z\alpha)^2$ relative to those shown. 

Comparing eqs.~\pref{running+0x} and \pref{lambda+0} reveals something interesting. Although the integration constant $(\msD_+/\msC_+)^{(0)}$ appears in both $B$ and $D$, it completely cancels out of the differential evolution equations, which depend only on the ratios $A/C$ and $B/D$. This shows that $(\msD_+/\msC_+)^{(0)}$ can also be regarded as the integration constant obtained when integrating \pref{universalDE}, and so carries the same information as do the parameters $\epsilon_{\star+}$ and $y_{\star+}$. Rewriting \pref{lambda+0} to have the form eq.~\pref{running+0x} makes this explicit:
\bea  \label{ca+0def}
\left( \frac{\msD_+}{\msC_+} \right)^{(0)} &=& - y_{\star+} \left(\frac{c-a}{c-a'}\right) (2\kappa\epsilon_{\star+})^{2\zeta} 
\simeq -\frac{16y_{\star+}(m\epsilon_{\star+})^2}{n(n+1)} \left( \frac{2Z\alpha m\epsilon_{\star+}}{n} \right)^{2\zeta-2} + \cdots \nn\\
 &=:& - \frac{\mfc}{n(n+1)} + \cO[(\za)^2] \,,
\eea
where the last equality defines the $n$-independent constant $\mfc = 16y_{\star+}(m\epsilon_{\star+})^2$. Since observable quantities (like electron energy shifts) depend on the boundary conditions only through the value $\msD/\msC$, eq.~\pref{ca+0def} shows that observables also only depend on the renormalization-group invariant parameters, like $(\epsilon_\star,y_\star)$, that characterize the effective couplings.

Another important property of \eqref{lambda+0x} is its $n$-independence, at least at leading order in $(Z\alpha)^2$. This considerably simplifies the $\cO(\exx)$ renormalization story, starting with being responsible for the vanishing of $\Lambda_+$ as defined in \pref{Lambda+form}. In particular, notice that the contents of the final braces in \pref{Lambda+form} have the form $\bar{\lambda}_+^{(0)} - \widetilde{\bar{\lambda}}_+^{(0)}$ where  $\widetilde{ \bar{\lambda}}_+^{(0)}$ and $\bar{\lambda}_+^{(0)}$ differ only by being evaluated using different quantum numbers -- such as $\widetilde{n}$ and $\widetilde j$ as opposed to $n$ and $j$ and so on. But because \pref{lambda+0x} is independent of these state quantum numbers it follows that $\widetilde{\bar{\lambda}}_+^{(0)} = \bar{\lambda}_+^{(0)}$ to the accuracy to which we work. Furthermore, as mentioned above (and is seen explicitly below), the $n$-dependence appearing in \pref{ca+0def} is precisely what is required to reproduce the proper energy shifts found in the literature for non-Coulomb nuclear structure related energy shifts \cite{friar, eides}.

 \subsubsection*{Negative parity}
 
We next repeat the above story for nuclear-spin independent interactions, but for $j=\frac12$ negative-parity states. At zeroth order in $\exx$ the relevant boundary condition \cite{ppeft3, ppeftA} is as given in \pref{running-0}
\be \label{lambda0-def}
\bar{\lambda}_-^{(0)} := \hat{c}_s^{(0)} + \hat{c}_v^{(0)}   =  -\frac{1}{\chi}\left[ \frac{\left(c  - a \right) +\left( c - a' \right) \( {\msD_-}/{\msC_-} \)^{(0)} (2\kappa\epsilon)^{-2\zeta} }{\left(c + a  \right) +\left( c  + a'  \right) \( {\msD_-}/{\msC_-} \)^{(0)} (2\kappa\epsilon)^{-2\zeta} } \right] \,,
\ee
which defines the variable $\bar{\lambda}^{(0)}_-$. This again shows how $\hat{c}_s^{(0)} + \hat{c}_v^{(0)}$ must depend on $\epsilon$ to ensure that physical quantities do not. It also falls into the category of evolution considered in \pref{gvseps}, with $g = - \bar{\lambda}_-^{(0)} \chi = - \(\hat{c}_s^{(0)} + \hat{c}_v^{(0)}\)\chi $ and
\bea
   \frac{A}{C} &=& \frac{c-a}{c+a} = \frac{1-\zeta - (m-\omega)Z\alpha/\kappa}{1+\zeta -(m+\omega)Z\alpha/\kappa} \simeq -\frac{1}{4n}(Z\alpha)^2 + \cdots  \nn\\
   \frac{B}{D} &=& \frac{c-a'}{c+a'} =  \frac{1+\zeta - (m-\omega)Z\alpha/\kappa}{1-\zeta -(m+\omega)Z\alpha/\kappa}  \simeq -\frac{1}{n} + \cdots \,.
\eea

The universal evolution equivalent to \pref{lambda0-def} then is
\bea
\label{lambda-0}
 \bar{\lambda}_-^{(0)} &=&- \frac{1}{2\chi} \left( \frac{c-a}{c+a} - \frac{c-a'}{c+a'} \right) v_-^{(0)}(\epsilon) - \frac{1}{2\chi} \left( \frac{c-a}{c+a} + \frac{c-a'}{c+a'} \right)  \\
 &=&- \frac{1}{2\chi}  \left( \frac{c-a}{c+a} - \frac{c-a'}{c+a'} \right) \left[ \frac{\left( \epsilon /\epsilon_{\star-} \right)^{2\zeta} + y_{\star-}}{\left(\epsilon / \epsilon_{\star-}\right)^{2\zeta} - y_{\star-}} \right]- \frac{1}{2\chi}\left( \frac{c-a}{c+a} + \frac{c-a'}{c+a'} \right)\,,\nn
\eea
whose leading form for small $Z\alpha$ is
\be 
\label{lambda-0x}
 \bar{\lambda}_-^{(0)} \simeq - \frac{1}{Z\alpha} \Bigl(v_-^{(0)} - 1 \Bigr) 
 = - \frac{1}{Z\alpha} \left\{ \left[ \frac{\left( \epsilon /\epsilon_{\star-} \right)^{2\zeta} + y_{\star-}}{\left(\epsilon / \epsilon_{\star-}\right)^{2\zeta} - y_{\star-}} \right] -1 \right\} \,,
\ee
which again drops terms suppressed by $(Z\alpha)^2$ relative to those shown. Rewriting \pref{lambda-0} to have the form eq.~\pref{lambda0-def} allows $(\msD_-/\msC_-)^{(0)}$ to be expressed in terms of $\epsilon_{\star-}$ and $y_{\star-}$, giving
\bea  \label{ca-0def}
\left( \frac{\msD_-}{\msC_-} \right)^{(0)} &=& - y_{\star-} \left(\frac{c+a}{c+a'}\right) (2\kappa\epsilon_{\star-})^{2\zeta} \\
&\simeq& -4y_{\star-} \left(\frac{n-1}{n^3}\right) (Z\alpha m \epsilon_{\star-})^2 \left( \frac{2Z\alpha m\epsilon_{\star-}}{n} \right)^{2\zeta-2} + \cdots \,.\nn
\eea
Note the $\cO[(Z\alpha)^2]$ suppression of this relative to the corresponding result \pref{ca+0def} in the parity-even case, as expected due to the near-nucleus suppression of $l = 1$ electronic states relative to $l = 0$ wave functions.

\subsubsection{Nuclear-spin dependent contributions}
 
The arguments made to this point are special cases of those used for spinless nuclei in \cite{ppeftA}. This section now extends these considerations to include terms at linear order in $\exx$, focussing in turn on $j=\frac12$ states with even and odd parity.

 \subsubsection*{Positive parity}
 
Consider first the apparent $\epsilon$-dependence coming from the $\cO(\exx)$ part of the boundary condition \pref{bc++}, as given explicitly in \pref{cFfromBC}. As in previous sections the $\epsilon$-dependence of the couplings can be read off directly from the boundary condition, which in this case states that
\bea 
\bar{\lambda}_+^{(1)} &:=& \hat{c}_s^{(1)} - \hat{c}_v^{(1)} +\frac38 \left( \frac{ \hat c_\ssF}{\exx} \right) \\
&=& - \chi \left( \frac{\msD_+}{\msC_+} \right)^{(1)}  \frac{2c\left(a'-a\right) (2\kappa \epsilon)^{-2\zeta}}{\left[\left(c - a\right) + \left(c - a'\right)\left( \msD_+ / \msC_+ \right)^{(0)}  (2\kappa \epsilon)^{-2\zeta} \right]^2}\,.\nn
\eea
This expression uses the positive-parity results $\varpi = +1$ and $2\varpi + 1 = 3$. It is useful to trade the dependence on $(\msD_+/\msC_+)^{(0)}$ in this expression for $\epsilon_{\star+}$ using \pref{ca+0def}, leading to
\be \label{lambda1-star+}
\bar{\lambda}_+^{(1)}   = - \left[ \frac{2 \chi  c(a'-a)}{(c-a)^2} \right] \frac{(\epsilon/\epsilon_{\star+})^{2\zeta}}{\left[(\epsilon/\epsilon_{\star+})^{2\zeta} - y_{\star+} \right]^2} \left( \frac{\msD_+}{\msC_+} \right)^{(1)}   (2\kappa \epsilon_{\star+})^{-2\zeta}  \, .
\ee

\begin{figure}[h!]
\includegraphics{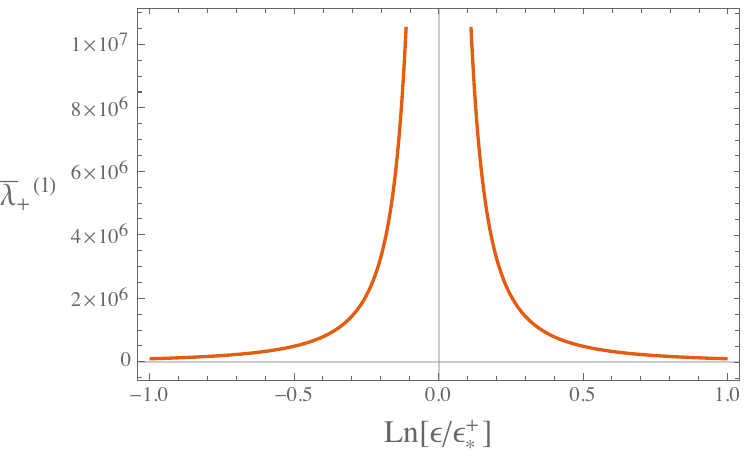}
\caption{A plot of how $\bar{\lambda}_+^{(1)} (\epsilon, \eta)$ runs as a function of $\epsilon$ (with $\eta$ fixed), based on \eqref{lambda1-star+}. The RG-invariant scale $\epsilon_{\star+}$ appearing here is the same one that also labels the running of the spin-independent interactions in $\bar{\lambda}_+^{(0)}$. The vertical scale is arbitrary because this depends on choices made for the value of $\eta$, as described in the main text.}
\label{figure2}
\end{figure}

Just like in previous sections, the requirement that observables remain $\epsilon$-independent requires $\bar{\lambda}_+^{(1)}$ (and so also its particular combination of $\hat{c}_s^{(1)}, \hat{c}_v^{(1)}$ and $\hat{c}_\ssF$) must vary with $\epsilon$ as indicated in this expression, with $(\msD_+/\msC_+)^{(1)}$ held fixed. Notice, in particular, that this coupling evolution does not require any new invariant parameters beyond $\epsilon_{\star+}$ and $y_{\star+}$ already encountered in the running of $\bar{\lambda}_+^{(0)}$. With these definitions the expression \pref{DC1vscF+} for $(\msD_+/\msC_+)^{(1)}$ becomes
\bea \label{DC1vscF+x}
 \exx \left( \frac{\msD_+}{\msC_+} \right)^{(1)} \!\!\! &=&  -  \frac{(c-a)^2(2\kappa\epsilon_{\star+})^{2\zeta}}{2\chi c \left(a'-a\right)} \left[ \left( \frac{\epsilon}{\epsilon_{\star+}} \right)^{2\zeta} - y_{\star+} \right]^2 \left(\frac{\epsilon_{\star+}}{\epsilon} \right)^{2\zeta}   \left[ \exx \(\hat{c}_s^{(1)} - \hat{c}_v^{(1)}\) +\frac{3\hat c_\ssF}{8} \right] , \notag \\
 &\simeq&  -  \frac{8 Z\alpha (m\epsilon_{\star+})^2}{n(n+1)} \left( \frac{2Z\alpha m\epsilon_{\star+}}{n} \right)^{2\zeta-2}\left[ \left( \frac{\epsilon}{\epsilon_{\star+}} \right)^{2\zeta} - y_{\star+} \right]^2 \left(\frac{\epsilon_{\star+}}{\epsilon} \right)^{2\zeta}   \\
 && \qquad \qquad \qquad \qquad \qquad \qquad \qquad \qquad   \times \left[ \exx \(\hat{c}_s^{(1)} - \hat{c}_v^{(1)}\) +\frac{3\hat c_\ssF}{8} \right] \,. \nn 
\eea

There is an important difference between this expression and previous discussions. The difference is that $\exx \(\hat{c}_s^{(1)} - \hat{c}_v^{(1)}\) + \frac38\, \hat{c}_\ssF = \bar{\lambda}^{(1)}_+ = \bar{\lambda}^{(1)}_+ (\epsilon, \eta)$ must also depend on the other regularization parameter, $\eta$, in order to cancel the explicit $\eta$-dependence hidden in $C_\eta$ in expressions like \pref{E1}. This $\eta$-dependence is hidden in the couplings $\hat{c}_s^{(1)}$, $\hat c_v^{(1)}$ and $\hat{c}_\ssF$ since these are the parameters whose renormalization absorbs this particular dependence. So although the $\epsilon$-dependence of $ \exx \(\hat{c}_s^{(1)} - \hat{c}_v^{(1)}\) +\frac38 \,\hat c_\ssF$ cancels the explicit $\epsilon$-dependence in \pref{DC1vscF+x}, the same cannot be true for the $\eta$ dependence, implying $(\msD_+/\msC_+)^{(1)}$ is implicitly a function of $\eta$. The ultimate cancellation of this $\eta$-dependence is described below after first summarizing how the $\epsilon$-dependence cancels for parity-odd states at $\cO(\exx)$.

\subsubsection*{Negative parity}
 
Consider first the apparent $\epsilon$-dependence coming from the $\cO(\exx)$ part of the boundary condition \pref{bc--}, as given explicitly in \pref{cFfromBC-}. For negative parity this states that
\bea \label{lambda1-}
\bar{\lambda}_-^{(1)} &:=& \hat{c}_s^{(1)} + \hat{c}_v^{(1)} + \frac38\left( \frac{\hat{c}_\ssF}{\exx} \right)\\
 &=& \frac{1}{\chi} \left( \frac{\msD_-}{\msC_-} \right)^{(1)}  \frac{2c\left(a'-a\right) (2\kappa \epsilon)^{-2\zeta}}{\left[\left(c + a\right) + \left(c + a'\right)\left( \msD_- / \msC_- \right)^{(0)}  (2\kappa \epsilon)^{-2\zeta} \right]^2}\, .\nn
\eea
Trading the dependence on $(\msD_-/\msC_-)^{(0)}$ in this expression for $\epsilon_{\star-}$ using \pref{ca-0def} then gives
\be \label{lambda1-star}
\bar{\lambda}_-^{(1)}   =  \left[ \frac{2 c(a'-a)}{\chi (c+a)^2} \right] \frac{ (\epsilon/\epsilon_{\star-})^{2\zeta} }{\left[ (\epsilon/\epsilon_{\star-})^{2\zeta}  - y_{\star-} \right]^2} \left( \frac{\msD_-}{\msC_-} \right)^{(1)}   (2\kappa \epsilon_{\star-})^{-2\zeta}  \, .
\ee

\begin{figure}[h!]
\centering
\includegraphics{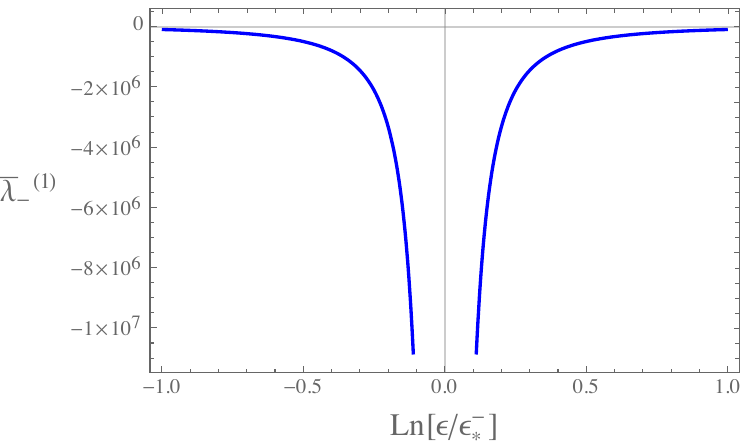}
\caption{The running of $\bar{\lambda}_-^{(1)}$ as a function of $\epsilon$, based on \eqref{lambda1-star}. The RG-invariant scale $\epsilon_{\star-}$ is the same as the scale controlling the running of $\bar{\lambda}_-^{(0)}$. For plotting purposes we hold $\eta$ fixed and so the vertical axis has an arbitrary scale that depends on the precise values chosen for $\eta$.}
\label{figure5}
\end{figure}

As before, this shows how $\bar{\lambda}_-^{(1)}$ (and so also the particular combination of $\hat{c}_s^{(1)}, \hat{c}_v^{(1)}$ and $\hat{c}_\ssF$ characteristic of negative parity states) must vary with $\epsilon$, with $(\msD_-/\msC_-)^{(1)}$ held fixed, in order to keep observables $\epsilon$-independent. The solution, \pref{DC1vscF-}, for $(\msD_-/\msC_-)^{(1)}$ then becomes
\bea \label{DC1vscF+xx}
 \exx \left( \frac{\msD_-}{\msC_-} \right)^{(1)}  &=&   \frac{\chi (c+a)^2(2\kappa\epsilon_{\star-})^{2\zeta}}{2 c \left(a'-a\right)} \left[ \left( \frac{\epsilon}{\epsilon_{\star-}} \right)^{2\zeta} - y_{\star-} \right]^2 \left(\frac{\epsilon_{\star-}}{\epsilon} \right)^{2\zeta}  \left\{ \exx \left[ \hat{c}_s^{(1)} + \hat{c}_v^{(1)} \right] +\frac{3\hat{c}_\ssF}8  \right\} , \notag \\
&\simeq&   \frac{2(n-1)(Z\alpha)^3 (m\epsilon_{\star-})^2}{n^3} \left( \frac{2Z\alpha m\epsilon_{\star-}}{n} \right)^{2\zeta-2}  \\
&& \qquad \qquad \qquad \qquad   \times \left[ \left( \frac{\epsilon}{\epsilon_{\star-}} \right)^{2\zeta} - y_{\star-} \right]^2 \left(\frac{\epsilon_{\star-}}{\epsilon} \right)^{2\zeta} \left\{ \exx \left[ \hat{c}_s^{(1)}  + \hat{c}_v^{(1)}  \right] + \frac{3\hat{c}_\ssF}{8}  \right\} , \nn
\eea
again exhibiting a $P$-wave $(Z\alpha)^2$ suppression relative to the parity-even case.  This evolution is shown in Figure \ref{figure5}.

 \subsection{Cancellation of $\eta$-dependence}
 
We now return to describing how divergent $\eta$-dependence found in earlier sections gets renormalized.  

To this end recall that to linear order in $\exx$ the nuclear-structure contribution to electronic energy levels has the form given in \pref{energylevelanswer}, of which the main focus in this section is on the first two terms on the right-hand side:
\be \label{energylevelanswerx}
     \delta \omega_{n\ssF j \varpi}  + \varepsilon_{n\ssF j \varpi}^{(1)} = \Bigl( \delta \omega_{n\ssF j \varpi}^{(0)} +  \delta\omega_{n\ssF j \varpi}^{(1)} \Bigr) + \Bigl( \varepsilon^{\rm{hfs}}_{n\ssF j \varpi}  + \delta\varepsilon_{n\ssF j \varpi}^{(1)} \Bigr) \,.
\ee
Here $\delta \omega_{n\ssF j \varpi}^{(0)}$ and $\delta \omega_{n\ssF j \varpi}^{(1)}$ are the spin-independent and $\cO(\exx)$ contributions to $\delta \omega_{n\ssF j\varpi}$, given by using $\msD/\msC = (\msD/\msC)^{(0)} + \exx (\msD/\msC)^{(1)}$  in the energy expression \pref{normshift}, while $\varepsilon^{\rm{hfs}}_{n\ssF j \varpi}$ is the hyperfine (point-nucleus but nuclear-spin dependent) energy shift of eq.~\pref{Efermi} and 
\bea
\label{E1NS}
 \delta \varepsilon^{(1)}_{n \ssF j \varpi} &=& \varepsilon^{\rm{hfs}}_{n\ssF j \varpi} \left[ \frac{1 + (\msD/\msC)(\mfN_1/\mfN_{\rm pt}) + (\msD/\msC)^2 (\mfN_2/\mfN_{\rm pt} )}{1 + (\msD/\msC)(\mfD_1/\mfD_{\rm pt}) + (\msD/\msC)^2 (\mfD_2/\mfD_{\rm pt} )} -1 \right] \nn\\
&=& \varepsilon^{\rm{hfs}}_{n\ssF j \varpi} \left[  \left( \frac{\msD}{\msC} \right)^{(0)} \left( \frac{\mfN_1}{\mfN_{\rm pt}} - \frac{\mfD_1}{\mfD_{\rm pt}}  \right) + \cdots \right] \\
&\simeq&  \varepsilon^{\rm{hfs}}_{n\ssF j \varpi}  \left[  C_\eta - \frac{\mfc}{n} +\cdots \right] \,,\nn
\eea
is the nuclear-structure part of the contribution to $\varepsilon_{n\ssF j \varpi}^{(1)}$ of eq.~\pref{E1} once $\varepsilon^{\rm{hfs}}_{n\ssF j \varpi}$ has been subtracted out.  

The issue to be addressed arises because the integrations appearing in $\delta\varepsilon_{n\ssF j \varpi}^{(1)}$ diverge in the near-nucleus ($r \to 0$) limit; a divergence that is dealt with using a regulation parameter $\eta$. This section traces how this unphysical $\eta$-dependence is renormalized by the effective couplings $\hat{c}_s^{(1)}, \hat{c}_v^{(1)}$ and $\hat{c}_\ssF$  that appear in $\delta \omega_{n\ssF j \varpi}^{(1)}$ through its dependence on $(\msD/\msC)^{(1)}$ (such as, {\it e.g.}, eq.~\pref{DC1vscF+x}). 

To display this cancellation explicitly we require the dependence of $\delta \omega_{n\ssF j \varpi}^{(1)}$ on $\(\msD_+/\msC_+ \)^{(1)}$. This is found by expanding the general result \eqref{normshift} of Appendix \ref{AppendixC} -- found by solving \pref{normal} for nonzero $\msD/\msC$. Specializing \eqref{normshift} to positive-parity, $j=\frac12$ states one finds
\bea \label{omega0exp}
\delta \omega_{n\frac{1}{2}+}^{(0)}  &\simeq& - \frac{\kappa_\ssD^3 {n(n+1)} \left( {\msD_+}/{\msC_+} \right)^{(0)} \Big[ 1-(2 - 2\zeta)\( H_{n+1} + \gamma \)  \Big]}{2m^2Z\alpha\,\mfH} \nn\\
&\simeq&  \frac{8 y_{\star+} (Z\alpha)^2 m (m \epsilon_{\star+})^2}{n^3\mfH}  \left( \frac{2Z\alpha m\epsilon_{\star+}}{n} \right)^{2\zeta-2}  + \cdots\,,
\eea
and
\bea \label{omega1exp}
  \delta \omega_{n\ssF \frac{1}{2}+}^{(1)} &\simeq&   - \exx X_\ssF\left[ \frac{\kappa_\ssD^3 n(n+1)  }{2 m^2Z\alpha}  \left( \frac{\msD_+}{\msC_+} \right)^{(1)}+ \cdots \right] \nn\\
  &\simeq&    \frac{ 4(Z\alpha)^3 m (m \epsilon_{\star+})^2 }{n^3}  \left( \frac{2Z\alpha m\epsilon_{\star+}}{n} \right)^{2\zeta-2}  \\
  && \qquad\qquad \times \left[ \left( \frac{\epsilon}{\epsilon_{\star+}} \right)^{\zeta} - y_{\star+} \left(\frac{\epsilon_{\star+}}{\epsilon} \right)^{\zeta} \right]^2 \left[ \exx \left( \hat{c}_s^{(1)} - \hat{c}_v^{(1)} \right) + \frac{3\hat{c}_\ssF}{8} \right] X_\ssF + \cdots \,,\nn
\eea
where the second lines of  \pref{omega0exp} and \pref{omega1exp} respectively use \pref{ca+0def} and \pref{DC1vscF+x}, as well as
\bea \label{mfHeq}
  \mfH &:=& 1-  \frac{n(n+1)\left( {\msD_+}/{\msC_+} \right)^{(0)} \left[1 -4(1-\zeta) \left(H_{n+1} + \gamma \right) \right]}{4(1-\zeta)}  - 5 (1-\zeta) + \cdots\nn \\
  &\simeq& 1+  \frac{ 4y_{\star+}(m\epsilon_{\star+})^2}{(Z\alpha)^2}   + \cdots \,,
\eea
with $H_n = \sum_{k=1}^n 1/k$ being the harmonic numbers and $\gamma$ being Euler's constant. As we shall see, matching implies $m \epsilon_{\star+} \propto mRZ\alpha$ and so  the ratio $m\epsilon_{\star+}/(Z\alpha)$ proves to be small for electrons (though only order $\frac12$ for muons). The ellipses in these expressions represent terms that involve additional powers of one or both of the small quantities $(\msD_+/\msC_+)^{(0)}$ or $(Z\alpha)^2$. 

The $\eta$-dependence of the couplings is now determined by requiring physical quantities not depend on $\eta$. This implies 
\be
   \frac{\exd}{\exd \eta} \left[ - \exx X_\ssF\;\frac{(Z\alpha)^2 m (n+1)  }{2 n^2} \left( \frac{\msD_+}{\msC_+} \right)^{(1)}+ \delta\varepsilon_{n\ssF \frac12 +}^{(1)} + \cdots \right] = 0\,,
\ee
and so, using \pref{E1NS} 
\be
\label{dc+cond}
\frac{\exd}{\exd \eta}\left( \frac{\msD_+}{\msC_+} \right)^{(1)}X_\ssF \simeq \frac{2n^2}{(Z\alpha)^2  (n+1)} \left( \frac{\varepsilon_{n\ssF \frac12 +}^{\rm hfs} }{\exx  \, m}\right)\; \frac{\exd C_\eta}{\exd \eta}  
 \simeq  \frac{2X_\ssF Z\alpha}{n (n+1)}  \; \frac{\exd C_\eta}{\exd \eta} \,.
\ee

Since $C_\eta$ is $n$-independent (at leading order in $\za$) and since eq.~\pref{DC1vscF+x} implies the same is also true of $n(n+1) (\msD_+/\msC_+)^{(1)}$, it follows that the $\eta$-dependence can be cancelled by performing an $\eta$-dependent but $n$-independent shift of the combination $\exx \left( \hat{c}_s^{(1)} - \hat{c}_v^{(1)} \right) + \frac38\,\hat{c}_\ssF $. The integral of \pref{dc+cond} 
then is
\bea
\label{dc+1sol}
\left(\frac{\msD_+}{\msC_+} \right)^{(1)} &=&  \left(\frac{\msD_+}{\msC_+} \right)^{(1)}_{phys} + \frac{2n^2}{(Z\alpha)^2  (n+1)}  \left( \frac{ \varepsilon_{n\ssF \frac12 +}^{\rm hfs}}{\exx X_\ssF m} \right) C_\eta \nn\\
&\simeq&  \left(\frac{\msD_+}{\msC_+} \right)^{(1)}_{phys} + \frac{2  Z\alpha}{n (n+1)}  \; C_\eta \,,
\eea
where the first term is both $\epsilon$- and $\eta$-independent and inversely proportional to $n(n+1)$. 

Although the above discussion cancels the $\eta$-dependent part of $\delta\varepsilon_{n\ssF \frac12 +}^{(1)}$, there is (as always) clearly considerable freedom in choosing the finite parts of the counterterms. We here use this freedom to define the new $n$-independent RG-invariant parameter $\epsilon_{\ssF+}$, through the definition (notice the resemblance to \pref{ca+0def}, apart from overall sign)
\be
\label{dc+1physdef}
\left( \frac{\msD_+}{\msC_+} \right)^{(1)}_{phys} := \(\frac{c-a}{c-a'}\) \left( \frac{2Z\alpha m\epsilon_{\ssF+}}{n} \right)^{2\zeta} \simeq \frac{16 (m \epsilon_{\ssF+})^2}{n(n+1)} \left( \frac{2Z\alpha m\epsilon_{\ssF+}}{n} \right)^{2\zeta-2} \,.
\ee
With this definition the net spin-dependent energy shift at this order simply becomes,
\bea 
\label{energy+phys}
\delta \omega_{n\ssF \frac{1}{2}+}^{(1)} + \delta \varepsilon_{n\ssF \frac{1}{2}+}^{(1)} &\simeq&  - \exx X_\ssF \; \frac{\kappa_\ssD^3 n(n+1)  }{2 m^2Z\alpha}  \left( \frac{\msD_+}{\msC_+} \right)^{(1)}  + \varepsilon_{n\ssF \frac{1}{2}+}^{\rm hfs} \left( C_\eta - \frac{\mfc}{n}  \right) \notag \\
&=&-\exx X_\ssF \left( \frac{\kappa_\ssD^3}{m^2} \right) \left[ \frac{ n(n+1)  }{2Z\alpha}   \left( \frac{\msD_+}{\msC_+} \right)^{(1)}_{phys}  + \frac{\mfc}{n} \right] \\
&\simeq& -\exx X_\ssF \left[ \frac{  (Z\alpha)^2 m}{n^3 } \right]   8 (m \epsilon_{\ssF+})^2 \left( \frac{2Z\alpha m\epsilon_{\ssF+}}{n} \right)^{2\zeta-2}  + \cdots,\nn
\eea
where the last line drops the $\mfc/n$ term, as appropriate at the order we work. This is to be compared, say, with \pref{omega0exp} for $\delta \omega^{(0)}_{n\ssF \frac{1}{2}+}$.

A similar story goes through as well for $j = \frac12$ negative parity states, in principle involving the definition of a new parameter $\epsilon_{\ssF-}$, however we do not pursue this further because the additional $(Z\alpha)^2$ suppression of $P$-wave states makes the contribution of this new parameter to atomic energy shifts too small to be relevant to the order we work. We therefore drop the parity label and in what follows simply use $\epsilon_\ssF := \epsilon_{\ssF+}$ to denote the parameter relevant to nuclear spin-dependence.

\subsection{Matching to nuclear moments}
 
The previous sections show that without loss of generality all nuclear finite-size effects can be described -- at least out to contributions with dimension (length)${}^3$ --  in terms of the three RG-invariant parameters $\epsilon_{\star+}$, $\epsilon_{\star-}$ and $\epsilon_\ssF$. 

In this section our goal is to illustrate how the values of these parameters can in principle be calculable in terms of known nuclear moments from an underlying nuclear model. Later sections describe how our parameters can also be obtained from precision measurements of atomic energy levels. Besides providing some intuition for how big our parameters should be, relating them to nuclear moments allows a check on our energy-level calculations, which must reproduce those of specific models once specialized to the model's assumptions.

\subsubsection{Nuclear moments}
 
Perhaps the simplest nuclear models replace the nucleus with specified charge and magnetization distributions, $\rho_c$ and $\rho_m$, and although these are over-simplifications of the real quantum systems, they do allow explicit calculation of finite-size effects in the nucleus' electromagnetic response. These distributions are normalized such that
\begin{equation}
Ze = \int \exd^3\mathbf{x}' \, \rho_{c}(\mathbf{x}'), \hspace{18pt} \mu_\ssN = \int \exd^3\mathbf{x}' \, \rho_{m}(\mathbf{x}'),
\end{equation}
where (as in earlier sections) $Ze$ is the total nuclear charge and $\mu_\ssN$ is the nuclear magnetic moment (including the $g$-factor). Because atoms are so much larger than nuclei, atomic observables tend to sample only the first few moments of these distributions, defined by
\begin{equation}
\langle r^k \rangle_c = \frac{1}{Ze} \int \exd^3\mathbf{x}' \, r^k \rho_{c}(\mathbf{x}') \,, \qquad \langle r^k \rangle_m = \frac{1}{\mu_\ssN} \int \exd^3\mathbf{x}' \, r^k \rho_{m}(\mathbf{x}') \,.
\end{equation}

For instance, the first model-independent parameterization of a nuclear-size atomic energy  shift was written down by Karplus, Klein and Schwinger for Hydrogen in \cite{kks} for the $nS_{1/2}$\footnote{Here, and in the rest of the paper we use the spectroscopic notation, $nL_j^\ssF$, where $n$ is the principal quantum number, $L$ is the orbital angular momentum quantum number, $j$ is the total angular momentum quantum number of the orbiting lepton and $F$ is the total atomic angular momentum quantum number (if appropriate and necessary).} state, giving
\be 
\delta \omega_{n \frac{1}{2} +}^{(0)} \simeq \frac{2}{3} (Z\alpha)^4 m_r^3 \langle r^2 \rangle_c \,,
\ee 
where $m_r$ is the reduced mass. 

A further step was taken by Zemach in \cite{zemach}, who computed the influence of the magnetization distribution to  find (for total atomic spin $F = 0,1$)
\be 
\label{zemachmoment}
\delta \omega_{1 \ssF \frac{1}{2} +}^{(1)} \simeq  - 2\, \exx X_\ssF   \left( \frac{m_r}{m} \right)^3 (Z\alpha)^4 m^2\langle r \rangle_{cm} + \cdots \,,
\ee
with $\exx$ as defined in \pref{exxdef} and the first Zemach moment, $\langle r \rangle_{cm}$, being one of many such moments defined by
\begin{equation}
\langle r^k \rangle_{cm} = \frac{1}{Ze\, \mu_\ssN}  \int \exd^3 \mathbf{x}{}' \int \exd^3\mathbf{y}{}'  \, r^k \rho_{c}(\mathbf{x}{}') \rho_{m}(\mathbf{y}{}') \,.
\ee

Friar categorized finite-size effects \cite{friar} out to third-order perturbation theory in $Z\alpha$ as a function of nuclear moments and showed (among other things) that for positive-parity $j=\frac12$ ({\it i.e.}~$nS_{1/2}$) states they can be written as,
\bea
\label{friarShalf}
\delta \omega_{n \frac{1}{2} +}^{(0)} &=& \frac{2}{3}(Z\alpha)^4 \frac{m_r^3}{n^3} \left\{ \langle r^2 \rangle_c - \frac{1}{2} m_r(Z\alpha)\langle r^3 \rangle_{cc} - (Z\alpha)^2 \left[ \frac{\langle r^3 \rangle_c \langle r^{-1} \rangle_c}{3} - I^{\rm REL}_2 - I^{\rm REL}_3  \right. \right. \notag \\
&& \left. + \langle r^2 \rangle_c \left( H_{n-1} +  \gamma - \frac{13n^2 + 4n - 9}{4n^2} + \left\langle \ln\left[ \frac{2m_r (Z\alpha) r}{n}\right] \right\rangle_c  \right) \right] \nn\\
&& \quad+ m_r^2(Z\alpha)^2 \left[ I^{\rm NR}_2 + I^{\rm NR}_3 + \frac{2}{3} \langle r^2 \rangle_c \left( \left\langle r^2\ln\left[ \frac{2m_r (Z\alpha) r}{n}\right] \right\rangle_c \right. \right.   \\
&&\quad \bigg. \left. \left. + \langle r^2 \rangle_c \left(H_{n-1} + \gamma - \frac{4n+3}{3n} \right) \right) + \langle r^3 \rangle_c \langle r \rangle_c + \frac{\langle r^4 \rangle_c}{10n^2}  + \langle r^5 \rangle_c \langle r^{-1} \rangle_c  \right] \bigg\} + \ldots \, ,\nn
\eea
where $H_n$ are again the harmonic numbers -- defined below \pref{mfHeq} -- while $\gamma$ is Euler's constant, the ellipses represent terms of order $(Z\alpha)^7$ or higher, and $I^{\rm NR}_2, I^{\rm NR}_3, I^{\rm REL}_2, I^{\rm REL}_3$ are parametric integrals whose detailed form plays no role in what follows (but, for those interested, can be found in \cite{friar}).

Calculations like these based on fixed distributions of charge and magnetization miss dynamical effects, such as those due to nucleon motion and polarizability (that are not included in \cite{friar}). These effects are included in the modern approaches to precision atomic calculations that dominate the more recent literature \cite{pachucki2015, karshenboim2015-2, dorokhov2018, pachucki2018}, which involve more detailed modelling of nucleon substructure, nucleon motion and inter-nucleon interactions. Some of the results of these more sophisticated calculations nonetheless overlap with eqs.~\eqref{friarShalf} and \eqref{friarPhalf}, typically when describing `elastic' contributions (for which the nucleus is assumed to remain unexcited within internal lines when evaluating the relevant Feynman graphs -- such as those describing virtual photon exchange with the orbiting lepton). Other contributions fall outside the above expression, such as those inelastic contributions that sum over excited nuclear states and are related to the electric and magnetic polarizabilities of the nucleus. As can be seen in calculations for deuterium in \cite{friar2013, pachucki2018}, the split between elastic and inelastic contributions can be artificial, and they are better considered together. 

Combining these dynamical effects \cite{pachucki2018} with the  results given above then gives (for the nuclear-spin independent shift of $nS_{1/2}$ states),
\bea
\label{pacShalf}
 \delta \omega_{n \frac{1}{2} +}^{(0)} &=&  \frac{2}{3}(Z\alpha)^4 \frac{m_r^3}{n^3} \left\{ \langle r^2 \rangle_c - \frac{1}{2} m_r(Z\alpha)\langle r^3 \rangle_{cc}^{\rm eff} - (Z\alpha)^2 \langle r^2 \rangle_c \left( H_{n-1} +  \gamma + \frac{9}{4n^2} - \frac{1}{n}  \right. \right. \notag \\
&& \left. - 3  +  \ln\left[ \frac{2m_r (Z\alpha)}{n}\right] + \ln[\langle r_{\scriptscriptstyle C2} \rangle]  \right) + m_r^2(Z\alpha)^2 \left[ \frac{\langle r^4 \rangle_c}{15n^5}  \right.   \\
&& \quad\left. \left.  + \frac{2}{3} \langle r^2 \rangle_c \langle r^2 \rangle_c   \left( H_{n-1} + \gamma - \frac{1}{n} + \, 2 +  \ln\left( \frac{2Z\alpha m_r \langle r_{\scriptscriptstyle C1} \rangle}{n} \right) \right) \right] \right\} + \ldots,\nn
\eea
where the ellipses now only denote terms of order $\cO\left( (Z\alpha)^7 \)$ or higher, $\langle r^3 \rangle_{cc}^{\rm eff}$ is an effective radius that takes into account the inelastic contributions of the two-photon Coulomb exchange\footnote{There is a cancellation between the original $\langle r^3 \rangle_{cc}$ term in \eqref{friarShalf} (known as the Friar moment) and a certain part of the polarizability \cite{friar2013} but since the inelastic contributions at this order depend on the lepton quantum numbers the same way as the elastic contributions they can be combined to define the effective nuclear moment, $\langle r^3 \rangle_{cc}^{\rm eff}$.}, and $\langle r_{\scriptscriptstyle C2} \rangle, \langle r_{\scriptscriptstyle C1} \rangle$ are again other nuclear moments, whose definitions from \cite{pachucki2018} we do not repeat here as they do not qualitatively contribute to our discussion. 

For some purposes it is also necessary to know similar results for the nucleus-induced energy shift for parity-negative $j =\frac12$ states ($nP_{1/2}$). These are suppressed by the small size of the wave-function at the nucleus, which introduces two more powers of $Z\alpha$, with the result coming from a static nuclear charge distribution \cite{friar} to order $\cO\left[ (Z\alpha)^6 \right]$ given by
\be 
\label{friarPhalf}
 \delta \omega_{n \frac{1}{2} -}^{(0)} = \frac{n^2-1}{3n^5} (Z\alpha)^4 m_r^3 \left\{ \frac{1}{2} (Z\alpha)^2 \langle r^2 \rangle_c + \frac{1}{15} (Z\alpha)^2 m_r^2 \langle r^4 \rangle_c \right\} + \ldots \,,
\ee
where the ellipses again denote terms higher order in $(Z\alpha)$. Nuclear finite-size related polarizability contributions for Hydrogen do not yet contribute at order $(Z\alpha)^6$ for $P$-states and so can be ignored. (Even if present, such terms would not change the arguments made below.)

Energy shifts sensitive to nuclear spin -- such as \pref{zemachmoment} -- are also relevant at order $(Z\alpha)^6$, since -- {\it c.f.}~eq.~\pref{exxdef} --  for Hydrogen $\exx \sim (m/M) (Z\alpha)$. Since $m/M \sim Z\alpha$ for electrons the result \pref{zemachmoment} suffices to present experimental accuracy for ordinary Hydrogen, but the larger muon mass (and high experimental precision) implies that corrections involving both nuclear structure and spin can also be important for muonic Hydrogen. These are written in terms of momentum-space integrals over the proton form-factors in \cite{martynenko2004}, and a position-space equivalent is calculated in \cite{kalinowski2018} as a part of elastic nuclear-structure corrections, leading (in our notation -- see eq.~\pref{energylevelanswer} and \pref{energylevelanswerx}) to the total spin-dependent result  
\bea
\label{fshfs}
\delta \omega_{n \ssF \frac{1}{2} +}^{(1)} + \delta\varepsilon_{n\ssF \frac{1}{2} +}^{(1)}   &\simeq& - \frac{\exx X_\ssF}{n^3}     \bigg\{ 2 \left( \frac{m_r}{m} \right)^2 (Z\alpha)^4 m_r^2 \langle r \rangle_{cm} -  \frac{4}{3} (Z\alpha)^5 m_r^3 \langle r^2 \rangle_c   \left[-\frac{1}{n}  + \gamma + H_{n-1} \phantom{\frac12}\right. \bigg. \nn\\
&&\qquad\qquad\qquad \left. \left.  + \ln\left( \frac{2Z\alpha m \langle r_{pp} \rangle}{n} \right) + \frac{\langle r^2 \rangle_m}{4n^2 \langle r^2 \rangle_c}\right] \right\}   + \cdots  
\eea
with atomic spin $F = 0,1$ and with ellipses representing terms of $(Z\alpha)^7$ or higher. Here $\langle r_{pp} \rangle$ is yet another nuclear parameter (whose detailed form is found in \cite{kalinowski2018}, but whose precise definition is not needed in what follows).

 \subsubsection{Matching to RG-invariants}
 
Eqs.~\pref{pacShalf},  \eqref{friarPhalf}  and \pref{fshfs} seem to involve a lot of nuclear parameters. But while it is true that these parameters all capture something different (and in principle measurable) about the electromagnetic properties of nuclei, a major point in this paper (and of \cite{ppeftA}) is that these nuclear parameter do not all appear independently if one's interest is only the very low energies accessed by atomic energy shifts.

The formalism used in this paper captures nuclear effects using dramatically fewer parameters, and can do so because it expands from the get-go in powers of the small ratio of nuclear to atomic size. It is the timely use of this low-energy approximation that underlies its simplicity. Furthermore, it does {\it not} make assumptions about the validity of any particular nuclear models (including dynamical effects, such as polarizabilities). It is therefore guaranteed to capture all possible nuclear effects for atomic levels, and must in particular include the predictions of any specific model. In particular, this means that the energy-shift formulae \pref{pacShalf},  \eqref{friarPhalf}  and \pref{fshfs} must agree with those computed in earlier sections, for some choice of the parameters $\epsilon_{\star+}$, $\epsilon_{\star-}$ and $\epsilon_\ssF$.

In this section we compare our predictions for the nucleus-generated $nS_{1/2}$ and $nP_{1/2}$ energy shifts to the above results and by doing so identify (or `match') how the parameters $\epsilon_{\star+}$, $\epsilon_{\star-}$ and $\epsilon_\ssF$ are related to the various moments appearing in \pref{pacShalf},  \eqref{friarPhalf}  and \pref{fshfs}. Doing so also shows that the traditional moments always appear together in these three combinations, so for the purposes of calculating atomic energy shifts there are fewer independent `effective' nuclear moments than one might naively think.  

To this end, the energy shift computed for $nS_{1/2}$ and $nP_{1/2}$ states using the steps above starting from \pref{normshift}, for the spin-independent nuclear-size contribution -- accurate to order $(Z\alpha)^5 m^4 R^3$ and $(Z\alpha)^6 m^3 R^2$  -- written using the RG-invariant $\epsilon_{\star\pm}$ is given by \cite{ppeftA}
\bea
\label{ppeftAenergy}
\delta \omega_{n \frac12 +}^{(0)} &=& \frac{8(Z\alpha)^2}{n^3}\left( \frac{m_r}{m} \right)^2 m_r^3\, y_{\star+} \epsilon_{\star+}^2 \left\{ 1 + (Z\alpha)^2 \left[ \frac{12n^2-n-9}{4n^2(n+1)} \right. \right.  \\
&&\qquad\qquad\qquad\qquad\qquad\qquad \left. \left. - \ln\left( \frac{2Z\alpha m_r \epsilon_{\star+}}{n} \right) + 2  - \gamma - H_{n+1} \right] \right\} + \cdots  \nn\\
\delta \omega_{n \frac12 -}^{(0)} &=&+ 2\left( \frac{n^2-1}{n^5} \right) (Z\alpha)^4 \left( \frac{m_r}{m} \right)^2 m_r^3  \epsilon_{\star-}^2 + \ldots, \label{ppeftAenergy2}
\eea
where the ellipses represent terms of higher order in $(Z\alpha)^2$ and $(m\epsilon_\star)$ than those written. In this expression we evaluate $\kappa^\ssD_{nj} = Z\alpha m_r/\cN  \simeq Z\alpha m_r/n + \cdots$, with $m_r = mM/(m+M)$ being the system's reduced mass. 

Earlier sections in this paper also compute the nuclear-size-related effects at linear order in the nuclear spin -- {\it i.e.}~linear in $\exx$ -- using the effective couplings with dimension (length)${}^2$ in $S_p$. The result obtained from \eqref{energy+phys} and \eqref{dc+1physdef} gives,
\be 
\label{spinenergy}
\delta \omega_{n \ssF \frac12 +}^{(1)} =  - \exx X_\ssF \frac{8 (Z\alpha)^2}{n^3} \left( \frac{m_r}{m} \right)^2 m_r^3 \, \epsilon_\ssF^2  + \cdots \,,
\ee
where ellipses represent terms higher order in $m\epsilon_\ssF$ and $(Z\alpha)^2$ than those written (such as the mixed hyperfine, finite-size effects for negative-parity, $j=\frac12$ states found in \cite{karshenboim1997}). To this same accuracy no finite-size corrections enter into the $j=\frac12$ negative-parity energy shift. 

We now can compare formulae to read off expressions for the RG-invariant scales $\epsilon_{\star\pm}$ and $\epsilon_\ssF$. We do so starting with the parity-even spin-independent shifts -- {\it i.e.}~equating \eqref{pacShalf} to \eqref{ppeftAenergy}, excluding the terms suppressed relative to the leading one by $(R/a_\ssB)^2$, -- which requires the following terms to agree for all $n$:
\bea
\label{scalarmatch}
&&\frac{8(Z\alpha)^2}{n^3} \left( \frac{m_r}{m} \right) ^2 m_r^3 \,y_{\star+} \epsilon_{\star+}^2 \left\{ 1 + (Z\alpha)^2 \left[ \frac{12n^2-n-9}{4n^2(n+1)}  \right. \right. \notag \\
&&  \qquad \qquad \qquad \qquad \qquad \qquad \qquad \qquad \left. \left.  + 2 - \gamma  - H_{n+1} - \ln\left( \frac{2Z\alpha m_r \epsilon_\star^+}{n} \right)  \right] \right\} \notag \\
&&=  \frac{2}{3}(Z\alpha)^4 \frac{m_r^3}{n^3} \langle r^2 \rangle_c \left\{ 1  + (Z\alpha)^2 \left[ \frac{1}{n} + 3 - H_{n-1} -  \gamma - \frac{9}{4n^2}  \right.\right.  \\
&&\qquad\qquad\qquad\qquad\qquad \left.\left. -  \ln\left( \frac{2Z\alpha m_r\langle r_{\scriptscriptstyle C2} \rangle }{n}\right)  \right] \right\} - \frac{m_r^4}{3n^3} (Z\alpha)^5 \langle r^3 \rangle_{cc}^{\rm eff} \,. \nn
\eea
First off, agreement on the overall sign requires $y_{\star+} = +1$. Second, although these at first sight appear to differ in their $n$ dependence, writing $H_{n+1}$ in terms of $H_{n-1}$ (and a little algebra) shows this to be an illusion. They agree provided only that the RG-invariant $\epsilon_{\star+}$ is chosen to be (at this order)
\bea
\label{scalarmoments}
\epsilon_{\star+}^2 &:=& \frac{(Z\alpha)^2}{12} \left( \frac{m}{m_r}\right)^2 \left\{ \langle r^2 \rangle_c\left[ 1 + (Z\alpha)^2 \left( 1 + \frac{1}{2}\ln \left[ \left( \frac{m}{m_r} \right)^2 \frac{(Z\alpha)^2 \langle r^2 \rangle_c}{12 \langle r_{\scriptscriptstyle C 2} \rangle^2} \right] \right) \right] - \right. \notag \\
&& \qquad \qquad \qquad \qquad \qquad \qquad \qquad  \left.-\frac{1}{2} m_r(Z\alpha) \langle r^3 \rangle_{cc}^{\rm eff} \right\} \,.
\eea
A similar exercise for the parity odd states compares the leading term in \eqref{friarPhalf} to \eqref{ppeftAenergy2}, giving agreement when  
\be \label{scalarmoments-}
\epsilon_{\star-}^2 := \frac{(Z\alpha)^2}{12}  \left( \frac{m}{m_r}\right)^2  \langle r^2 \rangle_c \,,
\ee
to the order required. Several things are noteworthy about these expressions
\begin{itemize}
\item First, notice that to the order we work this also implies 
\be
   \epsilon_{\star+} = \epsilon_{\star-} =: \epsilon_\star \,,
\ee
as perhaps might have been expected for a parity-preserving nucleus. In particular, to the order we work only a single parameter controls the entire contribution to spin-independent nuclear-size-related energy shifts. 

\item Second, nuclear effects ultimately enter through so few parameters because they can only influence atomic properties by changing the value of the integration constant $\msD/\msC$ arising when integrating the radial mode equation. The total number of independent parameters necessary is therefore given by the number of integration constants available. Although each partial wave contributes an independent integration constant, each partial wave is also suppressed at the nuclear position by additional powers of $Z\alpha$. Constants associated with higher partial waves can only enter energy shifts once a minimal precision is required in powers of $Z\alpha$. 

\item Third, notice that matching implies the overall size $\epsilon_\star \sim Z \alpha R$, where $R \sim 1$ fm is a typical nuclear-physics scale (that arises from $\langle r^2 \rangle_c \sim R^2$).  This shows how $\epsilon_\star$ encodes both the nucleus'  intrinsic size, but also the strength with which this size is probed. Because electromagnetic forces are weak it follows that $\epsilon_\star \ll R$. 

\item Finally, matching shows that the independent parameters $\epsilon_\star$ and $\epsilon_\ssF$ depend explicitly on the lepton mass, and so parameters as measured using atomic Hydrogen do not directly apply to muonic Hydrogen. Although in principle this mass dependence can be computed, its calculation involves more detailed information about the nuclear moments. From the point of view of minimizing nuclear-physics related errors we therefore treat parameters like $\epsilon_\star$ as being independent for muonic and electronic Hydrogen.
\end{itemize}

A similar comparison can be done for the spin-dependent nuclear-size contributions -- {\it i.e.}~eqs.~\pref{spinenergy} and \pref{fshfs}. Since \pref{spinenergy} only works to leading-order accuracy we restrict to comparing only with the first term in \eqref{fshfs} when inferring the value for $\epsilon_\ssF^+$. Doing so shows they agree provided we identify
\be 
\label{spinmoments}
\epsilon_\ssF^2 = \frac{(Z\alpha)^2 \langle r \rangle_{cm}}{4 m_r}  + \cdots \,.
\ee
The ellipses show that this comparison can receive corrections once the matching is done at higher orders in the ratio of nuclear to atomic size ($R/a_\ssB$), or of $\alpha$.

\section{Predictions for energy shifts}
\label{energies}

This section takes the previous section's results for how atomic energy levels depend on finite nuclear size, and uses them to identify observables from which nuclear effects can be eliminated. Most of the discussion is aimed at electrons orbiting a proton (ordinary Hydrogen), but (motivated by the prospects of improving measurements) implications are also explored for muonic Hydrogen.

\subsection{Isolating finite-nucleus effects}
\label{observables}

The main idea is simple: electronic energy levels are given (to the accuracy needed here) by \pref{energylevelanswer} -- and the discussion immediately following \pref{energylevelanswer} -- using also \pref{energylevelanswerx}. The result is rewritten here as
\be \label{energylevelanswer4}
   \omega_{n\ssF j \varpi}  = \omega_{n\ssF j \varpi}^{\rm pt} + \omega_{n\ssF j \varpi}^{\NS} \,,
\ee
where
\bea \label{energylevelanswer4x}
  \omega_{n\ssF j \varpi}^{\rm pt} &=& \omega^{\ssD}_{nj} +  \varepsilon_{n\ssF j \varpi}^{\rm hfs} +  \varepsilon_{n\ssF j \varpi}^{(ho)} + \varepsilon^{\QED}_{n\ssF j\varpi} + \varepsilon_{n\ssF j \varpi}^{{\rm pt-rec}}  \nn\\
    \omega_{n\ssF j \varpi}^{\NS}  &=&   \delta \omega_{n j \varpi}^{(0)}  + \delta \omega_{n\ssF j \varpi}^{(1)}  + \delta \varepsilon_{n\ssF j \varpi}^{(1)} + \varepsilon^{\ssN-\QED}_{n\ssF j\varpi} + \varepsilon_{n\ssF j \varpi}^{{\rm N-rec}}   \,.
\eea
Here $\omega_{n\ssF j \varpi}^{\rm pt}$ contains all terms that would be present for a spinning point nucleus: $\omega^\ssD_{nj}$ being the Dirac-Coulomb energy levels of \pref{Diracomega} and \pref{Diracomega2}; $\varepsilon_{n\ssF j \varpi}^{\rm hfs}$ given by the hyperfine structure \pref{Efermi} caused by the point-nucleus's magnetic moment; $ \varepsilon_{n\ssF j \varpi}^{(ho)}$ containing the higher-order magnetic moment effects for a point nucleus; $\varepsilon^{\QED}_{n\ssF j\varpi}$ describing the series in powers of $\alpha$ that give all the QED corrections to the first two (including the Lamb shift), also computed for a point nucleus; and $ \varepsilon_{n\ssF j \varpi}^{{\rm pt-rec}}$ containing all point-nucleus recoil corrections \cite{yerokhin2015} to the assumed order. All of these contributions are under good theoretical control and can be calculated in principle with very high precision. For our purposes all these contributions are conveniently summarized to our assumed precision in \cite{hh} for electronic Hydrogen and in \cite{aldo2013th} for its muonic counterpart.

All of the influence of nuclear size resides in $ \omega_{n\ssF j \varpi}^{\NS} $ of \pref{energylevelanswer4x}. Of these, the first two terms, $\delta \omega_{n j \varpi}^{(0)} + \delta \omega_{n\ssF j \varpi}^{(1)}$, give the change of energy -- {\it c.f.} eq.~ \pref{normshift} -- induced by the change in the small-$r$ boundary condition that the presence of the nucleus generates. The superscripts on these expressions indicate their dependence on nuclear spin, with $\delta \omega^{(k)}$ being proportional to $\exx^k$, with $\exx$ given in \pref{exxdef}. For $nS_{1/2}$ and $nP_{1/2}$ states these are given explicitly by eqs.~\pref{ppeftAenergy}, \pref{ppeftAenergy2} and \pref{spinenergy}. The contribution $\delta \varepsilon_{n\ssF j \varpi}^{(1)}$, on the other hand, describes the nuclear-structure modifications to the hyperfine energy, given by the $\epsilon_\ssF$-dependent (and, in principle, $\mfc$-dependent) terms in \pref{energy+phys}. 

The rest of this section exploits the fact that nuclear effects enter into these quantities (to the order computed here) only through the two independent parameters $\epsilon_\star$ and $\epsilon_\ssF$. 

The remaining nuclear terms, $\varepsilon^{\ssN-\QED}_{n\ssF j\varpi}$ and $\varepsilon_{n\ssF j \varpi}^{{\rm N-rec}}$, complicate the details (but not the logic) of this exploitation, by complicating the formulae involved. These terms represent the non-pointlike nuclear finite-size contributions to the QED corrections and to recoil corrections, which are calculable (see below) but only depend on the value of $\epsilon_\star$ (or change the relationship between $\epsilon_\star$, $\epsilon_\ssF$ and nuclear moments), but do not introduce any new parameters of principle.

\subsubsection{Nuclear corrections to QED contributions}

There are several ways that QED corrections enter into the above story. The most direct way is as the perturbative expansion in the bulk interaction $\cL_{\QED\,{\rm int}} = i e A_\mu \ol{\Psi} \gamma^\mu \Psi$ of the bulk lagrangian \pref{QEDlep}. For graphs involving only electrons and photons these may be evaluated in the usual way, with the usual results. 

What is unusual about the QED Feynman rules obtained from the action given in \pref{QEDlep} and \pref{Spdefnospin} is the Feynman rules for the nuclear degrees of freedom. In the effective theory used here the only nuclear degrees of freedom are its first-quantized center-of-mass position, $y^\mu(\tau)$, and its spin, $\xi^\mu(\tau)$. In deriving this action all other nuclear degrees of freedom are integrated out, leaving them to contribute to low-energy observables only through their contributions to effective interactions like $c_s$, $c_v$, $c_\ssF$ and so on. But the graphs can nonetheless be evaluated, with the functional integration over $y^\mu$ capturing in particular nuclear-recoil effects associated with the nucleus' motion in response to electron/photon interactions. 

From this point of view the nuclear finite-size corrections to long-distance QED effects are calculated by evaluating Feynman graphs involving the nuclear effective couplings $c_s$, $c_v$ and so on. No new independent constants enter in these corrections because they are explicitly built from the same couplings that are used to define the parameters $\epsilon_\star$ and $\epsilon_\ssF$, implying the existence of a formula of the form $ \varepsilon^{\ssN-\QED}_{n\ssF j\varpi} =  \varepsilon^{\ssN-\QED}_{n\ssF j\varpi}(\epsilon_\star, \epsilon_\ssF)$, whose explicit form we require in the steps outlined below. 

Rather than evaluating the $c_i$-dependent Feynman graphs (with $i = s,v,F$) to compute this function ab-initio, we instead are able to infer the result using standard evaluations of nuclear corrections to QED effects found in the literature. The procedure is very different for electrons and muons, so we treat them separately in what follows.

\subsubsection*{Electrons}
 
For atomic Hydrogen one-loop QED corrections involve a vacuum polarization loop as well as one-loop vertex corrections for both the electron and nuclear couplings. The energy shift obtained by evaluating these graphs in a second-quantized theory of nucleons coupled to QED gives the following nucleus-dependent QED energy corrections  \cite{eides},
\bea \label{FSQEDgraph} 
 \varepsilon_{n \ssF \frac12 +}^{\ssN-\QED}(e) &=& \frac{2}{3} \left( 4\ln 2 - 5 \right) \alpha (Z\alpha)^5 \frac{m_r^3}{n^3} \langle r^2  \rangle _c   \\
&& \qquad\qquad + \exx X_\ssF \left( \frac{m_r}{m} \right)^2 \frac{\alpha(Z\alpha)^4}{\pi n^3} m_r^2 \langle r \rangle_{cm} \left\{\frac{5}{2} - \frac{4}{3} \left[ \ln \left( \frac{\Lambda^2}{m^2} \right) - \frac{317}{105} \right] \right\} \,.\nn
\eea
Here $\Lambda$ is a nuclear energy scale related to the dipole parameterization of the nuclear form factors used when evaluating the nuclear electromagnetic vertices in these graphs. To translate this into a useful form for the present purposes, all of the model-dependent variables -- like $\Lambda$ and the moments $\langle r^2 \rangle_c$ and $\langle r \rangle_{cm}$ -- must be traded for a dependence on the existing variables $\epsilon_\star$ and $\epsilon_\ssF$ (as we know must be possible). 

This is a particularly simple process for electrons, and it is simple because the important scales circulating within the QED loops have energies of order the electron mass. As a result they involve very high energies relative to the scales allowed in our low-energy effective description. Because of this any QED loop-generated effects that explicitly involve nuclear properties can only influence physics within a Compton wavelength of the nucleus, and so from the point of view of the EFT can be described by a local operator localized at the nuclear position. But because the action $S_p$ of \pref{Spdefnospin} and \pref{SpForm} contains the most general local interactions involving the given degrees of freedom, {\it any} nucleus-dependent QED loops can simply be regarded as shifting the values of the effective couplings that are conceived to be functions of nuclear properties, and so correcting the formulae \pref{scalarmoments} and \pref{spinmoments}. 

Because of this the nucleus-dependent energy shift contributed by QED loops are given by precisely the same formulae as above -- {\it i.e.}~eqs.~\pref{ppeftAenergy} and \pref{fshfs}. Within this picture the spin-independent and spin-dependent parts of \pref{FSQEDgraph} are completely captured by {\it omitting} $ \varepsilon_{n \ssF \frac12 +}^{\ssN-\QED}$ and then simply by using the modified results
\bea \label{eepsstarQED}
\epsilon_\star^2 &=& \frac{(Z\alpha)^2}{12} \left( \frac{m}{m_r}\right)^2 \Bigg\{ \langle r^2 \rangle_c\left[ 1 + (Z\alpha)^2 \left( 1 + \frac{1}{2}\ln \left[ \left( \frac{m}{m_r} \right)^2 \frac{(Z\alpha)^2 \langle r^2 \rangle_c}{12 \langle r_{\scriptscriptstyle C 2} \rangle^2} \right] \right) + \right. \Bigg. \notag \\
&&\qquad\qquad\qquad \Bigg. \Bigg. \qquad + \alpha(Z\alpha) \Big(4\ln2 -5 \Big) \Bigg]  -\frac{1}{2} m_r(Z\alpha) \langle r^3 \rangle_{cc}^{\rm eff} +  \ldots \Bigg\},
\eea
and
\bea \label{eepsFQED}
\epsilon_\ssF^2 &:=& \frac{(Z\alpha)^2 \langle r \rangle_{cm}}{4m_r}  \left\{ 1 + \frac{\alpha}{\pi} \left( \frac23 \left[ \ln \left( \frac{\Lambda^2}{m^2} \right) - \frac{317}{105} \right] - \frac54 \right) + \cdots \right\} 
\eea
instead of \pref{scalarmoments} and \pref{spinmoments} in the remainder of the energy shifts: eqs.~\pref{ppeftAenergy}, \pref{ppeftAenergy2} and \pref{spinenergy}.

 \subsubsection*{Muons}
 
Incorporation of nucleus-dependent one-loop QED corrections can be done in a similar way for muonic Hydrogen, though with an important difference. The explicit one-loop calculation has been done for muonic Hydrogen, with the result \cite{karshenboim1997, pachucki1996, borie2012, eides}
\bea
\label{muradmoments}
  \varepsilon_{n \ssF \frac12 +}^{\ssN-\QED}(\mu) &=& \frac{2}{3} \left( 4\ln2 - 5 \right) \alpha (Z\alpha)^5 \frac{m_{r}^3}{n^3} \langle r^2  \rangle _c   \\
&& \qquad +X_\ssF \exx \left( \frac{m_{r}}{m} \right)^2 \frac{\alpha(Z\alpha)^4}{\pi n^3}  m_{r}^2 \langle r \rangle_{cm}\left\{ \frac{5}{2} - \frac{4}{3} \left[ \ln \left( \frac{\Lambda^2}{m^2} \right) - \frac{317}{105} \right] \right\}   \notag \\
&& \qquad \qquad  +  \frac{4 }{9 n^3} \left[ \frac{\alpha(Z\alpha)^4}{\pi} \right]m_{r}^3 \left\{ \langle r^2  \rangle _c - \frac{1}{2}Z\alpha \,m_{r} \langle r^3 \rangle_{cc}^{\rm eff}\right\}\Xi_{n\frac12+},\nn
\eea
and  
\be
\label{muradmoments-}
 \varepsilon_{n \ssF \frac12 -}^{\ssN-\QED}(\mu) = +  \frac{4 }{9 n^3} \left[ \frac{\alpha(Z\alpha)^4}{\pi} \right]m_{r}^3 \left\{ \langle r^2  \rangle _c - \frac{1}{2}Z\alpha\,m_{r} \langle r^3 \rangle_{cc}^{\rm eff}\right\}\Xi_{n\frac12-} \,.
\ee
The quantity $\Xi_{nj\varpi}$ appearing here vanishes (to the order we work) for $j \neq \frac12$, and is given for $j = \frac12$ by 
\bea
\label{Xidef}
(Z\alpha)^2\; \Xi_{n \tfrac12 \varpi} &:=& \left\{ \left( \frac{\mfm_e}{m_{r}} \right)^2  \frac{n(n-l-1)!}{\left[ (n+l)! \right]^3} \int_0^\infty \exd \rho_0 \, e^{-\rho_0} \rho_0^{2l+1} \left[ L^{2l+1}_{n-l-1} (\rho_0) \right]^2 \right. \notag \\
&&\qquad\qquad\qquad \left. \times  \int_1^\infty \exd x \, e^{-\left(\frac{\mfm_e}{m_{r}} \right) \frac{n x}{(Z\alpha)}\rho_0} \left(1 + \frac{1}{2x^2} \right) \sqrt{x^2 -1} \right\} \nn\\
&& \quad + \left\{ \frac{(n-1)!}{n\left[ (n)! \right]^3} \left( \frac{n}{2m_{r}} \right)^3  \left( \frac{4Z\alpha \, m_{r}}{n} \right) \right. \notag \\
&&\qquad\qquad \times \int \frac{\exd^3 \rho_0}{\rho_0} \, e^{-\frac{\rho_0}{2}} L^1_{n-1}(\rho_0) G'(\rho_0, 0) L^1_{n-1}(0)   \\
&& \qquad \qquad \qquad \left. \times \int_1^\infty \exd x \;  e^{-\left(\frac{\mfm_e}{m_{r}} \right) \frac{n x}{(Z\alpha)} \rho_0} \left(1 + \frac{1}{2x^2} \right) \frac{\sqrt{x^2 -1}}{x^2}  \,  \right\} \delta_{\varpi +}   \,, \nn
\eea
in which the factor of $(Z\alpha)^2$ is extracted so that $\Xi_{n\frac12\varpi}$ is order unity.\footnote{This factor of $(Z\alpha)^2$ can be displayed more explicitly by rescaling the integration variable $x \to \hat x := x/(Z\alpha)$.} 

In these expressions $m$ and $m_r$ denote, as usual, the muon mass and the muon-proton reduced mass, $m_r = mM/(m+M)$. The electron mass is here (and only here) denoted $\mfm_e$ to emphasize that it is {\it not} the orbital lepton's mass, and enters through the contribution of electrons in virtual loops. The orbital-angular momentum quantum number $l$ is the unique one consistent with $j = l \pm \frac12$ and $\varpi = (-)^l$, while $\rho_0 = 2m_r Z\alpha r/n$ is the non-relativistic dimensionless radial variable of the Schr\"odinger-Coulomb problem and $L^k_n(x)$ are the associated Laguerre polynomials,
\begin{equation}
L^k_n(x) = \sum_{p=0}^{n} (-1)^p \frac{(n+k)!}{(n-p)!(k+p)! p!} \;x^p \,.
\end{equation}
Finally $G'(x, 0)$ is the reduced Schr\"odinger-Coulomb Green's function for $nS_{1/2}$ states, which is not known for general $n$ but is computed for $n=2$ in \cite{pachucki1996} to calculate the above radiative corrections for the $2P-2S$ Lamb shift in muonic Hydrogen.

All of the terms in \pref{muradmoments} that do not involve $\Xi_{n\frac12\varpi}$ come from the contributions of virtual muon loops, and so have the same functional form as did the electron loops for electronic Hydrogen. Because the important loop momenta for these graphs is set by the muon mass, their contribution to nucleus-dependent effects can also be captured by modifying the effective nuclear couplings. They consequently contribute to a shift in $\epsilon_\star$ and $\epsilon_\ssF$ of the same form as in eqs.~\pref{eepsstarQED} and \pref{eepsFQED}, but with $m$ and $m_r$ denoting the muon mass and the muon-proton reduced mass. 

It is the $\Xi_{n\frac12\varpi}$ terms that are the key difference between the result \pref{eepsstarQED} for electronic Hydrogen and \pref{muradmoments} for muonic Hydrogen. These terms come from the vacuum polarization graph, in which virtual electrons circulate within the loop. Although the dominant momenta in this loop still have magnitudes of order the electron mass, the electron Compton wavelength is not much smaller than the muonic Bohr radius. This precludes absorbing this  graph into the value of an effective coupling like $\epsilon_\star$ or $\epsilon_\ssF$. 

To summarize, the full calculation of nucleus-induced energy shifts in muonic Hydrogen, including QED contributions \cite{pachucki1996, eides, karshenboim2018}, is captured by using the modified parameters $\epsilon_\star$ and $\epsilon_\ssF$ of \pref{eepsstarQED} and \pref{eepsFQED} in the energy shift \pref{ppeftAenergy}, \pref{ppeftAenergy2} and \pref{spinenergy}, {\it and} including only the electron loop separately, using
\be
\label{murad}
   \varepsilon_{n \ssF \frac12 +}^{\ssN-\QED}(\mu)  = \frac{16}{3n^3} \left[ \frac{\alpha(Z\alpha)^2}{\pi} \right]  \left( \frac{m_{r}}{m}\right)^2 m_{r}^3 \,\epsilon_{\star}^2 \; \Xi_{n \frac12 +} \,.
\ee
For electronic Hydrogen this last contribution is not separately required.

\subsubsection{Structure-dependent recoil corrections}
 
The arguments used above for QED corrections apply equally well to recoil corrections. Recoil corrections for a point nucleus are well-known, $\varepsilon_{n\ssF j \varpi}^{{\rm pt-rec}} $, $ \varepsilon_{n\ssF j \varpi}^{{\rm N-rec}}$ \cite{Fell, PKGrotch, bodwin1987, faustov2017-2} as are many explicit nuclear-size contributions, the leading ones of these that are not simply due to substitutions of reduced mass into the charge-radius term give \cite{borie2012, friar, karshenboim2015-2, yerokhin2015}
\be 
   \varepsilon_{n\ssF j \varpi}^{{\rm N-rec}}  = - \frac{(Z\alpha)^5}{n^3} \left( \frac{m_r^3}{M} \right) \langle r \rangle_{cm} \,,
\ee
where we see the Zemach radius, $\langle r \rangle_{cm}$ emerge in a spin-independent context.

What matters is that to this order the $n$-dependence of this result is precisely the same as that of eqs.~\pref{ppeftAenergy}, and so it can be absorbed into $\delta \omega_{n\ssF j \varpi}^{(0)}$ by shifting $\epsilon_\star$ from the value given in \pref{eepsstarQED} by adding the new contribution 
\be \label{recoilepsshift}
  \delta \epsilon_\star^2 = - \frac{(Z\alpha)^3}{12} \left( \frac{m}{m_r} \right)^2 \frac{\langle r \rangle_{cm}}{M} \,.
\ee
The upshot is that these nuclear recoil terms, though numerically significant, only modify the relationship between $\epsilon_\star$ and nuclear properties; what they do not change is the functional form of \pref{ppeftAenergy} as a function of $\epsilon_\star$. 

 \subsubsection{Observables}
\label{sssec:Observables}
 
The above sections use a first-quantized EFT to compute all spin-independent nuclear-size contributions to atomic energy levels that arise at orders $(Z\alpha)^4 m^3 R^2$, $(Z\alpha)^5 m^4 R^3$ and $(Z\alpha)^6 m^3 R^2$, plus all contributions linear in nuclear spin out to order $\exx \,m^2R (Z\alpha)^4$. We show that the results agree with the extant calculations in the literature \cite{eides, aldo2013th, hh}. Most importantly these calculations show that, on very general grounds, the many nuclear moments that arise in standard calculations  are all captured as contributions to only two independent effective parameters, our $\epsilon_\star$ and $\epsilon_\ssF$. In this section we use these results to predict the size of numerous atomic transition energies. By fitting the two parameters $\epsilon_\star$ and $\epsilon_\ssF$ themselves from well-measured atomic transitions we remove the usual nuclear uncertainties from these calculations. 

We focus first on atomic Hydrogen, since for this many more transitions are measured, but these same techniques can equally well be applied to muonic Hydrogen. The main difference for muonic Hydrogen is the relative scarcity of measured transitions, though those that have been measured have been done with spectacular accuracy. We provide a single prediction for a soon-to-be-measured muonic Hydrogen transition at the end of this section.

As ever, our starting point is the energy-level expressions \pref{energylevelanswer4} and \pref{energylevelanswer4x}, reproduced again here:
\be 
   \omega_{n\ssF j\varpi}  =  \omega^{\rm pt}_{n\ssF j\varpi}   +  \omega_{n\ssF j \varpi}^\NS    \,,
\ee
where \pref{energylevelanswer4x} gives the point-nucleus contribution, and is regarded here to be a known quantity (since our focus is on nuclear contributions), evaluated to any desired accuracy.  Eq.~\pref{energylevelanswer4x} also gives the nuclear-size part of the level shifts as 
\be
    \omega_{n\ssF j \varpi}^\NS :=     \delta \omega_{n j \varpi}^{(0)}  + \delta \omega_{n\ssF j \varpi}^{(1)}  + \delta \varepsilon_{n\ssF j \varpi}^{(1)} + \varepsilon^{\ssN-\QED}_{n\ssF j\varpi}  \,,
\ee
where the first three terms are considered in detail in this paper given by \pref{ppeftAenergy}, \pref{ppeftAenergy2}, \pref{spinenergy} and \pref{E1NS}, in which the parameters $y_{\star+} = +1$, as well as $\epsilon_\star$ and $\epsilon_\ssF$ first arise. The final term is the nuclear QED radiative correction $\varepsilon^{\ssN-\QED}_{n\ssF j\varpi}$ of \pref{murad}, that only need be considered for muonic Hydrogen (because, as shown above, all of the other nucleus-dependent QED effects can be absorbed into \pref{ppeftAenergy}, \pref{ppeftAenergy2} and \pref{spinenergy} when one uses \pref{eepsstarQED} and \pref{eepsFQED} for $\epsilon_\star$ and $\epsilon_\ssF$ rather than \pref{scalarmoments} and \pref{spinmoments}). $\varepsilon_{n\ssF j \varpi}^{{\rm N-rec}} $ is omitted from this formula because it can also be absorbed into \pref{ppeftAenergy} through the shift \pref{recoilepsshift}.  In practice the nuclear corrections of interest here are only important for $j=\frac12$ states; the case to which we specialize below.

Of course, the observables of experimental interest are not atomic energies, they are the frequencies, $\nu\left(nL_{j}^{\ssF} - n' L'_{j'}{}^{\ssF'}\right)$,  of radiation emitted in a transition between an initial state $nL_j^\ssF$ and a final state $n'L'_{j'}{}^{\ssF'}$. These are what can be measured with great precision, and are given in terms of energy differences of the initial and final states,
\be
\nu\left(nL_{j}{}^{\ssF} - n'L'_{j'}{}^{\ssF'}\right) = \left( \omega_{n \ssF j \varpi}^{\rm pt} - \omega_{n' \ssF' j' \varpi'}^{\rm pt} \right)+ \left(\delta \omega_{n\ssF j \varpi}^{\rm \ssN \ssS} - \delta \omega_{n' \ssF' j' \varpi'}^{\rm \ssN \ssS}\right) +\left( \varepsilon_{n\ssF j \varpi}^{\ssN-\QED}   -  \varepsilon_{n' \ssF' j' \varpi'}^{\ssN-\QED} \right).
\ee

The next two sections use the two best-measured values for these frequencies to determine $\epsilon_\star$ and $\epsilon_\ssF$, from which predictions can be made for all other levels without introducing any uncertainties associated with the model-dependence of explicit nuclear calculations. What makes this strategy work is the observation that only these two parameters are needed, to the accuracy we work. All of the many potentially relevant nuclear moments that are naively needed to this accuracy actually only appear in atomic energy shifts through the two formulae \pref{eepsstarQED} and \pref{eepsFQED} that predict the values for the two parameters $\epsilon_\star$ and $\epsilon_\ssF$. But one never need compute these two parameters from moments if one instead infers them directly from atomic experiments.

Precisely how many parameters are required for any given accuracy? The answer depends on the number of powers of $Z\alpha$ and in $R/a_\ssB$ one wishes to keep (where $a_\ssB$ is the relevant Bohr radius and $R \sim 1$ fm is a typical nuclear length scale), since this controls when the values of new integration constants like $\msD/\msC$ for higher spherical harmonics become needed. Table \ref{table1} gives the size of various nuclear contributions to energy levels, obtained by estimating the size of nuclear moments to be $\langle r^k \rangle \sim R^k$. For each order in the expansion in powers of $Z\alpha$ and $R/a_\ssB \sim mRZ\alpha$, this table also shows how many parameters are in principle required. 

Estimates for the numerical size of each contribution is given in Table \ref{table1} for both atomic Hydrogen (for which $mRZ\alpha \ll Z\alpha$), and for muonic Hydrogen (for which $mRZ\alpha \sim Z\alpha$). We consider each of these two cases in turn.

 \subsection{Atomic Hydrogen}
\label{electronPunchline}
 
In atomic Hydrogen the three most accurately measured transitions are \cite{kramida, hh, parthey2011}
\begin{eqnarray}
\label{ah}
\nu \left( 1S^{F=1}_{\frac{1}{2}} - 1S^{F=0}_{\frac{1}{2}} \right) &=:&  \nu_{1S_{hfs}} = 1\,420\,405.751\,768\, (1) \hspace{6pt} \mathrm{kHz}, \notag \\
\nu \left( 2S^{F=1}_{\frac{1}{2}} - 2S^{F=0}_{\frac{1}{2}} \right) &=:& \nu_{2S_{hfs}} = 177 \, 556.834 \, 3 \, (67) \hspace{6pt} \mathrm{kHz},  \\
\nu \left( 2S^{F=1}_{\frac{1}{2}} - 1S^{F=1}_{\frac{1}{2}} \right) &=:& \nu_{21} = 2\,466\,061\,102\,474.806 \, (10) \hspace{6pt} \mathrm{kHz},\nn
\end{eqnarray} 
which have experimental errors of size $10^{-6} \, {\rm kHz}$, $6.7 \times 10^{-3} \, {\rm kHz}$ and $10^{-2} \, {\rm kHz}$ respectively. A large library of other measured transitions having experimental errors of 1 kHz or worse is given in \cite{kramida}, and similar precision also arises in more recent experiments, such as the recent $3S_{\frac{1}{2}}^{F=1} - 1S_\frac{1}{2}^{F=1}$ and Lamb shift measurements of atomic Hydrogen in \cite{fleurbaey2018} and \cite{hessels2019} respectively. Many of these transitions are reproduced here in Tables \ref{table2} through \ref{table4}. 

As Table \ref{table1} shows, for atomic Hydrogen the two parameters $\epsilon_\star$ and $\epsilon_\ssF$ suffice to describe nuclear contributions to atomic energy shifts down to an accuracy of about $10^{-3}$ kHz, which is much smaller than the $\sim 1$ kHz experimental accuracy listed \cite{kramida} for most of the transitions appearing in Tables \ref{table2} through \ref{table4}. For ease of comparison, contributions of order $10^{-3}$ kHz are shaded in green in Table \ref{table1}. Notice in particular that these estimates show that the contributions of $\delta \varepsilon_{n\ssF j \varpi}^{(1)}$ coming from $\mfc$ in \pref{E1}, which first arise at order $\exx\, m^3R^2(Z\alpha)^5$, are too small to be relevant at an accuracy of a Hz, and so can be neglected in what follows.

 \subsubsection*{Fitting for $\epsilon_\star$ and $\epsilon_\ssF$}
 
Before providing more precise numerical estimates for nuclear transitions we provide the explicit formulae to be used to obtain them. Recalling that all $\epsilon_\star$ and $\epsilon_\ssF$ dependence appears in eqs.~\pref{ppeftAenergy}, \pref{ppeftAenergy2} and \pref{spinenergy} and that \pref{E1NS} can be neglected, and that we can set $y_{\star+} = +1$ and (for electrons) $\varepsilon^{\ssN-\QED}_{n\ssF j\varpi} = 0$, the nucleus-dependence of the parity-even $j=\frac12$ levels is
\bea
\label{efstofit}
\delta \omega^{\rm \ssN \ssS}_{n \ssF \frac12+}(e)  &=&\frac{8}{n^3}\left( \frac{m_{r}}{m_e} \right)^2 (Z\alpha)^2 m_{r}^3  \left\{ \epsilon_{\star, e}^2 \left[ 1 + (Z\alpha)^2 \left( 2 -\gamma  - H_{n+1} \phantom{\frac12} \right. \right.\right. \notag \\
&&\qquad\qquad\qquad\qquad\qquad \left. \left. - \ln \left( \frac{2Z\alpha m_r \epsilon_{\star, e}}{n} \right) + \frac{12n^2 -n -9}{4n^2(n+1)}\right) \right]  \nn\\
&& \qquad\qquad\qquad\qquad\qquad\qquad  \left. -  \left( \frac{g_\ssN m_e  }{2M} \right) Z \alpha \,\epsilon_{\ssF, e}^2   X_\ssF \right\} 
\eea
where the subscript `$e$' again emphasizes that this expression applies only for electronic (and not muonic) Hydrogen-like atoms with spin-half nuclei. Here $X_\ssF$ is defined in \pref{amatrix} and we leave factors of nuclear charge, $Z$, explicit in the answer (for applications to general spin-half nuclei), although take $Z = 1$ for our numerical applications to Hydrogen. As above, $M$ denotes the nuclear mass and we write its magnetic moment as $\mu_\ssN = Ze g_\ssN/(2M)$ (specializing to Hydrogen via the replacement $g_\ssN \to g_p$).  The analogous formula for parity-odd $j=\frac12$ states is
\be 
\label{efstofit2}
\delta \omega^{\rm \ssN \ssS}_{n \ssF \frac12 -}(e)   = 2 (Z\alpha)^4\left( \frac{n^2-1}{n^5} \right)  \left( \frac{m_{r}}{m_e} \right)^2 m_{r}^3\, \epsilon_{\star, e}^2 \,.
\ee 

Since terms not proportional to $X_\ssF$ cancel from any hyperfine interval, we fix $\epsilon_{\ssF, e}$ using the experimentally measured $2S$ hyperfine splitting frequency \eqref{efstofit}. Using (for $j = \frac12$ states) $X_\ssF = +\frac23$ for $F=1$ and $X_\ssF = -2$ for $F=0$, gives 
\be 
 \nu_{2\ssS_{hfs}}     =     \left(\omega^{\rm pt}_{2 1 \frac{1}{2} +} - \omega^{\rm pt}_{2 0 \frac{1}{2} +} \right) - \frac{8}{3} \left( \frac{m_{r}}{m_e} \right)^2 \left( \frac{  g_\ssN m_e}{2M} \right) (Z\alpha)^3 m_{r}^3 \epsilon_{\ssF, e}^2 \,.
\ee
It is convenient to group together the experimentally measured value and the theoretical point-nucleus effects, since these are both regarded as given when effects related to nuclear size are of interest. Define therefore the precisely known quantity
\be
  \widehat{\Delta \omega}_{n\ssS_{hfs}} := \left( \omega^{\rm pt}_{n 1 \frac{1}{2} +} - \omega^{\rm pt}_{n 0 \frac{1}{2} +} \right) -  \nu_{n\ssS_{hfs}} \,,
\ee
in terms of which $\epsilon_{\ssF,e}$ is accurately determined by
\begin{equation}
\label{epsilonffit}
 \widehat{\Delta \omega}_{2\ssS_{hfs}} = \frac83 \left( \frac{m_{r}}{m_e} \right)^2 \!\! \left( \frac{g_\ssN m_e }{2M} \right) (Z\alpha)^2 m_{r}^3 \,\epsilon_{\ssF, e}^2  \,.
\end{equation}
Notice also that these formulae also predict the nuclear-size contributions to the two hyperfine transitions given in \pref{ah} are related by  
\be \label{1is2x8}
  \widehat{\Delta \omega}_{1\ssS_{hfs}} = 8\; \widehat{\Delta \omega}_{2\ssS_{hfs}} \,.
\ee

We similarly use $\nu_{21}$ to fix $\epsilon_{\star, e}$, while using \pref{epsilonffit} to eliminate $\epsilon_{\ssF,e}$, and so
\bea
\nu_{21} &=& \left( \omega^{\rm pt}_{21 \frac{1}{2} +} - \omega^{\rm pt}_{1 1 \frac{1}{2} +}\right) \nn\\
&&\quad + (Z\alpha)^2 \left( \frac{m_{r}}{m_e} \right)^2 m_{r}^3\epsilon_{\star, e}^2 \left\{ 1 + (Z\alpha)^2 \left[2 -\gamma - \frac{17}{16} - \ln \Bigl(Z\alpha m_r \epsilon_{\star, e} \Bigr) \right]  \right\} \nn\\
&&\quad - 8(Z\alpha)^2 \left( \frac{m_{r}}{m_e} \right)^2 m_{r}^3\epsilon_{\star, e}^2 \left\{ 1 + (Z\alpha)^2 \left[2 - \gamma - \frac{5}{4} - \ln \Bigl( 2Z\alpha m_r\epsilon_{\star, e} \Bigr) \right]  \right\}  \notag \\
&&\qquad\qquad - \frac{1}{4} \, \widehat{\Delta \omega}_{2\ssS_{hfs}} + 2 \, \widehat{\Delta \omega}_{2\ssS_{hfs}},
\eea
which can be rewritten
\bea \label{SolveMeForEpsilon*}
 \widehat{\Delta \omega}_{21} &=& - \frac{7}{4} \, \widehat{\Delta \omega}_{2\ssS_{hfs}} \\
&&\quad + 7(Z\alpha)^2  \left( \frac{m_{r}}{m_e} \right)^2 m_{r}^3 \epsilon_{\star, e}^2 \left\{ 1+(Z\alpha)^2 \left[ \frac{81}{112} - \gamma - \frac{8}{7}\, \ln 2 - \ln \Bigl(Z\alpha\, m_r \epsilon_{\star, e} \Bigr)  \right] \right\} \,, \nn
\eea
which defines the precisely known quantity 
\be
   \widehat{\Delta \omega}_{21} :=   \left( \omega^{\rm pt}_{21 \frac{1}{2} +} - \omega^{\rm pt}_{1 1 \frac{1}{2} +}\right)  - \nu_{21} \,.
\ee
For numerical purposes it is useful to have numerical values for these quantities, which are \cite{hh}  
\be
\widehat{\Delta\omega}_{\scriptscriptstyle 1S_{hfs}} = 58.07(57) \;\hbox{kHz} \,, \quad 
\widehat{\Delta\omega}_{\scriptscriptstyle 2S_{hfs}} = 7.22(57) \;\hbox{kHz} \quad \hbox{and} \quad
\widehat{\Delta\omega}_{\scriptscriptstyle 21} = 955.31(57) \;\hbox{kHz} \,,
\ee
which also shows that \pref{1is2x8} is satisfied, within the errors. 

Eq.~\pref{SolveMeForEpsilon*}, has the form
\be 
\mathfrak{x}_e = z_e \left( \mathfrak{y}_e - \ln z_e \right)  = -z_e \ln\left( e^{-\mathfrak{y}_e}z_e\right) \,, 
\ee
with $z_e := (m_{r}\, \epsilon_{\star, e})^2$ and
\bea
\mathfrak{x}_e &:=& 2 \left( \frac{m_e}{m_{r}} \right)^2 \frac{\left( \widehat{\Delta \omega}_{21} + \frac{7}{4} \, \widehat{\Delta \omega}_{2\ssS_{hfs}} \right)}{7m_{r}(Z\alpha)^4}, \notag \\
\mathfrak{y}_e &:=&  \frac{2}{(Z\alpha)^2} - 2\ln(Z\alpha)+ \frac{81}{56} -2 \gamma  - \frac{16}{7} \ln 2 \,.
\eea
This is to be solved for $z$, and so has solutions given by branches of the Lambert $W$-function\footnote{$W(z)$ is defined as the solution to $W(z) \, e^{W(z)} = z$, and is multiple-valued with branches labelled by an integer $k$. The branches relevant for real $z$ are $W_0(z)$, which is defined for $z > 0$, and $W_{-1}(z)$, whose argument satisfies $e^{-1} < z < 0$.}
\be 
\label{epstar}
(m\, \epsilon_{\star, e})^2 = e^\cW \quad \hbox{with} \quad
 \cW := W_{-1} \left( -\mathfrak{x}_e \, e^{-\mathfrak{y}_e} \right) + \mathfrak{y}_e \,.
\ee

The Lambert $W$-function returns real values only for real arguments in the range $x > - e^{-1}$, and is double valued for arguments $-e^{-1} < x < 0$. One of the branches takes values $-1 < W_0(x) < 0$ while the other satisfies $W_{-1} < -1$ in this range. We choose the branch, $W_{-1}(x)$, here because $(m_{r}\epsilon_{\star, e})^2$ is both real and small, and because $Z\alpha \ll 1$ implies $\mathfrak{y}_e \gg 1$. These two statements are only consistent with one another, and with eq.~\pref{epstar}, if $W(x)$ is order $-\mathfrak{y}_e$ for $x$ near zero. The numerical values inferred in this way for $\epsilon_\star$ and $\epsilon_\ssF$ are given in Table \ref{sat}.

\begin{table}
\centering
\begin{tabular}{|c|c|c|}
\hline
Transition & $\left(2S_{1/2}^{\ssF=1} - 2S_{1/2}^{\ssF=0}\right)$ & $\left(2S_{1/2}^{\ssF=1} - 1S_{1/2}^{\ssF=1}\right)$ \\
\hline
Experimental value & $177 \, 556.834 \,3$ (kHz) & $2\, 466 \, 061 \, 102 \, 474.806$ (kHz) \\
\hline
Experimental error & $0.0067$ (kHz) & $0.010$ (kHz) \\
\hline
\hline
Pt. nucl. theory & $177 \, 564.05$ (kHz) & $2 \, 466 \, 061 \, 103 \, 430.12$ (kHz)\\
\hline
Pt. nucl. error & $0.57$ (kHz) & $0.57$ (kHz) \\
\hline
\hline
Inferred param. & $(m \epsilon_{\ssF, e})^2$ & $(m \epsilon_{\star, e})^2$ \\
\hline
Fitted value & $3.71\times 10^{-8} $  & $2.1020 \times10^{-11}$ \\
\hline
Prop. exp. error & $0.0035 \times 10^{-8}$  & $0.000034 \times 10^{-11}$ \\
\hline
Prop. theory error & $0.29 \times 10^{-8}$  & $0.0025 \times 10^{-11}$ \\
\hline 
\end{tabular}
\caption{The experimental values (row 2), the experimental errors (row 3) and the point-nucleus theoretical values (row 4) and errors (row 5) for the reference transitions in atomic Hydrogen used for fixing the values of the two nuclear parameters listed in row 6. The last 3 rows give the values inferred for these parameters (row 7) and the errors they inherit due to the experimental uncertainty (row 8) and the precision of the point-nucleus calculation (row 9).}
\label{sat}
\end{table}

Given these explicit solutions for $\epsilon_{\star, e}$ and $\epsilon_{\ssF, e}$ as functions of the two well-measured energy differences (combined with well-understood point-nucleus theory contributions) we may now use these to directly express predictions for the nuclear part of the energy shift for any other energy levels, without direct reference to nuclear physics. For parity-even $j=\frac12$ states this gives
\bea
\label{prediction}
\delta \omega^{\rm \ssN\ssS}_{n \ssF  \frac12 +}(e)   &=&   \frac{8}{n^3} (Z\alpha)^2 \left( \frac{m_{r}}{m_e} \right)^2 m_{r}\,  e^{\cW} \notag \\ 
&&\quad\times  \left\{ 1 + (Z\alpha)^2 \left[2 - \gamma - H_{n+1} - \ln\left( \frac{2Z\alpha}{n} \right) - \frac{\cW}{2}  + \frac{12n^2 - n - 9}{4n^2(n+1)} \right] \right\}  \nn\\
&& \qquad-  \frac{3}{n^3}X_\ssF \widehat{\Delta \omega}_{2\ssS_{hfs}},
\eea
while for parity-odd $j=\frac12$ states one instead finds 
\be
\label{prediction-}
\delta \omega^{\rm \ssN\ssS}_{n \ssF  \frac12 -}(e)  =  2\left( \frac{n^2 -1}{n^5} \right) (Z\alpha)^4  \left( \frac{m_{r}}{m_e} \right)^2 m_{r} \, e^{\cW }  \,.
\ee
Transition frequencies are then simply given by differences of the above, for different choices for $n$ and $F$.

\begin{table}[h!]
\centering
\begin{tabular}{|c|c|c|c|c|c|}
\hline
Transition & $\nu_{\rm exp}$ &  $\Delta E^{\rm fs}$ & $\Delta E^{\rm exp}$ & $\Delta E^{\rm th}$ & $\Delta E^{\rm trunc}$\\ [3pt]
\hline
$2P_{1/2}^{\ssF=1} - 2S_{1/2}^{\ssF=0}$ & 909 871.7(3.2)  & $-143.70$ & 0.0069 & 0.57 & 0.00031 \\ [5pt]
\hline
$1S_{1/2}^{\ssF=1} - 1S_{1/2}^{\ssF=0}$ & 1 420  405.751 768(1)  & $-57.8$ & 0.054 & 4.5 & 0.0033 \\ [5pt]
\hline
$8S_{1/2}^{\ssF=1} - 2S_{1/2}^{\ssF=1}$ & 770 649  350  012(9)  & $-134.348$ & 0.0014 & 0.080 & 0.00010\\ [5pt]
\hline
$8D_{3/2}^{\ssF=2, 1} - 2S_{1/2}^{\ssF=1}$ & 770  649 504  450(8)  & $-136.481$ & 0.0014 & 0.081 & 0.00010 \\ [5pt]
\hline
$8D_{5/2}^{\ssF=3, 2} - 2S_{1/2}^{\ssF=1}$ & 770  649  561  584(6)  & $-136.481$ & 0.0014 & 0.081 & 0.00010\\ [5pt]
\hline
$12D_{3/2}^{\ssF=2, 1} - 2S_{1/2}^{\ssF=1}$ & 799  191  710  473(9)  & $-136.481$ & 0.0014 & 0.081 & 0.00010\\ [5pt]
\hline
$12D_{5/2}^{\ssF=3, 2} - 2S_{1/2}^{\ssF=1}$ & 799 191  727  404(7)  & $-136.481$ & 0.0014 & 0.081 & 0.00010\\ [5pt]
\hline
$3S_{1/2}^{\ssF=1} - 1S_{1/2}^{\ssF=1}$ & 2 922 743 278 671.5 (2.6)  & $-1051.35$ & 0.011 & 0.62 & 0.00079 \\ [5pt]
\hline
\end{tabular}
\caption{Transitions from \cite{hessels2019} (row 2), from \cite{kramida} (rows 3-8) and \cite{fleurbaey2018} (row 9) that are measured with better than 10 kHz accuracy in atomic Hydrogen.  Column 2 gives their experimental values (with experimental errors in brackets); all values given in kHz. Column 4 gives the nuclear-finite-size contribution to the transition energy predicted by eqs.~\pref{prediction} and \pref{prediction-}. Columns 5--7 give the uncertainties in this prediction: column 5 is the error from measurement errors in the reference transitions; column 6 gives the error due to theoretical uncertainty in the point-nucleus finite-size effects \cite{hh}; while column 7 is the error due to neglect of higher orders in $\exx$, $Z\alpha$ and $R/a_\ssB = mRZ\alpha$ beyond those given by green squares in Table \ref{table1}. Uncertainty in values for $\alpha$ and Ry give errors significantly smaller than those listed.}
\label{ahprecise}
\end{table}

There are three main sources of error when using expressions \pref{prediction} and \pref{prediction-} for transition frequencies. One of these -- the `truncation' error -- arises because the above expressions drop terms beyond a fixed order in $Z\alpha$ and $mRZ\alpha$. For electronic Hydrogen this truncation puts a floor of about $0.001 \, {\rm kHz}$ to the nuclear contribution to atomic energy shifts. To this must also be added two other sources of error: the experimental accuracy with which the input quantities $\nu_{21}$ and $\nu_{2\ssS_{hfs}}$ are measured (with current values given in \pref{ah}), and the uncertainty with which the point-nucleus prediction for $\omega^{\rm pt}_{n\ssF j \varpi}$ is known (which is limited in principle by the persistence of theorists). All three sources of error can be much smaller than is permitted for explicit calculations of the nuclear moments using nuclear models, and can also expect to improve into the future unconstrained by limitations in nuclear modelling. 

Table \ref{ahprecise} lists several atomic levels (taken from \cite{kramida} and \cite{fleurbaey2018, hessels2019}) that are measured to better than 10 kHz accuracy, and compares for each the overall size of the nuclear-structure prediction of eqs.~\pref{prediction} and \pref{prediction-}, as well as the three sources of error in this prediction described above. Tables \ref{table2} through \ref{table4} list these similar information for a larger class of measured transitions given in \cite{kramida}. In all cases the nuclear error is much smaller than the current experimental uncertainties. Many rows of these tables share the same values because the only shift that is larger than $10^{-3}$ kHz in size arises from the shift of the $S$-wave state, which is common to many transitions in the list.

In our numerical evaluations of these formulae we use the values for binding energies as given in \cite{hh} to fit the two parameters $\epsilon_{\star, e}, \epsilon_{\ssF, e}$. The errors given in \cite{hh} for the point-nucleus parts of the theory are at the 0.1 kHz level (which as these authors report, is satisfactory for the current experimental precision). The implied uncertainty for transition energies (and so also for $\epsilon_{\ssF, e}$) then is effectively twice as large  because transition frequencies involve energy differences. The same large uncertainty inherently exists for $\epsilon_{\star, e}$, which is found through a different interval of binding energies. Currently this error dominates both the experimental error and the `truncation' uncertainty mentioned earlier.  The good news is that the theoretical error in the point-nucleus part of the energy differences can be made much smaller simply by including higher-order calculations.

 \subsection{Muonic Hydrogen}
\label{muonPunchline}
 
Lastly, we comment on how the above calculations are adapted for muonic Hydrogen. The most accurate measurements of transitions in this system are \cite{pohlnature, pohl2016, karshenboim2015}
\bea
\label{muexp}
\nu_t :=  \nu \left( 2P^{F=2}_{{3}/{2}} -  2S^{F=1}_{{1}/{2}} \right) &=&  206.292 \, 7 \, (27) \hspace{6pt} \mathrm{meV},\notag \\
\nu_s := \nu \left( 2P^{F=1}_{{3}/{2}} -  2S^{F=0}_{{1}/{2}} \right) &=&  225.853 \, 6 \, (43) \hspace{6pt} \mathrm{meV},
\eea
with an experimental uncertainty of approximately $10^{-3}$ meV. 

The last column of Table \ref{table1} shows the size for muonic Hydrogen of each term in the expansions in powers of $Z\alpha$ and $mRZ\alpha$, and in particular shows that the same orders considered above for electronic Hydrogen -- {\it i.e.} spin-independent contributions at order $m^3 R^2 (Z\alpha)^4$, $m^4 R^3 (Z\alpha)^5$ and $m^3 R^2 (Z\alpha)^6$ together with the Zemach moment contribution at $m^2 R \exx (Z\alpha)^4$ -- also control nuclear effects in muonic Hydrogen down to a precision of about 0.01 meV. To this accuracy there are therefore only two nuclear parameters relevant, $\epsilon_{\star,\mu}$ and $\epsilon_{\ssF,\mu}$, whose values can be inferred using the experimental results \pref{muexp}. Once a third transition frequency is measured nucleus-independent predictions can in principle be tested. 

Calculations reaching the experimental precision of $10^{-3}$ meV, however, likely also require including contributions at order $m^5 R^4 (Z\alpha)^6$ (indicated in yellow in Table \ref{table1}). Although these can be computed using the methods in this paper, we do not do so here, for several reasons.  First, proper treatment of boundary conditions to this accuracy also requires generalizing $S_p$ to include spin-independent effective couplings out to dimension (length)${}^4$, and spin-dependent couplings out to order (length)${}^3$. This in turn involves analyzing the running of the existing couplings out to higher accuracy in $\rho_\epsilon$ than was performed here. Furthermore, as Table \ref{table1} shows, this new term introduces the additional complication that corrections to the $j=\frac32$ modes first become relevant at this order, potentially introducing a new integration constant, $\msD/\msC$, and possibly requiring the addition of a third RG invariant parameter. Although this requires nothing new conceptually, it is a considerable complication that we defer to future work.

In what follows we instead work only to the 0.01 meV accuracy that our calculations above already capture, and identify how the two independent parameters $\epsilon_{\star,\mu}$ and $\epsilon_{\ssF,\mu}$ are determined by existing observations, and sketch how to use these to predict the nuclear-structure part of the predictions for any other muonic Hydrogen levels that might be measured in the future.

 \subsubsection*{Determining $\epsilon_\star$ and $\epsilon_\ssF$}
 
The finite-size effects in muonic Hydrogen, written in terms of RG-invariants and approximately accurate to order $10^{-3}$ meV in the spin-independent sector but only to $10^{-2}$ meV  in the spin-dependent sector are captured by the sum of the contributions from \eqref{ppeftAenergy}, \eqref{ppeftAenergy2} and \eqref{spinenergy}, as applied to the muon, as well as adding \eqref{murad}, giving
\bea
\label{mutofit}
&&\delta \omega_{n \ssF \frac12 +}^{ \rm \ssN\ssS}(\mu) + \varepsilon^{\ssN-\QED}_{n\ssF \frac12 +}(\mu) =  \frac{8}{n^3}\left( \frac{m_{r}}{m_\mu} \right)^2 (Z\alpha)^2 m_{r}^3 \left\{\epsilon_{\star, \mu}^2 \left[ 1 +\left( \frac{2\alpha}{3\pi } \right)  \Xi_{n \frac12 +} \right. \right. \notag \\
&&\qquad\qquad\qquad\qquad \left.  + (Z\alpha)^2 \left( 2 -\gamma  - H_{n+1}  - \ln \left( \frac{2Z\alpha m_r \epsilon_{\star, \mu}}{n} \right) + \frac{12n^2 -n -9}{4n^2(n+1)}\right)  \right]  \notag \\
&& \qquad\qquad\qquad\qquad \qquad  \left. -     \left( \frac{g_\ssN m_\mu}{2M} \right) (Z\alpha) \epsilon_{\ssF, \mu}^2 X_\ssF   \right\}
\eea
which is of almost exactly the same form as in the electronic case, except for the term proportional to $\Xi_{nj\varpi}$ (defined in \eqref{Xidef}) which encodes the radiative corrections to finite-size effects due to electron vacuum polarization. Similarly
\bea
\label{mutofit-}
\delta \omega_{n \ssF \frac12 -}^{ \rm \ssN\ssS}(\mu) + \varepsilon^{\ssN-\QED}_{n\ssF \frac12 -}(\mu) &=&  2 \left( \frac{n^2-1}{n^5} \right) (Z\alpha)^4 \left( \frac{m_{r}}{m_\mu} \right)^2 m_{r}^3 \epsilon_{\star, \mu}^2   \\
&&\qquad + \frac{16}{3n^3}  \left[ \frac{\alpha(Z\alpha)^2}{\pi} \right]  \left( \frac{m_{r}}{m_\mu}\right)^2 m_{r}^3 \epsilon_{\star, \mu}^2\, \Xi_{n\frac{1}{2}-}. \nn
\eea

We follow ref.~\cite{aldo2013th} and define two useful combinations of the two measurements of \pref{muexp}, which help isolate the complications due to the electronic vacuum polarization.  The first linear combination of measurements in \cite{aldo2013th} is largely dominated by the hyperfine energy contributions and is useful for extracting $\epsilon_{\ssF, \mu}$ directly, much in the same way as was done for the $2S$ hyperfine splitting in the electronic case, leading to
\bea 
\nu_s - \nu_t &=& \left(\omega^{\rm pt}_{2 1 \frac{3}{2} +} - \omega^{\rm pt}_{2 0 \frac{1}{2} +} \right) - \left( \omega^{\rm pt}_{2 2 \frac{3}{2} +} - \omega^{\rm pt}_{2 1 \frac{1}{2} +} \right) \nn\\
&& \qquad\qquad\qquad  - \frac{8}{3} \left( \frac{m_{r}}{m_\mu} \right)^2 \left( \frac{m_\mu e \mu_p}{4\pi} \right) (Z\alpha)^2 m_{r}^3 \epsilon_{\ssF, \mu}^2  \,.
\eea
In this expression the point-nuclear theory terms combine into the hyperfine splitting combination for the $2S_{{1}/{2}}$ and the $2P_{{3}/{2}}$ states, motivating the definition
\be 
\widehat{\Delta \omega}_{hfs} := \left( \omega^{\rm pt}_{2 1 \frac{1}{2} +} - \omega^{\rm pt}_{2 0 \frac{1}{2} +} \right) - \left( \omega^{\rm pt}_{2 2 \frac{3}{2} +} - \omega^{\rm pt}_{2 1 \frac{3}{2} +} \right) - \Bigl(\nu_s - \nu_t \Bigr),
\ee
in terms of which a numerical value for $\epsilon_{\ssF, \mu}$ can be obtained, since
\be 
\label{epfmufit}
 \widehat{\Delta \omega}_{\scriptscriptstyle hfs} = \frac83 \left( \frac{m_{r}}{m_\mu} \right)^2 \left( \frac{m_\mu e \mu_p}{4\pi} \right) (Z\alpha)^2 m_{r}^3\,\epsilon_{\ssF, \mu}^2  \,.
\ee

For later convenience we record the numerical value for the point-nucleus theoretical expressions, as collected by \cite{aldo2013th}. For the transitions $\nu_s$ and $\nu_t$ of \pref{muexp} they are
\be
\label{num00}
\omega^{\rm pt}_{2 1 \frac{3}{2} +} - \omega^{\rm pt}_{2 0 \frac{1}{2} +} =  209.9450\, (26) \, \mathrm{meV}\quad\hbox{and} \quad
\omega^{\rm pt}_{2 2 \frac{3}{2} +} - \omega^{\rm pt}_{2 1 \frac{1}{2} +} =  229.6813\, (34) \, \mathrm{meV},
\ee
while for the hyperfine intervals one has
\be 
\label{num1}
\omega^{\rm pt}_{2 1 \frac{1}{2} +} - \omega^{\rm pt}_{2 0 \frac{1}{2} +} =  22.9858\, (26) \, \mathrm{meV}\quad\hbox{and} \quad
\omega^{\rm pt}_{2 2 \frac{3}{2} +} - \omega^{\rm pt}_{2 1 \frac{3}{2} +} =  3.2480\, (2) \, \mathrm{meV},
\ee
which also include the state-mixing $\delta$ contribution\footnote{This is a point-nucleus mixing of the $F=1$ levels for $j=\frac12$ and $j=\frac32$ that arises at second order in the nuclear magnetic field.} \cite{aldo2013th} that is part of $\varepsilon_{n\ssF j \varpi}^{(ho)}$. 

The second useful linear combination, $\frac{1}{4}(\nu_s + 3\nu_t)$ is defined so that the mixed hyperfine, finite-size effects tracked by the variable $\epsilon_{\ssF, \mu}$ cancel -- to the accuracy used here -- and hence for our purposes also allows for a direct fit of the $\epsilon_{\star, \mu}$ parameter. The contributions that survive in this second combination are separated by the authors of \cite{aldo2013th} into various point-like theory effects including the traditional $(2P_{\frac{1}{2}} -2S_\frac{1}{2})$ Lamb shift, and the nuclear-size dependent piece. In our present notation this second variable becomes
\bea
\label{epsmufit}
\frac{1}{4}\Bigl( \nu_s + 3\nu_t\Bigr) &=& \frac{1}{4} \left( \omega^{\rm pt}_{2 1 \frac{3}{2} +} - \omega^{\rm pt}_{2 0 \frac{1}{2} +} \right) + \frac{3}{4} \left( \omega^{\rm pt}_{2 2 \frac{3}{2} +} - \omega^{\rm pt}_{2 1 \frac{1}{2} +} \right)  \notag \\
&& - \left( \frac{m_{r}}{m_\mu} \right)^2 (Z\alpha)^2 m_\mu^3 \epsilon_{\star, \mu}^2 \left\{ 1 + (Z\alpha)^2 \left[  \frac{133}{48}-\gamma - H_3 - \ln\Bigl(Z\alpha \, m_{r} \epsilon_{\star, \mu} \Bigr)  \right]  \right. \notag \\
&&\qquad\qquad \bigg. + \frac{2\alpha}{3\pi }  \left(\Xi_{2\frac{1}{2} +} - \Xi_{2\frac{3}{2} +} \right) \bigg\} \,.
\eea

Some useful numerical values as transcribed from \cite{aldo2013th} with the help of \cite{karshenboim2015} are also quoted here for later use,
\bea
\label{num2}
\frac{1}{4} \left( \omega^{\rm pt}_{2 1 \frac{3}{2} +} - \omega^{\rm pt}_{2 0 \frac{1}{2} +} \right) + \frac{3}{4} \left( \omega^{\rm pt}_{2 2 \frac{3}{2} +} - \omega^{\rm pt}_{2 1 \frac{1}{2} +} \right) &=&  214.8791 \, (25) \, \, {\rm meV}, \notag \\
\hbox{and} \quad \frac{2\alpha }{3\pi } \left(\Xi_{2\frac{1}{2} +} - \Xi_{2\frac{3}{2} +} \right) &=& 0.0038556.
\eea
These motivate the following definition
\be 
\widehat{\Delta \omega}_{\scriptscriptstyle Lamb} := \frac{1}{4} \left( \omega^{\rm pt}_{2 1 \frac{3}{2} +} - \omega^{\rm pt}_{2 0 \frac{1}{2} +}  - \nu_s \right) + \frac{3}{4} \left( \omega^{\rm pt}_{2 2 \frac{3}{2} +} - \omega^{\rm pt}_{2 1 \frac{1}{2} +} - \nu_t \right),
\ee
which simplifies solving \eqref{epsmufit} for the value of $\epsilon_{\star, e}$. From here on in the argument proceeds much as for electrons, defining
\begin{equation}
\mathfrak{x}_\mu = z_\mu \Bigl( \mathfrak{y}_\mu - \ln z_\mu \Bigr), 
\end{equation}
with parameters  
\bea
\mathfrak{x}_\mu &:=& 2\left( \frac{m_\mu}{m_{r}} \right)^2 \frac{\widehat{\Delta \omega}_{\scriptscriptstyle Lamb}}{(Z\alpha)^4 m_{r}}, \notag \\
\mathfrak{y}_\mu &:=&  \frac{2}{(Z\alpha)^2}\left[ 1 + \frac{2\alpha}{3\pi }  \left(\Xi_{2\frac{1}{2} +} - \Xi_{2\frac{3}{2} +} \right) \right] + \frac{15}{8} -2\gamma  -2 \ln(Z\alpha)  \,,
\eea
leads to a solution involving the Lambert $W$-function
\be 
(m_{r} \epsilon_{\star, \mu})^2 = e^{\cW} \quad \hbox{where}
\quad \cW := W_{-1}\left(-\mathfrak{x}_\mu e^{-\mathfrak{y}_\mu}\right) + \mathfrak{y}_\mu \,.
\ee

This last equation, with \pref{epfmufit}, give the required solution for both $\epsilon_{\star, \mu}$ and $\epsilon_{\ssF, \mu}$ in terms of well-understood point-nucleus parts of the theory and experimental values. Using these in \pref{mutofit} and \pref{mutofit-} for other energy levels allows other transition energies to be computed without the usual nuclear uncertainties.  Predictions made in this way are completely independent of nuclear models and their associated inaccuracies.

\begin{table}
\centering
\begin{tabular}{|c|c|c|}
\hline
Transition  & $\nu_s - \nu_t$ & $\frac14(\nu_s + 3\nu_t)$ \\
\hline
Exp. value (meV)  & 19.5609 & 211.1829  \\
\hline
Exp. error (meV) & 0.0051 & 0.0023 \\
\hline
Pt. nucl. theory (meV)  & 19.7363 & 214.8791 \\
\hline
Pt. nucl. error (meV)  & 0.0030 & 0.0025\\
\hline
\hline
Parameter  &  $(m \epsilon_{F, \mu})^2$ & $(m \epsilon_{\star, \mu})^2$\\
\hline
Inferred value & $3.51 \times 10^{-6}$ & $9.0068 \times 10^{-7}$ \\
\hline
Prop. exp. error & $0.10 \times 10^{-6}$ & $0.0056 \times 10^{-7}$ \\
\hline
Prop. theory error  & $0.060 \times 10^{-6}$ & $0.0052 \times 10^{-7}$ \\
\hline
\end{tabular}
\caption{Measured transitions in muonic Hydrogen and linear combinations of these measurements (row 2) that are useful for fitting finite-size effects. The experimental errors are given in row 3, the point-nucleus theoretical contributions in row 4 and the errors in these in row 5. The parameters that we fit for are given in row 6, their fitted values are in row 7, and their uncertainty coming from the propagated experimental error are in row 8, while that coming from propagated point-nucleus theoretical errors are in row 9.}
\label{mat}
\end{table}

As an application of the predictivity of these techniques consider the planned measurements of the ground state hyperfine splitting experiment of muonic Hydrogen, whose precision is expected to be $\sim 10^{-4}$ meV \cite{kanda, schmidt}, and whose value is expected to provide the Zemach moment of the proton to a higher accuracy. As discussed above, to obtain this same theoretical accuracy using the techniques pursued here requires including higher-order terms than have so far been computed. We nonetheless predict here, for illustrative purposes, the nuclear contribution to this amplitude to the accuracy possible with the calculations given above, and find
\be \label{epsformnumb}
\delta \omega_{1 1 \frac{1}{2} +} ^{ \rm \ssN\ssS}(\mu) +\varepsilon^{\ssN-\QED}_{1 1 \frac{1}{2} +}(\mu) - \delta \omega_{1 0 \frac{1}{2} +}^{ \rm \ssN\ssS}(\mu) - \varepsilon^{\ssN-\QED}_{1 0 \frac{1}{2} +} (\mu) = -1.415(48) \, {\rm meV},
\ee
with the total uncertainty resulting from a net experimental error of $(0.041) \, {\rm meV}$, net point-like theoretical error of $(0.021) \, {\rm meV}$ and a truncation error of $(0.013) \, {\rm meV}$. For comparison, a similar calculation directly using \eqref{fshfs} and \eqref{muradmoments}, and simply quoting the values of (and errors for) nuclear moments (and the parameter $\Lambda$) estimated from nuclear models \cite{aldo2013th} instead gives  
\be \label{momformnumb}
\delta \omega_{1 1 \frac{1}{2} +} ^{ \rm \ssN\ssS}(\mu) +\varepsilon^{\ssN-\QED}_{1 1 \frac{1}{2} +}(\mu) - \delta \omega_{1 0 \frac{1}{2} +}^{ \rm \ssN\ssS}(\mu) - \varepsilon^{\ssN-\QED}_{1 0 \frac{1}{2} +} (\mu) = -1.385(47) \, {\rm meV},
\ee
which are consistent with comparable quoted errors. Even if one accepts that the errors in nuclear models used in \pref{momformnumb} are well-understood, because the errors in \pref{epsformnumb} are controlled only by experiments and theory calculations using point nuclei, they can improve dramatically as these are improved, without needing new approaches to nuclear theory.   

Although not yet at an accuracy of $10^{-4}$ meV for muonic Hydrogen, we regard the above exercise to be a proof of principle that nuclear-modelling uncertainties can be banished for muonic Hydrogen using essentially the same steps as for atomic Hydrogen.

\section{Summary and Outlook}
\label{conc}
 
To summarize, this paper extends earlier arguments based on first-quantized EFTs for spinless nuclei (PPEFTs) \cite{ppeft3, ppeftA} to include nuclear spin. The response of `bulk' electromagnetic fields and a Dirac lepton field to this point nucleus is computed in order to capture how nuclear structure alters leptonic energy levels. 

Spin is included by supplementing the nuclear center-of-mass coordinate, $y^\mu(\tau)$, with Grassmann (anti-commuting) classical variables, $\xi^\mu(\tau)$, that are also localized on the nuclear world-line. Once quantized, the Grassmann variables $\xi^\mu$ fill out a finite-dimensional quantum state space that represents spatial rotations (and thereby encodes the finite-dimensional space of nuclear spin-states). General EFT principles ensure that such a first-quantized effective action can capture the low-energy behaviour of {\it any} spinning nucleus provided one includes all possible interactions in the first-quantized nuclear effective action, subject to the other symmetries of the problem and unitarity. The effects of the effective nuclear couplings get transferred to electromagnetic and lepton degrees of freedom through a set of matching boundary conditions \cite{ppeft1, ppeft2, ppeft3}  that govern the behaviour of bulk modes in the near-nucleus regime $r = \epsilon \ll a_\ssB$, where $a_\ssB$ is the lepton's Bohr radius.

Experience with spinless nuclei \cite{ppeft3, ppeftA} shows that although there are many effective couplings (or nuclear moments) these only turn out to contribute to atomic energy shifts through a limited number of combinations. Ref.~\cite{ppeftA} showed that when computing atomic energy shifts for spinless nuclei of size $R$, and if one is expanding energies in a powers series in $Z\alpha$ and $R/a_\ssB \sim mRZ\alpha$, then up to and including effects of order $m^4 R^3 (Z\alpha)^5$ or $m^3 R^2 (Z\alpha)^6$ all nuclear moments contribute only through a single parameter, $\epsilon_\star$, that has dimensions of length. Although $\epsilon_\star$ can depend in a complicated way on nuclear moments -- typically with $\epsilon_\star \sim (Z\alpha) R$ -- the leptonic energies themselves are functions of these moments only through their dependence on $\epsilon_\star$. 

We here show that a similar statement also holds once nuclear spin is included. Working to the same order in $Z\alpha$ and $mRZ\alpha$, and also including similar sized nuclear-spin-dependent terms, shows that all nuclear moments appear in atomic energies only through two parameters, $\epsilon_\star$ and $\epsilon_\ssF$. We verify that we reproduce the explicit nuclear calculations in the literature (to the order we work) and provide explicit expressions for how these parameters depend on nuclear moments.

A technical issue that arises in these calculations concerns the divergences that one finds when computing matrix elements found when perturbing in the nuclear magnetic fields. These divergences arise because the presence of nonzero nuclear size makes modes external to the nucleus more singular near the origin. Strictly speaking this divergent behaviour stops once nuclear structure intervenes, but nuclear structure is not present to do so within the PPEFT formalism, wherein nuclei are replaced by point objects with many effective couplings. We show that sensible predictions can nonetheless be made, because the near-nucleus divergences can be renormalized into the values of the effective nuclear couplings. 

We have carried through this EFT program and applied it to compute nuclear effects in atomic Hydrogen. We correctly capture existing results for the energy shifts due to the charge radius, nuclear polarizabilities, Friar and Zemach moments and others, and thereby verify that these all contribute through only the two independent parameters $\epsilon_\star$ and $\epsilon_\ssF$, down to contributions at the $10^{-2}$ kHz order in atomic Hydrogen. By fitting these two parameters to two particularly well-measured transitions, we can predict the nuclear-size contributions to a large number of energy levels listed in \cite{kramida, fleurbaey2018, hessels2019}. Our uncertainties are independent of nuclear models, and are currently dominated by the precision with which pure QED corrections have been computed for point nuclei. Our errors are reduced by at least one order of magnitude compared to what is reported in \cite{hh} for such finite-size effects. Our results are summarised in Tables \ref{sat} and \ref{ahprecise} for the best measured transitions, and in Tables \ref{table2} through \ref{table4} for a broader class of transitions. 

We repeat the exercise for muonic Hydrogen, with results given in Table \ref{mat}. Again two parameters suffice to capture finite-size nuclear effects down to errors of order 0.01 meV, although this is not yet competitive with the accuracy of current (and upcoming) measurements. The required improvement is a straightforward extension of the methods used here, making them much easier to perform than are traditional nuclear methods. 

We remark that the same techniques apply equally well to nuclear-structure contributions to energy shifts for heavier and more complicated spinning nuclei such as deuterium, tritium, various helium isotopes, lithium and beryllium, with convergence of the low-energy EFT expansion expected to be quickest for those nuclei with the largest internal gap to exciting internal nuclear degrees of freedom.

In future work we hope to carry out a meta-analysis of available data for most low-Z electronic and muonic atoms and make predictions of the finite-size effects in transitions of these system that are relevant for future experiments \cite{schmidt, kanda} that are equally well unclouded by inaccuracies of nuclear moments as they only depend on non-finite-size theory and experimental measurements.

\section*{Acknowledgements}
We thank Marko Horbatsch, Eric Hessels, Krzysztof Pachucki and Randolph Pohl for helpful discussions. CB and PH thank the Mainz Institute for Theoretical Physics for its hospitality during the workshop {\it Precision Measurements and Fundamental Physics: The Proton Radius Puzzle and Beyond}. This work was partially supported by funds from the Natural Sciences and Engineering Research Council (NSERC) of Canada. Research at the Perimeter Institute is supported in part by the Government of Canada through NSERC and by the Province of Ontario through MRI. 

\newpage

\begin{table}[!h]
\centering

\begin{tabular}{|l|r|r|r|r|}
\hline
Transition & $\Delta E^{\rm fs}$ (kHz) & $\Delta E^{\rm exp}$ (kHz) & $\Delta E^{\rm th}$ (kHz) & $\Delta E^{\rm trunc}$ (kHz) \\
\hline
$2P_{1/2}^{F=1} - 2S_{1/2}^{F=0}$ &  $-143.70$ & 0.0069 & 0.57 & 0.00031  \\
\hline
$2P_{3/2}^{F=1} - 2S_{1/2}^{F=0}$ &  $-143.70 $ & 0.0069 & 0.57 & 0.00031  \\
\hline
$3P_{1/2}^{F=1} - 2S_{1/2}^{F=1}$ & $-136.480 $ & 0.0014 & 0.081 & 0.00010 \\
\hline
$3P_{1/2}^{F=0} - 2S_{1/2}^{F=1}$ &  $-136.480 $ & 0.0014 & 0.081 & 0.00010 \\
\hline
$3P_{3/2}^{F=2} - 2S_{1/2}^{F=1}$ &  $-136.481$ & 0.0014 & 0.081 & 0.00010 \\
\hline
$3P_{3/2}^{F=1} - 2S_{1/2}^{F=1}$ &  $-136.481$ & 0.0014 & 0.081 & 0.00010 \\
\hline
\hline
$8S_{1/2}^{F=1} - 2S_{1/2}^{F=1}$ &  $-134.348$ & 0.0014 & 0.080 & 0.00010 \\
\hline
$8D_{3/2}^{F=2} - 2S_{1/2}^{F=1}$ &  $-136.481$ & 0.0014 & 0.081 & 0.00010 \\
\hline
$8D_{3/2}^{F=1} - 2S_{1/2}^{F=1}$ & $-136.481$ & 0.0014 & 0.081 & 0.00010 \\
\hline
$8D_{5/2}^{F=3} - 2S_{1/2}^{F=1}$ &  $-136.481$ & 0.0014 & 0.081 & 0.00010 \\
\hline
$8D_{5/2}^{F=2} - 2S_{1/2}^{F=1}$ &  $-136.481$ & 0.0014 & 0.081 & 0.00010 \\
\hline
$10D_{5/2}^{F=3} - 2S_{1/2}^{F=1}$ &  $-136.481$ & 0.0014 & 0.081 & 0.00010 \\
\hline
$10D_{5/2}^{F=2} - 2S_{1/2}^{F=1}$ & $-136.481$ & 0.0014 & 0.081 & 0.00010 \\
\hline
$12D_{3/2}^{F=2} - 2S_{1/2}^{F=1}$ &  $-136.481$ & 0.0014 & 0.081 & 0.00010 \\
\hline
$12D_{3/2}^{F=1} - 2S_{1/2}^{F=1}$ &  $-136.481$ & 0.0014 & 0.081 & 0.00010 \\
\hline
$12D_{5/2}^{F=3} - 2S_{1/2}^{F=1}$ &  $-136.481$ & 0.0014 & 0.081 & 0.00010 \\
\hline
$12D_{5/2}^{F=2} - 2S_{1/2}^{F=1}$ & $-136.481$ & 0.0014 & 0.081 & 0.00010 \\
\hline
\hline
$3P_{1/2}^{F=1} - 3S_{1/2}^{F=0}$ &  $-42.58$ & 0.0020 & 0.17 & 0.000092 \\
\hline
$3P_{3/2}^{F=2} - 3S_{1/2}^{F=1}$ &  $-40.439$ & 0.00042 & 0.024 & 0.000031  \\
\hline
$3P_{3/2}^{F=2} - 3S_{1/2}^{F=0}$ & $-42.58$ & 0.0020 & 0.17 & 0.000092 \\
\hline
$3P_{3/2}^{F=1} - 3S_{1/2}^{F=1}$ &  $-40.439$ & 0.00042 & 0.024 & 0.000031 \\
\hline
$3P_{3/2}^{F=1} - 3S_{1/2}^{F=0}$ & $-42.58$ & 0.0020 & 0.17 & 0.000092 \\
\hline
$3D_{3/2}^{F=2} - 3S_{1/2}^{F=1}$ & $-40.439$ & 0.00042 & 0.024 & 0.000031 \\
\hline
$3D_{3/2}^{F=2} - 3S_{1/2}^{F=0}$ & $-42.58$ & 0.0020 & 0.17 & 0.000092 \\
\hline
$3D_{3/2}^{F=1} - 3S_{1/2}^{F=1}$ &  $-40.439$ & 0.00042 & 0.024 & 0.000031 \\
\hline
$3D_{3/2}^{F=1} - 3S_{1/2}^{F=0}$ & $-42.58$ & 0.0020 & 0.17 & 0.000092 \\
\hline
$3D_{5/2}^{F=2} - 3S_{1/2}^{F=0}$ & $-42.58$ & 0.0020 & 0.17 & 0.000092 \\
\hline
\end{tabular}
\caption{Finite-nuclear-size effects (column 2) with three sources of errors (columns 3--5) for Hydrogen transitions listed in ref.~\cite{kramida} that can be measured at the 0.01kHz level. See Table \ref{table3} (and main text) for more detailed descriptions of the column entries.}
\label{table2}
\end{table}

\newpage

%\newpage
\begin{table}[!ht]
\centering

\begin{tabular}{|l|r|r|r|r|}
\hline
Transition &  $\Delta E^{\rm fs}$ (kHz) & $\Delta E^{\rm exp}$ (kHz) & $\Delta E^{\rm th}$ (kHz) & $\Delta E^{\rm trunc.}$ (kHz)  \\
\hline
$4P_{1/2}^{F=1} - 4S_{1/2}^{F=1}$ &  $-17.060$ & 0.00018 & 0.010 & 0.000013 \\
\hline
$4P_{1/2}^{F=1} - 4S_{1/2}^{F=0}$ &  $-17.962$ & 0.00086 & 0.071 & 0.000039 \\
\hline
$4P_{1/2}^{F=0} - 4S_{1/2}^{F=1}$ &   $-17.060$ & 0.00018 & 0.010 & 0.000013 \\
\hline
$4P_{1/2}^{F=0} - 4S_{1/2}^{F=0}$ &  $-17.962$ & 0.00086 & 0.071 & 0.000039 \\
\hline
$4D_{3/2}^{F=2} - 4S_{1/2}^{F=1}$ &   $-17.060$ & 0.00018 & 0.010 & 0.000013 \\
\hline
$4D_{3/2}^{F=2} - 4S_{1/2}^{F=0}$ &  $-17.962$ & 0.00086 & 0.071 & 0.000039 \\
\hline
$4D_{3/2}^{F=1} - 4S_{1/2}^{F=1}$ & $-17.060$ & 0.00018 & 0.010 & 0.000013 \\
\hline
$4D_{3/2}^{F=1} - 4S_{1/2}^{F=0}$ &  $-17.962$ & 0.00086 & 0.071 & 0.000039 \\
\hline
$4P_{3/2}^{F=2} - 4S_{1/2}^{F=1}$ &  $-17.060$ & 0.00018 & 0.010 & 0.000013 \\
\hline
$4P_{3/2}^{F=2} - 4S_{1/2}^{F=0}$ &  $-17.962$ & 0.00086 & 0.071 & 0.000039 \\
\hline
$4P_{3/2}^{F=1} - 4S_{1/2}^{F=1}$ & $-17.060$ & 0.00018 & 0.010 & 0.000013 \\
\hline
$4P_{3/2}^{F=1} - 4S_{1/2}^{F=0}$ &  $-17.962$ & 0.00086 & 0.071 & 0.000039 \\
\hline
$4D_{5/2}^{F=3} - 4S_{1/2}^{F=1}$ &   $-17.060$ & 0.00018 & 0.010 & 0.000013 \\
\hline
$4D_{5/2}^{F=3} - 4S_{1/2}^{F=0}$ &  $-17.962$ & 0.00086 & 0.071 & 0.000039 \\
\hline
$4D_{5/2}^{F=2} - 4S_{1/2}^{F=1}$ &  $-17.060$ & 0.00018 & 0.010 & 0.000013 \\
\hline
$4D_{5/2}^{F=2} - 4S_{1/2}^{F=0}$ &  $-17.962$ & 0.00086 & 0.071 & 0.000039 \\
\hline
\hline
$5P_{1/2}^{F=1} - 5S_{1/2}^{F=1}$ &  $-8.7346$ & 0.000091 & 0.0052 & 0.0000066  \\
\hline
$5P_{1/2}^{F=1} - 5S_{1/2}^{F=0}$ & $-9.197$ & 0.00044 & 0.037 & 0.000020 \\
\hline
$5P_{1/2}^{F=0} - 5S_{1/2}^{F=1}$ &  $-8.7346$ & 0.000091 & 0.0052 & 0.0000066\\
\hline
$5P_{1/2}^{F=0} - 5S_{1/2}^{F=0}$ &  $-9.197$ & 0.00044 & 0.037 & 0.000020 \\
\hline
$5P_{3/2}^{F=2} - 5S_{1/2}^{F=1}$ &  $-8.7348$ & 0.000091 & 0.0052 & 0.0000066 \\
\hline
$5P_{3/2}^{F=2} - 5S_{1/2}^{F=0}$ &  $-9.197$ & 0.00044 & 0.037 & 0.000020 \\
\hline
$5P_{3/2}^{F=1} - 5S_{1/2}^{F=1}$ & $-8.7348$ & 0.000091 & 0.0052 & 0.0000066 \\
\hline
$5P_{3/2}^{F=1} - 5S_{1/2}^{F=0}$ &  $-9.197$ & 0.00044 & 0.037 & 0.000020 \\
\hline
\end{tabular}
\caption{More nuclear-size effects listed in \cite{kramida}. Column 2 gives the nuclear-finite-size contribution to the transition energy predicted by eq.~\pref{prediction}. Columns 3--5 give errors inherent in column 2: column 3 is the error due to uncertainty in $\msD_{\ssL}/\msC_{\ssL}$ due to measurement errors in the reference transitions; column 4 gives the uncertainty due to uncertainty in the unperformed parts of the calculation not associated with nuclear finite-size effects; column 5 is the error due to neglect of higher orders in $\exx$, $Z\alpha$ and $R/a_\ssB = mRZ\alpha$ beyond those given by green squares in Table \ref{table1}. Uncertainty in values for $\alpha$ and $Ry$ give errors significantly smaller than those listed.}
\label{table3}
\end{table}

\newpage

\phantom{+}
\vspace{5mm}

\begin{table}[htb!]
\centering
\begin{tabular}{|l|r|r|r|r|}
\hline
Linear combination of transitions & $\Delta E^{\rm fs}$& $\Delta E^{\rm exp}$& $\Delta E^{\rm th}$ & $\Delta E^{\rm trunc}$\\
  &  (kHz) &  (kHz) &  (kHz) & (kHz)\\
\hline
$\left(4P_{1/2}^{F=1} - 2S_{1/2}^{F=1} \right) -\frac14\left( 2S_{1/2}^{F=1} - 1S_{1/2}^{F=1} \right)$ & $102.35$ & 0.0029 & 0.16 & 0.00021 \\
\hline
$\left(4P_{1/2}^{F=0} - 2S_{1/2}^{F=1} \right) -\frac14\left( 2S_{1/2}^{F=1} - 1S_{1/2}^{F=1} \right)$ & $102.35$ & 0.0029 & 0.16 & 0.00021 \\
\hline
$\left(4S_{1/2}^{F=1} - 2S_{1/2}^{F=1} \right) -\frac14\left( 2S_{1/2}^{F=1} - 1S_{1/2}^{F=1} \right)$ & $119.41$ & 0.0028 & 0.16 & 0.00020  \\
\hline
$\left(4P_{3/2}^{F=2} - 2S_{1/2}^{F=1} \right) -\frac14\left( 2S_{1/2}^{F=1} - 1S_{1/2}^{F=1} \right)$ & $102.35$ & 0.0029 & 0.16 & 0.00021 \\
\hline
$\left(4P_{3/2}^{F=1} - 2S_{1/2}^{F=1} \right) -\frac14\left( 2S_{1/2}^{F=1} - 1S_{1/2}^{F=1} \right)$ & $102.35$ & 0.0029 & 0.16 & 0.00021 \\
\hline
$\left(4D_{5/2}^{F=3} - 2S_{1/2}^{F=1} \right) -\frac14\left( 2S_{1/2}^{F=1} - 1S_{1/2}^{F=1} \right)$ & $102.35$ & 0.0029 & 0.16 & 0.00021 \\
\hline
$\left(4D_{5/2}^{F=2} - 2S_{1/2}^{F=1} \right) -\frac14\left( 2S_{1/2}^{F=1} - 1S_{1/2}^{F=1} \right)$ & $102.35$ & 0.0029 & 0.16 & 0.00021 \\
\hline
$\left(6S_{1/2}^{F=1} - 2S_{1/2}^{F=1} \right) -\frac14\left( 3S_{1/2}^{F=1} - 1S_{1/2}^{F=1} \right)$ &$131.41$ & 0.0031 & 0.17 & 0.00022 \\
\hline
$\left(6D_{5/2}^{F=3} - 2S_{1/2}^{F=1} \right) -\frac14\left( 3S_{1/2}^{F=1} - 1S_{1/2}^{F=1} \right)$ & $126.36$ & 0.0031 & 0.18 & 0.00022 \\
\hline
$\left(6D_{5/2}^{F=2} - 2S_{1/2}^{F=1} \right) -\frac14\left( 3S_{1/2}^{F=1} - 1S_{1/2}^{F=1} \right)$ & $126.36$ & 0.0031 & 0.18 & 0.00022 \\
\hline
\end{tabular}
\caption{Contribution of nuclear-size effects and the errors in this prediction for specific linear combintions of transition energies (whose motivation comes from experimental considerations), as taken from \cite{kramida}, that are observable at the 0.001 kHz level. See Table \ref{table3} for more details on the definitions of each column.}
\label{table4}
\end{table}

\begin{appendices}

\section{Spin formalism}
\label{AppendixA}
 
This appendix summarizes the quantization procedure for the Grassmann fields, $\xi^\mu(s)$, and sketches the derivation of the final form for the nuclear action described in the main text.

 \subsection*{Quantization}
 
The free spinning particle has action
\begin{equation}
S = \int \exd s\, \mathcal{L} = - \int \exd s \, \left[ M \sqrt{-\dot{y}^2} + i \xi^\mu \dot{\xi}_\mu \right] \,,
\end{equation}
where the configuration variables are the bosonic coordinate $y^\mu(s)$ and the Grassmann variables $\xi^\mu(s)$. This proves to be a constrained system because the symmetries of the problem (such as reparameterization invariance along the world-line) imply that these variables and their canonical momenta are not independent.

As described in detail in \cite{henneaux} the canonical quantization procedure for systems with constraints proceeds as follows. First identify the conjugate momenta and the Hamiltonian, using  
\be
   \delta S = \int \exd s \; \Bigl( \delta \dot y^\mu \, p_\mu + \delta \dot \xi^\mu \, \pi_\mu \Bigr) \,,
\ee
and so 
\be 
p_\mu = \frac{\partial \mathcal{L}}{\partial \dot{y}^\mu} =  \frac{M\dot{y}_\mu}{\sqrt{-\dot{y}^2}}  \qquad \hbox{and} \qquad 
\pi_\mu = \frac{\partial \mathcal{L}}{\partial \dot{\xi}^\mu} = i\xi_\mu \,.
\ee
In principle one wishes to invert these expressions to write the velocities, $\dot y^\mu$ and $\dot \xi^\mu$ as functions of the momenta, and to use these to construct the Hamiltonian from the Lagrangian. For constrained systems, like the one considered here, this inversion cannot be done. For instance, for the Grassmann field the $\dot \xi^\mu$ does not even appear in $\pi_\mu$, while the bosonic momentum satisfies the identity
\begin{equation}
p^\mu p_\mu = -M^2 \,,
\end{equation}
(which is the correct dispersion relation for a relativistic massive particle). The inability to solve for velocities in terms of positions and momenta arises because the system's positions and velocities are related by the following two \textit{primary} constraints,
\be \label{AppA:constraints}
\phi_1 := p^2 + M^2 = 0, \qquad \hbox{and} \qquad
\Phi_\mu := \pi_\mu -i\xi_\mu = 0 \,.
\ee

It is useful to incorporate the primary constraints into the Lagrangian,
\begin{equation}
\mathcal{L}_c = -  M\sqrt{-\dot{y}^2} - i\xi^\mu \dot{\xi}_\mu - \theta \phi_1 - \Theta^\mu \Phi_\mu \,,
\end{equation}
where $\theta$ and $\Theta^\mu$ are Lagrange multipliers. The variation of $\cL$ with respect to $y^\mu$ and $\xi^\mu$ subject to the constraints \pref{AppA:constraints} is equivalent to the unconstrained variation of $\cL_c$ provided that the new variables $\theta$ and $\Phi_\mu$ are also varied. The Hamiltonian of this theory including the constraints is then:
\begin{eqnarray}
H_c &=& \dot{y}^\mu p_\mu + \dot{\xi}^\mu \pi_\mu - \mathcal{L}_c, \notag \\
&=& \theta \phi_1 + \Theta^\mu \Phi_\mu.
\end{eqnarray}

Primary constraints like \pref{AppA:constraints} need not exhaust all of the constraints because even if the primary constraints are imposed on any initial conditions, additional constraints might be necessary to ensure that \pref{AppA:constraints} remain true for all times. The time evolution of any function of canonical variables, $A(q, p, t)$, is given by
\begin{equation}
\frac{\exd A}{\exd t}  = \frac{\partial A}{\partial t} + \left( A, H \right)_P = 0,
\end{equation}
where $(\cdots , \cdots )_P$ denotes the Poisson bracket, defined for Grassmann even and odd variables by \cite{casalbuoni}:
\begin{eqnarray}
\left( E_1, E_2 \right)_{P} &=& \left( \frac{\partial E_1}{\partial q^\alpha} \frac{\partial E_2}{\partial p^q_\alpha} - \frac{\partial E_2}{\partial q^\alpha} \frac{\partial E_1}{\partial p^q_\alpha} \right) + \left( \frac{\partial E_1}{\partial \xi^\alpha} \frac{\partial E_2}{\partial \pi^\xi_\alpha} - \frac{\partial E_2}{\partial \xi^\alpha} \frac{\partial E_1}{\partial \pi^\xi_\alpha} \right), \notag \\
\notag \\
\left( E, O \right)_{P} &=& \left( \frac{\partial E}{\partial q^\alpha} \frac{\partial O}{\partial p^q_\alpha} - \frac{\partial O}{\partial q^\alpha} \frac{\partial E}{\partial p^q_\alpha} \right) + \left( \frac{\partial E}{\partial \xi^\alpha} \frac{\partial O}{\partial \pi^\xi_\alpha} + \frac{\partial O}{\partial \xi^\alpha} \frac{\partial E}{\partial \pi^\xi_\alpha} \right), \notag \\
\left( O, E \right)_{P} &=& \left( \frac{\partial O}{\partial q^\alpha} \frac{\partial E}{\partial p^q_\alpha} - \frac{\partial E}{\partial q^\alpha} \frac{\partial O}{\partial p^q_\alpha} \right) - \left( \frac{\partial O}{\partial \xi^\alpha} \frac{\partial E}{\partial \pi^\xi_\alpha} + \frac{\partial E}{\partial \xi^\alpha} \frac{\partial O}{\partial \pi^\xi_\alpha} \right), \notag \\
\notag \\
\left( O_1, O_2 \right)_{P} &=& \left( \frac{\partial O_1}{\partial q^\alpha} \frac{\partial O_2}{\partial p^q_\alpha} + \frac{\partial O_2}{\partial q^\alpha} \frac{\partial O_1}{\partial p^q_\alpha} \right) - \left( \frac{\partial O_1}{\partial \xi^\alpha} \frac{\partial O_2}{\partial \pi^\xi_\alpha} + \frac{\partial O_2}{\partial \xi^\alpha} \frac{\partial O_1}{\partial \pi^\xi_\alpha} \right) \,.
\end{eqnarray}
Any further constraints required to ensures that primary constraints hold for all times are called  \textit{secondary} constraints. 

For the constraints of \pref{AppA:constraints} we find:
\be 
\frac{\exd\phi_1}{\exd s} = (\phi_1, H)_P =  \theta \left( p^2 + M^2, p^2 + M^2 \right)_P + \Theta^\mu  \left( p^2 + M^2, \pi_\mu -i\xi_\mu \right)_P = 0 \,,
\ee
and
\bea 
\frac{\exd\Phi_\mu}{\exd s} &=& (\Phi_\mu , H)_P =  \theta \left( \pi^\mu - i \xi^\mu,  p^2 + M^2 \right)_P - \Theta^\nu  \left( \pi^\mu - i \xi^\mu, \pi_\nu -i\xi_\nu \right)_P \notag \\
&=& \Theta^\nu \left( -i \delta^\mu_\alpha \delta^\alpha_\nu - i \eta_{\beta \nu} \delta^\beta_\alpha \eta^{\mu\gamma} \delta_\gamma^\alpha \right)= -2i \Theta^\mu \,,
\end{eqnarray}
and so the evolution of the bosonic constraint yields no new restrictions while preservation of the fermionic constraint in time constrains the Grassmann Lagrange multiplier to vanish. 

The primary constraints have the following Poisson brackets with one another
\be
\left( \phi_1, \phi_1 \right)_P =  
\left( \phi_1, \Phi^\mu \right)_P =
\left( \Phi^\mu, \phi_1 \right)_P = 0 \quad \hbox{and} \quad
\left( \Phi_\mu, \Phi_\nu \right)_P = 2i \eta_{\mu\nu} \,.
\ee
Writing these constraints as a 5-component column vector, $\phi_\alpha = \{ \phi_1, \Phi_\mu \}$, these brackets can be arranged into a matrix,
\begin{equation}
\Delta_{\alpha\beta} := \left( \phi_\alpha, \phi_\beta \right)_P = \left[ \begin{array}{cc}
0 & \mathbf{0}^T \\
\mathbf{0} & 2i \eta_{\mu\nu} \end{array} \right] \,.
\end{equation}
Zero eigenvectors of this matrix are called first-class constraints, and are obstructions to the program of quantizing by using commutators to replace Dirac brackets, defined by 
\begin{equation}
\left(A, B \right)_D = \left( A, B \right)_P - \left( A, \phi_\alpha \right)_P \left( \Delta^{-1} \right)_{\alpha\beta} \left(\phi_\beta, B \right)_P \,.
\end{equation}

Zero eigenvectors are associated with local symmetries for which gauge conditions must be chosen as supplementary constraints. In the above example $\Delta$ is diagonal and so its only zero vector corresponds to the bosonic constraint $\phi_1$, corresponding to the freedom to redefine the world-line parameterization. This symmetry can be removed by choosing a gauge condition and checking its time evolution. The freedom to reparameterize time can be removed by fixing a coordinate condition like 
\begin{equation}
\varphi := y^0 - s= 0 \,,
\end{equation}
and the evolution of this new condition now fixes the final Lagrange multiplier, since
\begin{equation}
\frac{\exd \varphi}{\exd s} = -1 + 2\theta p^0 = 0 \,.
\end{equation}
With this choice the variable $y^0$ is no longer dynamical and only the spatial components of the position-vector need be quantized. Their conjugate momenta are
\begin{equation}
p_i = \frac{M\dot{y}_i}{\sqrt{1-\dot{y}^2}},
\end{equation}
which can now be inverted for the velocities:
\begin{equation}
\dot{y}^i = \frac{p^i}{\sqrt{p^ip_i + M^2}} \,.
\end{equation}

Finally, quantization proceeds by replacing Dirac brackets with commutators and anticommutators, so
\be
i\left( E_1, E_2 \right)_D \to \left[ \hat{E}_1, \hat{E}_2 \right] \,, \quad 
i \left( O, E \right)_D \to \left[ \hat{O}, \hat{E} \right] \quad \hbox{and} \quad
i \left( O_1, O_2 \right)_D \to \left\{ \hat{O}_1, \hat{O}_2 \right\} \,.
\ee
Using this for the variables $\left\{ y^i, \xi^\mu, p^i \right\}$ in the present instance leads to
\be 
\left[ \hat{x}^i, \hat{p}_j \right] = i \delta^i_j \qquad \hbox{and} \qquad
\left\{ \hat{\xi}^\mu, \hat{\xi}^\nu \right\} = - \frac{1}{2} \eta^{\mu\nu} \,,
\ee
as used in the main text.

\subsection*{Representations}
 
The bosonic commutators in the previous section are easily represented using position and derivative operators, but it remains to choose how to represent the anti-commutator. Defining $\hat{\xi}^\mu := \frac{i}{2} \Gamma^\mu$, we see that the anti-commutator goes over to the Clifford algebra,
\begin{equation}
\left\{ \Gamma^\mu, \Gamma^\nu \right\} = 2 \eta^{\mu\nu},
\end{equation}
and any representation of this Clifford algebra provides a quantization of the Grassmann fields. 

In the main text we work in the rest-frame of the nucleus, making it convenient to choose a basis for the matrices that make it simple to distinguish particles from anti-particles, and so use the $2n \times 2n$ matrices
\be \label{NuclearGamma}
\Gamma^0 = -i \left[ \begin{array}{cc} \mathds{1} & 0 \\
0 & -\mathds{1} \end{array} \right] \qquad \hbox{and} \qquad \Gamma^k = (-i) \left[ \begin{array}{cc} 0 & \tau^k \\ -\tau^k & 0 \end{array} \right]
\ee
and so defining $\Gamma_5 := -i \Gamma^0 \Gamma^1 \Gamma^2 \Gamma^3$ gives
\be 
\Gamma_5 \Gamma^k = (-i) \left[ \begin{array}{cc} \tau^k & 0 \\ 0 & -\tau^k \end{array} \right] 
 \quad \hbox{and} \quad
\Gamma_5 = - \left[ \begin{array}{cc} 0 & \mathds{1} \\ \mathds{1} & 0 \end{array} \right] \,,
\ee
while $\Gamma^{\mu\nu} := -\frac{i}{4} \left[ \Gamma^\mu, \Gamma^\nu \right]$ implies
\be 
 \Gamma^{0k} = \frac{i}{2} \left[ \begin{array}{cc} 0 & \tau^k \\ \tau^k & 0 \end{array} \right]  \quad \hbox{and} \quad \Gamma^{jl} = \frac{1}{2} \epsilon^{jlk} \left[ \begin{array}{cc} \tau^k & 0 \\ 0 & \tau^k \end{array} \right] \,.
\ee

In the above expressions $\mathds{1}$ denotes the $n \times n$ unit matrix and $\tau^i$ denotes the $n \times n$ representation of the rotation generators, whose choice determines how nuclear spin is represented. For spin-half nuclei the $\tau^k$ are Pauli matrices,
\begin{equation}
\tau^x = \left( \begin{array}{cc} 0 & 1 \\ 1 & 0 \end{array} \right)\,, \hspace{6pt} \tau^y = \left( \begin{array}{cc} 0 & -i \\ i & 0 \end{array} \right)\,, \hspace{6pt} \tau^z = \left( \begin{array}{cc} 1 & 0 \\ 0 & -1 \end{array} \right) \,, \hspace{6pt}
\end{equation}
while for spin-one nuclei the matrices
\begin{equation}
\tau^x_{(3)} = \frac{1}{\sqrt{2}}\left[ \begin{array}{ccc} 0 & 1 & 0 \\ 1 & 0 & 1  \\ 0 & 1 & 0 \end{array} \right], \hspace{6pt} \tau^y_{(3)} = \frac{1}{\sqrt{2}}\left[ \begin{array}{ccc} 0 & -i & 0 \\ i & 0 & -i \\ 0 & i & 0 \end{array} \right], \hspace{6pt} \tau^z_{(3)} = \left[ \begin{array}{ccc} 1 & 0 & 0 \\ 0 & 0 & 0 \\ 0 & 0 & -1 \end{array} \right], \hspace{6pt}
\end{equation}
are instead used, and so on. 

In this basis the particle and antiparticle states in the particle rest frame are given by
\begin{equation}
\ket{\psi} = e^{ip\cdot x} \left[ \begin{array}{cc} \alpha \\ \beta \end{array} \right],
\end{equation}
where $\alpha$ and $\beta$ represent the particle- and anti-particle solutions respectively. It is the state $\alpha$ for spin-half nuclei that appears in the main text, and for these only $\Gamma^0$ and $\Gamma^{jl}$ have nonvanishing matrix elements for nuclei at rest.

\subsection*{Comparison with second-quantized nuclei}

It is instructive to compare the first-quantized action found above with what a second-quantized field theory with two fermion species would yield. Let us write down the lowest order terms in each case, assuming the fermions are now also charged under electromagnetism.

The lowest-order second-quantized effective action that respects all the previously mentioned symmetries (and a U(1) gauge symmetry) for two charged fermions is:
\begin{equation}
S = - \int   \exd^4 x \, \left\{ \frac{1}{4} F_{\mu\nu} F^{\mu\nu} + \ol{\Psi} \left[ \slashed{D} + m_e \right] \Psi + \ol{\Phi} \left[ \slashed{D} + M \right] \Phi + a_\ssN \left( \ol{\Phi} \,\Gamma^{\mu\nu} \Phi \right) F_{\mu\nu} + \cdots \right\}, 
\end{equation}
where $\slashed{D}\Psi = \gamma^\mu \left( \partial_\mu + ieA_\mu \right) \Psi$ and $\slashed{D} \Phi =\Gamma^\mu \left( \partial_\mu - iZe A_\mu \right)\Phi$, for a nucleus with charge $+Ze$. This action contains two parameters for each fermion species, the mass and the electric charge, just as does the leading first-quantized action
\begin{equation}
S = - \int   \exd s \, \left\{ M \sqrt{-\dot{y}^2} + i \xi^\mu \dot{\xi}_\mu - q \dot{y}^\mu A_\mu - i \mu_\ssN \xi^\mu \xi^\nu F_{\mu\nu} + \cdots \right\},
\end{equation}
and it contains the same number of parameters.

Notice that writing $F_{jk} = \epsilon_{jkl} B^l$ for a magnetic field $\bfB$ turns the last term into
\be
- i \mu_\ssN \xi^j \xi^l F_{jl} =  - \frac{\mu_\ssN}{2} \Gamma^{jl} F_{jl} = - \frac{\mu_\ssN}{4} \, \epsilon^{jlk} \epsilon_{jlm} B^m \left[ \begin{array}{cc} \tau_k & 0 \\ 0 & \tau_k \end{array} \right] = \left[ \begin{array}{cc} -\bm{\mu} \cdot \bfB & 0 \\ 0 & -\bm{\mu} \cdot \bfB\end{array} \right] \,,
\ee
when the Hamiltonian is computed, confirming the identification of $\mu_\ssN$ as the nuclear magnetic moment (and once the magnetic-moment contribution is extracted from $\ol{\Phi} (\slashed{D} + M)\Phi$ it transpires that $a_\ssN$ contains the contribution $g-2$ to the nuclear magnetic moment $\mu_\ssN$).

\section{Fermionic boundary conditions}
\label{AppendixFBC}
 
In the text, the boundary condition \eqref{fermionbc1} is described as arising as in the classic delta-function potential: by integrating the fermion field equations over a sphere of radius $\epsilon$, and dropping all but the derivative and delta-function terms. This does not mean that it requires an explicit extrapolation of $\Psi$ right into the nucleus, however. Indeed, the PPEFT formalism is designed expressly to \textit{avoid} dealing with the physics in the core. Though qualitatively correct, the delta-function description is not really precisely defined. This appendix outlines the more detailed derivation of this boundary condition, following the discussion in Appendix A of \cite{ppeft1} (and fleshed out in \cite{ppeft2, ppeft3, falltocenter, ppeftA, EFTBook}), focussing specifically on the special issues that arise with first-order fermionic field equations.

Within the PPEFT approach used here all of the internal degrees of freedom for the nucleus are integrated out, leaving only the centre-of-mass position, $y^\mu(s)$, and spin, $\xi^\mu(s)$. These variables are regarded as collective coordinates: {\it i.e.} modes that appear in the low-energy theory because they are related to the action of Poincar\'e symmetries on the nuclear state (which in general is neither translation nor rotation invariant). The coupling between these two modes and the bulk fields given in the text is found by writing down the most general action that involves them all while properly realizing the symmetries, organized in an expansion in interactions of successively higher dimension -- eq.~\eqref{SpForm}. We reproduce the important interactions from that action for the electron field $\Psi$ here for convenience: 
\begin{eqnarray}
\label{eq:app:BCs:Sp}
S_p^\text{int} &=& - \int \exd s \; \ol{\Psi}(y(s)) \left[ \sqrt{-\dot{y}^2} \left( c_s + i c_2 \epsilon_{\alpha\beta\gamma\delta}\xi^\alpha \xi^\beta \xi^\gamma \xi^\delta \gamma_5 + i c_\ssF \xi^\mu \xi^\nu \gamma_{\mu\nu} \right) \right.  \notag  \\
			   && \qquad\qquad\qquad + i \dot{y}^\mu \left( c_v \gamma_\mu + c_3  \epsilon_{\alpha\beta\gamma\delta}\xi^\alpha \xi^\beta \xi^\gamma \xi^\delta \gamma_5\gamma_\mu \right) \Big] \Psi(y(s)) + \cdots \,.
\end{eqnarray}
In particular, $\Psi$ is evaluated `on the world-line of the nucleus', but in an EFT sense wherein spatial resolutions are limited to be only over distances  $L  \gg R$, where $R \sim 1$ fm is a representative size of the nucleus. Notice that to the dimensions of interest in this paper only terms bilinear in $\Psi$ are required, which simplifies the discussion because it allows the neglect of any two-  or higher-body contact interactions. 

The task is to make precise how the effective couplings in \pref{eq:app:BCs:Sp} can be translated into the correct near-nucleus behaviour of $\Psi$. To this end define the world-tube swept out by a ball $\mathcal{B}_\epsilon(y)$ of radius $\epsilon$ that is instantaneously centred on the nucleus. The radius of this ball is chosen so that $R \ll \epsilon \ll a_\ssB$ (where, as in the main text,  $a_\ssB$ is the electronic Bohr radius). In principle one could imagine specifying the value of $\Psi$ itself, or of its radial derivative, on the surface of this world-tube, but this is too prescriptive because the precise value of a bulk field at any particular position on this world-tube depends not only on the sources situated inside it, but also on any other sources or fields that are outside (though with an influence that falls off with that source's distance from the ball). What is sought is a construction that is dynamical, in that it can respond to the presence of all sources that play a role in the path integral. 

The required dynamical boundary condition is found by defining a `boundary' action, $I_\cB$, on the surface of a world-tube swept out by $\mathcal{B}_\epsilon(y)$, defined by the property that the path integral over $\Psi$, $y^\mu$ and $\xi^\mu$ exterior to $\cB_\epsilon(y)$ reproduces all of the results of the full theory, and thereby makes precise the implications of an action like \pref{eq:app:BCs:Sp}. The formulation of such an action is simplest in the limit where recoil corrections are neglected, because in this limit the position of $\cB_\epsilon(y)$ does not move. 

Concretely, writing the field as a sum over a basis of modes (as in the main text) $\Psi = \sum_\beta \Psi_\beta$, in the nuclear centre-of-mass frame the interactions in the boundary action required to work to the same accuracy as \pref{eq:app:BCs:Sp} is
\be
	\label{eq:app:BCs:Sb}
	I_\cB^\text{int} = \int \exd^2 \Omega \, \epsilon^2\sum_\beta \ol\Psi_\beta(\epsilon)\left( \hat c_s(\beta; \epsilon) - i\hat{c}_v(\beta; \epsilon)\gamma^0 + \hat{c}_F(\beta;\epsilon)\mathbf{I}\cdot\bm{\Sigma} \right)\Psi_\beta(\epsilon)
\ee
where we discard $c_2$ and $c_3$ as in section \ref{section:lepModes} (since they are not relevant for a nucleus at rest after projecting out the anti-particle solution), and as in the text $\mathbf{I} = \frac 12 \bm{\tau}$ and $\bm\Sigma$ satisfies $\gamma^{ij}=\epsilon^{ijk}\Sigma_k$. For applications to atoms we take $\beta = \{n, j, F, \ldots\}$ to run over the mode labels described in the main text. 

In principle there is an independent \textit{boundary} coupling, $\hat{c}_s, \hat{c}_v,$ and $\hat{c}_\ssF$, for every mode $\beta$ \cite{falltocenter}, and this is required because each eigenmode satisfies slightly different boundary conditions in the nuclear region. These all separately depend on $\epsilon$ because the boundary condition required to capture the effects of a nucleus depend on the size of the ball $\cB_\epsilon$ that is used. In general the couplings in \pref{eq:app:BCs:Sb} are found by matching to nuclear properties (as usual for EFTs), but for $S$-wave modes the connection between the couplings of \pref{eq:app:BCs:Sb} and \pref{eq:app:BCs:Sp} is simply given by dimensional reduction: schematically $4\pi \epsilon^2 \hat c_i = c_i$.

As usual the $\epsilon$-dependence of these couplings is chosen to ensure nothing physical depends on the value of $\epsilon$, and so changes in $\epsilon$ generate a renormalization-group (RG) flow amongst these couplings. What is important is that the RG-invariant parameters (like $\epsilon_\star$ and $\epsilon_\ssF$ of the main text, on which physical observables depend) do {\it not} depend on the mode label $\beta$, for the reasons described in more detail in Appendix \ref{ssec:AppEvoEq}. This independence of $\beta$ expresses the fact that the physical effective properties of the nucleus should not depend on the quantum numbers of the electrons that are used as probes.

The boundary conditions implied by the action $I_\cB$ are found when evaluating the path integral over $\Psi$, with the nucleus replaced by $I_\cB$. In a semiclassical evaluation this involves computing the saddle point, against which the total action is stationary against variations of $\Psi$ both away from and on the ball $\cB_\epsilon$. Stationarity with respect to variations that vanish at $\epsilon$ leads to the standard bulk field equations, with mode solutions as given in section \ref{section:lepModes}. Stationarity with respect to variations on the boundary $\cB_\epsilon$ then gives boundary conditions for each mode, of the form
\be
	\label{eq:app:BCs:mainBC}
	\Bigl[ \gamma^r + \hat c_s(\beta; \epsilon) - i\hat{c}_v(\beta; \epsilon)\gamma^0 + \hat{c}_F(\beta;\epsilon)\mathbf{I}\cdot\bm{\Sigma} \Bigr] \Psi_\beta(\epsilon) = 0,
\ee
where the $\gamma^r$ term comes from an integration by parts in the bulk action. 

For a second-order field equation (like the Schr\"odinger or Klein-Gordon equations discussed in \cite{ppeft1, ppeft2}) this would be the whole story, since the analog of \pref{eq:app:BCs:mainBC} then gives a relation between the field and its radial derivative at $r=\epsilon$. Interpreting \pref{eq:app:BCs:mainBC} is trickier for fermions because it is not a differential condition, and has nontrivial solutions for $\Psi_\beta(\epsilon)$ only if the matrix in the square brackets has a zero eigenvalue.
To see the implications of this observation consider how \eqref{eq:app:BCs:mainBC} constrains the radial eigenmodes found in the main text.  

\subsection*{Radial Boundary Conditions}
\label{app:BCs:rad}

For convenience, we restate here the eigenmodes given in \eqref{FModeFuncs}. We do so working in the same Dirac basis for leptons as for nucleons \pref{NuclearGamma}, reproduced here for ease of reference: 
 \begin{equation}
	\label{eq:app:BCs:defGammas}
	\gamma^0 = -i
	\begin{pmatrix} 
		\mathds{1}_2 & 0  \\ 0 & -\mathds{1}_2
	\end{pmatrix},
	\qquad \text{and} \qquad 
	\gamma^r = -i
	\begin{pmatrix} 
	0 & \sigma^r \\ -\sigma^r & 0
	\end{pmatrix},
\end{equation}
where $\mathds{1}_2$ is the $2\times2$ identity matrix. Writing $\Psi_\beta = e^{-i\omega_\beta t}\psi_\beta$, we define:
\begin{equation}
	\label{eq:app:BCs:FModeFuncs}
	\psi_{n\ssF j \varpi} = \left( \begin{array}{r} \mathcal{Y}_{F f_z}^{j, \varpi} \, \mff_{nj \varpi}(r) \\ i \mathcal{Y}_{F f_z}^{j, -\varpi} \, \mfg_{nj \varpi}(r) \end{array} \right) \,.
\end{equation}

These modes satisfy the useful identity for the action of $\sigma^r$,
\begin{equation}
	\label{eq:app:BCs:evalSigR}
	\sigma^r \mathcal{Y}^{j,\varpi}_{F,f_z} = -\mathcal{Y}^{j,-\varpi}_{F,f_z} \,.
\end{equation}
The action of $\mathbf{I} \cdot \boldsymbol{\Sigma}$ appearing in \pref{eq:app:BCs:mainBC}, restricted to a degenerate subspace with specific electronic angular momentum $j$, can be evaluated using the projection identity \cite{bs}:
\bea
	\label{eq:app:BCs:projId}
  \cZ_{\ssF j\varpi}. &:=& \Braket{\mathbf{I}\cdot\bm{\Sigma}} = \frac{\Braket{\mathbf{J}\cdot\mathbf{I}}\Braket{\mathbf{J}\cdot\bm{\Sigma}}}{\Braket{\mathbf{J}\cdot\mathbf{J}}},  \\
										&=& \frac 1{4j(j+1)} \left[ F(F+1)-I(I+1) - j(j+1) \right] \left[ j(j+1) - l (l +1) + s(s+1) \right], \notag \\
										  &=& \frac{1+\varpi(2j +1)}{8j(j+1)} \Bigl[ F(F+1) - j(j+1) - I(I+1)\Bigr]  \qquad \hbox{($s=\frac12$)} \notag \\
										&=& \left[ \frac{ 1+\varpi (2j +1)}{8} \right] X_\ssF \,. \nn
\eea
Here the first line defines the constant $\cZ_{\ssF j\varpi}$, the second-last line specializes to $s = \frac12$ and $l = j -  \frac12 \,\varpi$ and the last line uses the definition \pref{amatrix} of $X_\ssF$. Specialized to states with $j = \frac12$ this gives
\be
  \cZ_{\ssF\varpi} := \cZ_{\ssF \frac12 \varpi} = \frac{2\varpi + 1}{8} \; X_\ssF \,,
\ee
which for $I = \frac12$ becomes
\be
  \cZ_{\ssF\varpi}  = \frac{2\varpi + 1}{6}  \left[ F(F+1) - \frac32 \right] \,.
\ee

In general, $\mathbf{I}\cdot\boldsymbol{\Sigma}$ is not diagonalized by the states in \eqref{eq:app:BCs:FModeFuncs} because it turns out that $\mathbf{I} \cdot \bm{\Sigma}$ mixes the same opposite parity states as does the hyperfine interaction, \textit{i.e.}~states whose angular momentum quantum numbers only differ in their value for $j$. Consequently the boundary condition needs to be handled with care. However, for the $j = \frac12$ states relevant for this paper, the off-diagonal elements are suppressed by additional factors of $m R Z\alpha$ relative to the diagonal elements, and this puts them beyond the precision with which we work in this paper.\footnote{This suppression arises because the negative-parity $j =1/2$ Dirac-Coulomb mode-functions go as $\rho^{\zeta-1} \approx \rho^{0}$, which yields diagonal expectation values of $\langle \mathbf{I} \cdot \bm{\Sigma} \rangle_{\rm d} \sim (mRZ\alpha)^{2(j+1/2)-2} \sim 1$ on the $\mathbf{I} \cdot \bm{\Sigma}$ operator, but leads to matrix elements mixing this state with the positive-parity $j=3/2$ state that go as $\langle \mathbf{I} \cdot \bm{\Sigma} \rangle_{\rm off-d} \sim (mRZ\alpha)^{j+j'-1} \sim (mRZ\alpha)$. As such, this mixing effect arises at the next order in the $R/a_\ssB$ expansion of the EFT action and is therefore not considered here.} 

Restricted to the $j = \frac12$ eigenspace we may treat $\mathbf{I}\cdot\bm{\Sigma}$ as if it were diagonal in the basis \eqref{eq:app:BCs:FModeFuncs}. Using the identity in \eqref{eq:app:BCs:projId} the boundary condition \eqref{eq:app:BCs:mainBC} for $\Psi$ modes reads\footnote{It might seem unusual to assign an $F$-dependence to the Dirac mode functions, however this dependence arises because the integration constants $\msD/\msC$ differ for different $F$, as we see from the boundary condition derived below.}:
\begin{equation}
	\label{eq:app:BCs:BC12}
	\left( \begin{matrix} \hat{c}_s^{\varpi f} + \cZ_{\ssF , \varpi} \, \hat{c}_\ssF^{\varpi f}  -  \hat{c}_v^{\varpi f}  & -i\sigma^r  \\
	i\sigma^r & \hat{c}_s^{\varpi f}  + \cZ_{\ssF ,-\varpi} \, \hat{c}_\ssF^{\varpi f}  + \hat{c}_v^{\varpi f}   \end{matrix} \right) 
	\left( \begin{array}{r} \mathcal{Y}_{F f_z}^{\frac 12, \varpi} \, \mff_{n\ssF \frac 12 \varpi}(\epsilon) \\ i \mathcal{Y}_{F f_z}^{\frac 12, -\varpi} \, \mfg_{n\ssF \frac 12 \varpi}(\epsilon) \end{array} \right) = 0.
\end{equation}
The superscript on the couplings $\hat{c}_{s,v,\ssF}^{\varpi f}$ is meant as a reminder that they depend in principle on the mode's parity $\varpi = \pm$ and the atom's total spin\footnote{We use lower-case $f$ to denote dependence on nuclear spin $F$ due to the unfortunate notational choice that already uses capital $F$ to label the coupling $\hat c_\ssF$.} $F = j \pm \frac12 = 0, 1$.  

\subsubsection*{Coupling constraint}

For generic couplings the boundary condition \pref{eq:app:BCs:BC12} implies $\mathfrak{f}_{n\ssF\frac 12 \varpi}(\epsilon) = \mathfrak{g}_{n\ssF\frac 12 \varpi}(\epsilon) = 0$ whenever the pre-multiplying matrix is invertible. So having a nonvanishing spinor at $r=\epsilon$ requires the boundary couplings must satisfy 
\begin{equation}
	\label{eq:app:BCs:couplConst}
	1 + \left(\hat{c}_v^{\varpi f} - \frac \varpi3\left[ F(F+1) - \frac 32 \right]  \, \hat{c}_\ssF^{\varpi f} \right)^2 =  \left(\hat{c}_s^{\varpi f} + \frac 16\left[ F(F+1) - \frac 32 \right]\hat{c}_\ssF^{\varpi f}\right)^2 ,
\end{equation}
for both $F = 0$ and $F = 1$. This shows that the couplings $c_{s,v}^{\varpi f}$ and $c_\ssF^{\varpi f}$ are not all independent of one another.

This relationship amongst the effective couplings can be made explicit order-by-order in $\exx$, keeping in mind that $\hat c_\ssF^{\varpi f}$ starts at $\cO(\exx)$ while $\hat{c}_{s,v}^{\varpi, f} = \left( \hat{c}_{s,v}^{\varpi} \right)^{(0)} + \exx \,\left(\hat{c}_{s,v}^{\varpi, f} \right)^{(1)} + \cdots$. At $\cO(\exx^0)$ the leading constraint shows that $\left( \hat{c}_{s,v}^{\varpi} \right)^{(0)}$ satisfy the $F$-independent constraint found for spinless nuclei in \cite{ppeftA}  
\bea
1 = \left[ \left( \hat{c}_s^{\varpi} \right)^{(0)} \right]^2 - \left[ \left( \hat{c}_v^{\varpi} \right)^{(0)} \right]^2 \,,
\eea
while at first-order,
\be
  \exx \left[ \left(\hat{c}_s^{\varpi}\right)^{(0)}  \left( \hat{c}_s^{\varpi f} \right)^{(1)} - \left(\hat{c}_v^{\varpi}\right)^{(0)}   \left( \hat{c}_v^{\varpi f} \right)^{(1)} \right] = - \, \frac{\hat{c}_\ssF^{\varpi f}}{6} \left[ F(F+1) - \frac{3}{2} \right] \left[ \left(\hat{c}_s^{\varpi}\right)^{(0)}  + 2 \varpi  \left(\hat{c}_v^{\varpi}\right)^{(0)} \right] \,.
\ee
This last expression is consistent with $c_\ssF^\varpi$ being $F$-independent while $\left( \hat{c}_s^{\varpi f} \right)^{(1)}$ and $ \left( \hat{c}_v^{\varpi f} \right)^{(1)}$ are proportional to $F(F+1) - \frac32$.

\subsubsection*{Boundary condition}

To identify more explicitly the implications of the boundary condition for the radial functions, rewrite \pref{eq:app:BCs:BC12} as the two conditions
\bea
	\label{eq:app:BCs:BC12Eval}
	\left[ \left( \hat{c}_s^{\varpi f} + \cZ_{\ssF, \varpi} \,\hat{c}_\ssF^{\varpi f} -  \hat{c}_v^{\varpi f} \right) \mathfrak{f}_\varpi - \mathfrak{g}_\varpi  \right]\mathcal{Y}_{F f_z}^{\frac 12, \varpi} &=& 0 \\
	 \left[
	-i\mathfrak{f}_\varpi  + i \left( \hat{c}_s^{\varpi f} + \cZ_{\ssF, -\varpi} \,\hat{c}_\ssF^{\varpi f} + \hat{c}_v^{\varpi f} \right)\mathfrak{g}_\varpi  \right] \mathcal{Y}_{F f_z}^{\frac 12, -\varpi} &=& 0,\nn
\eea
which for brevity writes $\mathfrak{f}_\varpi := \mathfrak{f}_{n\ssF\frac 12 \varpi}(\epsilon)$ and $\mathfrak{g}_\varpi := \mathfrak{g}_{n\frac 12 \varpi}(\epsilon)$. Although this looks like two conditions for each choice of $\varpi$ and $F$, they are not independent because of condition \eqref{eq:app:BCs:couplConst}. The implications for the radial functions then are the ones used in eqs.~\pref{bc++} and \pref{bc--} of the main text:
\be 
	\label{eq:app:BCs:bothUsualBC}
	\left( \hat{c}_s^+ + \cZ_{\ssF}\, \hat{c}_\ssF^+ -  \hat{c}_v^+ \right)\mathfrak{f}_+ - \mathfrak{g}_+ = 0, \quad \text{and} \quad
	\left(\hat{c}_s^- +\cZ_{\ssF}\, \hat{c}_\ssF^- +  \hat{c}_v^- \right)\mathfrak{g}_- - \mathfrak{f}_- = 0,
\ee
where in both boundary conditions $\cZ_{\ssF} := \cZ_{\ssF +} = \frac 12 \left[ F(F+1) - \frac 32 \right]$.

 \section{Finite-size energy shift}
\label{AppendixC}
 
In this section we compute the finite-size energy shifts in terms of $\msD/\msC$, including its $(Z\alpha)^2$ corrections, which allows us to capture finite-size energy shifts in electronic atoms of $\cO \( m^3R^2 (Z\alpha)^6 \)$ magnitude.

\subsection*{Energy shift in the single-zero, single-pole approximation to $\mathcal{O} \left[( Z\alpha)^2 \right]$}

As described in the main text, the normalizability of the zeroth order wave-functions requires that the ratio of integration constants in the radial solutions satisfy:
\begin{equation}
\label{energyapp1}
-\left( \frac{\msD}{\msC} \right) = \frac{\Gamma\left[ 1 + 2\zeta \right] \Gamma\left[-\zeta - \frac{Z\alpha \omega}{\kappa} \right]}{\Gamma\left[1-2\zeta \right] \Gamma\left[ \zeta - \frac{Z\alpha\omega}{\kappa}\right]}.
\end{equation}
Our goal in this Appendix is to solve this equation for $\omega$ as a function of $\msD/\msC$, following the steps taken in \cite{ppeftA}. In particular, we seek solutions that are nearby to the standard Coulomb-Dirac expression for $\omega_\ssD$ and $\kappa_\ssD = \sqrt{m^2 - \omega_\ssD^2}$, for a point nucleus, \pref{Diracomega} and \pref{Diracomega2}, that are the solutions when $\msD/\msC = 0$.

Since $\zeta = \sqrt{\mfK ^2 - (Z\alpha)^2}$, it is close to the value of $|\mfK |$ and so $|\mfK | - \zeta \approx \mathcal{O} \left[\left( Z\alpha\right)^2 \right] \ll 1$. At the same time, the value of the poles of the gamma functions in the denominator are slightly shifted due to the nucleus having a finite-size, which we implement by taking $\omega = \omega^\ssD_{n j} + \delta \omega$ and it is $\delta \omega \ll \omega^\ssD_{n j} $ that this normalizability condition allows us to find as a function of the ratio of the integration constants. Additionally, as we have noted in the main text, the Dirac energies have the property that $\zeta - \frac{\omega^\ssD_{n j}}{\kappa_\ssD} = -N$, where $N$ is a non-negative integer, related to the principal quantum number, $n$, through 
\be
   N = n - |\mfK| \,. 
\ee
Then, to first order in $\delta \omega$ this combination without the subscripts becomes
\begin{equation}
\zeta - \frac{Z\alpha \omega}{\kappa} \approx -N - \left( \frac{Z\alpha m^2}{\kappa_\ssD^3} \right) \delta \omega,
\end{equation}
which allows us to write the condition in \eqref{energyapp1} as a function of the small quantities 
\be
   \delta y = 2|\mfK |-2\zeta  \simeq \cO\left[ \left( Z \alpha\right)^2 \right] 
   \quad \hbox{and} \quad \delta{x} = - \left( \frac{Z\alpha m^2}{\kappa_\ssD^3} \right) \delta \omega \,,
\ee
as 
\begin{equation}
\label{energyapp2}
-\left( \frac{\msD}{\msC} \right) = \frac{\Gamma\left[2|\mfK |+1 -\delta y \right] \Gamma\left[-(N+2|\mfK |) + \delta x + \delta y \right] }{\Gamma\left[ -(2|\mfK |-1) + \delta y \right] \Gamma\left[ -N + \delta x \right]}.
\end{equation}

Now, to capture the energy shift to an accuracy of $\mathcal{O} \left[\left( Z\alpha \right)^2 \right]$   we need to expand this formula for small $\delta x, \delta y$ and keep terms of order $\mathcal{O} \left( \delta x, \delta y, \delta x \delta y \right)$ and potentially $\cO \(\delta y^2 \)$ but not higher. In general, using the special property of gamma functions that $x\Gamma[x] = \Gamma[x+1]$ we can expand them around their poles in the following way,
\begin{eqnarray} \label{GammaPoleRes}
\Gamma\left[ -N + \delta z \right] &=& \frac{\Gamma\left[ -N +1 + \delta z \right]}{\left( -N + \delta z \right)} =  \frac{\Gamma\left[ 1 + \delta z \right]}{\left( -N + \delta z \right)\left(-N +1 + \delta z \right) \cdots \delta z}  \notag \\
&\approx& \frac{\Gamma\left[ 1\right] \left(1 + \delta z \frac{\Gamma'[1]}{\Gamma[1]} + \cdots\right)}{(-1)^N N! \,\delta z \left(1 - \delta z H_\ssN + \cdots \right)}\, ,
\end{eqnarray}
where the ellipses of \pref{GammaPoleRes} contain terms higher order in $\delta z$. Carrying out this expansion for each gamma function\footnote{Note that the arguments of the gamma functions in \eqref{energyapp2} that depend on $N$ in both numerator and denominator simultaneously approach negative integers in the limit $\delta x, \delta y \to 0$  and this necessitates expanding both gamma functions around their poles, hence the name ``single-zero, single-pole'' approximation.}  in \eqref{energyapp2} we end up with (after some algebra),
\bea
-\left( \frac{\msD}{\msC} \right) 
&=&- \frac{(2|\mfK |-1)! (2|\mfK |)! N! \left(\delta x \delta y \right) \left(1 - \frac{\delta y}{2|\mfK |} \left(1 + 4|\mfK | H_{2|\mfK |-1}\right) + \cdots \right)}{(N+2|\mfK |)! \left( \delta x \left[ 1 - 2 \delta y \(H_{N+2|\mfK|} + \gamma \) \right] + \delta y -  \delta y^2 \left[ H_{N+2|\mfK|} + \gamma \right] \right) },\nn\\
&&
\eea
which uses the identities:
\begin{equation}
\frac{\Gamma'[1]}{\Gamma[1]} = -\gamma, \hspace{12pt} \frac{\Gamma'[2|\mfK |+1]}{\Gamma[2|\mfK |+1]} = H_{2|\mfK |} -\gamma, \hspace{12pt} H_{2|\mfK |} + H_{2|\mfK |-1} = \frac{1}{2|\mfK |} \left( 1 + 4|\mfK | H_{2|\mfK |-1} \right). 
\end{equation}
Pulling out a factor of $\delta y$ from the denominator yields
\begin{equation}
\left( \frac{\msD}{\msC} \right) = \frac{(2|\mfK |-1)! (2|\mfK |)! N! \left(\delta x \right) \left[1 - \frac{\delta y}{2|\mfK |} \left(1 + 4|\mfK | H_{2|\mfK |-1}\right) + \cdots \right]}{(N+2|\mfK |)! \left(1 + \delta x \left[\frac{1}{\delta y} -2\left(H_{N+2|\mfK |} + \gamma \right)\right] - \delta y \left(H_{N+2|\mfK |} + \gamma \right) + \cdots \right)},
\end{equation}
and rearranging this for $\delta \omega$ (hidden inside $\delta x$) and writing it in terms of the principal quantum number $n$ gives the desired result for the finite-size energy shift to order $\mathcal{O} \left( Z\alpha^2 \right)$ as a function of the small parameter $\msD/\msC$:
\begin{equation}
\label{normshift}
\delta \omega \simeq - \frac{\kappa_\ssD^3 \mfB \left( {\msD}/{\msC} \right) \left[ 1 - \delta y \left(H_{n+|\mfK |} + \gamma \right) \right] }{m^2(Z\alpha)\left[1-  {\mfB\left({\msD}/{\msC} \right) \left[(\delta y)^{-1} -2 \left(H_{n+|\mfK |} + \gamma \right) \right]}  - \frac{\delta y}{2|\mfK |} \left(1 + 4|\mfK | H_{2|\mfK |-1} \right) + \cdots \right]} \,,
\end{equation}
where the ellipses represent terms that involve higher powers of $\delta x$  and
\be \label{mfBdef}
\mfB :=   \frac{(n+|\mfK |)!}{(n-|\mfK |)! (2|\mfK |)! (2|\mfK |-1)!}  \,.
\ee
 
 \section{Perturbing in the magnetic dipole}
\label{AppendixB}
 
This appendix derives the contributions to lepton-mode energy shifts due to the nuclear magnetic-dipole electromagnetic field. This is to be combined with the effects of finite-size nuclear effects in the main text. For the applications there we work to first order in the leptonic wave-functions, and
energy shifts.

\subsection*{Degenerate perturbation theory}

The discussion of the main text shows that the leptonic mode functions satisfy the equation of motion \pref{modeEQ}, reproduced here (after multiplying through by $i \gamma^0$) as
\begin{equation} \label{App:modeEQ}
\omega \psi = \left[ i\gamma^0 \boldsymbol{\gamma} \cdot \nabla + i\gamma^0 m +  eA_0^{\rm nuc}  \right] \psi - \left( e \gamma^0  \boldsymbol{\gamma} \cdot \bfA_{\rm nuc} \right) \psi \,,
\end{equation}
where $A_0^{\rm nuc}$ and $\bfA_{\rm nuc}$ are given by \pref{NucEMFs}, also reproduced here:
\be \label{App:NucEMFs}
   A_0^{\rm nuc} \simeq -\frac{Ze}{4\pi r} , \hspace{18pt} \bfA_{\rm nuc} \simeq \frac{\boldsymbol{\mu} \times \bfr}{4\pi r^3}  \,.
\ee

This is to be solved perturbatively in $\bfA_{\rm nuc}$. To do so we regard \pref{App:modeEQ} as a special instance of the eigenvalue condition,
\begin{equation}
\left[ H_0 + \lambda V \right] \ket{\psi_\ssA} = \omega_\ssA \ket{\psi_\ssA},
\end{equation}
with 
\be 
H_0 =   i\gamma^0 \boldsymbol{\gamma} \cdot \nabla + i\gamma^0 m +  eA_0^{\rm nuc}  
\quad \hbox{and} \quad
 V = -  e \gamma^0  \boldsymbol{\gamma} \cdot \bfA_{\rm nuc}  \,,
\ee
and $\lambda$ being a parameter that formally helps keep track of the order in $V$ (but that is set to unity at the end) \cite{sakurai, griffithsqm}. Seeking eigenstates and eigenvalues order by order in $\lambda$,
\be 
\omega = \omega^{(0)} + \lambda \omega^{(1)} + \lambda^2 \omega^{(2)} + \ldots \quad \hbox{and} \quad
\ket{\psi} = \ket{\psi}_{0} + \lambda \ket{\psi}_{1} + \lambda^2 \ket{\psi}_{2} + \ldots \,,
\ee
gives the hierarchy of conditions,
\begin{eqnarray}
\mathcal{O}(1):\quad\qquad\qquad H_0 \ket{\psi}_{0} &=& \omega^{(0)} \ket{\psi}_{0} \notag \\
\mathcal{O}\left(\lambda \right):\quad  H_0 \ket{\psi}_{1} + V\ket{\psi}_{0} &=& \omega^{(0)} \ket{\psi}_{1} + \omega^{(1)} \ket{\psi}_{0}, \notag \\
\mathcal{O}\left( \lambda^2 \right): \quad H_0 \ket{\psi}_{2} + V \ket{\psi}_{1} &=& \omega^{(0)} \ket{\psi}_{2} + \omega^{(1)} \ket{\psi}_{1} + \omega^{(2)} \ket{\psi}_{0} \,, \notag 
\end{eqnarray}
and so on.

\subsubsection*{Zeroth order}

The leading equation is:
\begin{equation}
 H_0 \ket{\psi}_{0} = \omega^{(0)} \ket{\psi}_{0} \,,
\end{equation}
which in the present instance is the Dirac-Coulomb equation, whose eigenvalues, $\omega^{(0)}_{nj} = \omega^\ssD_{nj}$ are given in \pref{Diracomega}, and whose eigenstates are labelled by the electronic principal, angular-momentum and parity quantum numbers described in the main text, and by the nuclear spin. That is, the zeroeth-order eigenstates are
\be
  | n j j_z \varpi \,; I I_z \rangle_0 =  |n j j_z \varpi  \rangle_0 \otimes |  I I_z \rangle_0 \,,
\ee
where $j = \frac12, \frac32, \cdots$ and $j_z = -j, -j+1,\cdots, j-1,j$ and $I_z = -I , -I +1, \cdots, I-1,I$ and $\varpi =\pm$.

Notice that these energy levels are degenerate, with $(2I+1)(2j+1)$ states distinguished by $j_z$ and $I_z$ sharing the same energy. This makes it necessary to use degenerate perturbation theory in what follows. That is, within any degenerate eigenspace a basis $|E, a \rangle_0$ of energy $E$ the zeroth-order eigenstates should be chosen to ensure that $V$ is diagonal: 
\be
   {}_0\langle E, b\,| V | E, a \rangle_0 = \cV(E,a) \delta_{ab} \,.
\ee
In practice the required basis are the states that are eigenstates of the total (combined nuclear and leptonic) angular momentum $\bfF = \bfJ + \bfI$ (see below for details). 

\subsubsection*{First order}

At first order in $\lambda$ the eigenvalue equation is
\begin{equation} \label{RaySchPt1}
\left[ H_0 - \omega^{(0)} \right] \ket{\psi}_{1}  =  \left[ \omega^{(1)}  - V \right] \ket{\psi}_{0}.
\end{equation}
Following the usual steps this implies the first-order energy shift for a state $|\omega^{(0)} ,a\rangle_0$ is
\be \label{eshift1}
\omega^{(1)}  = \cV(\omega^{(0)} ,a) = \frac{ {}_0\langle \omega^{(0)}, a | V | \omega^{(0)}, a \rangle_0}{{}_0\langle \omega^{(0)}, a |  \omega^{(0)}, a \rangle_0}\,,
\ee
and the corresponding zeroth-order energy eigenstate at this order is $| \omega^{(0)},a \rangle_0$. 

The first-order correction to this energy eigenstate implied by \pref{RaySchPt1} is then
\begin{equation}
\label{RaySchState1}
  \ket{\omega^{(0)}, a}_{1} = \bar{D} \frac{1}{\left[ \omega^{(0)} - H_0 \right]} \bar{D} V  \ket{\omega^{(0)},a}_{0}\,,
\end{equation}
where $\bar{D}$ denotes the projection matrix onto all zeroth-order states that are {\it not} degenerate with the original state $\ket{\omega^{(0)},a}_{0}$.

\subsubsection*{Eigenstates of total atomic spin}

Although the nuclear magnetic moment splits some of the degeneracy of Dirac-Coulomb levels, rotational invariance ensures that the resulting states retain a residual $(2F+1)$-dimensional degeneracy where $F$ is the total angular momentum quantum number for the entire atom (nucleus plus lepton): $\bfF = \bfJ + \bfI$.

This section writes these states out for the special case of a spin-half nucleus, as is relevant for our main application to muonic and atomic Hydrogen. We would like these functions to satisfy
\begin{align}\label{spinconds}
& \bfF^2 \mathcal{Y}^{j, \varpi}_{\ssF, f_z} = F(F+1) \mathcal{Y}^{j, \varpi}_{\ssF, f_z}, \hspace{16pt}  F_z \mathcal{Y}^{j, \varpi}_{\ssF, f_z} = f_z \mathcal{Y}^{j, \varpi}_{\ssF, f_z}, \hspace{16pt} \bfI^2 \mathcal{Y}^{j, \varpi}_{\ssF, f_z}= I(I+1) \mathcal{Y}^{j, \varpi}_{\ssF, f_z}, \notag \\
&\bfJ^2 \mathcal{Y}^{j, \varpi}_{\ssF, f_z} = j(j+1) \mathcal{Y}^{j, \varpi}_{\ssF, f_z}, \hspace{16pt}  \bfS^2 \mathcal{Y}^{j, \varpi}_{\ssF, f_z} = s(s+1) \mathcal{Y}^{j, \varpi}_{\ssF, f_z}, \hspace{16pt}  \bfL^2 \mathcal{Y}^{j, \varpi}_{\ssF, f_z} = l(l+1) \mathcal{Y}^{j, \varpi}_{\ssF, f_z}.
\end{align}
Any such state also diagonalizes $\bfI \cdot \bfJ$ and $\bfS \cdot \bfL$, with 
\bea
 2 \bfI \cdot \bfJ &=&  \Bigl( \bfI + \bfJ \Bigr)^2 - \bfI^2 - \bfJ^2 = F(F+1) - I(I+1) - j(j+1) \nn\\
 \hbox{and} \quad 2 \bfS \cdot \bfL &=&  \Bigl( \bfS + \bfL \Bigr)^2 - \bfS^2 - \bfL^2 = j(j+1) - s(s+1) - l(l+1) \,.
\eea

The electron spinor harmonics, $\Omega_{jlj_z\varpi}$ satisfy the last three of conditions \pref{spinconds} for $s = \frac12$, as well as $J_z \Omega_{jlj_z\varpi}  = j_z \Omega_{jlj_z\varpi}$. These are defined in eqs.~\pref{OmegaDef1} and \pref{OmegaDef2}, repeated here for convenience:
\begin{equation} \label{appOmegaDef}
\psi = \left( \begin{array}{c} \Omega_{jlj_z\varpi}(\theta,\phi) \; \mff_{nj}(r) \\ i\Omega_{jl'j_z\varpi}(\theta,\phi) \; \mfg_{nj}(r) \end{array} \right)  \quad \hbox{with} \quad
\Omega_{jlj_z\varpi} := \left( \begin{array}{c} \varpi \sqrt{\frac{l + \varpi\, j_z + \frac{1}{2}}{2l + 1}} \;Y_{l, j_z-\frac{1}{2}} (\theta, \phi) \\ \\  \sqrt{\frac{l - \varpi\, j_z + \frac{1}{2}}{2l+1}} \;Y_{l, j_z + \frac{1}{2}}(\theta, \phi) \end{array} \right) \,,
\end{equation}
where the left-hand equality gives the 4-component electron spinor -- in a basis for which $\gamma^0$ is diagonal, see \pref{eq:app:BCs:defGammas}  -- in terms of the 2-component electron spinor harmonics $\Omega_{jlj_z}$ defined in terms of ordinary scalar spherical harmonics in the right-hand equality. In the right-hand equality $\varpi = \pm 1$ is the parity quantum number and in the left-hand equality $l$ and $l'$ are related to $j$ and parity by $l =  j - \frac12 \varpi $ and $l' = j + \frac12 \varpi$. 

Similarly the nuclear $I$-states, 
\be \label{app:etadefs}
  \eta_{\frac{1}{2}, +\frac{1}{2}} = \left[ \begin{array}{c} 1 \\ 0 \end{array} \right] \quad \hbox{and} \quad
  \eta_{\frac{1}{2}, -\frac{1}{2}} = \left[ \begin{array}{c} 0 \\ 1 \end{array} \right] \,,
\ee
satisfy
\be
\bfI^2 \eta_{\ssI \ssI_z} = I(I+1) \eta_{\ssI \ssI_z} \qquad \hbox{and} \qquad
I_z \eta_{\ssI \ssI_z} = I_z \eta_{\ssI \ssI_z} \,.
\ee
We adopt the convention where square brackets denote nuclear-spin spinors while round brackets denote spinors in electron-spin space. 

In general, states of definite total spin are built from product states with given $j$ and $I$ by
\begin{equation}
\ket{F, f_z} = \sum_{j} \sum_{j_z} \sum_{I_z}  \braket{j,j_z;I, I_z | F, f_z } \ket{j,j_z ; I, I_z},
\end{equation}
for an appropriate set of Clebsch-Gordan coefficients, $\braket{j,j_z;I, I_z | F, f_z }$. For a spin-half nucleus, $I = \frac{1}{2}$, this reduces to \cite{sakurai},
\begin{eqnarray}
 \left| F = j \pm \frac{1}{2} ,f_z \right \rangle &=& \pm \sqrt{\frac{j+\frac{1}{2} \pm f_z}{2j + 1}} \;\left| j, f_z - \frac{1}{2}; \frac{1}{2}, +\frac{1}{2} \right\rangle + \notag \\
&& \qquad  + \sqrt{\frac{j+\frac{1}{2} \mp f_z}{2j + 1}} \;\left| j, f_z + \frac{1}{2}; \frac{1}{2}, -\frac{1}{2} \right \rangle \,.\nn\\
\end{eqnarray}

Using the explicit position-space representation given above the new basis of spinor harmonics with definite $F$ are the 4-component mixed electron/nuclear spin quantites
\begin{equation} \label{cYdef1}
\mathcal{Y}_{F = j \pm \frac{1}{2}, f_z}^{j, \varpi} = \pm \sqrt{\frac{j + \frac{1}{2} \pm f_z}{2j + 1}} \;\Omega_{jlf_z-\frac{1}{2},\varpi} \; \eta_{\frac{1}{2}, +\frac{1}{2}} +  \sqrt{\frac{j+\frac{1}{2} \mp f_z}{2j + 1}} \; \Omega_{jlf_z +\frac{1}{2},\varpi} \;\eta_{\frac{1}{2}, -\frac{1}{2}} \,,
\end{equation}
and so using \pref{app:etadefs} the explicit 4-component spinors with fixed $F$ are  
\begin{equation}\label{cYdef2}
\mathcal{Y}_{F = j +  \frac{\nu}{2}, f_z}^{j, \varpi} = \left[ \begin{array}{c}
\nu \sqrt{\frac{j + \frac{1}{2} + \nu f_z}{2j + 1}} \; \Omega_{j, l, f_z - \frac{1}{2},\varpi} \\
\sqrt{\frac{j + \frac{1}{2} -\nu f_z}{2j + 1}}\;  \Omega_{j, l, f_z + \frac{1}{2},\varpi} \end{array} \right],
\end{equation}
where $\nu = \pm$ corresponds to the choice for $F = j \pm \frac{1}{2} = j +  \frac\nu2$ and $\varpi$ is the parity of the electron spinor harmonic, and $l =  j - \frac12 \varpi $.

As a concrete example, consider $F=1, f_z = 0, \pm1$ and $j = \frac12$ states with positive and negative parity, for which the above give the explicit positive-parity ($S$-wave) angular functions,
\be 
\mathcal{Y}_{1, 0}^{\frac{1}{2}, +}  = \frac{1}{\sqrt{2}} \left[ \begin{array}{c}
\Omega_{\frac{1}{2}, 0, -\frac{1}{2},+} \\
\Omega_{\frac{1}{2}, 0, + \frac{1}{2},+} \end{array} \right], \hspace{12pt} \mathcal{Y}_{1, +1}^{\frac{1}{2}, +} =  \left[ \begin{array}{c}
\Omega_{\frac{1}{2}, 0, \frac{1}{2},+} \\
0 \end{array} \right], \hspace{12pt} \mathcal{Y}_{1, -1}^{\frac{1}{2}, +} = \left[ \begin{array}{c}
0 \\
\Omega_{\frac{1}{2}, 0, - \frac{1}{2},+} \end{array} \right] \,,
\ee
while the negative-parity ($P$-wave) states instead are
\be 
\mathcal{Y}_{1, 0}^{\frac{1}{2}, -} = \frac{1}{\sqrt{2}} \left[ \begin{array}{c}
\Omega_{\frac{1}{2}, 1, -\frac{1}{2},-} \\
\Omega_{\frac{1}{2}, 1, + \frac{1}{2},-} \end{array} \right], \hspace{12pt} \mathcal{Y}_{1, +1}^{\frac{1}{2}, -} =  \left[ \begin{array}{c}
\Omega_{\frac{1}{2}, 1, \frac{1}{2},-} \\
0 \end{array} \right], \hspace{12pt}  \mathcal{Y}_{1, -1}^{\frac{1}{2}, -} = \left[ \begin{array}{c}
0 \\
\Omega_{\frac{1}{2}, 1, - \frac{1}{2},-} \end{array} \right] \,.
\ee
The orthonormality of the spherical spinors and of the nuclear spin states,
\begin{equation}
\int \exd^2 \Omega_2 \, \, \Omega_{j', l', j'_z}^\dagger \Omega_{j, l, j_z} = \delta_{jj'} \delta_{ll'} \delta_{j_z j'_z }\qquad \hbox{and} \qquad \eta_{\ssI, \ssI_z}^\dagger \eta _{\ssI', \ssI'_z} = \delta_{\ssI_z, \ssI'_z}
\end{equation}
ensure the above spinor harmonics are orthonormal
\begin{equation}
\int \exd^2 \Omega_2 \, \, \left( \mathcal{Y}_{\ssF', f'_z}^{j', \varpi'} \right)^\dagger \mathcal{Y}_{\ssF, f_z}^{j, \varpi} = \delta_{\ssF \ssF'} \delta_{jj'} \delta_{f'_z f_z} \delta_{\varpi\varpi'} \,.
\end{equation}
 
 \section{Evaluation of matrix elements}
\label{ssec:MatrixElementApp}
 
This section evaluates the radial integrals that arise when evaluating the magnetic-moment contributions to energy shifts. Some of these integrals diverge due to singularities in the integrands as $r \to 0$, and for these we also evaluate the regularization procedure we use when separating out the divergent and finite parts. The divergences all have a specific dependence on the principal quantum number $n$, that is consistent with their being renormalized into shifts of the effective coupling $\hat c_\ssF$. As a result the main content of the finite contributions is restricted to those terms that depend differently on $n$ than do the divergent ones.

We consider in turn the integrals associated with both the first-order energy shift and the first-order state change.

\subsection*{Energy shift}
\label{Appssec:EShift}

The first order energy shift due to the nuclear dipole field is given by \pref{E1partial2},
\be
\label{E1partial2App}
 \varepsilon^{(1)}_{n \ssF j \varpi} =  - \frac{\mfK\, \exx X_\ssF}{m} \frac{(2\kappa)^3}{2m} \left( \frac{\mfN}{\mfD} \right) =  - 4\exx \mfK \, X_\ssF \frac{ \kappa^3}{m^2} \left( \frac{\mfN}{\mfD} \right) = -4\exx \mfK\, X_\ssF \,m\left( \frac{\za}{\cN} \right)^3 \left( \frac{\mfN}{\mfD} \right) \,,
\ee
where
\be  \label{Diracomega2:app}
  \cN = n \sqrt{1 - \frac{2(n-|\mfK  |) (Z\alpha)^2}{n^2(\zeta + |\mfK  |)}} \to n \sqrt{1 - \frac{2(n-1) (Z\alpha)^2}{n^2 (\zeta + 1)}} \,,
\ee
and
\be \label{exxdef:app}
  \exx := \frac{m e \mu_\ssN}{4\pi}  \to  \frac{\za }{2} \left( \frac{m }{M} \right)  g_p  \,,
\ee
where $g_p$ is the proton $g$-factor and
\bea
\label{amatrix:app}
   X_\ssF &:=&     \frac{F(F+1) - j(j+1) - I(I+1)}{j(j+1)} \\
   &=& \left\{ {(j+1)^{-1}\quad \hbox{if $F = j+\frac12$}\atop  - j^{-1} \quad \hbox{if $F = j-\frac12$}} \right. 
   \to  \left\{ { 2/3 \quad \hbox{if $F = 1$}\atop  - 2 \quad \hbox{if $F = 0$}} \right. \,. \nn
\eea
In these expressions the arrows specialize to the positive parity $\varpi = +$, $j = \frac12$ states of Hydrogen. 

The numerator and denominator functions are obtained as matrix elements of the interaction Hamiltonian (as described in the main text) and so contain the integrals we seek to evaluate. They both depend on the integration constant ratio, $\msD/\msC$, and so can be written
\bea
\label{NumDenom}
  \mfN &=& \mfN_{\rm pt} + \left( \frac{\msD}{\msC} \right) \mfN_1 +  \left( \frac{\msD}{\msC} \right)^2 \mfN_2 \nn\\ 
  \mfD &=& \mfD_{\rm pt} + \left( \frac{\msD}{\msC} \right) \mfD_1 +  \left( \frac{\msD}{\msC} \right)^2 \mfD_2 \,.
\eea
where eq.~\pref{ca+0def} gives the integration constant for parity-even $j=\frac12$ states as
\bea  \label{App:ca+0def}
\left( \frac{\msD_+}{\msC_+} \right)^{(0)}  &\simeq& -\frac{16y_{\star+}(m\epsilon_{\star+})^2}{n(n+1)} \left( \frac{2Z\alpha m\epsilon_{\star+}}{n} \right)^{2\zeta-2} + \cdots\nn\\
&\simeq& - \frac{\mfc}{n(n+1)} + \cO[(\za)^2] \,.
\eea
which defines $\mfc = 16y_{\star+}(m\epsilon_{\star+})^2$. Since matching reveals that $\epsilon_{\star+} \sim \cO(R\za)$ where $R \sim 1$ fm is a typical nuclear scale, we see that $\mfc \sim  \cO[(m R \za)^2]$. In particular $\mfN_{\rm pt}/\mfD_{\rm pt}$ is revealed to be the point-nucleus contribution to the hyperfine energy (which provides a useful check). 

The functions $\mfN_{\rm pt}$, $\mfN_1$, $\mfN_2$, $\mfD_{\rm pt}$, $\mfD_1$ and $\mfD_2$ are given in terms of the following class of integrals, that the rest of this section evaluates in detail:
\be \label{AppIntDef}
\cI_{\mathfrak{i} \mathfrak{j}}^{(p)} := \int_0^\infty \exd \rho \, e^{-\rho} \rho^{p} \cM_\mathfrak{i} \cM_\mathfrak{j} \,,
\ee
where we use the notation $\cM(a;b;z) := {}_1F_1(a;b;z)$ for confluent hypergeometric functions and the integrands are
as given in \pref{hypergeos} 
\begin{eqnarray}\label{hypergeosApp}
&&\mathcal{M}_1 :=  \mathcal{M}\left(a, b; \rho \right)\,, \;\; \mathcal{M}_2 :=  \mathcal{M} \left( a+1, b;\rho \right)\,,\nn\\ \;\;
&&\mathcal{M}_3 :=  \mathcal{M}\left( a', b'; \rho\right)\,, \;\;  \mathcal{M}_4 := \mathcal{M}\left( a'+1, b'; \rho \right) \,.
\end{eqnarray}
with parameters defined as in \pref{hyperparam}
\begin{eqnarray}
\label{hyperparam0}
a &:=& \zeta-\frac{\zas  \omega}{\kappa}, \hspace{12pt} a' :=  -\left( \zeta + \frac{\zas  \omega}{\kappa} \right), \hspace{12pt} b := 1+ 2\zeta, \hspace{12pt} b' := 1-2\zeta, \hspace{12pt}  \notag \\
c &:=& \mfK  - \frac{\zas   m}{\kappa}, \quad \rho := 2\kappa r, \hspace{24pt} \kappa := \sqrt{m^2 - \omega^2}, \hspace{24pt} \zeta := \sqrt{\mfK ^2 - (\za )^2} \,,
\end{eqnarray}
and $\mfK = - \varpi(j + \frac12)$ where $\varpi = \pm 1$ is the state's parity. 

Our main interest is in $j=\frac12$ and $\varpi = +1$ states for which $\mfK = -1$. For this choice we have 
\be
   \zeta =\sqrt{1 - (\za )^2} = 1 + \mfz  
\ee
where $\mfz$ is order $(\za)^2$. Similarly, for bound states one has
\be
   \kappa = \sqrt{m^2 - \omega^2} =\frac{\zas m}{\cN} \simeq \frac{\zas m}{n} + \cO[(\za)^2] \,,
\ee
where $n = 1,2,...$ is the principal quantum number, and so 
\be
   \frac{\zas\omega}{\kappa} = n + \lambda \quad \hbox{and} \quad
   \frac{\zas m}{\kappa} = n + \mu 
\ee
 with $\lambda$ and $\mu$ also order $(\za)^2$. In this regime the parameters of \pref{hyperparam0} are
\begin{eqnarray}
\label{hyperparamApp}
&& a = 1-n + \mfz - \lambda \,, \quad
a' = -1-n - \mfz - \lambda  \,, \quad
 b  = 3 +2 \mfz \nn\\
&&\qquad\qquad b' = -1 - 2\mfz \quad\hbox{and} \quad
c  = -1-n  - \mu \,.
 \end{eqnarray}
 
 In terms of these quantities we have
\bea \label{Nforms}
 \mfN_{\rm pt}  &=& \cI_{11}^{(2\zeta-2)} - \left( \frac{a}{c}\right)^2 \cI_{22}^{(2\zeta-2)}  \nn\\
  \mfN_1 &=& 2  \left[ \cI_{13}^{(-2)} - \left(\frac{aa'}{c^2}\right) \cI_{24}^{(-2)} \right] \\
 \mfN_2 &= &   \cI_{33}^{(-2\zeta-2)} - \left( \frac{a'}{c} \right)^2  \cI_{44}^{(-2\zeta-2)}  \,,\nn
\eea
and
\bea \label{Dforms}
 \mfD_{\rm pt} &=& 2\cI_{11}^{(2\zeta)} - \frac{4\omega}{m}  \left( \frac{a}{c} \right) \cI_{12}^{(2\zeta)} + 2\left( \frac{a}{c} \right)^2 \cI_{22}^{(2\zeta)} \nn\\
 \mfD_1 &=& 2  \left[ 2\cI_{13}^{(0)} -  \frac{2\omega}{m}  \left( \frac{a'}{c} \right) \cI_{14}^{(0)} -  \frac{2\omega}{m}  \left( \frac{a}{c} \right) \cI_{23}^{(0)} +  2\left(\frac{aa'}{c^2} \right) \cI_{24}^{(0)} \right]   \\
  \mfD_2 &=&  2  \left[ \cI_{33}^{(-2\zeta)}  - \frac{2\omega}{m}  \left( \frac{a'}{c} \right)  \cI_{34}^{(-2\zeta)} + \left( \frac{a'}{c}\right)^2 \cI_{44}^{(-2\zeta)} \right]  \,.\nn
\eea

\subsubsection*{Hypergeometric facts}

The relevant integrands depend on generalized hypergeometric functions, some of whose definitions and properties -- taken from \cite{slater}, \cite{gr} and online libraries such as \cite{dlmf}  -- are summarized here. These functions are defined within the domain of convergence by the following infinite series
\bea
\label{chfseries}
 {}_{\ssA} \cF_{\ssB} \left[  {a_1, \cdots a_{\ssA} \atop  b_1 \cdots b_{\ssB}} ;  z \right] &:=& \sum_{k=0}^\infty \frac{(a_1)_k \cdots  (a_{\ssA})_k}{ (b_1)_k \cdots (b_\ssB)_k} \; \frac{z^k}{k!} \nn\\
 &=& 1 + \frac{a_1 \cdots a_\ssA}{b_1 \cdots b_\ssB} \; z +  \frac{a_1(a_1 + 1) \cdots a_\ssA(a_\ssA+1)}{b_1(b_1+1) \cdots b_\ssB(b_\ssB+1)} \; \frac{z^2}{2} + \cdots
\eea
and are extended to general complex $z$ by analytic continuation. Here $(a)_\mathfrak{i}$ are the Pochhammer symbols defined by 
\be \label{PochDef}
   (a)_\mathfrak{i}:= a(a+1) \cdots (a+\mathfrak{i}-1) \quad \hbox{when} \quad \mathfrak{i} \geq1 \,,
\ee
and $(a)_0 :=1$. 

These definitions show that if any of the $a$-type arguments is a non-positive integer then the series expansion terminates after a finite number of terms. Similarly, the coefficients of the series are not well-defined if any of the $b$-type arguments is a non-positive integer. Both of these situations actually arise in Dirac-Coulomb wave-functions, which involve confluent hypergeometric functions (given by the special case $\cA = \cB =1$). In particular 
\be\label{-1result}
  {}_3\cF_2\left[{a, b, -1 \atop c, d}; \rho \right] =  1 - \left( \frac{ab}{cd} \right)  \rho 
  \quad\hbox{and so} \quad
  {}_3\cF_2\left[{a, b, -1 \atop c, d}; 1 \right] = \frac{cd-ab}{cd} \,.
\ee
Also notice that
\be
  {}_3\cF_2\left[{a, b, d \atop c, d}; \rho \right] = {}_2\cF_1\left[{a , b \atop c }; \rho \right] = {}_2\cF_1(a,b;c;\rho)\,,
\ee
is a standard hypergeometric function, and so
\be \label{hyper21at1}
   {}_3\cF_2\left[{a, b, d \atop c, d}; 1 \right]  =  {}_2\cF_1\left[{a , b \atop c }; 1 \right] = {}_2\cF_1(a,b;c;1) = \frac{\Gamma(c) \Gamma(c-a-b)}{\Gamma(c-a) \Gamma(c-b)}   \,.
\ee
Strictly speaking the last equality only holds for $\mathfrak{R}(c-a-b) > 0$, and is defined for other values by analytic continuation.

\subsubsection*{Basic integrals}

Because $\cM(a;c;\rho) \to 1$ as $\rho \to 0$, the basic integral of interest, \pref{AppIntDef}, converges at $\rho = 0$ if Re $p > -1$. Although the exponential factor $e^{-\rho}$ might seem to ensure automatic convergence as $\rho \to \infty$, the large-$\rho$ asymptotic expansion 
\be
  \cM(a;b;\rho) \sim \frac{e^\rho \, \rho^{a-b}}{\Gamma(a)} \sum_{k=0}^\infty \frac{(1-a)_k (1-b)_k}{k!} \; \rho^{-k} \,,
\ee
shows that convergence actually depends on the values of $a$, $b$, $a'$ and $b'$ and $p$.

The integral can be evaluated by expanding one of the hypergeometric functions and formally integrating term-by-term \cite{slater}:
\bea \label{type1a}
  \cI_d(a,b;a',b') &:=&  \int_0^\infty \exd \rho \, e^{-\rho} \rho^{d-1} \cM\left[a; b; \rho \right] \cM \left[a'; b';\rho \right], \notag \\
   &=& \frac{\Gamma(b')}{\Gamma(a')} \sum_{k=0}^\infty \frac{\Gamma(a'+k)}{k!\, \Gamma(b'+k)} \int_0^\infty \exd \rho\; e^{-\rho} \rho^{d-1+k} \cM(a,b;\rho) \nn\\
   &=& \frac{\Gamma(d) \Gamma(b') \Gamma(b' - d - a') }{\Gamma(b' - a') \Gamma(b' - d)} \, {}_3\cF_2\left[{a, d, 1 + d - b' \atop b, 1+d+a'-b'}; 1 \right] \\
&=& \frac{\Gamma(d) \Gamma(b) \Gamma(b - d - a) }{\Gamma(b - a) \Gamma(b - d)} \, {}_3\cF_2\left[{a', d, 1 + d - b \atop b', 1+d+a-b}; 1 \right]\,,\nn
\end{eqnarray}
where the last equality uses the manifest symmetry of the original integrand under $(a,b) \leftrightarrow (a',b')$. A useful special case that arises sometimes is $d=b'$, in which case \pref{type1a} simplifies to
\bea \label{type1aspec}
 \cI_d(a,b;a',b') &=&   \frac{\Gamma(b) \Gamma(b') \Gamma(b-a-b') }{ \Gamma(b-a)\Gamma(b-b') }  \;{}_3\cF_2\left[ {a' , b', 1+b'-b \atop b' , 1+b'+a-b} ;1\right]\nn\\
 &=&  \frac{\Gamma(b) \Gamma(b') \Gamma(b-a-b') }{ \Gamma(b-a)\Gamma(b-b') }  \;{}_2\cF_1\left[ {a' , 1+b'-b \atop  1+b'+a-b} ;1\right]\\
 &=&  \frac{\Gamma(b) \Gamma(b') \Gamma(b-a-b')\Gamma(1+a +b'-b) \Gamma(a-a') }{ \Gamma(b-a)\Gamma(b-b') \Gamma(1+b'-b + a-a') \Gamma(a)} \,.\nn
\eea

\subsubsection*{Integrals appearing in $\mfN_{\rm pt}$ and $\mfD_{\rm pt}$}

Consider first the convergent integrals that give the standard hyperfine structure. This tests that we are evaluating things properly.

\bigskip\noindent{\it The integral $\cI_{11}^{p}$}

\smallskip\noindent
We start with the integral
\bea
  \cI_{11}^{(p)} &=& \int_0^\infty \exd \rho \, e^{-\rho} \rho^{p} \cM(a,b;\rho) \cM(a,b;\rho) = \cI_{p+1}(a,b;a,b) \nn \\
  &=&   \frac{\Gamma(p+1) \Gamma(b) \Gamma(b  - a - p - 1) }{\Gamma(b - a) \Gamma(b - p -1)} \, {}_3\cF_2\left[{a, p+1, p+2 - b \atop b, p+2+a-b}; 1 \right] \,,
\eea
which with
\be \label{App:aavals}
  a = \zeta - \frac{\zas\omega}{\kappa} \,, \quad b = 1 + 2\zeta \,, \quad b-a = 1+\zeta + \frac{\zas\omega}{\kappa} \,,
\ee
gives
\be
  \cI_{11}^{(p)}  = \frac{\Gamma(p+1) \Gamma(1+2\zeta) \Gamma(\zeta -p + \zas\omega/\kappa ) }{ \Gamma(2\zeta - p )\Gamma(1 + \zeta + \zas\omega/\kappa)} \, {}_3\cF_2\left[{\zeta - \zas\omega/\kappa, p+1, p+1 - 2\zeta \atop 1+2\zeta, p+1-\zeta-\zas\omega/\kappa}; 1 \right]  \,. 
\ee
This integral arises in $\mfN_{\rm pt}$ and $\mfD_{\rm pt}$ with the two cases $p = 2\zeta$ and $p = 2\zeta - 2$. We consider each of these cases in turn.

Specializing to $p=2\zeta$ and simplifying the result using \pref{hyper21at1} gives
\bea
  \cI_{11}^{(2\zeta)} &=& \frac{[ \Gamma(1+2\zeta) ]^2\Gamma(-\zeta  + \zas\omega/\kappa ) }{ \Gamma(0 )\Gamma(1 + \zeta + \zas\omega/\kappa)} \, {}_3\cF_2\left[{\zeta - \zas\omega/\kappa, 1+2\zeta, 1  \atop 1+2\zeta, 1+\zeta-\zas\omega/\kappa}; 1 \right]  \nn\\
  &=&\;   \frac{[ \Gamma(1+2\zeta) ]^2\Gamma(1-\zeta  + \zas\omega/\kappa ) }{ \Gamma(1 + \zeta + \zas\omega/\kappa)  }   
\eea
which survives despite the $\Gamma(0)$ in the denominator of the prefactor\footnote{More precisely, the expression contains the ill-defined quantity $\Gamma(a^\prime - a)/\Gamma(b^\prime - b) \to  \Gamma(0)/\Gamma(0)$. This quantity requires regulating, and there is freedom in choosing how to do this. The easiest way is to choose $a^\prime = a + \delta_a$ and  $b^\prime = b + \delta_b$ then take the limit that  $\delta_a, \delta_b \to 0$, so that $\Gamma(\delta_a)/\Gamma(\delta_b) \to \delta_b/\delta_a = \pm 1$. The sign on the ratio depends on how $\delta_a$ and $\delta_b$ are taken to $0$ and can be fixed by ensuring the results for the point-like integrals align with the standard Dirac-Coulomb wavefunctions, which turns out to require the negative sign.} because of a compensating factor of $\Gamma(0)$ in the numerator coming from using \pref{hyper21at1} in the limit $c=b+a$. Expanding to lowest order in $(\za)^2$ then gives the result
\be
  \cI_{11}^{(2\zeta)} = \; \frac{[ \Gamma(3+2\mfz) ]^2\Gamma(n-\mfz  + \lambda ) }{ \Gamma(2 + n+ \mfz + \lambda) } = \frac{4 }{ n(n+1)} + \cO[(\za)^2]    \,.
\ee

Specializing next to $p=2\zeta-2$ and using \pref{-1result} to simplify the result leads to 
\bea
  \cI_{11}^{(2\zeta-2)} &=& \frac{\Gamma(2\zeta-1) \Gamma(1+2\zeta) \Gamma(2-\zeta  + \zas\omega/\kappa ) }{ \Gamma(2 )\Gamma(1 + \zeta + \zas\omega/\kappa)} \, {}_3\cF_2\left[{\zeta - \zas\omega/\kappa, 2\zeta-1, -1 \atop 1+2\zeta,-1+\zeta-\zas\omega/\kappa}; 1 \right] \nn\\
  &=& \frac{\Gamma(2\zeta-1) \Gamma(1+2\zeta) \Gamma(1-\zeta  + \zas\omega/\kappa ) }{ ( 1+2\zeta )\Gamma(1 + \zeta + \zas\omega/\kappa)} \left(1+ \frac{2\zas\omega}{\kappa} \right) \\
  &=&
   \frac{2(2n+1) }{ 3n(n+1)} + \cO[(\za)^2]  \,.\nn
\eea

\bigskip\noindent{\it The integral $\cI_{12}^{p}$}

\medskip\noindent
In this case we have the integral
\bea
  \cI_{12}^{(p)} &=& \int_0^\infty \exd \rho \, e^{-\rho} \rho^{p} \cM(a,b;\rho) \cM(a+1,b;\rho) = \cI_{p+1}(a,b;a+1,b) \nn \\
  &=&   \frac{\Gamma(p+1) \Gamma(b) \Gamma(b  - a - p - 1) }{\Gamma(b - a) \Gamma(b - p -1)} \, {}_3\cF_2\left[{a+1, p+1, p+2 - b \atop b, p+2+a-b}; 1 \right]  \,,
\eea
in which we again use \pref{App:aavals}, leading to 
\be
  \cI_{12}^{(p)}  = \frac{\Gamma(p+1) \Gamma(1+2\zeta) \Gamma(\zeta -p + \zas\omega/\kappa ) }{ \Gamma(2\zeta - p )\Gamma(1 + \zeta + \zas\omega/\kappa)} \, {}_3\cF_2\left[{1+\zeta - \zas\omega/\kappa, p+1, p+1 - 2\zeta \atop 1+2\zeta, p+1-\zeta-\zas\omega/\kappa}; 1 \right]  \,. 
\ee

Specializing to the case $p=2\zeta$ gives in this case
\bea
  \cI_{12}^{(2\zeta)} &=& \frac{[ \Gamma(1+2\zeta) ]^2\Gamma(-\zeta  + \zas\omega/\kappa ) }{ \Gamma(0 )\Gamma(1 + \zeta + \zas\omega/\kappa)} \, {}_3\cF_2\left[{1+\zeta - \zas\omega/\kappa, 1+2\zeta, 1  \atop 1+2\zeta, 1+\zeta-\zas\omega/\kappa}; 1 \right]   \nn\\
  &=& \frac{[ \Gamma(1+2\zeta) ]^2\Gamma(-\zeta  + \zas\omega/\kappa ) }{ \Gamma(0 )\Gamma(1 + \zeta + \zas\omega/\kappa)} \left(- \zeta + \frac{\zas\omega}{\kappa}\right) = 0\,,
\eea
which vanishes because of the uncanceled factor of $\Gamma(0)$ in the denominator.

By contrast, in the case $p=2\zeta-2$ one instead finds
\bea
  \cI_{12}^{(2\zeta-2)} &=& \frac{\Gamma(2\zeta-1) \Gamma(1+2\zeta) \Gamma(2-\zeta  + \zas\omega/\kappa ) }{ \Gamma(2 )\Gamma(1 + \zeta + \zas\omega/\kappa)} \, {}_3\cF_2\left[{1+\zeta - \zas\omega/\kappa, 2\zeta-1, -1 \atop 1+2\zeta,-1+\zeta-\zas\omega/\kappa}; 1 \right] \nn\\
   &=& \frac{2\Gamma(2\zeta-1) \Gamma(1+2\zeta) \Gamma(1-\zeta  + \zas\omega/\kappa ) }{(1+2\zeta) \Gamma(1 + \zeta + \zas\omega/\kappa)}  \left( \zeta + \frac{\zas\omega}{\kappa} \right) \\
  &=& 
  \frac{4 }{ 3n} + \cO[(\za)^2]  \,.\nn
\eea

\bigskip\noindent{\it The integral $\cI_{22}^{p}$}

\medskip\noindent
Next up is
\bea
  \cI_{22}^{(p)} &=& \int_0^\infty \exd \rho \, e^{-\rho} \rho^{p} \cM(a+1,b;\rho) \cM(a+1,b;\rho) = \cI_{p+1}(a+1,b;a+1,b) \nn \\
  &=&   \frac{\Gamma(p+1) \Gamma(b) \Gamma(b  - a - p - 2) }{\Gamma(b - a-1) \Gamma(b - p -1)} \, {}_3\cF_2\left[{a+1, p+1, p+2 - b \atop b, p+3+a-b}; 1 \right] \,,
\eea
in which we still use \pref{App:aavals}, finding
\be
  \cI_{22}^{(p)}  = \frac{\Gamma(p+1) \Gamma(1+2\zeta) \Gamma(-1+\zeta -p + \zas\omega/\kappa ) }{ \Gamma(2\zeta - p )\Gamma(\zeta + \zas\omega/\kappa)} \, {}_3\cF_2\left[{1+\zeta - \zas\omega/\kappa, p+1, p+1 - 2\zeta \atop 1+2\zeta, p+2-\zeta-\zas\omega/\kappa}; 1 \right]  \,. 
\ee

In the case $p=2\zeta$ this becomes 
\bea
  \cI_{22}^{(2\zeta)} &=& \frac{[ \Gamma(1+2\zeta) ]^2\Gamma(-1-\zeta  + \zas\omega/\kappa ) }{ \Gamma(0 )\Gamma( \zeta + \zas\omega/\kappa)} \, {}_3\cF_2\left[{1+\zeta - \zas\omega/\kappa, 1+2\zeta, 1  \atop 1+2\zeta, 2+\zeta-\zas\omega/\kappa}; 1 \right]  \nn\\
  &=& -\frac{[ \Gamma(1+2\zeta) ]^2\Gamma(-1-\zeta  + \zas\omega/\kappa ) }{ \Gamma( \zeta + \zas\omega/\kappa)}  \left(1+\zeta - \frac{\zas\omega}{\kappa} \right)  \\
  &=& \frac{4  }{ n(n-1)} + \cO[(\za)^2] \nn
\eea
where again the $\Gamma(0)$ in the denominator cancels a similar factor in the numerator. 

When $p=2\zeta-2$ the result instead is
\bea
  \cI_{22}^{(2\zeta-2)} &=& \frac{\Gamma(2\zeta-1) \Gamma(1+2\zeta) \Gamma(1-\zeta  + \zas\omega/\kappa ) }{ \Gamma(2 )\Gamma( \zeta + \zas\omega/\kappa)} \, {}_3\cF_2\left[{1+\zeta - \zas\omega/\kappa, 2\zeta-1, -1 \atop 1+2\zeta,\zeta-\zas\omega/\kappa}; 1 \right] \nn\\
  &=& \frac{\Gamma(2\zeta-1) \Gamma(1+2\zeta) \Gamma(-\zeta  + \zas\omega/\kappa ) }{( 1+2\zeta)\Gamma( \zeta + \zas\omega/\kappa)} \left( -1 + \frac{2\zas\omega}{\kappa } \right) \nn\\
  &=&  \frac{2(2n-1) }{ 3n(n-1)} + \cO[(\za)^2] \quad\hbox{(if $n\neq 1$)} \,. 
\eea
At face value the singularity when $n=1$ implies this goes like $1/(\za)^2$ for $n=1$, but this doesn't matter since this integral ultimately appears in the energy shifit premultiplied by factors of $(n-1)$.

\bigskip\noindent{\it Combining results for $\mfN_{\rm pt}$ and $\mfD_{\rm pt}$}

\medskip\noindent
Using the above integrals in \pref{Nforms} finally gives the expressions
\be \label{Nptform}
 \mfN_{\rm pt}  =  \cI_{11}^{(2\zeta-2)} - \left(\frac{1-n+\mfz - \lambda}{-1-n-\mu}\right)^2 \cI_{22}^{(2\zeta-2)}  =   \frac{4 }{ (n+1)^2} + \cO[(\za)^2]   \,,
\ee
and
\bea \label{Dptform}
 \mfD_{\rm pt}   &=&  2\cI_{11}^{(2\zeta)} - \frac{4\omega}{m}  \left(\frac{1-n+\mfz - \lambda}{-1-n-\mu}\right) \cI_{12}^{(2\zeta)} + 2\left(\frac{1-n+\mfz - \lambda}{-1-n-\mu}\right)^2 \cI_{22}^{(2\zeta)} \\
  &=& \frac{16}{(n+1)^2} +  \cO[(\za)^2]  \,,\nn
\eea
which together imply $\mfN_{\rm pt} /\mfD_{\rm pt} = \frac14 + \cO[(\za)^2]$. Using this in \pref{E1partial2} or  \pref{E1partial2App} gives the prediction for hyperfine splitting for a point nucleus,
\bea
\label{E1partial2x}
 \varepsilon^{\rm hfs}_{n \ssF \frac12 +} &=& -4\exx \mfK X_\ssF \,m\left( \frac{\za}{\cN} \right)^3 \left( \frac{\mfN_{\rm pt}}{\mfD_{\rm pt}} \right) = -\exx \mfK X_\ssF \,m\left( \frac{\za}{n} \right)^3 + \cO[(\za)^2] \notag \\
 &\to&  \frac{ g_p m^2}{M} \left[ \frac{(\za)^4}{2n^3} \right] X_\ssF + \cO[(\za)^2] \,,
\eea
in agreement with the literature. 

\subsubsection*{Integrals appearing in $\mfN_1$ and $\mfD_1$}

We next apply the result \pref{type1a} to the integrals appearing in $\mfN_1$ and $\mfD_1$. In this case it is the cases $p = 0$ and $p = -2$ that are of interest, and it proves useful to specialize the general integral to these two cases for general $a$, $b$, $a'$ and $b'$, keeping in mind that $b = 1+2\zeta$ and  $b' = 1-2\zeta$ imply that $b' = 2-b$. 

For instance, taking $p=0$ ({\it i.e.}~$d=1$) and $b' = 2-b$ in \pref{type1a}, and simplifying using \pref{hyper21at1}, gives
\bea \label{type1aqq2}
 \cI_1(a,b;a',2-b) &=&   \frac{\Gamma(b) \Gamma(1) \Gamma(b-a-1) }{ \Gamma(b-a)\Gamma(b-1) }  \;{}_3\cF_2\left[ {a' , 1, 2-b \atop 2-b , 2+a-b} ;1\right]\nn \\
   &=&  \frac{1-b}{ 1+a-a'-b  }  \,. 
\eea 
Similarly, taking $p = -2$ (and so $d = -1$) and $b' = 2-b$ in \pref{type1a} gives
\bea \label{type1ayy22}
 \cI_{-1}(a,b;a',2-b) &=&     \frac{\Gamma(b) \Gamma(d) \Gamma(b-a+1) }{ \Gamma(b-a)\Gamma(b+1) }  \;{}_3\cF_2\left[ {a' , -1, -b \atop 2-b , a-b} ;1\right] \\
  &=&   \frac{  \Gamma(-1) }{ b(b-2) }  \Bigl[  (2-b)(a-b) + a' b  \Bigr]   \,, \nn
\eea 
which diverges (for all $\za$) due to the $\Gamma(d)$ factor as $d \to -1$. We regulate this divergence dimensionally, which in this instance merely means writing $d = -1 + \eta$ with the regularization parameter $\eta$ taken to zero at the end, once the divergence has been renormalized away. With this in mind we write $\Gamma(-1) = \Gamma(-1+\eta) = \Gamma(\eta)/(-1+\eta) = - \Gamma(0)[1  + \cO(\eta)]$ in what follows, in practice typically dropping the $\cO(\eta)$ terms.

\bigskip\noindent{\it The integral $\cI_{13}^{p}$}

\medskip\noindent
The general formula applies directly to $\cI_{13}^{(p)}$, for which
\be
  \cI_{13}^{(p)} := \int_0^\infty \exd \rho \, e^{-\rho} \rho^{p} \cM(a,b;\rho) \cM(a',b';\rho) = \cI_{p+1}(a,b;a',b') \,,
\ee
in which we use
\be \label{App:aba'b'}
  a = \zeta - \frac{\zas\omega}{\kappa} \,, \quad b = 1 + 2\zeta \,, \quad  a' = -\zeta - \frac{\zas\omega}{\kappa} \,, \quad b' = 1 - 2\zeta  \,.
\ee
and so $b' = 2-b$ and $a-a' = 2\zeta$. Again we require the cases $p=0$ and $p=-2$ (or $d=1$ and $d = -1$). 

Specializing to $p=0$ in \pref{type1aqq2} gives a divergent result because $1+a-a' = 1+2\zeta = b$. Regularizing this divergence by deforming $a = \zeta - (\zas\omega/\kappa) + \eta_a$ (with $\eta_a \to 0$ at the end) we have
\be
  \cI_{13}^{(0)}   = -\; \frac{2\zeta  }{ \eta_a } =: - \Gamma_a(0) \, 2\zeta= -2 \Gamma_a(0) + \cO[(\za)^2] \,,
\ee
which defines $\Gamma_a(0) = \eta_a^{-1}[1+\cO(\eta_a)]$. Similarly using the $p=-2$ in \pref{type1ayy22} gives
\be
   \cI_{13}^{(-2)} =  \frac{  \Gamma(-1) }{ (1-2\zeta)(1+2\zeta) } \left(1 + \frac{2\zas\omega}{\kappa} \right) =  - \left( \frac{2n+1}{3} \right) \Gamma(-1) + \cO[(\za)^2] \,. 
\ee

\bigskip\noindent{\it The integral $\cI_{14}^{p}$}

\medskip\noindent
Next consider  
\be
  \cI_{14}^{(p)} := \int_0^\infty \exd \rho \, e^{-\rho} \rho^{p} \cM(a,b;\rho) \cM(a'+1,b';\rho) = \cI_{p+1}(a,b;a'+1,b')\,,
\ee
evaluated using \pref{App:aba'b'}, which implies $b' = 2-b$. In this case only $p=0$ (or $d = 1$) is required and so using \pref{type1aqq2} gives
\be
 \cI_{14}^{(0)} = \cI_1(a,b;a'+1,2-b) =  \frac{1-b}{ a-a'-b  }  = 2\zeta = 2 + \cO[(\za)^2] \,.
\ee

\bigskip\noindent{\it The integral $\cI_{23}^{p}$}

\medskip\noindent
The required integral in this case is
\be
  \cI_{23}^{(p)} := \int_0^\infty \exd \rho \, e^{-\rho} \rho^{p} \cM(a+1,b;\rho) \cM(a',b';\rho) = \cI_{p+1}(a+1,b;a',b')\,,
\ee
and only $p=0$ (or $d = 1$) is required. Using \pref{type1aqq2} this time gives
\be
 \cI_{23}^{(0)} = \cI_1(a+1,b;a',2-b) =  \frac{1-b}{2+ a-a'-b  }  = -2\zeta = -2 + \cO[(\za)^2] \,.
\ee

\bigskip\noindent{\it The integral $\cI_{24}^{p}$}

\medskip\noindent
The final integral in this section is $\cI_{24}^{(p)}$, for which
\be
  \cI_{24}^{(p)} := \int_0^\infty \exd \rho \, e^{-\rho} \rho^{p} \cM(a+1,b;\rho) \cM(a',b';\rho) = \cI_{p+1}(a+1,b;a'+1,b')\,,
\ee
and both $p=0$ (or $d = 1$) and $p=-2$ (or $d=-1$) are needed. In the case $p=0$, using \pref{type1aqq2} and noting that $\cI_1(a,b;a',2-b)$ depends on $a$ and $a'$ only through their difference, $a-a'$, implies $\cI_{24}^{(0)} = \cI_{13}^{(0)}$ and so
\be
 \cI_{24}^{(0)} = \cI_{13}^{(0)} =  - \Gamma_a(0) \, 2\zeta= -2 \Gamma_a(0) + \cO[(\za)^2]  \,.
\ee
For $p=-2$, on the other hand, using \pref{type1ayy22} gives
\bea  
 \cI_{24}^{(-2)} &=& \cI_{-1}(a+1,b;a'+1,2-b) =   \frac{  \Gamma(-1) }{ b(b-2) }  \Bigl[  (2-b)(1+a-b) + (1+a' )b  \Bigr] \nn\\
 &=&  \frac{  \Gamma(-1) }{ (1+2\zeta)(-1+2\zeta) }  \left( 1 - \frac{2\zas\omega}{\kappa} \right)   = - \left( \frac{2n-1}{3} \right)  \Gamma(-1)+\cO[(\za)^2]   \,. \nn
\eea

\bigskip\noindent{\it Combining results for $\mfN_1$ and $\mfD_1$}

These integrals when combined in \pref{Nforms}  give
\bea \label{Nformsv2}
  \mfN_1 
 &= & 2 \cI_{13}^{(-2)} -2 \left(\frac{1-n+\mfz - \lambda}{-1-n-\mu}\right)\left(\frac{-1-n-\mfz - \lambda}{-1-n-\mu} \right) \cI_{24}^{(-2)} \nn\\
 &=& - 2\left( \frac{2n+1}{3} \right) \Gamma(-1) + 2\left(\frac{n-1}{n+1}\right)  \left( \frac{2n-1}{3} \right) \Gamma(-1)+ \cO[(\za)^2] \\
  &=&- \left( \frac{4n}{n+1} \right) \Gamma(-1)  + \cO[(\za)^2] \,,\nn
\eea
while \pref{Dforms} similarly becomes
\bea  
 \mfD_1   &=& 4\cI_{13}^{(0)} -  \frac{4\omega}{m}  \left(\frac{-1-n-\mfz - \lambda}{-1-n-\mu} \right) \cI_{14}^{(0)} -  \frac{4\omega}{m}   \left(\frac{1-n+\mfz - \lambda}{-1-n-\mu}\right)  \cI_{23}^{(0)} \nn\\
  &&\qquad\qquad\qquad\qquad\qquad +  4 \left(\frac{1-n+\mfz - \lambda}{-1-n-\mu}\right)\left(\frac{-1-n-\mfz - \lambda}{-1-n-\mu} \right) \cI_{24}^{(0)} \nn   \\
  &=&- \; \frac{16 }{n+1} \Bigl[   n\, \Gamma_a(0) +1 \Bigr]      + \cO[(\za)^2]    \,.
\eea
Combining these with \pref{Nptform} and \pref{Dptform}, which say $\mfN_{\rm pt} = 4/(n+1)^2+ \cdots$ and $\mfD_{\rm pt} = 16/(n+1)^2 + \cdots$ we finally get
\bea
   \frac{\mfN_1}{\mfN_{\rm pt}} &=&  -  n(n+1) \; \Gamma(-1)  + \cO[(\za)^2]  \nn\\
   \frac{\mfD_1}{\mfD_{\rm pt}} &=&  - (n+1) \;  \Bigl[   n\, \Gamma_a(0) +1 \Bigr]      + \cO[(\za)^2]    \,.
\eea

Keeping in mind that 
\be
  \frac{\msD}{\msC} \simeq - \frac{\mfc}{n(n+1)} 
\ee
where $\mfc \propto (m \epsilon_\star)^2$ is defined in \pref{App:ca+0def}, we see
\be
 \left( \frac{\msD}{\msC}\right) \frac{\mfN_1}{\mfN_{\rm pt}} =   \mfc \; \Gamma(-1)  + \cO[(\za)^2]  
 \ee
 and
 \be
  \boxed{ \left( \frac{\msD}{\msC}\right)  \frac{\mfD_1}{\mfD_{\rm pt}} = \mfc \;  \left[  \Gamma_a(0) +\frac{1}{n} \right]      + \cO[(\za)^2]   } \,.
\ee
What is important here is the divergent terms are $n$-independent, as is required for them to be absorbed into the counter-term $c_\ssF$.  

The finite contribution in $\mfD_1$ does not cancel but it is small enough to be negligible for our purposes. To see why, recall that $\mfc \sim (m \epsilon_\star)^2 \sim (mR\za)^2$ while the point hyperfine splitting is order $(m^2/M) (\za)^4$. Taking $m/M \sim mR$, the finite part of the $\mfD_1$ piece contributes to the energy by an amount of order $m(\za)^3 (mR\za)^3$. Keeping in mind that the charge radius contributes at order $m (\za)^2 (mR\za)^2$ we see the finite part of $\mfD_1$ is suppressed relative to the charge radius by a factor of order $mR (\za)^2$. For electrons this is smaller than the $(\za)^2 \sim mR\za$ order to which we work, and for muons it is comparable to the other $(mR\za)^2$ terms that have been neglected (but whose size is of practical interest for some experiments).

\subsubsection*{Integrals appearing in $\mfN_2$ and $\mfD_2$}

Finally, consider the integrals in the $(\msD/\msC)^2$ part of the hyperfine energy. 

\bigskip\noindent{\it The integral $\cI_{33}^{p}$}

\medskip\noindent
The first integral of interest here is  
\begin{eqnarray}
\cI_{33}^{(p)} &:=&  \int_0^\infty \exd \rho \, e^{-\rho} \rho^{p} \cM\left[a'; b'; \rho \right] \cM \left[a'; b';\rho \right] = \cI_{p+1}(a',b'; a',b') \\
&=& \frac{\Gamma(p+1) \Gamma(b') \Gamma(b'  - a' - p -1) }{\Gamma(b' - a') \Gamma(b' - p-1)} \, {}_3\cF_2\left[{a', p+1, p+2 - b' \atop b', 2+p+a'-b'}; 1 \right] \,,\nn
\end{eqnarray}
in which we use
\be
  a' = - \zeta - \frac{\zas\omega}{\kappa} \,, \quad b' = 1 - 2\zeta \,, \quad b'-a' = 1-\zeta + \frac{\zas\omega}{\kappa} \,.
\ee
This is most easily obtained from the result for $\cI_{11}^{(p)}$ found above by making the replacement $\zeta \to - \zeta$, leading to
\be
  \cI_{33}^{(p)} = \frac{\Gamma(p+1) \Gamma(1-2\zeta) \Gamma(-\zeta -p + \zas\omega/\kappa ) }{ \Gamma(-2\zeta - p )\Gamma(1 - \zeta + \zas\omega/\kappa)} \, {}_3\cF_2\left[{-\zeta - \zas\omega/\kappa, p+1, p+1 + 2\zeta \atop 1-2\zeta, p+1+\zeta-\zas\omega/\kappa}; 1 \right] \,,
\ee
for which we require $p = -2\zeta$ and $p = -2\zeta - 2$.  

In the case $p=-2\zeta$ using \pref{hyper21at1} allows the integral to be written
\bea
  \cI_{33}^{(-2\zeta)}   &=& \frac{[\Gamma(1-2\zeta)]^2 \Gamma(\zeta  + \zas\omega/\kappa ) }{ \Gamma( 0 )\Gamma(1 - \zeta + \zas\omega/\kappa)} \, {}_3\cF_2\left[{-\zeta - \zas\omega/\kappa, 1 - 2\zeta, 1  \atop 1-2\zeta, 1-\zeta-\zas\omega/\kappa}; 1 \right]  \nn \\
 &=& - \;  \frac{[\Gamma(1-2\zeta)]^2 \Gamma(1+\zeta  + \zas\omega/\kappa ) }{ \Gamma(1 - \zeta + \zas\omega/\kappa)}   \,.\nn
\eea
Notice that in this expression the Gamma function $\Gamma(1-2\zeta)$ diverges when $\za \to 0$. What is important about this singularity is how it depends on $n$, since this allows it also to be absorbed into the effective coupling $c_\ssF$. To see this explicitly, notice that for small $\za$ the above becomes
 \be
  \cI_{33}^{(-2\zeta)} = -\Gamma(-1-2\mfz)]^2 \Bigl[ n(n+1) + \cO[(\za)^2]  \Bigr] \,.
\ee

The case $p=-2\zeta-2$ similarly gives
\bea
  \cI_{33}^{(-2\zeta-2)}   &=& \frac{\Gamma(-1-2\zeta) \Gamma(1-2\zeta) \Gamma(2+\zeta  + \zas\omega/\kappa ) }{ \Gamma(2 )\Gamma(1 - \zeta + \zas\omega/\kappa)} \, {}_3\cF_2\left[{-\zeta - \zas\omega/\kappa, -1-2\zeta, -1 \atop 1-2\zeta, -1-\zeta-\zas\omega/\kappa}; 1 \right]  \nn \\
    &=&  \frac{ [\Gamma(1-2\zeta) ]^2\Gamma(1 +\zeta  + \zas\omega/\kappa ) }{2\zeta (1+2\zeta)(1-2\zeta) \Gamma(1 - \zeta + \zas\omega/\kappa)} \left( 1+ \frac{2\zas\omega}{\kappa} \right)   \,.\nn
\eea
which as $\za \to 0$ becomes
\be
  \cI_{33}^{(-2\zeta-2)}  = [\Gamma(-1-2\mfz) ]^2 \left[- \frac16 \, n(n+1) (2n+ 1)  + \cO[(\za)^2]\right] \,.\nn
\ee

\bigskip\noindent{\it The integral $\cI_{34}^{p}$}

\medskip\noindent
Consider next  
\begin{eqnarray}
\cI_{34}^{(p)} &:=&  \int_0^\infty \exd \rho \, e^{-\rho} \rho^{p} \cM\left[a'; b'; \rho \right] \cM \left[a'+1; b';\rho \right] = \cI_{p+1}(a',b'; a'+1,b') \\
&=& \frac{\Gamma(p+1) \Gamma(b') \Gamma(b'  - a' - p -1) }{\Gamma(b' - a') \Gamma(b' - p-1)} \, {}_3\cF_2\left[{a'+1, p+1, p+2 - b' \atop b', 2+p+a'-b'}; 1 \right] \,,\nn
\end{eqnarray}
which becomes
\be
  \cI_{34}^{(p)}   = \frac{\Gamma(p+1) \Gamma(1-2\zeta) \Gamma(-\zeta -p + \zas\omega/\kappa ) }{ \Gamma(-2\zeta - p )\Gamma(1 - \zeta + \zas\omega/\kappa)} \, {}_3\cF_2\left[{1-\zeta - \zas\omega/\kappa, p+1, p+1 + 2\zeta \atop 1-2\zeta, p+1+\zeta-\zas\omega/\kappa}; 1 \right]   \,.
\ee
In the case $p=-2\zeta$ this gives
\bea
  \cI_{34}^{(-2\zeta)}   &=& \frac{[ \Gamma(1-2\zeta)]^2 \Gamma(\zeta  + \zas\omega/\kappa ) }{ \Gamma( 0 )\Gamma(1 - \zeta + \zas\omega/\kappa)} \, {}_3\cF_2\left[{1-\zeta - \zas\omega/\kappa, 1- 2\zeta, 1  \atop 1-2\zeta, 1-\zeta-\zas\omega/\kappa}; 1 \right] \nn \\
  &=& \frac{[ \Gamma(1-2\zeta)]^2 \Gamma(\zeta  + \zas\omega/\kappa ) }{ \Gamma( 0 )\Gamma(1 - \zeta + \zas\omega/\kappa)} \left(  \zeta + \frac{\zas\omega}{\kappa } \right) = 0 \,.\nn
\eea
which vanishes due to the uncanceled $\Gamma(0)$ in the denominator.

\bigskip\noindent{\it The integral $\cI_{44}^{p}$}

\medskip\noindent
Finally consider the case  
\begin{eqnarray}
\cI_{44}^{(p)} &:=&  \int_0^\infty \exd \rho \, e^{-\rho} \rho^{p} \cM\left[a'+1; b'; \rho \right] \cM \left[a'+1; b';\rho \right] = \cI_{p+1}(a'+1,b'; a'+1, b') \nn \\
&=& \frac{\Gamma(p+1) \Gamma(b') \Gamma(b'  - a' - p -2) }{\Gamma(b' - a'-1) \Gamma(b' - p-1)} \, {}_3\cF_2\left[{a'+1, p+1, p+2 - b' \atop b', 3+p+a'-b'}; 1 \right] \,,
\end{eqnarray}
which becomes
\be
  \cI_{44}^{(p)}  = \frac{\Gamma(p+1) \Gamma(1-2\zeta) \Gamma(-1-\zeta -p + \zas\omega/\kappa ) }{ \Gamma(-2\zeta - p )\Gamma( - \zeta + \zas\omega/\kappa)} \, {}_3\cF_2\left[{1-\zeta - \zas\omega/\kappa, p+1, p+1 + 2\zeta \atop 1-2\zeta, p+2+\zeta-\zas\omega/\kappa}; 1 \right]   \,.
\ee

When $p=-2\zeta$ the above formula becomes
\bea
  \cI_{44}^{(-2\zeta)}   &=& \frac{[ \Gamma(1-2\zeta)]^2 \Gamma(-1+\zeta  + \zas\omega/\kappa ) }{ \Gamma( 0 )\Gamma( - \zeta + \zas\omega/\kappa)} \, {}_3\cF_2\left[{1-\zeta - \zas\omega/\kappa, 1-2\zeta, 1  \atop 1-2\zeta, 2-\zeta-\zas\omega/\kappa}; 1 \right] \nn \\
  &=&  \frac{[ \Gamma(1-2\zeta)]^2 \Gamma(-1+\zeta  + \zas\omega/\kappa ) }{ \Gamma( - \zeta + \zas\omega/\kappa)} \left( 1-\zeta - \frac{\zas\omega}{\kappa } \right)   \,.
\eea
Expanding around $\za = 0$ shows the same divergent pole as for $\cI^{(-2\zeta)}_{33}$, leading to
\be
  \cI_{44}^{(-2\zeta)}  = [ \Gamma(-1-2\mfz)]^2 \Bigl[ -n(n-1) + \cO[(\za)^2] \Bigr]  \,.
\ee
Next take $p = -2\zeta -2$ in which case
\bea
  \cI_{44}^{(-2\zeta-2)}   &=& \frac{\Gamma(-1-2\zeta) \Gamma(1-2\zeta) \Gamma(1+\zeta  + \zas\omega/\kappa ) }{ \Gamma(2 )\Gamma( - \zeta + \zas\omega/\kappa)} \, {}_3\cF_2\left[{1-\zeta - \zas\omega/\kappa, -1-2\zeta, -1 \atop 1-2\zeta, -\zeta-\zas\omega/\kappa}; 1 \right] \nn \\
    &=&\frac{[\Gamma(1-2\zeta)]^2 \Gamma(\zeta  + \zas\omega/\kappa ) }{2\zeta(1+2\zeta)(1-2\zeta) \Gamma( - \zeta + \zas\omega/\kappa)} \left( -1+\frac{2\zas\omega}{\kappa} \right)   \,,
\eea
which expands out to give
\be
  \cI_{44}^{(-2\zeta-2)} = [\Gamma(-1-2\mfz)]^2 \left[ -\frac16  n(n-1 ) \left(2n -1 \right)  + \cO[(\za)^2] \right]  \,.
\ee

\bigskip\noindent{\it Combining results $\mfN_2$ and $\mfD_2$}

\medskip\noindent
These integrals combine to give
\bea  
 \mfN_2  &=&  \cI_{33}^{(-2\zeta-2)} - \left( \frac{-1-n-\mfz - \lambda}{-1-n-\mu}\right)^2  \cI_{44}^{(-2\zeta-2)} \\
 &=& [\Gamma(-1-2\mfz) ]^2  \Bigl[- n^2  + \cO[(\za)^2] \Bigr] \,,\nn
\eea
and
\bea
    \mfD_2
    &=&  2 \left\{  \cI_{33}^{(-2\zeta)}  - \frac{2\omega}{m}  \left( \frac{-1-n-\mfz - \lambda}{-1-n-\mu} \right)  \cI_{34}^{(-2\zeta)} + \left(\frac{-1-n-\mfz - \lambda}{-1-n-\mu}\right)^2 \cI_{44}^{(-2\zeta)}  \right\} \nn\\
      &=&2 [\Gamma(-1-2\mfz)]^2 \Bigl[  -2n^2 + \cO[(\za)^2] \Bigr] \,.
\eea
and so
\be
  \frac{\mfN_2}{\mfN_{\rm pt}} =  [\Gamma(-1-2\mfz) ]^2 \left[-  \frac{n^2(n+1)^2}{4} + \cO[(\za)^2] \right] \,,
\ee
and
\be
  \frac{\mfD_2}{\mfD_{\rm pt}} = - [\Gamma(-1-2\mfz)]^2 \left[   \frac{n^2(n+1)^2}{4} + \cO[(\za)^2] \right] \,.
\ee

Keeping in mind that 
\be
  \frac{\msD}{\msC} \simeq - \;\frac{\mfc}{n(n+1)} 
\ee
where $\mfc \propto (m \epsilon_\star)^2$ so
\be
  \left(  \frac{\msD}{\msC}  \right)^2 \frac{\mfN_2}{\mfN_{\rm pt}} =  [\Gamma(-1-2\mfz) ]^2 \left[-  \frac{\mfc^2}{4} + \cO[(\za)^2] \right]  \,,
\ee
and
\be
   \left(  \frac{\msD}{\msC}  \right)^2\frac{\mfD_2}{\mfD_{\rm pt}} =  -[\Gamma(-1-2\mfz)]^2 \left[   \frac{\mfc^2}{4} + \cO[(\za)^2] \right] \,.
\ee

\bigskip\noindent{\it Combined result}

\medskip\noindent
Finally combining all terms gives
\bea
  \frac{\mfN}{\mfD} 
  &=&  \frac{\mfN_{\rm pt} }{\mfD_{\rm pt} } \left[ \frac{1 + (\msD/\msC)(\mfN_1/\mfN_{\rm pt}) + (\msD/\msC)^2 (\mfN_2/\mfN_{\rm pt} )}{1 + (\msD/\msC)(\mfD_1/\mfD_{\rm pt}) + (\msD/\msC)^2 (\mfD_2/\mfD_{\rm pt} )}  \right] \nn\\
  &\simeq&  \frac{\mfN_{\rm pt} }{\mfD_{\rm pt} } \left[ 1 + C - \frac{\mfc}{n} +\cdots \right]  \,,
\eea
where $C$ is an $n$-independent but divergent constant whose precise value does not matter because it gets absorbed into the renormalization of $c_\ssF$. The prediction beyond the contribution of $c_\ssF$ is completely contained in the $\mfc/n$ term, which is smaller than the order to which we work. 

\subsection*{Radial matrix element of the first-order state correction}
\label{sec:StateShiftApp}

In the main text we have anticipated that the first-order wave-function corrections will lead to a more complicated RG behaviour of the combined PPEFT couplings $\hat{c}_s \pm \hat{c}_v + \langle \mathbf{I} \cdot \mathbf{S} \rangle \hat{c}_\ssF$ of $j=1/2$ states through the boundary conditions \eqref{eq:app:BCs:bothUsualBC} (worked out in detail in Appendix \ref{AppendixFBC}), repeated here for convenience
\begin{equation}
\left( \hat{c}_s^+ + \cZ_{\ssF}\, \hat{c}_\ssF^+ -  \hat{c}_v^+ \right) = \frac{\mfg_+}{\mff_+}, \quad \text{and} \quad
\left(\hat{c}_s^- + \cZ_{\ssF}\, \hat{c}_\ssF^- +  \hat{c}_v^- \right) = \frac{\mff_-}{
\mfg_-}.
\end{equation}
This occurs because the radial functions on the right-hand sides of these conditions change under the perturbation of the nuclear magnetic dipole field and can be expanded in a perturbation series as $\mff_\varpi = \mff_\varpi^{(0)} + \exx \mff_\varpi^{(1)} + \cdots$ and $\mfg_\varpi = \mfg_\varpi^{(0)} + \exx \mfg_\varpi^{(1)} + \cdots$, where the ellipses stand for the second- and higher-order corrections.

These corrections to the radial functions come about as a result of the state corrections, which we had formally calculated in \eqref{RaySchState1} and as applied to the eigenstates of total atomic angular momentum $\ket{n F F_z j \varpi}$ reads
\bea
\label{statecorr1}
\ket{n F F_z j \varpi}_1 &=& \frac{\cC_{n n \ssF \ssF_z j' j (-\varpi) \varpi}}{\omega_{nj\varpi}^\ssD - \omega_{n j' (-\varpi)}^\ssD} \ket{n F F_z j' (-\varpi)}_0 + \sum_{\widetilde{n} \not = n}  \left[ \frac{\cC_{\widetilde{n} n \ssF \ssF_z j j \varpi \varpi}}{\omega_{nj\varpi}^\ssD - \omega_{\widetilde{n} j \varpi}^\ssD} \ket{\widetilde{n} F F_z j \varpi}_0  \right. \notag \\
&+& \left.  \frac{\cC_{\widetilde{n} n \ssF \ssF_z j' j (-\varpi) \varpi}}{\omega_{nj\varpi}^\ssD - \omega_{\widetilde{n} j' (-\varpi)}^\ssD} \ket{\widetilde{n} F F_z j' (-\varpi)}_0 \right],
\eea
where the coefficients are defined as
\bea
\label{statecoeff}
\cC_{\widetilde{n} n \ssF \ssF_z \widetilde{j} j \widetilde{\varpi} \varpi} &:=& {}_0\braket{\widetilde{n} F F_z \widetilde{j} \widetilde{\varpi} | V |n F F_z j \varpi}_0 = -\left( \frac{e \mu_\ssN}{4\pi \widetilde{\cD}} \right) \int \exd^3 x \, r^{-2} \widetilde{\psi}^\dagger \gamma^0 \bm{\gamma} \cdot (\mathbf{I} \times \hat{\mathbf{r}}) \psi, \notag \\
&=& - \left( \frac{\exx}{m} \right) \int \exd \Omega_2\, \left( \cY_{\ssF, f_z}^{\widetilde{j} \widetilde{\varpi}} \right)^\dagger \left( I^\theta \sigma^\theta + I^\phi \sigma^\phi \right) \cY_{\ssF, f_z}^{j \varpi} \frac{\int_0^\infty \exd r \, \left( \widetilde{\mff} \mfg + \widetilde{\mfg} \mff \right)}{\widetilde{\cD}}, \notag \\
&=& \frac{(2\widetilde{\kappa})^3 \exx}{m^2} \left( \frac{\msC}{\widetilde{\msC} }\right) \int \exd \Omega_2\, \left( \cY_{\ssF, f_z}^{\widetilde{j} \widetilde{\varpi}} \right)^\dagger \left( I^\theta \sigma^\theta + I^\phi \sigma^\phi \right) \cY_{\ssF, f_z}^{j \varpi} \left( \frac{\mfN^s}{\widetilde{\mfD}} \right),
\eea
with
\be
\widetilde{\cD} = \int_0^\infty \exd r \, r^2 \left( \widetilde{\mff}^2 + \widetilde{\mfg}^2 \right) = \frac{\widetilde{\msC}^2m}{ (2\widetilde{\kappa})^3} \widetilde{\mfD}.
\ee

Notice that a total of three types of terms appear in the first-order wave-function correction: there are corrections coming from states that have the same angular momentum quantum numbers as the corrected state but that differ from it in their principal quantum number; there are corrections from states that have the same $F, F_z$ and $l$ quantum numbers but different $n, j$ values and opposite parity; lastly there is a contribution from a state that shares the same $n, F, F_z$ and $l$ value as the corrected state and differs from it only in its $j$-value and parity. This medley of corrections occurs due to the fact that although the $F$ eigenstates do diagonalize the degenerate subspaces of the combined nuclear and Dirac-Coulomb modes, they do \textit{not} diagonalize the actual perturbation, $\cL_{\rm int} = - e \gamma^0 \bm{\gamma} \cdot \mathbf{A}^{\rm nuc}$ (except for $S$-states for which $\mathbf{J} = \mathbf{S}$); the hyperfine perturbation is known to mix states that share all their quantum numbers except for $j$ and $\varpi$ \cite{brodsky, hh}. This effect manifests in the corrections proportional to $\cC_{n n \ssF \ssF_z j' j (-\varpi) \varpi}$ and $\cC_{\widetilde{n} n \ssF \ssF_z j' j (-\varpi) \varpi}$, which first appear for the negative-parity, $j =1/2$ states that receive a correction from the positive-parity $j=3/2$ states (and so we have $j =1/2$, $j'=3/2$ and $\varpi = -$ for these mixing corrections).

We can obtain a rough estimate for the sizes of these corrections knowing that the Dirac-Coulomb modes go as $\rho^{\zeta-1} \sim (mRZ\alpha)^{|\mfK| -1}$ and so $\ket{n, F, F_z, 1/2, -}_0 \sim \cO(1)$ whereas $\ket{n, F, F_z, 3/2, +}_0 \sim (mRZ\alpha)$, and assuming (as we will show below) $\cC_{\widetilde{n} n \ssF \ssF_z \widetilde{j} j \widetilde{\varpi} \varpi} \sim \exx (Z\alpha)^3$, while the energy differences are of size,
\begin{eqnarray*}
\left( \omega_{n j \varpi}^{\ssD} - \omega_{\widetilde{n} j \varpi}^\ssD \right) \sim (Z\alpha)^2, \quad  \left( \omega_{n j \varpi}^{\ssD} - \omega_{\widetilde{n} j' (-\varpi)}^\ssD \right) \sim (Z\alpha)^2,  \quad \left(\omega_{n j \varpi}^{\ssD} - \omega_{n j' (-\varpi)}^\ssD \right) \sim (Z\alpha)^4.
\end{eqnarray*}
The large size of the last energy difference is due to the Dirac-Coulomb modes having the same principal quantum number, $n$ but different angular momentum quantum numbers, $j, j'$.

Combining these estimates we find that the corrections coming from states with the same angular momentum quantum numbers but different principal quantum number are of order,
\be
\frac{\cC_{\widetilde{n} n \ssF \ssF_z j j \varpi \varpi}}{\omega_{nj\varpi}^\ssD - \omega_{\widetilde{n} j \varpi}^\ssD} \ket{\widetilde{n} F F_z j \varpi}_0 \sim \exx \frac{(Z\alpha)^3}{(Z\alpha)^2} = \exx(Z\alpha),
\ee
those coming from states with the same $F, F_z$ and $l$ value but different $n, j, \varpi$ values are of size,
\be
\frac{\cC_{\widetilde{n} n \ssF \ssF_z j' j (-\varpi) \varpi}}{\omega_{nj\varpi}^\ssD - \omega_{\widetilde{n} j' (-\varpi)}^\ssD} \ket{\widetilde{n} F F_z j' (-\varpi)}_0 \sim \exx \frac{(Z\alpha)^3}{(Z\alpha)^2} (mRZ\alpha) = \exx (Z\alpha) (mRZ\alpha).
\ee
and those coming from states with the same $n, F, F_z$ but different $j, \varpi$ quantum numbers as the corrected state yield a correction of size.
\be
\frac{\cC_{n n \ssF \ssF_z j' j (-\varpi) \varpi}}{\omega_{nj\varpi}^\ssD - \omega_{n j' (-\varpi)}^\ssD} \ket{n F F_z j' (-\varpi)}_0 \sim \exx \frac{(Z\alpha)^3}{(Z\alpha)^4} (mRZ\alpha) = \frac{\exx}{(Z\alpha)} (mRZ\alpha).
\ee

Looking at these sizes we can see that the corrections coming from mixing states with different angular momentum quantum numbers are proportional to $(mRZ\alpha)$, which pushes them outside of the scope of this paper as their calculation would require the PPEFT action to be computed to the next order in $R/a_\ssB$. Nevertheless, although we ignore these off-diagonal corrections to the negative parity, $j=1/2$ state in what follows, we compute the most general matrix element $\cC_{\widetilde{n} n \ssF \ssF_z \widetilde{j} j \widetilde{\varpi} \varpi}$ next.

In the $\cC_{\widetilde{n} n \ssF \ssF_z \widetilde{j} j \widetilde{\varpi} \varpi} $ coefficients above, $\widetilde{\mfD}$ has the same functional form as $\mfD$ defined in \eqref{NumDenom} and \eqref{Dforms} for the first-order energy shift with the parameters taken to be $n \to \widetilde{n}, \zeta \to \widetilde{\zeta}, \rho \to \widetilde{\rho}$, \textit{etc}. and so this part does not require further computation. What is new is the function in the numerator that integrates over the mixtures of the radial functions of the corrected and the correcting states that can be written in terms of the various integration constant ratios as,
\bea
\mfN^s &=& m (2\widetilde{\kappa})^{\widetilde{\zeta}-1} (2\kappa)^{\zeta-1} (\widetilde{\kappa}+\kappa)^{1-\widetilde{\zeta} -\zeta} \left\{ \mfN^s_{\rm pt} + \left( \frac{\msD}{\msC} \right) (2\kappa)^{-2\zeta} \mfN^s_1 + \left( \frac{\widetilde{\msD}}{\widetilde{\msC}} \right) (2\widetilde{\kappa})^{-2\widetilde{\zeta}} \widetilde{\mfN}^s_1 \right. \notag \\
&+& \left. \left( \frac{\msD \widetilde{\msD}}{\msC \widetilde{\msC}} \right) (2\widetilde{\kappa})^{-2 \widetilde{\zeta}} (2\kappa)^{-2\zeta} \mfN^s_2 \right\}.
\eea

Furthermore, similarly to the energy shift, the functions $\mfN^s_{\rm pt}, \mfN^s_1, \widetilde{\mfN}^s_1$ and $\mfN^s_2$ can all be written as instances of the integral,
\bea
\label{type2}
\cI_{\mfi \widetilde{\mfj}}^p := (\widetilde{\kappa}+\kappa)^{\widetilde{\zeta} + \zeta -1 } \int_0^\infty \exd r \, e^{-(\widetilde{\kappa} + \kappa)r} r^p \cM_\mfi \cM_{\widetilde{\mfj}},
\eea
where the tilde on the subscript of the hypergeometric function means that in all of its arguments the parameters are to be transformed as $n \to \widetilde{n}, \kappa \to \widetilde{\kappa}, \zeta \to \widetilde{\zeta}$ and so on. This integral differs from that of the energy shift in \eqref{AppIntDef} in that the integration variable is no longer the same as the argument of either of the hypergeometric functions and the exponential has a factor multiplying the integration variable and so is proportional to the more general form of \eqref{type1a}
\bea
 \cI_d (s; a, b, k; \widetilde{a}, \widetilde{b}, \widetilde{k}) &:=&  \int_0^\infty \exd r \, e^{-sr} r^{d-1} \cM[a; b ; k r ] \cM[\widetilde{a}; \widetilde{b};  \widetilde{k} r], \notag \\
&=& s^{-d} \Gamma[d] \sum_{q = 0}^\infty \frac{(a)_q (k)^q (d)_q}{(b)_q \, q! \,s^q} {}_2 \cF_1 \left[ {\widetilde{a}, d+q; \atop \widetilde{b}; } \frac{\widetilde{k}}{s} \right],
\eea
which after taking the limit $\widetilde{k} = k = s$ and redefining the integration variable to $\rho = kr$ becomes proportional to \eqref{type1a}. The equality on the second line can again be found by writing out one of the hypergeometric functions in its series form and carrying out the resulting integral term by term using standard techniques found in \cite{gr, slater}.

In terms of these integrals the functions of $\mfN^s$ can be written as,
\bea
\label{Nsform}
\mfN^s_{\rm pt} &=& \cI_{1 \widetilde{1}}^{(\widetilde{\zeta} + \zeta -2)} \cS_+ - \left( \frac{\widetilde{a}}{\widetilde{c}} \right) \cI_{1 \widetilde{2}}^{(\widetilde{\zeta} + \zeta -2)} \cS_- + \left( \frac{a}{c} \right) \cI_{2 \widetilde{1}}^{(\widetilde{\zeta} + \zeta -2)} \cS_- - \left( \frac{a\widetilde{a}}{c \widetilde{c}} \right) \cI_{2 \widetilde{2}}^{(\widetilde{\zeta} + \zeta -2)} \cS_+, \notag \\
\mfN^s_1 &=& \cI_{3 \widetilde{1}}^{(\widetilde{\zeta} - \zeta -2)} \cS_+ - \left( \frac{\widetilde{a}}{\widetilde{c}} \right) \cI_{3 \widetilde{2}}^{(\widetilde{\zeta} - \zeta -2)} \cS_- + \left( \frac{a'}{c} \right) \cI_{4 \widetilde{1}}^{(\widetilde{\zeta} - \zeta -2)} \cS_- - \left( \frac{a' \widetilde{a}}{c \widetilde{c}} \right) \cI_{4 \widetilde{2}}^{(\widetilde{\zeta} - \zeta -2)} \cS_+, \notag \\
\widetilde{\mfN}^s_1 &=&  \cI_{1 \widetilde{3}}^{(\zeta- \widetilde{\zeta} -2)} \cS_+ - \left( \frac{\widetilde{a}'}{\widetilde{c}} \right) \cI_{1 \widetilde{4}}^{(\zeta - \widetilde{\zeta} - 2)} \cS_- + \left( \frac{a}{c} \right) \cI_{2 \widetilde{3}}^{(\zeta - \widetilde{\zeta} - 2)} \cS_- - \left( \frac{a \widetilde{a}'}{c \widetilde{c}} \right) \cI_{2 \widetilde{4}}^{(\zeta - \widetilde{\zeta} - 2)} \cS_+, \notag \\
\mfN^s_2 &=&  \cI_{3 \widetilde{3}}^{(-\widetilde{\zeta} - \zeta -2)} \cS_+ - \left( \frac{\widetilde{a}'}{\widetilde{c}} \right) \cI_{3 \widetilde{4}}^{(-\widetilde{\zeta} - \zeta -2)} \cS_- + \left( \frac{a'}{c} \right) \cI_{4 \widetilde{3}}^{(-\widetilde{\zeta} - \zeta -2)} \cS_- - \left( \frac{a' \widetilde{a}'}{c \widetilde{c}} \right) \cI_{4 \widetilde{4}}^{(-\widetilde{\zeta} - \zeta -2)} \cS_+, \notag \\ 
\eea
where the dimensionless quantities, $\cS_\pm$ are defined as
\be
\cS_\pm := \sqrt{\left( 1+ \frac{\widetilde{\omega}}{m}\right) \left( 1 - \frac{\omega}{m} \right)} \pm  \sqrt{\left( 1+ \frac{\omega}{m}\right) \left( 1 - \frac{\widetilde{\omega}}{m} \right)} \sim \cO(Z\alpha).
\ee

In general, the integrals in $\mfN^s_1, \widetilde{\mfN}^s_1$ and $\mfN^s_2$ will diverge and a more careful analysis of their divergence structure and the energy shifts is required to see if they can be absorbed into the PPEFT couplings through boundary conditions such as  \eqref{eq:app:BCs:bothUsualBC}. However, the possible divergences appear along with ratios of integration constants such as $(\msD/\msC)$ and $(\widetilde{\msD}/\widetilde{\msC})$ and so are suppressed by factors of $(m\epsilon_\star)^{2\zeta} \sim (mRZ\alpha)^{2|\mfK|}$ and $(m\widetilde{\epsilon_\star})^{2\widetilde{\zeta}} \sim (mRZ\alpha)^{2|\widetilde{\mfK}|}$, which makes them negligible to the order we work here and so we do not explicitly calculate the integrals in $\mfN^s_1, \widetilde{\mfN}^s_1$ and $\mfN^s_2$.

Then, it is sufficient for our purposes (which is to capture the leading order corrections to the wave-functions) to calculate $\mfN^s_{\rm pt}$, which contains integrals that integrate over the point-nucleus parts of the radial functions of the Dirac-Coulomb modes. These evaluate to,
\bea
&&\cI_{1 \widetilde{1}}^{(\widetilde{\zeta} + \zeta - 2)} = (\widetilde{\kappa}+\kappa)^{\widetilde{\zeta}+ \zeta -1} \cI_{(\widetilde{\zeta} + \zeta - 1)} (\widetilde{\kappa} + \kappa ; a, b, 2\kappa; \widetilde{a}, \widetilde{b}, 2\widetilde{\kappa}) = \notag \\
&& \qquad \Gamma[\widetilde{\zeta} + \zeta -1] \sum_{q = 0}^\infty \frac{(a)_q (2\kappa)^q (\widetilde{\zeta} + \zeta -1)_q}{(b)_q \, q! \, (\widetilde{\kappa} + \kappa)^q} {}_2 \cF_1 \left[ {\widetilde{a}, \widetilde{\zeta} + \zeta -1 +q; \atop \widetilde{b}; } \frac{2\widetilde{\kappa}}{\widetilde{\kappa} + \kappa} \right], \notag\\
&&\cI_{1 \widetilde{2}}^{(\widetilde{\zeta} + \zeta - 2)} =  (\widetilde{\kappa}+\kappa)^{\widetilde{\zeta}+ \zeta -1} \cI_{(\widetilde{\zeta} + \zeta - 1)} (\widetilde{\kappa} + \kappa ; a, b, 2\kappa; \widetilde{a}+1, \widetilde{b}, 2\widetilde{\kappa}) = \notag \\
&& \qquad \Gamma[\widetilde{\zeta} + \zeta -1] \sum_{q = 0}^\infty \frac{(a)_q (2\kappa)^q (\widetilde{\zeta} + \zeta -1)_q}{(b)_q \, q! \, (\widetilde{\kappa} + \kappa)^q} {}_2 \cF_1 \left[ {\widetilde{a}+1, \widetilde{\zeta} + \zeta -1 +q; \atop \widetilde{b}; } \frac{2\widetilde{\kappa}}{\widetilde{\kappa} + \kappa} \right], \notag\\
&&\cI_{2 \widetilde{1}}^{(\widetilde{\zeta} + \zeta - 2)} =  (\widetilde{\kappa}+\kappa)^{\widetilde{\zeta}+ \zeta -1} \cI_{(\widetilde{\zeta} + \zeta - 1)} (\widetilde{\kappa} + \kappa ; a+1, b, 2\kappa; \widetilde{a}, \widetilde{b}, 2\widetilde{\kappa}) = \notag \\
&& \qquad \Gamma[\widetilde{\zeta} + \zeta -1] \sum_{q = 0}^\infty \frac{(a+1)_q (2\kappa)^q (\widetilde{\zeta} + \zeta -1)_q}{(b)_q \, q! \, (\widetilde{\kappa} + \kappa)^q} {}_2 \cF_1 \left[ {\widetilde{a}, \widetilde{\zeta} + \zeta -1 +q; \atop \widetilde{b}; } \frac{2\widetilde{\kappa}}{\widetilde{\kappa} + \kappa} \right], \notag\\
&&\cI_{2 \widetilde{2}}^{(\widetilde{\zeta} + \zeta - 2)} =  (\widetilde{\kappa}+\kappa)^{\widetilde{\zeta}+ \zeta -1} \cI_{(\widetilde{\zeta} + \zeta - 1)} (\widetilde{\kappa} + \kappa ; a+1, b, 2\kappa; \widetilde{a}+1, \widetilde{b}, 2\widetilde{\kappa}) = \notag \\
&& \qquad \Gamma[\widetilde{\zeta} + \zeta -1] \sum_{q = 0}^\infty \frac{(a+1)_q (2\kappa)^q (\widetilde{\zeta} + \zeta -1)_q}{(b)_q \, q! \, (\widetilde{\kappa} + \kappa)^q} {}_2 \cF_1 \left[ {\widetilde{a}+1, \widetilde{\zeta} + \zeta -1 +q; \atop \widetilde{b}; } \frac{2\widetilde{\kappa}}{\widetilde{\kappa} + \kappa} \right], \notag \\
\eea
and then combine in $\mfN^s_{\rm pt}$ to
\bea
&&\mfN^s_{\rm pt} = \Gamma[\widetilde{\zeta} + \zeta -1] \sum_{q = 0}^\infty \frac{(2\kappa)^q (\widetilde{\zeta} + \zeta - 1)_q}{(b)_q \, q! \, (\widetilde{\kappa} + \kappa)^q} \times \notag \\
&& \qquad   \left\{ \cS_+ \left( (a)_q {}_2 \cF_1 \left[ {\widetilde{a}, \widetilde{\zeta} + \zeta -1 +q; \atop \widetilde{b}; } \frac{2\widetilde{\kappa}}{\widetilde{\kappa} + \kappa} \right] -  \right. \right. \notag \\
&& \qquad \qquad \qquad \qquad \qquad \left. \left. \left( \frac{\widetilde{a} a}{\widetilde{c} c} \right) (a+1)_q {}_2 \cF_1 \left[ {\widetilde{a}+1, \widetilde{\zeta} + \zeta -1 +q; \atop \widetilde{b}; } \frac{2\widetilde{\kappa}}{\widetilde{\kappa} + \kappa} \right] \right) + \right. \notag \\
&& \qquad \left. \cS_- \left(  \frac{a(a+1)_q}{c}  {}_2 \cF_1 \left[ {\widetilde{a}, \widetilde{\zeta} + \zeta -1 + q; \atop \widetilde{b}; } \frac{2\widetilde{\kappa}}{\widetilde{\kappa} + \kappa} \right] - \right. \right. \notag \\
&& \qquad \qquad \qquad \qquad \qquad \left. \left. \frac{\widetilde{a} (a)_q}{\widetilde{c}} {}_2 \cF_1 \left[ {\widetilde{a}+1, \widetilde{\zeta} + \zeta -1 +q; \atop \widetilde{b}; } \frac{2\widetilde{\kappa}}{\widetilde{\kappa} + \kappa} \right] \right)\right\}.
\eea
A further simplification is possible thanks to $a = \zeta - Z\alpha \omega/\kappa = 1 - n$ and likewise $\widetilde{a} = \widetilde{\zeta} - Z\alpha \widetilde{\omega}/\widetilde{\kappa} = 1 - \widetilde{n}$ being non-positive integers for all values of $n, \widetilde{n}$ and hence terminating both the explicit series and the series representation of the hypergeometric functions. Then, using \eqref{chfseries} and the identities $a(a+1)_q = (a)_{q+1} = (a)_q (a+q)$ and $(a)_q (a+q)_t = (a)_{q+t}$ we can write after a little bit of algebra,
\bea
\mfN^s_{\rm pt} &=& \Gamma[\widetilde{\zeta} + \zeta -1] \sum_{q = 0}^\infty \sum_{t = 0}^\infty \frac{(2\kappa)^q (2\widetilde{\kappa})^t (\widetilde{\zeta} + \zeta - 1)_{q+t} (a)_q (\widetilde{a})_t}{(\widetilde{\kappa} + \kappa)^{q+t} \, q! \, t! \, (b)_q \, (\widetilde{b})_t \, \widetilde{c} c} \times \notag \\
&& \qquad \Big\{ \cS_+ \left( \widetilde{c} c - (a+q)(\widetilde{a}+t) \right) + \cS_- \left( a - \widetilde{a} +q -t \right) \Big\}, \notag \\
&=&  \sum_{q = 0}^{n-1} \sum_{t = 0}^{\widetilde{n}-1} \frac{\Gamma[\widetilde{\zeta} + \zeta -1](2\kappa)^q (2\widetilde{\kappa})^t (\widetilde{\zeta} + \zeta - 1)_{q+t} (1-n)_q (1-\widetilde{n})_t}{(\widetilde{\kappa} + \kappa)^{q+t} \, q! \, t! \, (1+2\zeta)_q \, (1+2\widetilde{\zeta})_t \, \left(\widetilde{\mfK} - \widetilde{\cN} \right) \Big( \mfK - \cN \Big)} \times \notag \\
&& \qquad \Big\{ \cS_+ \left[ \left(\widetilde{\mfK} - \widetilde{\cN} \right) \Big( \mfK - \cN \Big) - (1-n+q)(1-\widetilde{n}+t) \right] + \cS_- \left( \widetilde{n} - n + q - t \right) \Big\}, \notag \\
\eea
where the last two lines make use of the definitions of the hypergeometric parameters in \eqref{hyperparam} and $\kappa = mZ\alpha/\cN$.

Lastly, let us estimate the size of the coefficients $\cC_{\widetilde{n} n \ssF \ssF_z \widetilde{j} j \widetilde{\varpi} \varpi}$. Assuming that the angular integral and the ratio $(\msC/\widetilde{\msC})$ in \eqref{statecoeff} are $\cO(1)$ numbers and estimating
\bea
\mfN^s &\simeq& m (2\widetilde{\kappa})^{\widetilde{\zeta}-1} (2\kappa)^{\zeta-1} \mfN^s_{\rm pt} \sim m (2\widetilde{\kappa})^{\widetilde{\zeta}-1} (2\kappa)^{\zeta-1} (\widetilde{\kappa} + \kappa)^{1- \widetilde{\zeta} - \zeta} \cS_\pm, \notag \\
&\sim& m (mZ\alpha)^{\widetilde{\zeta}-1} (mZ\alpha)^{\zeta-1} (mZ\alpha)^{1-\widetilde{\zeta} - \zeta}(Z\alpha) \sim \cO(1)
\eea
and also that $\widetilde{\mfD} \simeq \widetilde{\mfD}_{\rm pt} \sim \cO(1)$. In this case we obtain,
\bea
\cC_{\widetilde{n} n \ssF \ssF_z \widetilde{j} j \widetilde{\varpi} \varpi} &=& \frac{(2\widetilde{\kappa})^3 \exx}{m^2} \left( \frac{\msC}{\widetilde{\msC} }\right) \int \exd \Omega_2\, \left( \cY_{\ssF, f_z}^{\widetilde{j} \widetilde{\varpi}} \right)^\dagger \left( I^\theta \sigma^\theta + I^\phi \sigma^\phi \right) \cY_{\ssF, f_z}^{j \varpi} \left( \frac{\mfN^s}{\widetilde{\mfD}} \right), \notag \\
&\sim& m \exx (Z\alpha)^3.
\eea

It turns out that a more careful calculation of $\cC_{\widetilde{n} n \ssF \ssF_z \widetilde{j} j \widetilde{\varpi} \varpi}$ is overkill because the leading corrections to the diagonal first-order state corrections of interest here (\textit{i.e.} ignoring the corrections coming from mixing different angular momentum modes) vanish as is calculated in the next section and argued around \eqref{Lambda+form} in the main text. As such, the leading size of the state corrections (or at least the diagonal contributions) gets suppressed by $(Z\alpha)^2$, making them negligible at the orders we work.

\section{RG evolution}
\label{AppendixF}

This Appendix collects useful parts of the renormalization story told in the main text.

\subsection*{Universal evolution}
\label{ssec:AppEvoEq}

The boundary conditions of the main text provide examples where the effective couplings are found to satisfy equations of the form
\be \label{appgvseps}
  g(\epsilon) =   \frac{A \rho_\epsilon^{2\zeta} + B}{C \rho_\epsilon^{2\zeta} + D}  \,,
\ee
where $g$ is a representative coupling -- such as $g=-(\hat c_s - \hat c_v)/\chi$ in eq.~\pref{running+0} or $g=-(\hat c_s + \hat c_v)\chi$ in \pref{running-0} -- and $\epsilon$ appears on the right-hand side through $\rho_\epsilon = 2\kappa \epsilon$ with $\kappa = \sqrt{m^2 - \omega^2}$. The power of $\rho_\epsilon$ appearing here is $\zeta = \sqrt{\mfK^2 - (Z\alpha)^2}$ where $\mfK = - \varpi (j + \frac12)$. For $j=\frac12$ parity-even states, for example, comparison with \pref{running+0} shows that the parameters $A,B,C$ and $D$ are given explicitly by
\be
   A = c + a \,, \quad B = (c + a' ) \left( \frac{\msD_+}{\msC_+} \right)^{(0)} \,, \quad C = c - a \quad \hbox{and} \quad D = (c - a' )\left( \frac{\msD_+}{\msC_+} \right)^{(0)}\,,
\ee
with parameters $a,a'$ and $c$ given in \pref{hyperparam}, and repeated here:
\be 
\label{hyperparamapp}
a = \zeta-\frac{Z\alpha \omega}{\kappa}, \quad a' =  -\left( \zeta + \frac{Z\alpha\omega}{\kappa} \right),  \quad c = \mfK  - \frac{Z\alpha m}{\kappa},\,.
\ee
For later use, eq.~\pref{appgvseps} also inverts to give 
\be \label{useme1}
   \rho_\epsilon^{2\zeta} =  \frac{B-D g }{C g - A} \,.
\ee

The goal is to derive a universal differential version of this evolution (see, for example \cite{ppeft1, ppeft2, ppeft3, ppeftA} for more details). To start this off directly differentiate \pref{appgvseps} holding $A,B,C,D$ fixed, leading to
\be \label{appgdiffeq}
  \epsilon \, \frac{\exd g}{\exd \epsilon} = 2\zeta \left[ \frac{AD-BC}{(C \rho_\epsilon^{2\zeta} + D)^2} \right] \rho_\epsilon^{2\zeta}=  2\zeta \left[ \frac{(Cg - A)(B-Dg)}{AD-BC} \right] \,,
\ee
where the second equality uses \pref{useme1} to trade $\rho_\epsilon^{2\zeta}$ for $g$. This evolution equation has fixed points at $g = g_*$, where
\be 
   g_* = \frac{A}{C} \quad \hbox{or} \quad g_* = \frac{B}{D} \,,
\ee
which can also be seen as the $\rho_\epsilon \to 0$ and $\rho_\epsilon \to \infty$ limits of \pref{appgvseps}.

This equation can be put into a standard form by redefining $g$ to ensure that $g_* = \pm 1$. To this end write 
\be
  g(\epsilon) = u(\epsilon) + \frac12 \left( \frac{A}{C} + \frac{B}{D} \right) \,,
\ee
in terms of which the fixed points are
\be
  u_* = \pm \frac12 \left(\frac{A}{C} - \frac{B}{D}  \right) = \pm \left( \frac{AD-BC}{2CD} \right) \,,
\ee
and \pref{appgdiffeq} becomes
\be
  \epsilon \, \frac{\exd u}{\exd \epsilon}  
  =  - \frac{2\zeta CD}{AD-BC} \left[ u - \left( \frac{AD-BC}{2CD} \right) \right] \left[ u + \left( \frac{AD-BC}{2CD} \right) \right] \,.
\ee

Finally rescale
\be
   u = \left[ \frac{AD-BC}{2CD} \right] v 
\ee
to see that 
\be \label{appuniversalDE}
  \epsilon \, \frac{\exd v}{\exd \epsilon} =  \zeta (1-v^2) \,
\ee
is an automatic consequence of \pref{appgvseps} once one defines
\be
  g = u + \frac{AD+BC}{2CD}  
  =\frac12 \left( \frac{A}{C} - \frac{B}{D} \right) v + \frac12 \left( \frac{A}{C} + \frac{B}{D} \right) \,.
\ee

These expressions emphasize that although the positions of the fixed points for $g$ depend on the ratios $A/C$ and $B/D$, the speed of evolution along the RG flow depends only on $\zeta$. Indeed the general solution to \pref{appuniversalDE} is 
\be \label{appRGsolution}
  v(\epsilon) = \frac{ (v_0 + 1)(\epsilon/\epsilon_0)^{2\zeta} + (v_0 - 1)}{(v_0 + 1)(\epsilon/\epsilon_0)^{2\zeta} - (v_0 - 1)} 
\ee
where the integration constant is chosen to ensure $v(\epsilon_0) = v_0$. For $\zeta > 0$ this describes a universal flow that runs from $v = -1$ to $v = +1$ as $\epsilon$ flows from 0 to $\infty$. 

Since the trajectories given in \pref{appRGsolution} cannot cross the lines $v = \pm 1$ for any finite nonzero $\epsilon$ there are two categories of flow, distinguished by the flow-invariant sign of $|v| - 1$ (see Figure \ref{figureOldFlow}). That is, if $|v_0| - 1$ is negative (positive) for any $0 < \epsilon_0 < \infty$, then $|v(\epsilon)| -1$ is negative (positive) for all $0 < \epsilon < \infty$. Every trajectory is therefore uniquely characterized by a pair of numbers. These can equally well be chosen to be the pair $(\epsilon_0, v_0)$ that specifies an initial condition $v_0 = v(\epsilon_0)$, or it can be taken to be the pair $(\epsilon_\star, y_\star)$ where $y_\star = \hbox{sign}(|v|-1) = \pm 1$ distinguishes the two classes of trajectories, and $\epsilon_\star$ is defined as the value of $\epsilon$ for which $v(\epsilon_\star) = 0$ (if $y_\star = -1$) or the value for which $v(\epsilon_\star) = \infty$ (if $y_\star = + 1$). The parameterization using $(\epsilon_\star, y_\star)$ is useful because physical observables turn out to have particularly transparent expressions in terms of these variables.  

For the specific case of $j= \frac12$ parity-even states these parameter combinations become $\zeta = \sqrt{1-(Z\alpha)^2}$ and
\bea
   \frac{A}{C} &=& \frac{c+a}{c-a} = \frac{1-\zeta +(m+\omega)Z\alpha/\kappa}{1+\zeta + (m-\omega)Z\alpha/\kappa}  \nn\\
   \frac{B}{D} &=& \frac{c+a'}{c-a'} =  \frac{1+\zeta +(m+\omega)Z\alpha/\kappa}{1-\zeta + (m-\omega)Z\alpha/\kappa} 
\eea
Using, in these, the leading Coulomb expression $m - \omega \simeq (Z\alpha)^2m/(2n^2)$ and so $\kappa \simeq Z\alpha m/n$ as well as $\zeta \simeq 1 - \frac12(Z\alpha)^2$ then leads to the approximate forms
\be
   \frac{A}{C} \simeq n + \cdots \,, \quad
   \frac{B}{D} \simeq \frac{2n}{(Z\alpha)^2} + \cdots\,,
\ee
up to terms suppressed by $(Z\alpha)^2$ compared to those shown.

\subsection*{Evolution for positive-parity $j=\frac12$ states}

The importance of calculating the first-order state corrections above is that the alternative boundary conditions in \eqref{bc++} and \eqref{bc--} set various combinations of the PPEFT couplings equal to the ratios of the full radial functions, $\mfg_{n\frac{1}{2}+}(\epsilon) / \mff_{n\frac{1}{2}+}(\epsilon)$ and $\mff_{n\frac{1}{2}-}(\epsilon)/\mfg_{n\frac{1}{2}-}(\epsilon)$ when applied to $j=1/2$ positive- and negative parity states respectively. The new coupling $c_\ssF$ sits on the left-hand side of these equations, which we assume to be of size $\exx$ and we further anticipate that the couplings present in the case of spinless nuclei $c_s, c_v$ also receive spin-dependent corrections that first appear at this order. Matching powers of $\exx$ on both sides of the boundary condition then requires us to compute all $\cO(\exx)$ contributions to the radial function ratios, which is what we will do now for both parities, starting with the positive-parity state. In what follows we will suppress both the arguments and the quantum number labels of the functions, except for parity.

\subsubsection*{Evolution for positive parity $j=1/2$ states}

On the right-hand side of the positive-parity, $j=1/2$ states' boundary conditions in \eqref{bc++} sits the ratio $\( \mfg_+/\mff_+\)$, which can be expanded to first order in degenerate perturbation theory schematically as,
\begin{align}
\label{gfratioapp1}
&\frac{\mfg_+}{\mff_+} = \frac{\mfg_+^{(0)} + \exx \mfg_+^{(1)} + \cdots}{\mff_+^{(0)} + \exx  \mff_+^{(1)} + \cdots} \approx \frac{\mfg_+^{(0)}}{\mff_+^{(0)}} + \exx \left( \frac{\mfg_+^{(1)}}{\mff_+^{(0)}} - \frac{\mfg_+^{(0)}}{\mff_+^{(0)}} \frac{\mff_+^{(1)}}{\mff_+^{(0)}} \right) + \cO \( \exx^2 \),
\end{align}
where $\mfg_+^{(0)}$ and $\mff_+^{(0)}$ are given in \eqref{dc} and $\mfg_+^{(1)}$ and $\mff_+^{(1)}$ are given in \eqref{fg1} using appropriate substitutions for the quantum number labels. Before proceeding any further, it is important to remember that the superscripts on these functions refer to their order in degenerate perturbation theory and not necessarily whether or not they are complete in any order in $\exx$. To emphasize, we had \textit{defined} $\mfg_+^{(1)}$ and $\mff_+^{(1)}$ to be the corrections to the radial solutions of the Dirac-Coulomb problem, $\mfg_+^{(0)}, \mff_+^{(0)}$  that come about purely as a result of degenerate perturbation theory, but \textit{not} including the expansion of the integration constant ratios in \eqref{caexp} and as such both $\mfg_+^{(0)}, \mfg_+^{(1)}$ and $\mff_+^{(0)}, \mff_+^{(1)}$ are still functions of the full $\left( \msD_+ /\msC_+ \right)$. This means that in order to get all the contributions to $\cO\(\exx\)$ in the ratio $\( \mfg_+/\mff_+ \)$ we still need to use \eqref{caexp} in $\mfg_+^{(0)}$ and $\mff_+^{(0)}$, but not in $\mfg_+^{(1)}, \mff_+^{(1)}$ since these are already $\cO\(\exx\)$. Then, focusing on the first term on the right-hand side of \eqref{gfratioapp1} we find
\begin{align}
&\!\! \frac{\mfg_+^{(0)}}{\mff_+^{(0)}} = -\chi \frac{\left[\cM_1 + \frac{a}{c} \cM_2 \right] + \( \frac{\msD_+}{\msC_+} \)^{(0)} \rho^{-2\zeta} \left[ \cM_3 + \frac{a'}{c} \cM_4 \right] + \( \frac{\msD_+}{\msC_+} \)^{(1)} \rho^{-2\zeta} \left[ \cM_3 + \frac{a'}{c} \cM_4 \right]}{\left[\cM_1 - \frac{a}{c} \cM_2 \right] + \( \frac{\msD_+}{\msC_+} \)^{(0)} \rho^{-2\zeta} \left[ \cM_3 - \frac{a'}{c} \cM_4 \right] + \( \frac{\msD_+}{\msC_+} \)^{(1)} \rho^{-2\zeta} \left[ \cM_3 - \frac{a'}{c} \cM_4 \right]}, \notag \\
&\!\!\! \qquad =  -\chi \frac{\left[\cM_1 + \frac{a}{c} \cM_2 \right] + \( \frac{\msD_+}{\msC_+} \)^{(0)} \rho^{-2\zeta} \left[ \cM_3 + \frac{a'}{c} \cM_4 \right]}{\left[\cM_1 - \frac{a}{c} \cM_2 \right] + \( \frac{\msD_+}{\msC_+} \)^{(0)} \rho^{-2\zeta} \left[ \cM_3 - \frac{a'}{c} \cM_4 \right]}  \notag \\
& \qquad \qquad \qquad - \exx \frac{2\(\frac{\msD_+}{\msC_+} \)^{(1)} \chi \rho^{-2 \zeta} \( a'\cM_1\cM_4 - a\cM_2 \cM_3 \) }{c \(\left[\cM_1 - \frac{a}{c} \cM_2 \right] + \( \frac{\msD_+}{\msC_+} \)^{(0)} \rho^{-2\zeta} \left[ \cM_3 - \frac{a'}{c} \cM_4 \right]\)^2}.
\end{align}
Substituting this into \eqref{gfratioapp1} along with the explicit functional forms from \eqref{dc} and making use of \eqref{fg1} and \eqref{caexp} we can write the ratio of positive parity radial functions as
\begin{align}
& \frac{\mfg_+}{\mff_+} \approx -\chi \frac{\left[\cM_1 + \frac{a}{c} \cM_2 \right] + \( \frac{\msD_+}{\msC_+} \)^{(0)} \rho^{-2\zeta} \left[ \cM_3 + \frac{a'}{c} \cM_4 \right]}{\left[\cM_1 - \frac{a}{c} \cM_2 \right] + \( \frac{\msD_+}{\msC_+} \)^{(0)} \rho^{-2\zeta} \left[ \cM_3 - \frac{a'}{c} \cM_4 \right]}  \notag \\
 & - \exx \left(  \frac{2\(\frac{\msD_+}{\msC_+} \)^{(1)} \chi \rho^{-2 \zeta} \( a'\cM_1\cM_4 - a\cM_2 \cM_3 \) }{c \(\left[\cM_1 - \frac{a}{c} \cM_2 \right] + \( \frac{\msD_+}{\msC_+} \)^{(0)} \rho^{-2\zeta} \left[ \cM_3 - \frac{a'}{c} \cM_4 \right]\)^2}  \right. \notag \\
& + \widehat{\sum} \sqrt{\frac{m-\widetilde{\omega}}{m+\omega}} \left( \frac{\widetilde{\msC}_+ e^{-\widetilde{\rho}/2} \widetilde{\rho}^{\zeta-1} }{\msC_+ e^{-\rho/2} \rho^{\zeta-1}} \right) \frac{\left[\cM_{\widetilde{1}} + \frac{\widetilde{a}}{\widetilde{c}} \cM_{\widetilde{2}} \right] + \( \frac{\widetilde{\msD}_+}{\widetilde{\msC}_+} \)^{(0)} \widetilde{\rho}^{-2\zeta} \left[ \cM_{\widetilde{3}} + \frac{\widetilde{a}'}{\widetilde{c}} \cM_{\widetilde{4}} \right]}{\left[\cM_1 - \frac{a}{c} \cM_2 \right] + \( \frac{\msD_+}{\msC_+} \)^{(0)} \rho^{-2\zeta} \left[ \cM_3 - \frac{a'}{c} \cM_4 \right]}  \notag \\
&- \widehat{\sum}  \chi \frac{\left[\cM_1 + \frac{a}{c} \cM_2 \right] + \( \frac{\msD_+}{\msC_+} \)^{(0)} \rho^{-2\zeta} \left[ \cM_3 + \frac{a'}{c} \cM_4 \right]}{\left[\cM_1 - \frac{a}{c} \cM_2 \right] + \( \frac{\msD_+}{\msC_+} \)^{(0)} \rho^{-2\zeta} \left[ \cM_3 - \frac{a'}{c} \cM_4 \right]}  \notag \\
& \times \left. \sqrt{\frac{m+\widetilde{\omega}}{m+ \omega}}  \left( \frac{\widetilde{\msC}_+ e^{-\widetilde{\rho}/2} \widetilde{\rho}^{\zeta-1} }{\msC_+ e^{-\rho/2} \rho^{\zeta-1}} \right) \frac{\left[\cM_{\widetilde{1}} - \frac{\widetilde{a}}{\widetilde{c}} \cM_{\widetilde{2}} \right] + \( \frac{\widetilde{\msD}_+}{\widetilde{\msC}_+} \)^{(0)} \widetilde{\rho}^{-2\zeta} \left[ \cM_{\widetilde{3}} - \frac{\widetilde{a}'}{\widetilde{c}} \cM_{\widetilde{4}} \right]}{\left[\cM_1 - \frac{a}{c} \cM_2 \right] + \( \frac{\msD_+}{\msC_+} \)^{(0)} \rho^{-2\zeta} \left[ \cM_3 - \frac{a'}{c} \cM_4 \right]} \right) + \cO\(\exx^2\).
\end{align}
With an eye to the future progression of this calculation in the main text, where the terms including the sums will cancel (to leading order in $\rho$), we can massage this into the form,
\begin{align}
& \frac{\mfg_+}{\mff_+} \approx -\chi \frac{\left[c \cM_1 + a \cM_2 \right] + \( \frac{\msD_+}{\msC_+} \)^{(0)} \rho^{-2\zeta} \left[ c\cM_3 + a' \cM_4 \right]}{\left[c\cM_1 - a \cM_2 \right] + \( \frac{\msD_+}{\msC_+} \)^{(0)} \rho^{-2\zeta} \left[ c \cM_3 - a' \cM_4 \right]}  \notag \\
 & - \exx \left(  \frac{2\(\frac{\msD_+}{\msC_+} \)^{(1)} \chi \rho^{-2 \zeta} \( a'\cM_1\cM_4 - a\cM_2 \cM_3 \) c }{\(\left[c\cM_1 - a \cM_2 \right] + \( \frac{\msD_+}{\msC_+} \)^{(0)} \rho^{-2\zeta} \left[ c \cM_3 - a' \cM_4 \right]\)^2}  \right. \notag \\
& + \widehat{\sum} \sqrt{\frac{m+\widetilde{\omega}}{m+\omega}} \left( \frac{\widetilde{\msC}_+ e^{-\widetilde{\rho}/2} \widetilde{\rho}^{\zeta-1} c}{\msC_+ e^{-\rho/2} \rho^{\zeta-1} \widetilde{c}} \right) \frac{\left[\widetilde{c} \cM_{\widetilde{1}} - \widetilde{a}\cM_{\widetilde{2}} \right] + \( \frac{\widetilde{\msD}_+}{\widetilde{\msC}_+} \)^{(0)} \widetilde{\rho}^{-2\zeta} \left[ \widetilde{c} \cM_{\widetilde{3}}  - \widetilde{a}'\cM_{\widetilde{4}} \right]}{\left[c \cM_1 - a\cM_2 \right] + \( \frac{\msD_+}{\msC_+} \)^{(0)} \rho^{-2\zeta} \left[ c\cM_3  - a'\cM_4  \right]}  \notag \\
&\times \left\{ \widetilde{\chi} \frac{\left[\widetilde{c} \cM_{\widetilde{1}} + \widetilde{a} \cM_{\widetilde{2}} \right] + \( \frac{\widetilde{\msD}_+}{\widetilde{\msC}_+} \)^{(0)} \widetilde{\rho}^{-2\zeta} \left[ \widetilde{c} \cM_{\widetilde{3}} + \widetilde{a}' \cM_{\widetilde{4}} \right]}{\left[\widetilde{c} \cM_{\widetilde{1}} - \widetilde{a} \cM_{\widetilde{2}} \right] + \( \frac{\widetilde{\msD}_+}{\widetilde{\msC}_+} \)^{(0)} \widetilde{\rho}^{-2\zeta} \left[ \widetilde{c} \cM_{\widetilde{3}} - \widetilde{a}' \cM_{\widetilde{4}} \right]}   \right. \notag \\
&- \left. \left. \chi \frac{\left[c \cM_1 + a \cM_2 \right] + \( \frac{\msD_+}{\msC_+} \)^{(0)} \rho^{-2\zeta} \left[ c \cM_3 + a' \cM_4 \right]}{\left[c \cM_1 - a \cM_2 \right] + \( \frac{\msD_+}{\msC_+} \)^{(0)} \rho^{-2\zeta} \left[ c \cM_3 - a' \cM_4 \right]} \right\} \right),
\end{align}
where the leading small $\epsilon$ expansion (and identically small $\rho$ expansion) yields,
\begin{align}
& \frac{\mfg_+}{\mff_+} \approx -\chi \frac{\left[c + a \right] + \( \frac{\msD_+}{\msC_+} \)^{(0)} \rho^{-2\zeta} \left[ c + a' \right]}{\left[c - a \right] + \( \frac{\msD_+}{\msC_+} \)^{(0)} \rho^{-2\zeta} \left[ c - a' \right]} - \exx \left(  \frac{2\(\frac{\msD_+}{\msC_+} \)^{(1)} \chi \rho^{-2 \zeta} \( a' - a \) c }{\(\left[c - a \right] + \( \frac{\msD_+}{\msC_+} \)^{(0)} \rho^{-2\zeta} \left[ c  - a' \right]\)^2}  \right. \notag \\
& + \widehat{\sum} \sqrt{\frac{m+\widetilde{\omega}}{m+\omega}} \left( \frac{\widetilde{\msC}_+ \widetilde{\rho}^{\zeta-1} c}{\msC_+  \rho^{\zeta-1} \widetilde{c}} \right) \frac{\left[\widetilde{c}  - \widetilde{a} \right] + \( \frac{\widetilde{\msD}_+}{\widetilde{\msC}_+} \)^{(0)} \widetilde{\rho}^{-2\zeta} \left[ \widetilde{c} - \widetilde{a}' \right]}{\left[c - a \right] + \( \frac{\msD_+}{\msC_+} \)^{(0)} \rho^{-2\zeta} \left[ c - a' \right]}  \notag \\
& \left. \times \left\{ \widetilde{\chi} \frac{\left[\widetilde{c} + \widetilde{a} \right] + \( \frac{\widetilde{\msD}_+}{\widetilde{\msC}_+} \)^{(0)} \widetilde{\rho}^{-2\zeta} \left[ \widetilde{c} + \widetilde{a}' \right]}{\left[\widetilde{c} - \widetilde{a} \right] + \( \frac{\widetilde{\msD}_+}{\widetilde{\msC}_+} \)^{(0)} \widetilde{\rho}^{-2\zeta} \left[ \widetilde{c} - \widetilde{a}' \right]} - \chi \frac{\left[c + a \right] + \( \frac{\msD_+}{\msC_+} \)^{(0)} \rho^{-2\zeta} \left[ c + a' \right]}{\left[c - a  \right] + \( \frac{\msD_+}{\msC_+} \)^{(0)} \rho^{-2\zeta} \left[ c - a' \right]} \right\} \right) + \cO\(\exx^2\).
\end{align}

\subsection*{Evolution for negative-parity $j=\frac12$ states}

Moving on to the negative parity, $j=1/2$ states, the right-hand side of the boundary condition in \eqref{bc--} is equivalent to the ratio $\( \mff_-/\mfg_-\)$, which using \eqref{psi1} can be expanded to first order in $\exx$ schematically as,
\begin{align}
\label{gfratioapp-}
&\frac{\mff_-}{\mfg_-} = \frac{\mff_-^{(0)} + \exx \mff_-^{(1)} + \cdots}{\mfg_-^{(0)} + \exx  \mfg_-^{(1)} + \cdots} \approx \frac{\mff_-^{(0)}}{\mfg_-^{(0)}} + \exx \left( \frac{\mff_-^{(1)}}{\mfg_-^{(0)}} - \frac{\mff_-^{(0)}}{\mfg_-^{(0)}} \frac{\mfg_-^{(1)}}{\mfg_-^{(0)}} \right) + \cO \( \exx^2 \),
\end{align}
where $\mff_-^{(0)}$ and $\mfg_-^{(0)}$ are given in \eqref{dc} and $\mff_-^{(1)}$ and $\mfg_-^{(1)}$ are given in \eqref{fg1} using appropriate substitutions for the quantum number labels. A before, these functions still contain the full integration constant ratio $\(\msD_-/\msC_-\)$, therefore to complete the expansion of $\(\mff_-/\mfg_-\)$ to linear order in $\exx$ we need to make use of \eqref{caexp} in $\mff^{(0)}_-/\mfg^{(0)}_-$. Concentrating on this term on the right-hand side of \eqref{gfratioapp-} we find
\begin{align}
&\!\! \frac{\mff_-^{(0)}}{\mfg_-^{(0)}} = -\chi^{-1} \frac{\left[\cM_1 - \frac{a}{c} \cM_2 \right] + \( \frac{\msD_-}{\msC_-} \)^{(0)} \rho^{-2\zeta} \left[ \cM_3 - \frac{a'}{c} \cM_4 \right] + \( \frac{\msD_-}{\msC_-} \)^{(1)} \rho^{-2\zeta} \left[ \cM_3 - \frac{a'}{c} \cM_4 \right]}{\left[\cM_1 + \frac{a}{c} \cM_2 \right] + \( \frac{\msD_-}{\msC_-} \)^{(0)} \rho^{-2\zeta} \left[ \cM_3 + \frac{a'}{c} \cM_4 \right] + \( \frac{\msD_-}{\msC_-} \)^{(1)} \rho^{-2\zeta} \left[ \cM_3 + \frac{a'}{c} \cM_4 \right]}, \notag \\
&\!\!\! \qquad =  -\chi^{-1} \frac{\left[\cM_1 - \frac{a}{c} \cM_2 \right] + \( \frac{\msD_-}{\msC_-} \)^{(0)} \rho^{-2\zeta} \left[ \cM_3 - \frac{a'}{c} \cM_4 \right]}{\left[\cM_1 + \frac{a}{c} \cM_2 \right] + \( \frac{\msD_-}{\msC_-} \)^{(0)} \rho^{-2\zeta} \left[ \cM_3 + \frac{a'}{c} \cM_4 \right]}  \notag \\
& \qquad \qquad \qquad + \exx \frac{2\(\frac{\msD_-}{\msC_-} \)^{(1)} \rho^{-2 \zeta} \( a'\cM_1\cM_4 - a\cM_2 \cM_3 \) }{c\chi \(\left[\cM_1 + \frac{a}{c} \cM_2 \right] + \( \frac{\msD_-}{\msC_-} \)^{(0)} \rho^{-2\zeta} \left[ \cM_3 + \frac{a'}{c} \cM_4 \right]\)^2}.
\end{align}
Substituting this into \eqref{gfratioapp-} along with the explicit functional forms from \eqref{dc} and making use of \eqref{fg1} and \eqref{caexp} we can write the ratio of negative parity radial functions as
\begin{align}
& \frac{\mff_-}{\mfg_-} \approx -\chi^{-1} \frac{\left[\cM_1 - \frac{a}{c} \cM_2 \right] + \( \frac{\msD_-}{\msC_-} \)^{(0)} \rho^{-2\zeta} \left[ \cM_3 - \frac{a'}{c} \cM_4 \right]}{\left[\cM_1 + \frac{a}{c} \cM_2 \right] + \( \frac{\msD_-}{\msC_-} \)^{(0)} \rho^{-2\zeta} \left[ \cM_3 + \frac{a'}{c} \cM_4 \right]}  \notag \\
& + \exx \left(  \frac{2\(\frac{\msD_-}{\msC_-} \)^{(1)} \rho^{-2 \zeta} \( a'\cM_1\cM_4 - a\cM_2 \cM_3 \) }{c \chi \(\left[\cM_1 + \frac{a}{c} \cM_2 \right] + \( \frac{\msD_-}{\msC_-} \)^{(0)} \rho^{-2\zeta} \left[ \cM_3 + \frac{a'}{c} \cM_4 \right]\)^2}  \right. \notag \\
& - \widehat{\sum} \sqrt{\frac{m+\widetilde{\omega}}{m-\omega}} \left( \frac{\widetilde{\msC}_- e^{-\widetilde{\rho}/2} \widetilde{\rho}^{\zeta-1} }{\msC_- e^{-\rho/2} \rho^{\zeta-1}} \right) \frac{\left[\cM_{\widetilde{1}} - \frac{\widetilde{a}}{\widetilde{c}} \cM_{\widetilde{2}} \right] + \( \frac{\widetilde{\msD}_-}{\widetilde{\msC}_-} \)^{(0)} \widetilde{\rho}^{-2\zeta} \left[ \cM_{\widetilde{3}} - \frac{\widetilde{a}'}{\widetilde{c}} \cM_{\widetilde{4}} \right]}{\left[\cM_1 + \frac{a}{c} \cM_2 \right] + \( \frac{\msD_-}{\msC_-} \)^{(0)} \rho^{-2\zeta} \left[ \cM_3 + \frac{a'}{c} \cM_4 \right]}  \notag \\
&+ \widehat{\sum}  \chi^{-1} \frac{\left[\cM_1 - \frac{a}{c} \cM_2 \right] + \( \frac{\msD_-}{\msC_-} \)^{(0)} \rho^{-2\zeta} \left[ \cM_3 - \frac{a'}{c} \cM_4 \right]}{\left[\cM_1 + \frac{a}{c} \cM_2 \right] + \( \frac{\msD_-}{\msC_-} \)^{(0)} \rho^{-2\zeta} \left[ \cM_3 + \frac{a'}{c} \cM_4 \right]}  \notag \\
& \times \left. \sqrt{\frac{m-\widetilde{\omega}}{m- \omega}}  \left( \frac{\widetilde{\msC}_- e^{-\widetilde{\rho}/2} \widetilde{\rho}^{\zeta-1} }{\msC_- e^{-\rho/2} \rho^{\zeta-1}} \right) \frac{\left[\cM_{\widetilde{1}} + \frac{\widetilde{a}}{\widetilde{c}} \cM_{\widetilde{2}} \right] + \( \frac{\widetilde{\msD}_-}{\widetilde{\msC}_-} \)^{(0)} \widetilde{\rho}^{-2\zeta} \left[ \cM_{\widetilde{3}} + \frac{\widetilde{a}'}{\widetilde{c}} \cM_{\widetilde{4}} \right]}{\left[\cM_1 + \frac{a}{c} \cM_2 \right] + \( \frac{\msD_-}{\msC_-} \)^{(0)} \rho^{-2\zeta} \left[ \cM_3 + \frac{a'}{c} \cM_4 \right]} \right) + \cO\(\exx^2\).
\end{align}
With an eye to the future progression of this calculation in the main text, where the terms including the sums will cancel (to leading order in $\rho$), we can massage this into the form,
\begin{align}
& \frac{\mff_-}{\mfg_-} \approx -\chi^{-1} \frac{\left[c \cM_1 - a \cM_2 \right] + \( \frac{\msD_-}{\msC_-} \)^{(0)} \rho^{-2\zeta} \left[ c\cM_3 - a' \cM_4 \right]}{\left[c\cM_1 + a \cM_2 \right] + \( \frac{\msD_-}{\msC_-} \)^{(0)} \rho^{-2\zeta} \left[ c \cM_3 + a' \cM_4 \right]}  \notag \\
& + \exx \left(  \frac{2\chi^{-1}\(\frac{\msD_-}{\msC_-} \)^{(1)} \rho^{-2 \zeta} \( a'\cM_1\cM_4 - a\cM_2 \cM_3 \) c }{\(\left[c\cM_1 + a \cM_2 \right] + \( \frac{\msD_-}{\msC_-} \)^{(0)} \rho^{-2\zeta} \left[ c \cM_3 + a' \cM_4 \right]\)^2}  \right. \notag \\
& - \widehat{\sum} \sqrt{\frac{m-\widetilde{\omega}}{m-\omega}} \left( \frac{\widetilde{\msC}_- e^{-\widetilde{\rho}/2} \widetilde{\rho}^{\zeta-1} c}{\msC_- e^{-\rho/2} \rho^{\zeta-1} \widetilde{c}} \right) \frac{\left[\widetilde{c} \cM_{\widetilde{1}} + \widetilde{a}\cM_{\widetilde{2}} \right] + \( \frac{\widetilde{\msD}_-}{\widetilde{\msC}_-} \)^{(0)} \widetilde{\rho}^{-2\zeta} \left[ \widetilde{c} \cM_{\widetilde{3}}  + \widetilde{a}'\cM_{\widetilde{4}} \right]}{\left[c \cM_1 + a\cM_2 \right] + \( \frac{\msD_-}{\msC_-} \)^{(0)} \rho^{-2\zeta} \left[ c\cM_3  + a'\cM_4  \right]}  \notag \\
&\times \left\{ \widetilde{\chi}^{-1} \frac{\left[\widetilde{c} \cM_{\widetilde{1}} - \widetilde{a} \cM_{\widetilde{2}} \right] + \( \frac{\widetilde{\msD}_-}{\widetilde{\msC}_-} \)^{(0)} \widetilde{\rho}^{-2\zeta} \left[ \widetilde{c} \cM_{\widetilde{3}} - \widetilde{a}' \cM_{\widetilde{4}} \right]}{\left[\widetilde{c} \cM_{\widetilde{1}} + \widetilde{a} \cM_{\widetilde{2}} \right] + \( \frac{\widetilde{\msD}_-}{\widetilde{\msC}_-} \)^{(0)} \widetilde{\rho}^{-2\zeta} \left[ \widetilde{c} \cM_{\widetilde{3}} + \widetilde{a}' \cM_{\widetilde{4}} \right]}   \right. \notag \\
&- \left. \left. \chi^{-1} \frac{\left[c \cM_1 - a \cM_2 \right] + \( \frac{\msD_-}{\msC_-} \)^{(0)} \rho^{-2\zeta} \left[ c \cM_3 - a' \cM_4 \right]}{\left[c \cM_1 + a \cM_2 \right] + \( \frac{\msD_+}{\msC_+} \)^{(0)} \rho^{-2\zeta} \left[ c \cM_3 + a' \cM_4 \right]} \right\} \right),
\end{align}
where the leading small $\epsilon$ expansion (and identically small $\rho$ expansion) yields,
\begin{align}
& \frac{\mff_-}{\mfg_-} \approx -\chi^{-1} \frac{\left[c  - a \right] + \( \frac{\msD_-}{\msC_-} \)^{(0)} \rho^{-2\zeta} \left[ c - a' \right]}{\left[c + a  \right] + \( \frac{\msD_-}{\msC_-} \)^{(0)} \rho^{-2\zeta} \left[ c  + a'  \right]} + \exx \left(  \frac{2\chi^{-1}\(\frac{\msD_-}{\msC_-} \)^{(1)} \rho^{-2 \zeta} \( a' - a \) c }{\(\left[c + a \right] + \( \frac{\msD_-}{\msC_-} \)^{(0)} \rho^{-2\zeta} \left[ c + a'  \right]\)^2}  \right. \notag \\
& - \widehat{\sum} \sqrt{\frac{m-\widetilde{\omega}}{m-\omega}} \left( \frac{\widetilde{\msC}_- e^{-\widetilde{\rho}/2} \widetilde{\rho}^{\zeta-1} c}{\msC_- e^{-\rho/2} \rho^{\zeta-1} \widetilde{c}} \right) \frac{\left[\widetilde{c}  + \widetilde{a} \right] + \( \frac{\widetilde{\msD}_-}{\widetilde{\msC}_-} \)^{(0)} \widetilde{\rho}^{-2\zeta} \left[ \widetilde{c}  + \widetilde{a}' \right]}{\left[c + a \right] + \( \frac{\msD_-}{\msC_-} \)^{(0)} \rho^{-2\zeta} \left[ c  + a'  \right]}  \notag \\
& \left. \times \left\{ \widetilde{\chi}^{-1} \frac{\left[\widetilde{c} - \widetilde{a} \right] + \( \frac{\widetilde{\msD}_-}{\widetilde{\msC}_-} \)^{(0)} \widetilde{\rho}^{-2\zeta} \left[ \widetilde{c}  - \widetilde{a}'  \right]}{\left[\widetilde{c}  + \widetilde{a} \right] + \( \frac{\widetilde{\msD}_-}{\widetilde{\msC}_-} \)^{(0)} \widetilde{\rho}^{-2\zeta} \left[ \widetilde{c}  + \widetilde{a}'  \right]} -  \chi^{-1} \frac{\left[c  - a  \right] + \( \frac{\msD_-}{\msC_-} \)^{(0)} \rho^{-2\zeta} \left[ c  - a' \right]}{\left[c  + a \right] + \( \frac{\msD_+}{\msC_+} \)^{(0)} \rho^{-2\zeta} \left[ c + a'  \right]} \right\} \right).
\end{align}

\section{List of symbols}

\label{los}

\begin{tabular}{l p{0.68\linewidth}}
$Z$ & Atomic number of an element \\
\\
$\alpha = \frac{e^2}{4\pi}$ & Fine-structure constant \\ 
\\
$N_{\rm exp}, N_{\rm nuc}$ & The number of available experimental observables and the number of nuclear parameters\\
\\
$\varepsilon_n$ & Bohr energy level of a lepton\\
\\
$n$ & Principal quantum number of a leptonic energy level\\
\\
$m$ & Mass of the lepton orbiting the nucleus \\
\\
$M$ & Mass of the nucleus\\
\\
$m_r$ & Reduced mass of the nucleus-lepton system\\
\\
$e$ & Electric charge unit \\
\\
$\hbar$ & Reduced Planck's constant\\
\\
$c$ & Speed of light in vacuum\\
\\
$k_\ssB$ & Boltzmann constant\\
\\
$\mathbf{v}_e, \mathrm{v}_e \sim (Z\alpha)$ & Velocity and speed of the nucleus-orbiting lepton\\
\\
$R$ & A length-scale of approximately nuclear size, \textit{i.e.} ~1 fm \\
\\
$a_\ssB = (mZ\alpha)^{-1}$ & Bohr radius of the atom\\
\\
$\exx = \frac{m e \mu_N}{4\pi}$ & The small parameter controlling the effects of the hyperfine interaction\\
\\
$j, j_z$ & Quantum numbers of the total leptonic angular momentum $\bfJ = \bfL + \bfS$ and its projection
\end{tabular}

\noindent
\begin{tabular}{l p{0.68\linewidth}}
$S_\NQED$ & The renormalizable action of a theory treating the nucleus as a relativistic point-like particle interacting with photons and another lepton 
species\\
\\
$F_{\mu\nu}$ & Field strength of the U(1) gauge field, $A^\mu(x)$ \\
\\
$A^\mu(x)$ & U(1) vector field \\
\\
$D_\mu = \partial_\mu - iqA_\mu$ & Covariant derivative of a field charged under the U(1) gauge group with charge $q$ \\
\\
$\Psi, \ol{\Psi}$ & Leptonic Dirac field and its Dirac conjugate \\
\\
$\Phi, \ol{\Phi}$ & Nuclear Dirac field and its Dirac conjugate \\
\\
$\gamma^\mu, \gamma_5$ & Dirac gamma matrices \\
\\
$\gamma^{\mu\nu} = -\frac{i}{4} \left[ \gamma^\mu, \gamma^\nu \right]$ & Lorentz algebra generators for Dirac particles\\
\\
$\slashed{D} = \gamma^\mu D_\mu$ & Slashes indicate contraction with Dirac gamma matrices\\
\\
$S_{\rm nuc}$ & The higher-dimensional extension of $S_\NQED$ containing non-renormalizable interactions between the second-quantized nuclear and leptonic fields and a U(1) gauge field\\
\\
$\widetilde c_s, \widetilde c_v, \widetilde c_d$ & Generic EFT couplings in $S_{\rm nuc}$ that are related to nuclear properties \\
\\
$\msP$ & Curve mapping the real line, $\mathbb{R}$ to the position of the nucleus\\ 
\\
$x^\mu$ & Arbitrary position 4-vector\\
\\
$s$ & Arbitrary parameter along the world-line of the nucleus\\
\\
$y^\mu(s)$ & 4-vector trajectory of the nucleus, parameterized by $s$\\
\\
$S_\QED$ & The standard QED action describing the interaction between a Dirac particle and a U(1) gauge field\\
\\
$S_p$ & The 1-dimensional action of a point-particle
\end{tabular}

\noindent
\begin{tabular}{l p{0.68\linewidth}}
$S = S_\QED + S_p$ & Total action of the PPEFT\\
\\
$\mathbf{v} \sim m(Z\alpha)/M$ & Velocity of the nucleus\\
\\
$\gamma := (1-\mathbf{v}^2)^{-1/2}$ & The relativistic conversion factor\\
\\
$c_s, c_v, c_\ssF, c_{em}, c_2, c_3$ & Generic EFT couplings in the PPEFT arising at order $({\rm length})^2$\\
\\
$\tau$ & Proper time along the point-particle's trajectory\\
\\
$\eta_{\mu\nu}, \eta^{\mu\nu}$ & Minkowski metric and its inverse with signature (-, +, +, +) \\
\\
$\bfe_r$ & Radially pointing unit normal vector\\
\\
$u_{\ssL}(t,r,\theta,\varphi)$ & Separable solution to the leptonic field equations\\
\\
$\cR_{\ssL}(\kappa r)$ & Radial part of the solution to the leptonic field equations\\
\\
$\omega$ & Energy of the leptonic field mode\\
\\
$\kappa$ & A function of the leptonic mode's energy, $\omega$. It is often given by the dispersion relation $\kappa = \sqrt{m^2 - \omega^2}$\\
\\
$L$ & Collection of angular momentum labels specific to the solution of the leptonic field equations\\
\\
$Y_{\ssL}(\theta, \varphi)$ & The angular part of the solution to the leptonic field equations\\
\\
$l , l _z$ & Quantum numbers of the orbital angular momentum and its $z$-component in a solution to 3-dimensional field equations of spinless fields\\
\\
$\msC_\ssL, \msD_\ssL$ & Integration constants in the solution to the ordinary second-order differential equation satisfied by the radial component of the leptonic field multiplying the near-origin convergent and divergent solutions respectively
\end{tabular}

\noindent
\begin{tabular}{l p{0.68\linewidth}}
$\cR^{\msC (\msD)}_{\ssL}(\kappa r)$ & What are traditionally thought of as the near-origin convergent (divergent) radial solutions to leptonic field equations\\
\\
$\mu_\ssN$ & Nuclear magnetic moment ({\it not the nuclear magneton}) including the nuclear $g$-factor\\
\\
$g_N, g_l$ & The nuclear and leptonic $g$-factors\\
\\
$r_p,~ r_\ssZ$ & Charge and Zemach radii of the proton as measured by \cite{pohlnature}\\
\\
$\bfI$ & Nuclear spin vector\\
\\
$\mathbf{F} = \mathbf{J} + \mathbf{I}$ & Total atomic angular momentum operator\\
\\
$F$ & Quantum number of the total angular momentum of the atomic system, $\bfF = \bfI + \bfJ$\\
\\
$S_\ssB$ & The `bulk' part of the PPEFT action, which for our purposes is the same as the QED action, $S_\ssB = S_\QED$\\
\\
$\xi^\mu(s)$ & Classical Grassmann field\\
\\
$\left\{ A, B \right\} = AB + BA$ & The anticommutator\\
\\
$[A, B] = AB - BA$ & The commutator \\
\\
$S_{p0}$ & The lowest-order part of the PPEFT action that describes the kinematics of the point-particle\\
\\
$\Gamma^\mu, \Gamma_5$ & Dirac gamma matrices acting on the Hilbert space of the nucleus\\
\\
$\Gamma^{\mu\nu} = -\frac{i}{4} \left[ \Gamma^\mu, \Gamma^\nu \right]$ & The Lorentz algebra generators for the Hilbert space of the nucleus\\
\\
$\epsilon_{\mu_1\mu_2\cdots \mu_n}$ & n-dimensional totally antisymmetric tensor\\
\\
$\mathds{1}$ & Identity operator
\end{tabular}

\noindent
\begin{tabular}{l p{0.68\linewidth}}
$\tau^i$ & The spin-matrices acting on nuclear-spin space\\
\\
$\boldsymbol{\tau} = (\tau^1~ \tau^2~ \tau^3)^T$ & The vector of spin matrices acting on nuclear-spin space\\
\\
$\bfE, \bfB$ & The electric and magnetic fields\\
\\
$\boldsymbol{\mu} = \mu_N \bfI$ & Nuclear magnetic moment\\
\\
$j^\mu$ & Electromagnetic 4-current\\
\\
\\
$A_0^{\mathrm{nuc}}, \mathbf{A}^{\mathrm{nuc}}$ & Electromagnetic fields directly generated by the nucleus\\
\\
$A_0^{\mathrm{rad}}, \mathbf{A}_{\mathrm{rad}}$ & Operator valued quantum field interaction of the electromagnetic field\\
\\
$ \boldsymbol{\Sigma} =  \left[ \begin{array}{cc} \bfS & 0 \\ 0 & \bfS \end{array} \right]  $ & The spin-operator for a Dirac-particle with $\bfS = \frac{1}{2} \bm{\sigma}$ the spin vector \\
\\
$\sigma^i$ & Pauli matrices acting on lepton-spin space\\
\\
$\boldsymbol{\sigma} =  (\sigma^1~ \sigma^2~ \sigma^3)^T$ & Vector of spin-matrices acting on electron spin-space\\
\\
$\psi(\bfx)$ & Spatial part of the solution to the leptonic field equations\\
\\
$l, l'$ & Quantum numbers of the orbital angular momentum of the leptons for both parities\\
\\
$\varpi= \pm$ & The parity quantum number, $(-)^l $ with $l = j - \frac12 \, \varpi$\\
\\
$\Omega_{jlj_z}(\theta, \phi)$ & 2 component spherical spinors of the Dirac-Coulomb problem \\
\\
$Y_{ll_z}(\theta, \phi)$ & Scalar spherical harmonics\\
\\
$\mff_{nj\varpi}(r), \mfg_{nj\varpi}(r)$ & Solutions to the radial part of the Dirac-Coulomb field equations\\
\\
$\mathcal{M}[\beta, \gamma; z] = {}_1{}\mathcal{F}_1[\beta; \gamma; z]$ & Confluent hypergeometric function
\end{tabular}

\noindent
\begin{tabular}{l p{0.68\linewidth}}
$\mathcal{M}_\mathfrak{i}$  & Denotes one of the confluent hypergeometric functions in $\mff_{nj}(r)$ and $\mfg_{nj}(r)$ with $\mathfrak{i} = 1,2,3,4$\\
\\
$\rho = 2\kappa r$ & Dimensionless radial variable of the Dirac-Coulomb problem\\
\\
$\mfK= \varpi \(j +\frac{1}{2} \) $ & Eigenvalue of the operator $\boldsymbol{\sigma} \cdot \bfp$ in the Dirac-Coulomb problem, a.k.a. the Dirac quantum number (normally denoted by $K$ in the literature\\
\\
$\bfp$ & Momentum operator of the Dirac fields\\
\\
$\zeta = \sqrt{\mfK^2 - (Z\alpha)^2}$ & Dimensionless combination appearing in the radial differential equations of the Dirac-Coulomb problem \\
\\
$\begin{array}{l} a = \zeta - \frac{Z\alpha \omega}{\kappa}, \\ \, b = 1+2\zeta \end{array}$ & Arguments of the confluent hypergeometric functions that appear in the near-origin finite parts of $\mff_{nj}(r), \, \mfg_{nj}(r)$ \\
\\
$\begin{array}{l} a' = -\left( \zeta + \frac{Z\alpha \omega}{\kappa} \right), \\ b' = 1-2\zeta \end{array}$ & Arguments of the  confluent hypergeometric functions that appear in the near-origin divergent parts of $\mff_{nj}(r), \, \mfg_{nj}(r)$ \\
\\
$c = \mfK - \frac{Z\alpha m}{\kappa}$ & Factor appearing in both types of solutions (near-source convergent and divergent) of the radial functions $\mff_{nj}(r), \, \mfg_{nj}(r)$ \\
\\
$\mathcal{N} = n\sqrt{1 - \frac{2(n-|\mfK|)(Z\alpha)^2}{n^2(\zeta + |\mfK|)}}$ & Relativistic numerical factor appearing in the point-like source solutions to the Dirac-Coulomb problem \\
\\
$\omega^\ssD_{nj} = m\sqrt{1 - \frac{(Z\alpha)^2}{\mathcal{N}^2}}$ & Bound state energy eiegenvalue of Dirac particles in a Coulomb potential sourced by a point-like nucleus with charge $(Ze)$\\
\\
$\kappa^\ssD_{nj} = \frac{mZ\alpha}{\mathcal{N}}$ & Function of the bound state lepton energy of a Dirac particle in a Coulomb potential sourced by a point-like nucleus with charge $(Ze)$\\
\\
$\delta \omega_{n \ssF j \varpi}$ & Energy shifts of a nuclear origin to the leptonic mode functions with quantum numbers $n, F, j, \varpi$ and Dirac-Coulomb energy $\omega^\ssD_{nj}$\\
\end{tabular}

\noindent
\begin{tabular}{l p{0.68\linewidth}}
\\
$\varepsilon_{n \ssF j \varpi}^{\rm mag} = \!\!\!\! {\varepsilon_{n \ssF j \varpi}^{(1)}  \atop \qquad + \varepsilon_{n \ssF j \varpi}^{(ho)}} $ & The energy shifts generated by the magnetic dipole moment of the nucleus at first and higher-order in $\exx$ respectively\\
\\
$\varepsilon_{n \ssF j \varpi}^{\QED} = \!\!\!\! {\varepsilon_{n \ssF j \varpi}^{{\rm pt} - \QED} \atop \qquad +  \varepsilon_{n \ssF j \varpi}^{N - \QED}} $ & The  energy shifts coming from various QED processes in the point-nucleus limit and the radiative corrections to finite-size effects through loop processes respectively\\
\\
$\varepsilon_{n \ssF j \varpi}^{\rm rec} = \!\!\!\! {\varepsilon_{n \ssF j \varpi}^{\rm pt - rec } \atop \qquad +  \varepsilon_{n \ssF j \varpi}^{\rm N - rec}} $ & The  energy shifts coming from nuclear recoil processes in the point-nucleus limit and the recoil corrections to the finite-size effects respectively\\
\\
$\mathcal{Y}_{\ssF f_z}^{j, \varpi} (\theta, \phi) $ & The new spinors that incorporate the hyperfine structure. They obey the eigenvalue relation $\bfF^2 \mathcal{Y}_\ssF = F(F+1) \mathcal{Y}$ and others found in Appendix \ref{AppendixB}\\
\\
$\psi_{n \ssF j \varpi} := \ket{n F f_z; I, j}_0 $ & The correct zeroth-order atomic states that diagonalize the degenerate subspaces of the mixed electron and nuclear states under the hyperfine interaction\\ 
\\
$\eta_{\ssI, \ssI_z}$ & The nuclear spin states of a nucleus with spin, $I$\\
\\
$\varepsilon^{(1)}_{n \ssF j \varpi}$ & The first-order energy shift caused by the presence of the nuclear magnetic dipole field calculated in perturbation theory\\
\\
$\cD, \widetilde{\cD}$ & The explicit, normalization factor that emerges in the energy shift and state-corrections for unnormalized states\\
\\
$\Sigma = i (\mathbf{I} \times \hat{\mathbf{r}}) \cdot \boldsymbol{\sigma}$ & Angular operator acting on the hyperfine spinors $\cY_{\ssF, f_z}^{j, \varpi}$\\
\\
$X_F$ & A combination of angular momentum quantum numbers defined in \eqref{amatrix} that ubiquitously appears at first order in $\exx$ due to rotational invariance\\
\\
$\mfN, \mfN^s$ & Collection of dimensionless integrals in the radial matrix elements appearing in the numerator of the first-order energy shift and the first-order state corrections respectively, due to the hyperfine interaction
\end{tabular}

\noindent
\begin{tabular}{l p{0.68\linewidth}}
$\mfD, \widetilde{\mfD}$ & Collection of dimensionless integrals appearing in the denominator of the first-order energy shift and state-corrections respectively, due to the hyperfine interaction \\
\\
$\begin{array}{l l l} \mfN_{\rm pt}, & \mfN_1, & \mfN_2, \\ \mfD_{\rm pt}, & \mfD_1, & \mfD_2 \end{array}$ & Set of dimensionless integrals in $\mfN, \mfD$ split into three categories: integrals over only near-origin convergent functions (`pt' subscript); integrals accompanied by one power of the integration constant ratio $\msD/\msC$ (labelled by the `1' subscript); integrals accompanied by two powers $\msD/\msC$ (subscript `2')\\
\\
$\begin{array}{l l l l} \mfN^s_{\rm pt}, & \widetilde{\mfN}^s_1, & \mfN^s_1, & \mfN^s_2, \\ \widetilde{\mfD}_{\rm pt}, & \widetilde{\mfD}_1, & \widetilde{\mfD}_2 & \end{array}$ & Set of dimensionless integrals in $\mfN^s, \widetilde{\mfD}$ split into the same categories as those in $\mfN, \mfD$\\
\\
$E^{(0)}_{n, j} = \omega^\ssD_{n j} + \delta \omega^{(0)}_{nj} $ & The zeroth-order energies of the atom with degeneracy $(2I + 1)(2j+1)$\\
\\
$\cC_{\widetilde{n} n \ssF F_z \widetilde{j} j \widetilde{\varpi} \varpi}$ & Coefficient of the first-order state-correstions\\
\\
$\delta \omega^{(0)}_{nj\varpi}$ & The spin-independent, zeroth-order finite-size energy shift determined by the normalizability condition \pref{normal}. This is the part of the energy shift that appears in our earlier work \cite{ppeft3, ppeftA} given by the zeroth-order, scalar part of the integration constant ratio $(\msD/\msC)^{(0)}$ \\
\\
$\delta \omega^{(1)}_{n \ssF j}$ & The spin-dependent, first-order finite-size energy shift coming from the normalizability condition \pref{normal} through $(\msD/\msC)^{(1)}$ \\
\\
$\cI^{(p)}_{\mathfrak{i} \mathfrak{j}}$ & Integrals that appear in the radial matrix elements of the first-order energy shift with $\mathfrak{i}, \mathfrak{j} \in [1, 4] $ denoting the four confluent hypergeometric functions in $\mff, \mfg$\\
\\
$\varepsilon^{\rm hfs}_{n \ssF j \varpi}$ & The hyperfine-splitting energy shift with relativistic corrections included\\
\\
$C_\eta$ & A regularization-scale dependent function that needs to be absorbed into the effective couplings in order to keep $\varepsilon_{n \ssF j \varpi}^{(1)}$ physical
\end{tabular}

\newpage

\noindent
\begin{tabular}{l p{0.68\linewidth}}
$\mathfrak{c} = 16 (m\epsilon_{\star+})^2$ & A function of the small dimensionless quantity $m\epsilon_{\star+}$ that controls finite-size effects in the PPEFT language\\
\\
$g_p$ & g-factor of the proton\\
\\
$\mu_p$ & Magnetic moment of the proton\\
\\
$nL_j{}^\ssF$ & Spectroscopic notation of an energy level with quantum numbers $n, l, j, F$
\\
$\cI^{(p)}_{\mathfrak{i} \widetilde{\mathfrak{j}}}$ & Integrals that appear in the $\mfN^s$ matrix-elements of the first-order state corrections. Here, the tilde signals the fact that the quantum numbers are different for the two hypergeometric functions in the integrand but the other indices are defined the same way as in $\cI^{(p)}_{\mathfrak{i} \mathfrak{j}}$\\
\\
$\widehat{\sum}$ & The sum factor of \eqref{SigmaHatDef} over all values of the principal quantum number that lie outside the degenerate subspace of the state whose corrections we are looking at\\
\\
$\mff_{n j \varpi}^{(0)}, \mfg_{n j \varpi}^{(0)}$ & The Dirac-Coulomb wave-functions\\
\\
$\mff_{n j \varpi}^{(1)}, \mfg_{n j \varpi}^{(1)}$ & First-order corrections to the Dirac-Coulomb wave-functions calculated in degenerate perturbation theory\\
\\ 
$\left( \msD/\msC \right)^{(0)}$ & The ratio of integration constants found in the case of a scalar source\\
\\
$\left( \msD/\msC \right)^{(1)}$ & The first order correction to the ratio of integration constants introduced as a compensation for the lack of new large-$r$ normalizability conditions for the full states once the hyperfine interaction is turned on \\
\\
$\epsilon$ & Radius of the Gaussian sphere on which the alternative boundary conditions implied by the PPEFT are set up\\
\\
$\epsilon_\star, \epsilon_0$ & RG-invariant scales associated with the finite-size effects of a scalar source and controlling the running of the PPEFT couplings $c_s, c_v, c_\ssF$
\end{tabular}

\noindent
\begin{tabular}{l p{0.68\linewidth}}
$\epsilon_\ssF$ & RG-invariant scale associated with the mixed finite-size, hyperfine effects \\
\\
$\cZ_{\ssF j \varpi}, \cZ_{\ssF\varpi}$ & Another combination of angular momentum quantum numbers that appears in the matrix elements of the $\mathbf{I} \cdot \bm{\Sigma}$ operator.
\\
$\hat{c}_i = \frac{c_i}{4\pi\epsilon^2}$ & The generic EFT coupling divided by the surface area of the sphere on which the new boundary conditions are set up. Equivalently, these are the dimensionless couplings that appear in the boundary action of the PPEFT \\
\\
$\hat{c}_s^{(0)}, \hat{c}_v^{(0)}$ & The coupling coefficients appearing at zeroth order in $\exx$ and so whose running is controlled by the spin-independent parts of the boundary-condition\\
\\
$\hat{c}_s^{(1)}, \hat{c}_v^{(1)}$ & Corrections to the $(length)^2$ coupling coefficients of scalar nuclei appearing at first order in $\exx$\\
$\chi = \sqrt{\frac{m-\omega}{m+\omega}}$ & A numerical factor that appears in the ratios of radial functions $\mff(r), \mfg(r)$\\
\\
$\Lambda_\pm$ & Contributions to the boundary conditions of leptonic modes that come from the first-order state-corrections\\
\\
$g(\epsilon), u(\epsilon), v(\epsilon) $ & Functions in terms of which the zeroth-order RG-flow can be universally determined\\
\\
$A, B, C, D$ & Constants in the universal evolution of coupling constants\\
\\
$y_\star = \pm 1 $ & An RG-invariant that determines which type of curve the couplings flow on in the zeroth-order RG evolution\\
\\
$\bar{\lambda}_\pm^{(0)}$ & The $n$-independent linear combination of spin-independent PPEFT couplings $\hat{c}^{(0)}_s$ and $\hat{c}^{(0)}_v$ that follows from the leading-order RG behaviour of the couplings\\
\\
$\bar{\lambda}_\pm^{(1)}$ & The linear combination of spin-dependent PPEFT couplings $\hat{c}^{(1)}_s, \hat{c}^{(1)}_v$ and $\hat{c}_\ssF$ that follows from $cO(\exx)$ RG behaviour of the couplings
\end{tabular}

\noindent
\begin{tabular}{l p{0.68\linewidth}}
$(\msD/\msC)^{(1)}_{phys}$ & The physical part of the normalizability compensating expansion of the ratio of leptonic integration constants at first order. This quantity controls the actual mixed hyperfine, finite-size effects coming from the large-$r$ normalizability condition\\
$\delta \omega_{n\ssF j \varpi} = \!\!\!\! {\delta \omega_{n\ssF j \varpi}^{(0)} \atop \qquad + \delta \omega_{n\ssF j \varpi}^{(1)} }$ & Nuclear-size dependent energy shift coming from the large-$r$ normalizability condition at zeroth and first orders in $\exx$\\
\\
$H_n$ & Harmonic numbers\\
\\
$\mfH$ & A function of $\msD/\msC$ appearing in $\delta \omega_{n\ssF j \varpi}$\\
\\
$\gamma$ & Euler-Mascheroni constant\\
\\
$\rho_{c/m}(\bf{x}')$ & The electric charge and magnetization densities of the proton \\
\\
$\langle r^2 \rangle_c, \langle r^3 \rangle_{cc}, \langle r \rangle_{cm}, \cdots$ & The charge radius squared, the Friar and the Zemach moments and other nuclear moments that can be used to parameterize finite-size effects\\
\\
$\langle r^3 \rangle_{cc}^{\rm eff}$ & The effective Friar moment, incorporating the finite-size parts of the nuclear polarizability contributions\\ 
\\
$\langle r_{\scriptscriptstyle C1} \rangle, \langle r_{\scriptscriptstyle C2} \rangle, \langle r_{pp} \rangle$ & Various moments used to capture elastic parts of the nuclear-structure effects from \cite{pachucki2018} and \cite{kalinowski2018}\\
\\
$\omega_{n\ssF j \varpi}^{\rm pt}$ & Theoretical contributions to the energy shift of leptons in the point-nucleus limit\\
\\
$\omega_{n\ssF j \varpi}^{\ssN\ssS}$ & Nuclear-size related energy shifts to the lepton energies\\
\\
$\Lambda$ & An arbitrary scale in radiative corrections to the leading mixed finite-size, hyperfine effects first derived in \cite{karshenboim1997} \\
\\
\\
$\Xi_{nj \varpi}$ & The radiative corrections to finite-size effects for muonic Hydrogen appearing in the traditional Lamb shift coming from the electronic vacuum polarization
\end{tabular}

\noindent
\begin{tabular}{l p{0.68\linewidth}}
$L_n^k(x)$ & The associated Laguerre polynomials\\
\\
$\rho_0=\frac{2m_r(Z\alpha)r}{n}$ & The dimensionless radial variable in the Schroedinger-Coulomb problem\\
\\
$G'(x, 0)$ & The reduced Schr\"odinger-Coulomb Green's function for $nS_{1/2}$ states\\
\\
$\nu\left(nL_j{}^\ssF - n'L'_{j'}{}^{\ssF'} \right)$ & The experimentally measured value of a transition between two energy levels\\
\\
$\nu_{1\ssS_{hfs}}, \nu_{2\ssS_{hfs}}, \nu_{21}$ & The experimentally measured energies of the $1S^{\ssF=1}_{j=\frac{1}{2}} - 1S^{\ssF=0}_{j=\frac{1}{2}}$, the $2S^{\ssF=1}_{j=\frac{1}{2}} - 2S^{\ssF=0}_{j=\frac{1}{2}}$ and the $2S^{\ssF=1}_{j=\frac{1}{2}} - 1S^{\ssF=1}_{j=\frac{1}{2}}$ transitions in atomic Hydrogen respectively \\
\\
$\nu_t, \nu_s$ & The experimentally measured $2P^{\ssF=2}_{j=\frac{3}{2}} - 2S^{\ssF=1}_{j=\frac{1}{2}}$ and the $2P^{\ssF=1}_{j=\frac{3}{2}} - 2S^{\ssF=0}_{j=\frac{1}{2}}$ transitions in muonic Hydrogen respectively \\
\\
$z_\ell = (m_{r, (\ell)} \epsilon_{\star, \ell})^2$ & A dimensionless combination of the RG-invariant $\epsilon_{\star, \ell}$ and the lepton ($\ell = e, \mu$) mass. We fit for this parameter in our numerical calculations \\
\\
$\mathfrak{x}_\ell$ & An energy scale appearing in the fitting of $z$ for the lepton $\ell = e, \mu$ \\
\\
$\mathfrak{y}_\ell$ & A dimensionless constant appearing in the fitting of $z$ for the lepton $\ell = e, \mu$ \\
\\
$W(t)$ & The Lambert-$W$ function\\
\\
$\cW = W_{-1}\(-\mathfrak{x} e^{-\mathfrak{y}}\) + \mathfrak{y}$ & The function of the Lambert-W function that determines $(m\epsilon_\star)^2$ \\
\\
$\widehat{\Delta \omega}_{2\ssS_{hfs}}$ & The difference, $\left[\omega^{\rm pt}_{2 1 \frac{1}{2} +} - \omega^{\rm pt}_{2 0 \frac{1}{2} +} \right]  - \nu_{2\ssS_{hfs}}$ between the experimentally measured value of the hyperfine-splitting of the $2S$ state in atomic Hydrogen and the size-independent contributions to this transition
\end{tabular}

\noindent
\begin{tabular}{l p{0.68\linewidth}}
$\widehat{\Delta \omega}_{21}$ & The finite-size contribution to the $2S^{\ssF=1}_{j=\frac{1}{2}} - 1S^{\ssF=1}_{j=\frac{1}{2}}$ transition, expressed as the difference between point-like theory contributions and the experimentally measured value, $\left[\omega^{\rm pt}_{2 1 \frac{1}{2} +} - \omega^{\rm pt}_{1 1 \frac{1}{2} +} \right] - \nu_{21}$ \\
\\
$\widehat{\Delta \omega}_{hfs}$ & The first linear combination of experimental values and point-like theoretical combinations to their measured intervals, $\left[\omega^{\rm pt}_{2 1 \frac{1}{2} +} - \omega^{\rm pt}_{2 0 \frac{1}{2} +} \right] - \left[\omega^{\rm pt}_{2 2 \frac{3}{2} +} - \omega^{\rm pt}_{2 1 \frac{3}{2} +} \right] - \left[ \nu_s - \nu_t \right] $ that can be used to fit $\epsilon_{\ssF, \mu}$ in muonic Hydrogen\\
\\
$\widehat{\Delta \omega}_{\scriptscriptstyle Lamb}$ & The second linear combination of experimental values and point-like theoretical combinations to their measured intervals,  $\frac{1}{4}\left[\omega^{\rm pt}_{2 1 \frac{3}{2} +} - \omega^{\rm pt}_{2 0 \frac{1}{2} +} \right] -\nu_s + \frac{3}{4} \left[\omega^{\rm pt}_{2 2 \frac{3}{2} +} - \omega^{\rm pt}_{2 1 \frac{1}{2} +} - \nu_t \right]$ that can be used to fit $\epsilon_{\ssF, \mu}$ in muonic Hydrogen\\
\\
$\Delta E^{\rm fs}$ & Finite-size contribution to a given energy shift\\
\\
$\Delta E^{\rm exp}$ & Experimental error on the value of $\Delta E^{\rm fs}$\\
\\
$\Delta E^{\rm th}$ & Error on $\Delta E^{\rm fs}$ generated by the size-independent contributions to energy shifts\\
\\
$\Delta E^{\rm trunc}$ & The truncation error on $\Delta E^{\rm fs}$ coming from ignoring terms in our finite-size series expansion\\
\\
$\cL$ & The Lagrangian density\\
\\
$p^\mu$ & Conjugate momentum to $y^\mu$ \\
\\
$\pi^\mu$ & Conjugate momentum to $\xi^\mu$ \\
\\
$\phi_1$ & Scalar constraint on the relativistic spinning point-particle\\
\\
$\Phi^\mu$ & Grassmann constraint on the relativistic spinning point-particle\\
\\
$\cL_c, H_c$ & Constrained Lagrangian and Hamiltonian of the relativistic spinning point-particle
\end{tabular}

\noindent
\begin{tabular}{l p{0.68\linewidth}}
$\theta$ & The scalar Lagrange-multiplier for $\phi_1$\\
\\
$\Theta^\mu$ & The Grassmann Lagrange multiplier for $\Phi^\mu$\\
\\
$(A, B)_\ssP$ & The Poisson bracket, defined for a theory with only Grassmann-even quantities as $(A, B)_\ssP = \frac{\partial A}{\partial q^i} \frac{\partial B}{\partial p_i} -  \frac{\partial A}{\partial p^q_i} \frac{\partial B}{\partial q^i}$ \\
\\
$\Delta_{\alpha\beta} := (\phi_\alpha, \phi_\beta)_\ssP$ & A matrix built out of the Poisson brackets of constraints\\
\\
$(A, B)_\ssD$ & The Dirac bracket\\
\\
$\varphi := y^0 - s$ & Imposed gauge condition to get rid of $\phi_1$\\
\\
$\alpha, \beta$ & Spinors of the Dirac field that are interpreted as the particle and anti-particle solutions in the rest-frame of the particle with the Dirac representation assumed. $\beta$ can also be a set of generic angular momentum labels depending on context\\
\\
$S_{p}^{\rm int}$ & The PPEFT action with the lepton field interactions\\
\\
$I^{\rm int}_{\cB}$ & Boundary action of nucleus-lepton interacrions\\
\\
$\cB_\epsilon$ & Ball of radius, $\epsilon$ where the boundary conditions are set up\\
\\
$\delta x = - \delta \omega \frac{(Z\alpha)m^2}{(\kappa^\ssD_{nj})^3}$ & Convenient dimensionless quantity that appears when finding the energy shift through the normalizability condition\\
\\
$ \delta y = 2|\mfK| - 2\zeta$ & Another convenient difference of dimensionless quantities of size $\cO\((Z\alpha)^2\)$ that appears when finding the energy shift from the normalizability condition\\
\\
$\mfB$ & A function of the principal and Dirac quantum numbers appearing in the energy shift implied by the normalizability condition \eqref{normal}
\\
$H_0$ & The zeroth-order Hamiltonian that can be solved exactly
\end{tabular}

\noindent
\begin{tabular}{l p{0.68\linewidth}}
$V, \lambda$ & The perturbation to $H_0$ and a parameter that formally helps keep track of the orders in $V$ and is eventually sent to 1\\
\\
$\bar{D}$ & Projection operator out of the degenerate subspace of a given state\\
\\
$\mu, \lambda, \mathfrak{z}$ & $\cO\left[  (Z\alpha)^2 \right]$ parameters that help us keep track of the regularization of the divergent integrals in energy shifts and state-corrections\\
\\
${}_\cA \cF_\cB$ & Hypergeometric function with $\cA$ numerator-type and $\cB$ denominator-type parameters\\
\\
$(a)_\mathfrak{i}$ & Pochhammer symbols\\
\\
$\cI_d(a, b ; a', b')$ & Generic integral appearing in $\mfN, \mfD$ and $\widetilde{\mfD}$\\
\\
$\eta_a$ & The regularization parameter that controls the divergences in matrix elements\\
\\
$\cI_d(s; a, b, k; \widetilde{a}, \widetilde{b}, \widetilde{k})$ & Generic integral appearing in $\mfN^s$\\
\\
$\cS_\pm := \!\!\! {\sqrt{(1+\frac{\widetilde{\omega}}{m})(1 - \frac{\omega}{m})} 
\atop \pm  \sqrt{(1 + \frac{\omega}{m})(1 - \frac{\widetilde{\omega}}{m})} }$  & Frequently appearing numerical factors in the first order state corrections
\end{tabular}

\end{appendices}

\end{document}